\documentclass[natbib,smallextended]{svjour3}       

\usepackage{graphicx}
\usepackage{amssymb}
\usepackage{aas_macros}
\usepackage[colorlinks=true,linkcolor=blue,citecolor=blue,urlcolor=blue]{hyperref}

\newcommand{\msol}{\ensuremath{M_{\odot}}}
\newcommand{\mstar}{\ensuremath{M_{\star}}}
\newcommand{\as}{^{\prime\prime}}

\begin{document}

\title{Nuclear star clusters}


\author{Nadine Neumayer \and
        Anil Seth \and
        Torsten B{\"o}ker 
}


\institute{N. Neumayer \at
            Max Planck Institute for Astronomy \\
            K\"onigstuhl 17, 69121 Heidelberg, Germany \\
            \email{neumayer@mpia.de}
           \and
           A. Seth \at
           Department of Physics and Astronomy, University of Utah \\
           115 South 1400 East, Salt Lake City, UT 84112, USA  \\
           \email{aseth@astro.utah.edu} 
           \and
           T. B{\"o}ker \at
           European Space Agency, c/o STScI \\
           3700 San Martin Drive, Baltimore, MD 21218, USA \\
           \email{tboeker@cosmos.esa.int} 
}

\date{Received: date / Accepted: date}

\maketitle

\begin{abstract}
We review the current knowledge about nuclear star clusters (NSCs), the spectacularly dense and massive assemblies of stars found at the centers of most galaxies. 
Recent observational and theoretical work suggest that many NSC properties, including their masses, densities, and stellar populations vary with the properties of their host galaxies.
Understanding the formation, growth, and ultimate fate of NSCs therefore is crucial for a complete picture of galaxy evolution.  Throughout the review, we attempt to combine and distill the available evidence into a coherent picture of NSC evolution.  Combined, this evidence points to a clear transition mass in galaxies of $\sim10^9\,M_\odot$ where the characteristics of nuclear star clusters change.  We argue that at lower masses, NSCs are formed primarily from globular clusters that inspiral into the center of the galaxy, while at higher masses, star formation within the nucleus forms the bulk of the NSC. 
We also discuss the coexistence of NSCs and central black holes, and how their growth may be linked.  The extreme densities of NSCs and their interaction with massive black holes lead to a wide range of unique phenomena including tidal disruption and gravitational wave events.  Lastly, we review the evidence that many NSCs end up in the halos of massive galaxies stripped of the stars that surrounded them, thus providing valuable tracers of the galaxies' accretion histories.  
\end{abstract}

\setcounter{tocdepth}{3}
\tableofcontents

\section{Introduction}
\begin{figure}[h]
    \centering
    \includegraphics[width=0.95\textwidth]{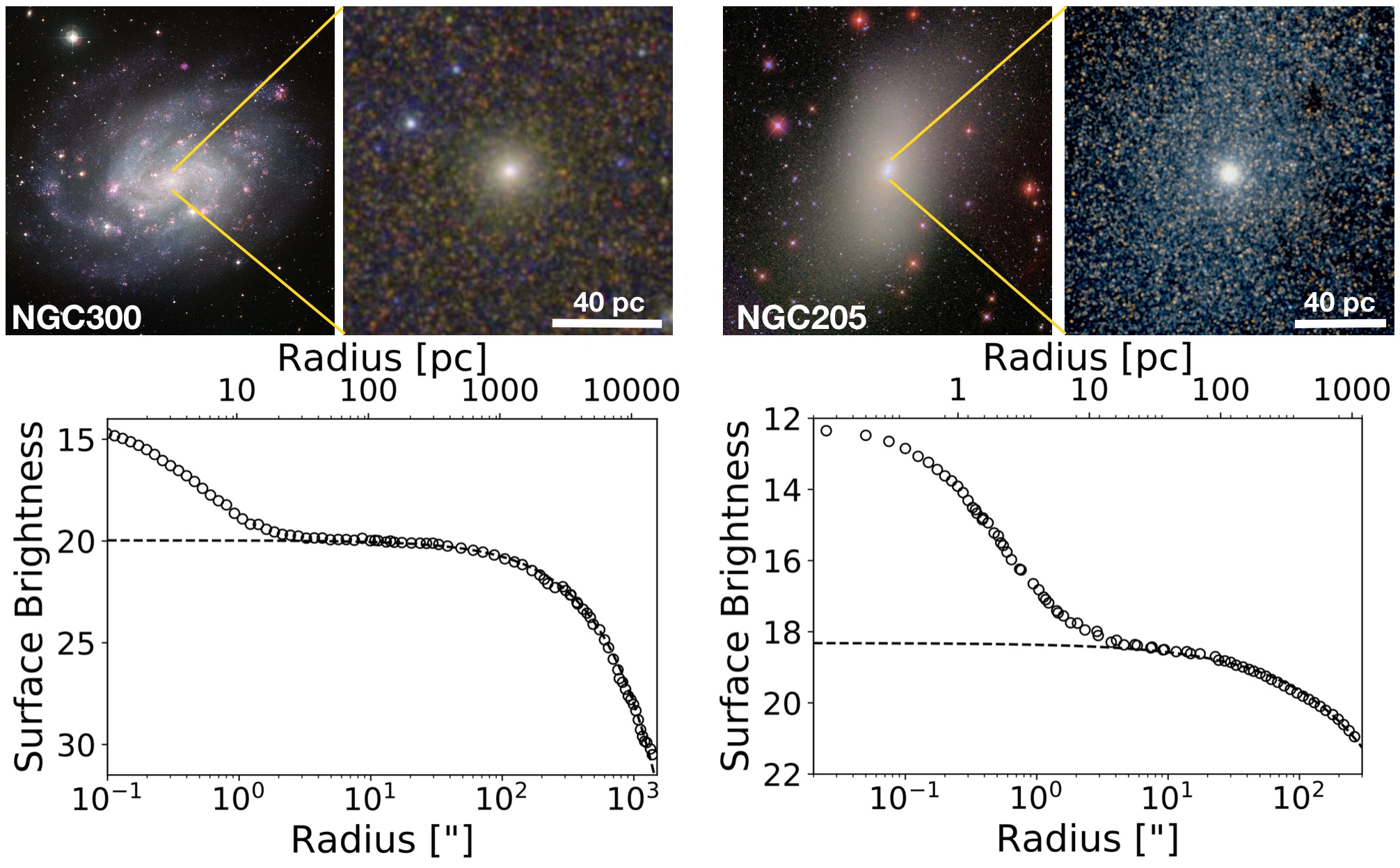}
    \caption{The NSCs in the late-type spiral NGC\,300 (left) and early-type galaxy \,NGC\,205 (M110, right). The top panels show galaxy-wide images with zoom-ins into the central regions of each galaxy. The bottom panels show surface brightness profiles of the two galaxies, which in both cases indicate the presence of an NSC by a rise above the light profile of the host galaxy body. Measured data (in units of $I$ band mag/arcsec$^2$) are shown with open circles, while the dashed line shows a fit to the galaxy profile outside of the NSC proper \citep[an exponential in NGC\,300 and a $n=1.4$ S\'ersic function in NGC\,205; data sources:][]{boker02,kim04,bland-hawthorn05,valluri05}. The large scale image of NGC\,300 is from \url{https://www.eso.org/public/images/eso1037a/}, while the inset HST image is from \citet{carson15}.  The large scale image of NGC\,205 is from Sloan imaging (retrieved from \url{https://cseligman.com/text/atlas/ngc2.htm}), while the inset HST image is from \url{https://www.nasa.gov/feature/goddard/2018/messier-110}. }
    \label{fig:NSC_demo}
\end{figure}

Nuclear star clusters (NSCs) are extremely dense and massive star clusters occupying the innermost region or `nucleus' of most galaxies. Observationally, NSCs are identified as luminous and compact sources that clearly `stand out' above their surroundings. This is illustrated in Fig.~\ref{fig:NSC_demo} which shows images and surface brightness profiles of two nearby NSCs in galaxies with very different morphologies: NGC\,300 is a disk-dominated late-type spiral without any discernible bulge component, while NGC\,205 is a low-mass early-type galaxy with only a minimal amount of gas and recent star formation. The comparison demonstrates that NSCs exist in very different host environments, which raises the question of whether NSC formation and evolution are governed by similar processes in all galaxy types, or whether NSCs follow evolutionary paths that depend on the properties of their host galaxy.

Providing an answer to this question is the goal of a very active research field. To date, more than 400 refereed articles have been written on NSCs, and the pace of discovery about these objects is accelerating, with $\sim$100 articles appearing in the past three years alone. The observational studies cover a wide range of topics, from measuring detailed properties of nearby NSCs (including the one in the Milky Way) to identifying large samples of nuclear star clusters in more distant galaxies of various morphological types. The literature also contains a large number of theoretical studies discussing the two main mechanisms for NSC formation that have been identified so far: the inward migration of 
star clusters on the one hand, and in-situ star formation triggered by high gas densities in the galaxy nucleus on the other. Other theoretical papers are focused on the unique and exotic events that can only occur in galaxy nuclei, e.g. tidal disruption events during which a supermassive black hole tears apart one of the stars of the nuclear star cluster, or gravitational wave emission by mergers of binary black holes.

Despite (or maybe because of) the vast amount of information available, a comprehensive compilation of our current knowledge is lacking, and a literature review that puts NSC properties in the context of their host galaxies is overdue. Specifically, we aim to shed light on the following three questions:
\begin{itemize}
    \item How do the stars of the NSC get to the nucleus?
    \item When do they get there, i.e., at what point in the evolution of the host galaxy?
    \item Are the answers to these questions the same across different galaxy masses and morphologies?
\end{itemize}

As will become clear, NSCs are very different from other massive star clusters, the properties of which have recently been reviewed by \citet{krumholz18} who explicitly exclude NSCs. We also note that the central star cluster of the Milky Way, the closest and best studied example for a typical NSC, will be discussed in some detail in Sect.~\ref{subsec:MWNSC} and whenever relevant in the context of the wider family of NSCs. However, we do not provide a comprehensive literature discussion for this object, as it is extensively covered by other recent reviews \citep{genzel10,schoedel14b,bland-hawthorn16}.  
We also point out that the black hole review of \citet{kormendy13} discusses NSCs, especially in the context of how their masses scale with the properties of their host galaxies, a topic we will also discuss in this review. Lastly, the formation of NSCs, and their relation to ultracompact dwarf galaxies are briefly reviewed by \citet{renaud18}.  

We have organized this review into a number of topical sections: after a short summary of early NSC studies in Sect.~\ref{sec:history}, we discuss in Sect.~\ref{sec:definition} how NSCs are identified observationally, including some of the challenges involved. In Sect.~\ref{sec:demographics}, we examine how frequently galaxies have NSCs at their centers, and how this varies with galaxy mass and environment.   The physical properties of NSCs, including their morphology, masses, densities, stellar populations, and kinematics are discussed in Sect.~\ref{sec:properties}. We then discuss in Sect.~\ref{sec:hosts} the evidence for any dependencies between NSC properties and those of their host galaxies, focusing especially on trends with galaxy stellar mass. Section~\ref{sec:formation} then discusses different formation scenarios for NSCs, and their implications for the evolution of galactic nuclei. The co-existence of NSCs and massive black holes in galactic nuclei is reviewed in Sect.~\ref{sec:nsc-mbh}. In Sect.~\ref{sec:stripped}, we examine the evidence that stripped NSCs are hidden amongst the populations of globular clusters (GCs) or ultra-compact dwarf galaxies (UCDs). Finally, we discuss the unique dynamics of NSCs in Sect.~\ref{sec:physics}, which can lead to transient tidal disruption and gravitational wave events.  We conclude in Sect.~\ref{sec:summary} by outlining some open issues and providing an outlook towards future observing facilities which will likely provide important new constraints on the nature of NSCs.  We note that the code and data used to make figures in this review is available at \url{https://github.com/anilseth/nsc_review}.

\section{Early studies}\label{sec:history}
In this section, we provide a short overview of `historical' (pre-2006) NSC studies. We (somewhat arbitrarily) consider 2006 as an inflection point, because in that year, three independent studies \citep{rossa06,wehner06,ferrarese06} concluded that NSCs are related to their host galaxies in much the same way as SMBHs, which triggered a renewed wave of interest in NSCs and their role in SMBH growth.

The fact that many galaxies, irrespective of their morphological type, exhibit a prominent and compact surface brightness peak in their morphological center has been known for a long time. High-quality photographic plate catalogs of nearby galaxies such as those of \cite{sandage61} and \cite{sandage94} clearly showed many of them to harbor a bright and unresolved central source. For many decades, however, detailed studies of these structures - and the central morphology of external galaxies in general - were difficult or impossible. One important reason is that galactic nuclei were often saturated or `burned out' in photographic plates, especially in high-surface brightness galaxies. In addition, the seeing-imposed resolution limit of early ground-based observatories smooths out small-scale features in the surface brightness profiles. This makes it difficult to distinguish between a steep, but continuous, rise towards the nucleus on the one hand, and a distinctive upturn or `kink' in the central few parsec on the other. The latter would indicate the presence of a separate morphological component, likely with structural and photometric properties different from those of the surrounding bulge and/or disk. 

The first study to unambiguously demonstrate that an extragalactic stellar nucleus is indeed a separate morphological component of its host galaxy was published by \cite{light74} who used a stratospheric balloon telescope to improve the resolution of their images of the Andromeda nucleus. Their decomposition of the surface brightness profile in the central arcminute of M\,31 led them to conclude that {\it ``... the nucleus of M\,31 is a distinct and separate feature of that galaxy''}, albeit without making any statement as to the nature of its emission. Their results motivated the initial theoretical work on NSC formation through dynamical friction by \citet{tremaine75}. Similar structures were soon confirmed in other nearby galaxies, e.g., M\,33 \citep{gallagher82}, M\,81, and a handful of others \citep{kormendy85}. Other early studies of spiral galaxy (disk) structure such as the one by \cite{romanishin83} clearly reveal the ubiquitous presence of unresolved central sources, but do not discuss these in any detail.

As for the Milky Way, the initial detection of its NSC required the development of infrared cameras because of the extreme extinction towards the Galactic center. The first literature reference to the Milky Way NSC was made by \citet{becklin68} who include the following note: {\it ``We have observed an extended source of infrared radiation which we believe to be at the nucleus of the Galaxy ... We believe that the infrared radiation most likely originates from stars ...''.} However, resolving individual stars within the Milky Way NSC became possible only several years later \citep{becklin75}.

Around 1980, high-quality photographic plates of nearby galaxy clusters became available which allowed the systematic study of large samples of low-surface brightness galaxies (whose centers were not saturated). Soon after, `nucleation' was recognized as a common morphological feature of (dwarf) galaxies \citep[e.g.,][]{reaves83,caldwell83,binggeli85,caldwell87}. 
Initially, these compact central structures were referred to as `star-like nuclei' or `semi-stellar nuclei' because of their morphological appearance, i.e., their resemblance to a central bright star. However, it was already suggested by \cite{caldwell83} that {\it ``Perhaps these nuclei are just tightly bound clusters of stars...''}, a notion that soon found support from spectroscopic observations. For example, \cite{bothun85} concluded that the nuclear emission was produced by a dense accumulation of stars, and so the term `stellar nucleus' began to describe both morphology and composition.

The review article by \cite{kormendy89} appears to be the first to clearly equate the terms `stellar nucleus' and `nuclear star cluster' while explicitly excluding  emission from an active galactic nucleus (AGN). Until recently, the terminology remained somewhat unsettled: `stellar nucleus' was traditionally used when describing (dwarf) elliptical galaxies, while studies of disk galaxies predominantly used `nuclear star cluster' (NSC). Most recent publications, however, have adopted the latter term, and we will also use it throughout this review, because it more clearly describes the nature of these objects.
\subsection{Imaging nuclear star clusters: the Hubble Space Telescope}\label{subsec:hst}
The rise of digital detectors and the launch of the Hubble Space Telescope (HST) revolutionized the study of NSCs.  By the mid-1990s, both ground-based observations using CCDs \citep[e.g.,][]{matthews97} and early HST-based studies \citep{phillips96,sarajedini96} had clearly demonstrated that NSCs are much more common than originally thought, and already revealed some early hints at scaling relations between NSC and host galaxy \citep{phillips96}. The enormous advance in spatial resolution of the refurbished HST triggered a number of large programs specifically aimed at the nuclear morphology of galaxies\footnote{We ignore here the plethora of HST studies aimed at characterizing the AGN phenomenon \citep{ravindranath01}. Many of these programs did provide important insight into the circum-nuclear morphology on scales of a few hundred pc, but the presence of a bright AGN usually outshines the stellar light emitted from the central few tens of pc.}. Much of our current knowledge about NSCs is based on three sets of early HST studies, which were divided by Hubble type and are briefly summarized below.
\begin{itemize}
    \item A large sample of NSCs in early-type spiral galaxies were observed in the optical \citep[WFPC2,][]{carollo97,carollo98a,carollo98b}, ultra-violet \citep[STIS,][]{scarlata04,hughes05}, and infrared \citep[NICMOS,][]{carollo02,seigar02}.  These studies found `superluminous star clusters' in the nuclei of more than half their sample galaxies, and provided early evidence for scaling relations between the luminosity of NSCs and that of their host galaxies. The latter result was soon refined by \cite{balcells03,balcells07} who pointed out a luminosity correlation between NSC and host galaxy \emph{bulge}. 
    \item In late-type spirals, the lack of bulges and lower dust content compared to early-type spirals makes the properties of NSCs easier to study.  An early study of `pure' disk galaxies by \citet{matthews99} observed four such galaxies and demonstrated the feasibility to derive structural parameters of their NSCs. A more systematic survey of late-type spirals was performed by \cite{boker02} who put on a firm statistical footing the notion that NSCs are ubiquitous even in these dynamically unevolved galaxies.
    \item In early-type galaxies, \citet{lotz01} examined the NSCs in dwarfs in the Virgo and Fornax cluster to test the dynamical friction paradigm for NSC formation.  This was followed by the systematic study of \citet{cote06} of 100 Virgo cluster galaxies across a wide range of luminosities which provided a clear view of NSC-galaxy scaling relations \citep{ferrarese06}. Their results demonstrated that the fraction of galaxies with NSCs varies with galaxy luminosity, and is nearly 100\% at the low-luminosity end of their sample.
\end{itemize}
 
All these studies agree that the vast majority of galaxies with stellar masses ($M_\star$) between $10^8$ and $10^{11}$\,\msol\ harbour a dominant, compact, and massive stellar cluster in their photometric center (discussed further in Sect.~\ref{sec:demographics}).   We note that following \citet{bullock17}, when we use low-mass or dwarf galaxy in this review, we refer to galaxies with $M_\star \lesssim 10^9$\,\msol.  

\subsection{Ground-based spectroscopic observations}\label{subsec:ground}
Given the modest aperture size of HST, its use for \emph{spectroscopic} NSC studies proves to be challenging. The survey of \cite{rossa06}, who obtained HST/STIS spectra of 40 NSCs from the Carollo and B\"oker samples, remains the only large HST-based effort to derive the stellar population of NSCs. While they were able to show that NSCs contain multiple stellar populations, and were among the first\footnote{Two additional papers \citep{wehner06,ferrarese06} arriving at similar conclusions were published in the same year.} to suggest a correlation between the stellar \emph{masses} of NSC and host galaxy bulge, the relatively poor signal-to-noise ratio in many of their spectra demonstrates the limited use of HST for NSC spectroscopy. 

In general, the superior light-collecting power of large ground-based telescopes is required to obtain deep spectroscopic data of individual NSCs. In low-mass spheroidal galaxies with shallow surface brightness profiles, as well as in bulge-less spiral galaxies with a high nucleus-to-disk contrast, spectroscopy of NSCs is possible even in seeing-limited conditions. This was demonstrated by the work of \cite{bothun85} and \cite{bothun88} who concluded that the stellar populations of dE,n nuclei in Virgo have relatively high metallicities, pointing to an extended star formation history. Similarly, \cite{walcher06} revealed the presence of young stellar populations in the NSCs of many late-type disk galaxies (see also Sect.~\ref{subsec:stelpop}). 

In most circumstances, however, seeing-limited observing conditions prevent a clean separation between NSC light and background emission from the host galaxy. Fortunately, over the same time span covered by the HST observations, advances in adaptive optics (AO) technology and laser guide star systems made it possible to overcome these seeing-imposed limitations. 

An especially powerful observational technique to study the structure, kinematics, and stellar populations of NSCs is integral field-spectroscopy (IFS). IFS not only allows to probe the extent to which NSCs truly occupy the dynamical center of their hosts, 
but also to reveal (kinematic and/or stellar population) sub-structures \emph{within} the NSC, provided it is nearby enough to be spatially resolved. 
Some examples for this type of analysis are highlighted in Sect.~\ref{subsec:kin}. Moreover, when combined with accurate surface brightness profiles, IFS data sets can provide strong constraints on the mass distribution, and in particular the presence of black holes in the mostly unexplored low-mass regime of galaxies with prominent NSCs. This topic is an especially important one which we will discuss further in Sect.~\ref{sec:nsc-mbh}.

\subsection{Early theoretical studies}\label{early_theory}
Although \citet{tremaine75} had already suggested that NSCs form through the infall and merging of globular clusters driven into the nucleus by dynamical friction, theories for the formation of nucleated low-mass early-type galaxies received renewed attention only following the Virgo cluster study of \cite{binggeli85}, which identified nucleation as a common feature in many low-mass (\mstar\,$\leq 10^9$\,\msol) early-type galaxies. 

The coalescence of globular clusters was further developed analytically by \cite{capuzzo93}, as well as by \cite{lotz01}, who also provided some observational evidence in support of the globular cluster infall scenario. The general validity of dynamical friction as a mechanism for nucleation was confirmed with $N$-body simulations by \cite{oh00}. On the other hand, \cite{milosavljevic04} concluded that in pure disk galaxies, the migration time scales are too long, and instead suggested gas infall and subsequent in-situ star formation as the more plausible alternative in disk galaxies.

Other early theoretical studies focused on gas infall and subsequent centralized star formation even in early-type galaxies include \cite{bailey80} who argues that inward-flowing stellar mass loss can lead to nuclear stellar disks, and \cite{mihos94} who point out that gas-rich mergers will produce dense stellar cores which could be identified as NSCs. In a similar vein, \cite{silk87} invoke late gas accretion and the ensuing star formation as a plausible mechanism to explain nucleation, and already point out that this may also explain the higher nucleation rate in more massive dwarf galaxies (see Sect.~\ref{sec:demographics}). 

We will provide an updated and more detailed summary of NSC formation mechanisms discussed in the recent literature in Sect.~\ref{sec:formation}.


\section{What is a nuclear star cluster?}\label{sec:definition}

Given the wealth of observational data on NSCs collected over the past two decades, and the wide range of host galaxy properties covered (i.e., Hubble type, stellar masses, environments, star formation activity, etc.), it seems useful to include a clarification of what a nuclear star cluster is.

A broad definition that encompasses the range of objects identified as nuclear star clusters in the literature is the following:  \\
{\it ``The detection of stellar excess light above the inward extrapolation of the host galaxy's surface brightness profile on scales of $\lesssim$50~pc.''}

While this definition leaves a number of gray areas, in the vast majority of cases, the NSC is unambiguously identified as a dense and massive compact stellar system located at the dynamical center of the host galaxy \citep[e.g.,][]{neumayer11}, as clearly illustrated in Fig.~\ref{fig:NSC_demo}.  

\begin{figure}[h]
    \centering
    \includegraphics[width=0.45\textwidth]{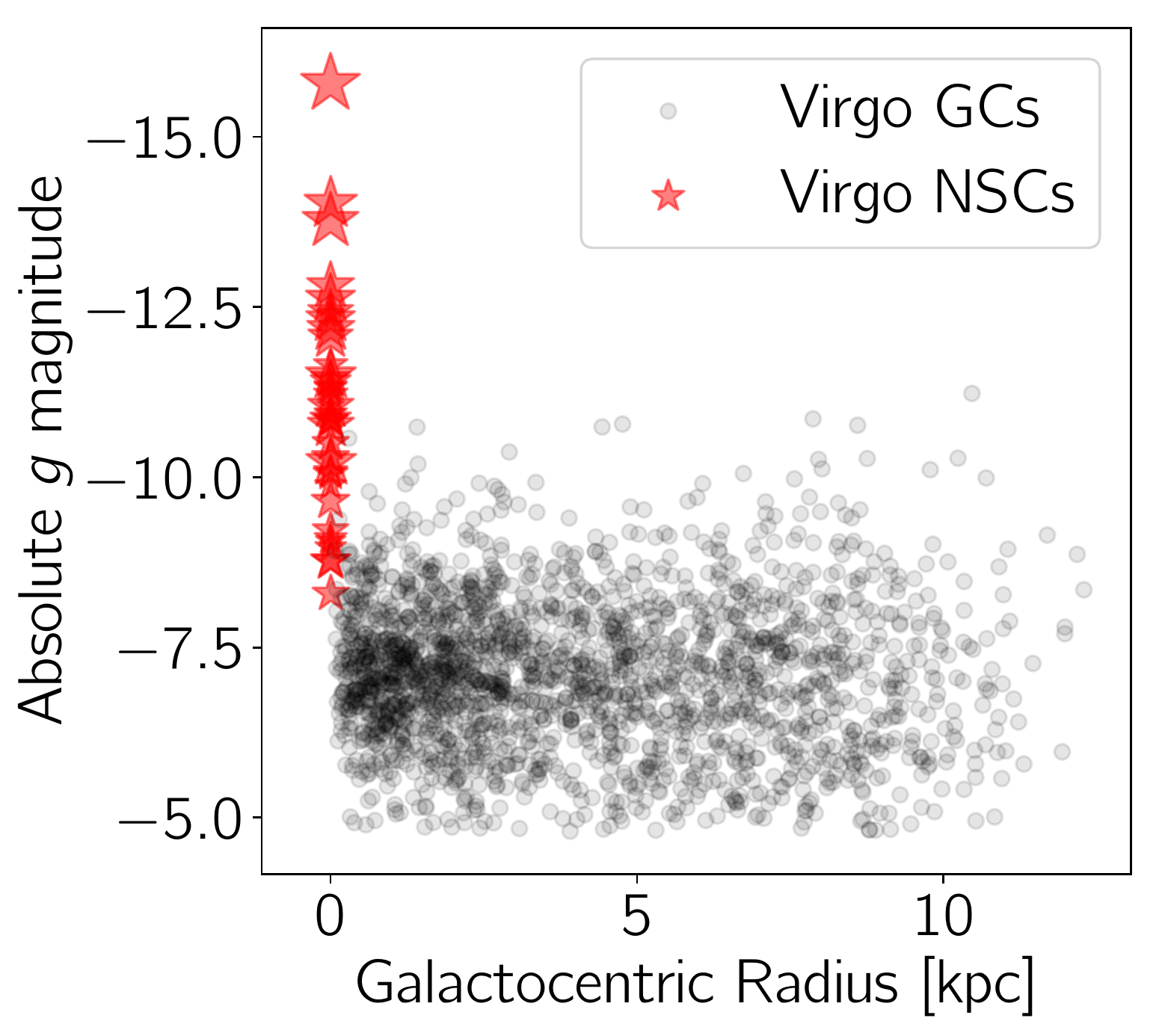}
    \includegraphics[width=0.45\textwidth]{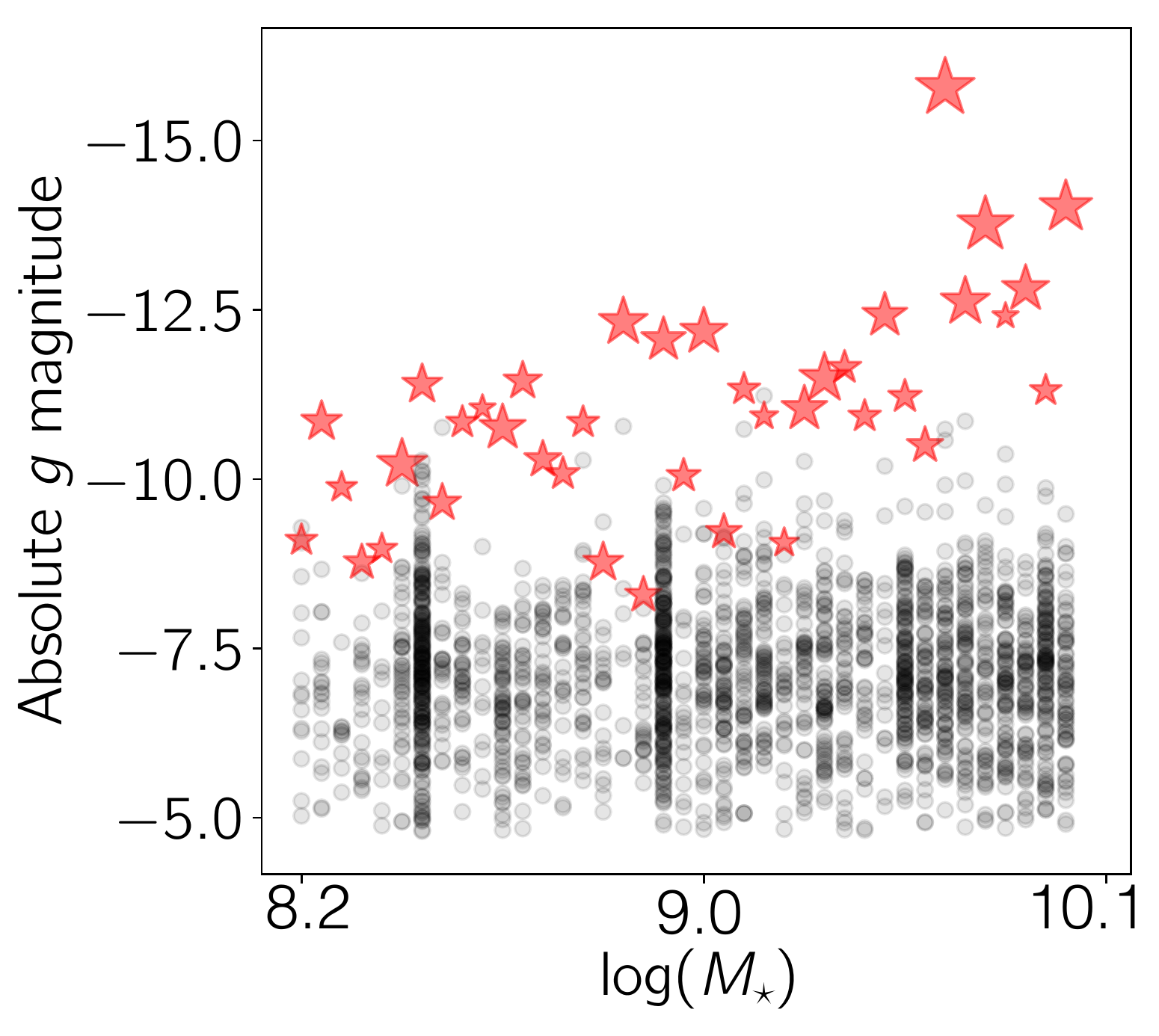}
    \caption{NSCs (red star symbols) are distinct from globular clusters (GCs, grey points).  In both panels, NSCs in the Virgo cluster \citep{cote06} are compared to the GC systems of their hosts, as derived by \citet{jordan09}.  The left panel shows the radial distribution of the clusters vs. their g-band luminosity (note that $M_g = -10$ corresponds to a luminosity of $1.1\times10^{6}\,L_\odot$).  The NSCs clearly have exceptional luminosities, and almost all are consistent with being at the photocenter of the galaxy. The right panel shows the distribution of GC and NSC absolute magnitudes separately for each galaxy, and ordered by host stellar mass \citep[from][]{spengler17}.  We note that here, the galaxies are evenly spaced in order of increasing stellar mass, and so their location on the x-axis is only approximate.  In both panels, the symbol size for each NSC is proportional to its effective radius, ranging from 2 to 40\,pc. Note that the increasing symbol size at higher luminosities is a result of a size-luminosity relation for these NSCs; this will be discussed further in Sect.~\ref{subsec:sizemass}.
     }
    \label{fig:special}
\end{figure}

In general, NSCs are very distinct, bright, and compact stellar clusters that are unrivalled in luminosity by other stellar clusters. This is clearly demonstrated in Fig.\,\ref{fig:special} which compares NSCs to the globular cluster systems of their hosts for the \cite{cote06} sample of early-type galaxies in Virgo. The right panel shows that independent of host galaxy mass, the NSC is nearly always the brightest cluster in the galaxy, often by a wide margin\footnote{A corresponding result for NSC candidates in low-mass dwarf galaxies was pointed out by \cite{georgiev09}.}. This result is remarkable, given that both populations have very similar effective radii \citep[e.g.,][see also Sect.~\ref{subsec:sizes}]{boker04}. 

Despite the prominence of NSCs, there are some complications in identifying them, both observationally and conceptually. These mostly make it difficult to \emph{exclude} their presence. In what follows, we list some of the challenges and ambiguities in identifying NSCs. 

\begin{itemize}
   \item{ {\bf Steep inner profiles:} it can sometimes be difficult to decide what exactly the ``inward extrapolation of the host galaxy's surface brightness profile'' is. This is especially true in more massive ellipticals and bulge-dominated early-type spirals with steep light profiles \citep[e.g.,][]{carollo02,cote07,kormendy09}.   Galaxies are typically fitted by S\'ersic profiles, and \citet{cote07} has shown that for a small range of early-type galaxy luminosities around $M_B=-20$mag, a single S\'ersic profile fits the galaxy all the way to the center, while at fainter luminosities most early-type galaxies show an excess within $0.02\,R_e$ (typically $\lesssim$20\,pc), which they refer to as NSCs.  However, this interpretation is controversial, as \citet{hopkins09} and \cite{kormendy13} further subdivide this group of lower-mass early-type galaxies into ellipticals with central light excess on the one hand, and diskier spheroidals with NSCs on the other.  For example, the center of M\,32 is described as having a central light excess by \citet{kormendy09}, but \citet{graham09} classify it as an NSC with an effective radius of 6\,pc, well within the distribution of sizes of typical NSCs.  In this review, we choose not to make a distinction between NSCs and central light excesses,  but will return to this issue when talking about the formation of NSCs in Sect.~\ref{sec:formation}. }
    \item{{\bf Nuclear disks:} flattened structures with disky morphologies are found in massive NSCs in both early-type \citep{cote06, nguyen18} and late-type galaxies including the Milky Way \citep{seth06,schoedel14b}. These structures show clear evidence for rotation \citep{seth08b,feldmeier14}. While the term ``nuclear star clusters'' may imply spheroidal objects, it is clear that any recent gas infall will likely occur along a preferred plane, and thus NSC growth via in-situ star formation will likely produce a disky morphology. At the same time, many galaxies have larger (circum)nuclear stellar disks on scales of $\approx$100\,pc that can be difficult to separate from the NSC proper. For example, the Milky Way has an extended nuclear stellar disk that is clearly distinct from the more compact NSC \citep{launhardt02}. On the other hand, some compact nuclear star-forming disks have been identified as a separate class of objects rather than being considered NSCs \citep[e.g.,][]{morelli10} despite having scales of only tens of parsecs. Clearly, some ambiguity exists, which has led some authors to exclude large flattened objects from being considered as NSCs \citep[e.g.,][]{scott13}. However, we stress that in most cases, there is a clear distinction between NSCs (with typical sizes of 5\,pc) and more extended circum-nuclear disks.  We note that in both items discussed so far, the ambiguity is primarily present in the nuclei of the most massive galaxies.}
    \item {\bf Ongoing starbursts:} some galactic nuclei are currently experiencing an episode of active star formation. In these cases, the obscuration by dense clouds of gas and dust, combined with the presence of young stars that dominate the light, can make it difficult to discern the true stellar mass distribution. A good example is NGC\,6946 which, despite its proximity, has eluded a characterization of its nuclear morphology \citep{schinnerer06,schinnerer07}.
    \item {\bf Ambiguous centers/non-nuclear clusters:} for higher mass galaxies with well-defined morphologies and organized rotation, the NSCs appear to be coincident with the photocenter \citep[e.g.,][]{boker02,cote06} as well as the dynamical center \citep{neumayer11}.  However, for many lower luminosity galaxies, the exact location of the galaxy center becomes more ambiguous, especially in irregular galaxies.  Nonetheless, the clusters living nearest the centers of these galaxies have similarly exceptional luminosities, suggesting they are in fact NSCs \citep{georgiev09}. Adding to the ambiguity, some star-forming galaxies harbor multiple bright clusters near their centers, making the identification of a bona-fide NSC difficult \citep[e.g.,][]{georgiev14}.
    \item {\bf Very low mass galaxies:} the recent discovery of low-mass star clusters ($<$6000~M$_\odot$) near the centers of very faint local group spheroidals \citep{crnojevic16,caldwell17} highlights an additional ambiguity.  In these galaxies, as well as in somewhat higher luminosity galaxies in Virgo and Fornax \citep{sanchez-janssen19,ordenes-briceno18}, the NSCs have lower masses than typical GCs. In contrast to the majority of NSCs, their properties or formation histories may be no different than those of ``normal'' star clusters.  
    \item {\bf Dust lanes:} the centers of massive spiral galaxies such as the Milky Way contain dense and thin dust-lanes that can make the identification of NSCs challenging. In the Milky Way, the NSC is obscured by $A_V \sim 30$ magnitudes \citep[e.g.,][]{fritz11,nogueras-lara19}, making it impossible to identify at optical wavelengths.  
\end{itemize}

To summarize this section, we stress again that in most cases, NSCs are distinct, well-defined objects, even though there are some ambiguous cases, particularly at the extreme ends of the NSC size and mass distribution.

\section{Nuclear star cluster demographics}\label{sec:demographics}
In this section, we focus on the question of how frequently NSCs are present in galaxies of different masses and types.  Over the last decade, significant additional data sets beyond the initial HST work have been obtained which characterize nuclear star clusters in a wide range of galaxy types and in different environments. A comprehensive list of large NSC studies, which are mostly focused on a specific range of Hubble type, is shown in Table~\ref{tab:samples}.  

\begin{figure}[h]
    \centering
    \includegraphics[width=0.49\textwidth]{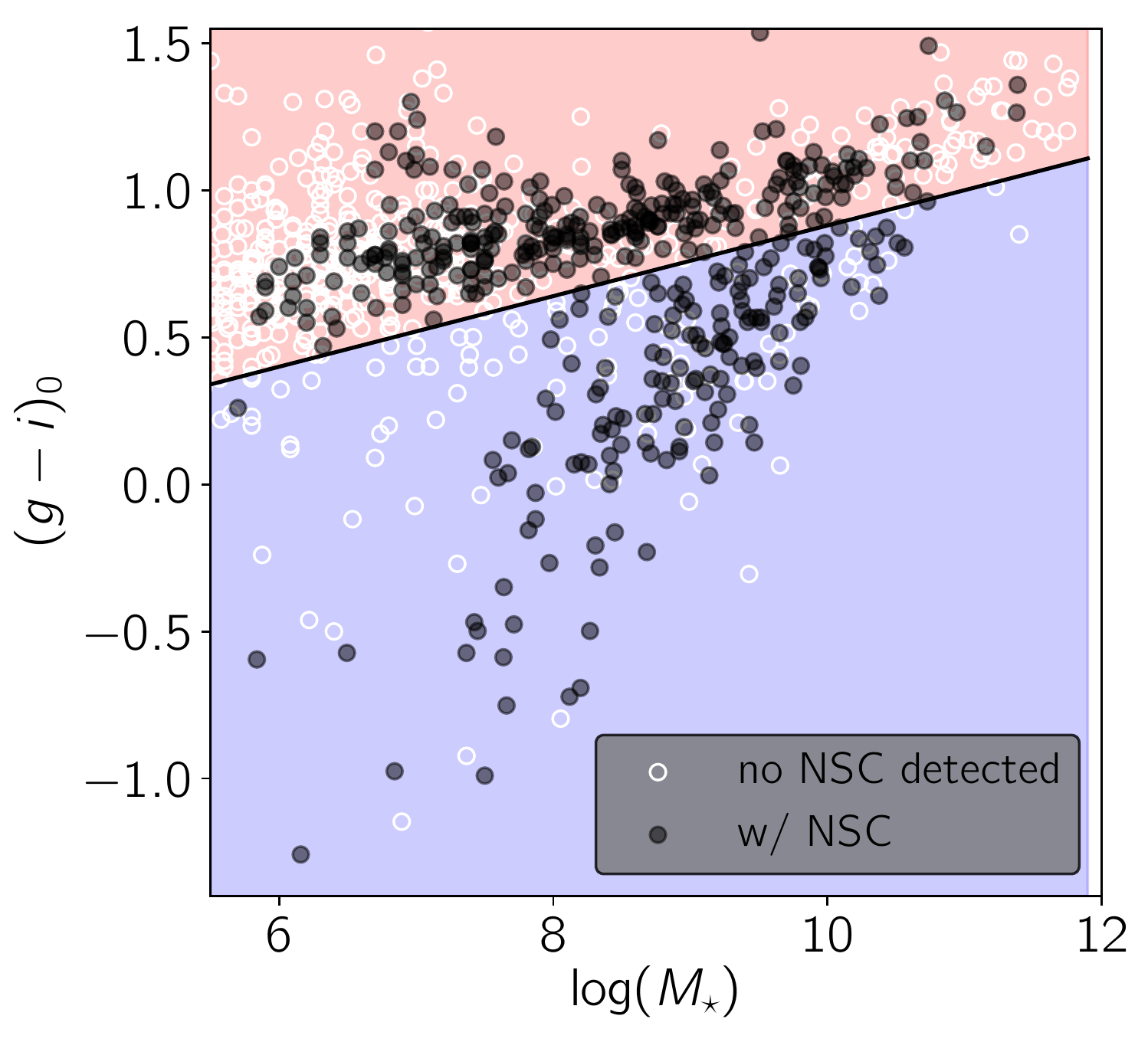}
    \includegraphics[width=0.47\textwidth]{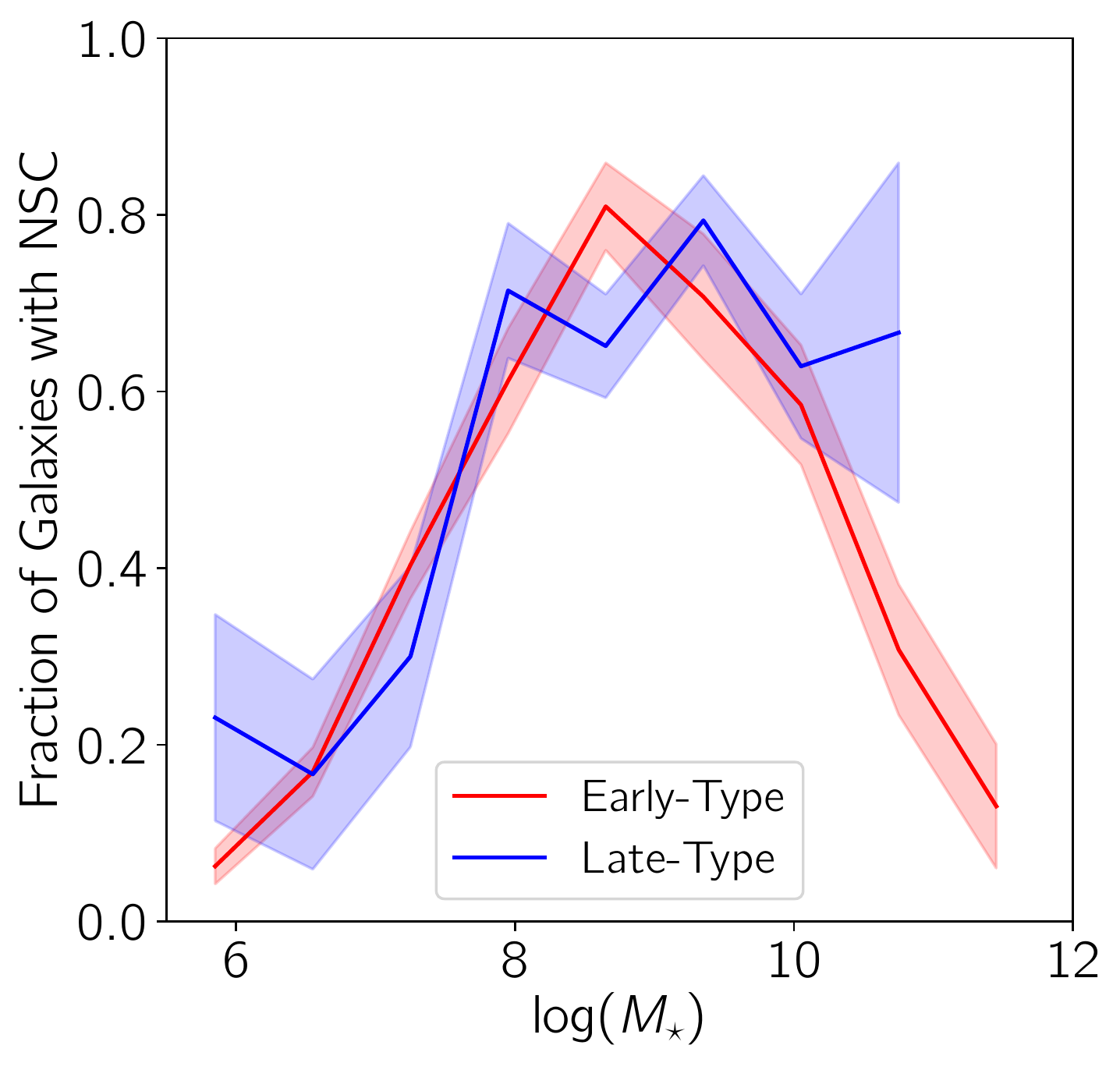}
    \caption{
    Nuclear star cluster are found in a wide range of galaxies, but preferentially those with stellar masses between $10^8$ and $10^{10}$\,\msol.  {\em Left:} A color-magnitude diagram of galaxies with (filled black circles) and without (open white circles) detected NSCs.  The background colors indicate the division between early-type galaxies (the red sequence) and late-type galaxies (the blue cloud).
    {\em Right:} The fraction of galaxies with identified NSCs as a function of stellar mass.  Because of various challenges in identifying NSCs (see Sect.~\ref{sec:definition}), these numbers can be considered lower limits for the NSC occupation fraction, especially in late-type galaxies.
    The data samples for which NSC searches have been conducted \citep{lauer05,ferrarese06b,georgiev09,georgiev14,eigenthaler18,sanchez-janssen19} comprise a total of 1191 galaxies. All have available color information that has been transformed to a common color of $(g-i)_0$. The values for $M_\star$, the stellar masses of the host galaxies, are derived using the \citet{bruzual03} models with a \citet{chabrier03} IMF. Where comparable masses were not listed in the original source, they were calculated using the $g-i$ color--M/L relations of \citet{roediger15}.     }
    \label{fig:demographics}
\end{figure}

\begin{table}
\caption[]{NSC literature samples\label{tab:samples} }
\begin{tabular}[t]{p{3cm} p{5cm} c c } 
\hline \hline 
 &  & &  \\
(1) & (2) & (3) & (4) \\ 
Authors & Galaxy Sample & Occ. Fraction & in Fig. 3? \\
\hline
\citet{carollo97,carollo98a,carollo98b,carollo02} & 94 early-type spirals (Sa-Sbc) with & & \\
& $\rm v_{\rm hel} < 2500\,km/s$, $i \leq 75^{\circ}$, and d $> 1'$ & 62\% (58/94) & no \\
\hline
\citet{boker02} & 77 face-on late-type spirals (Scd-Sm) & 77\% (59/77) & yes \\
\hline
\citet{lauer05} & 77 massive early-type spirals, 40 with & & \\
& color information & 20\% (8/40) & yes \\
\hline
\citet{cote06} & 100 early types in Virgo (E, S0, dE, dS0) & 66\% (66/100) &  yes \\
\hline 
\citet{seth06} & 14 edge-on late-type spirals & 64\% (9/14) & no \\
\hline
\citet{georgiev09} & 68 dwarf galaxies (mostly late-type) & 13\% (9/68) & yes \\
\hline
\citet{turner12} & 43 early types in Fornax & 72\% (31/43) &  no \\ 
\hline
\citet{georgiev14} & 323 non-active late-type spirals (Sbc-Sm) & 71\% (228/323) & yes \\ 
\hline
\citet{denbrok14} & 198 early types in Coma & 84\% (166/198) & no \\ 
\hline
\citet{baldassare14} & 23 early-type field galaxies & 26\% (6/23) & no \\
\hline
\citet{ordenes-briceno18} & 225 dwarf galaxies in Fornax & 32\% (71/225) & yes \\ 
\hline
\citet{sanchez-janssen19b} & 404 dwarf galaxies in Virgo & 26\% (107/404) & yes  \\
\hline
 &  & &  \\
\hline \hline 
\end{tabular}
\end{table}

\subsection{Trends with host galaxy mass, morphology, and color} 

We first quantify the fraction of galaxies with NSCs as a function of stellar mass, i.e., the `NSC occupation fraction'.  For early-type galaxies, the recent studies of hundreds of dwarf galaxies in galaxy clusters by \citet{denbrok14}, \citet{sanchez-janssen19}, and \citet{ordenes-briceno18} have made it clear that NSCs are found in $\gtrsim80\%$ of $\sim10^9\,M_\odot$ early-type galaxies, while the NSC occupation fraction drops steadily towards lower host masses, reaching nearly zero at galaxy stellar masses of $\sim10^6\,M_\odot$.  This data in clusters also appears to be consistent with the low $<25$\% nucleation fraction in early-type dwarf galaxies found around nearby (mostly spiral) galaxies within 12 Mpc by \citet{carlsten19}.  As for the high-mass end, earlier data from the HST surveys of Virgo and Fornax \citep{cote06,turner12} show a decline in the NSC occupation fraction at masses above 10$^9$~M$_\odot$, which is put in context with the lower mass galaxies in the more recent study by \citet{sanchez-janssen19}. In the highest mass ellipticals, NSCs are clearly rarer, but some do appear to exist. For example, \citet{lauer05} catalogs a number of  nuclear sources with nuclear absorption line spectra and colors similar to the main galaxy, suggesting they are in fact NSCs, although their classification does not universally agree with those of \citet{cote06}. Counting just these absorption line nuclei as bona-fide NSCs, the nucleation fraction in the \citet{lauer05} sample is 20\% (8 out of 40 galaxies with available colors).  

As pointed out by \cite{cote06}, a plausible explanation for the scarcity of NSCs in high-mass ellipticals is the merging of galactic nuclei with SMBHs. This process likely leads to the formation of binary black holes which, in the process of coalescence, transfer energy to the surrounding stars, thus decreasing the central density and effectively destroying the NSC \citep[e.g.,][]{quinlan97,milosavljevic01}.

In late-type galaxies, the available data sets are less extensive, with \citet{georgiev14} finding an NSC occupation fraction of 80\% in a sample of over 300 spirals of Hubble type Sbc and later, roughly consistent with earlier findings \citep{boker02,carollo02}.  A much lower NSC occupation fraction is found in a small sample of dwarf galaxies in \citet{georgiev09}.  Similarly, studies of satellite galaxies in \citet{habas19} suggest a low occupation fraction in dwarf irregular satellite galaxies in the nearby universe, lower than found for early-type galaxies at the same luminosity (but not mass).  
As discussed in Sect.~\ref{sec:definition}, NSCs can be more challenging to identify in late-type galaxies due to uncertain photocenters, dust obscuration, star formation, and young star clusters that create ambiguity when identifying the NSC. Because of this, the NSC occupation fraction in these studies are really lower limits. 

In Fig.~\ref{fig:demographics}, we combine the available information on the NSC occupation fraction from all major studies listed above in Table~\ref{tab:samples}. More specifically, we use all objects from the NSC samples of \cite{lauer05,cote06,georgiev09,georgiev14,ordenes-briceno18,sanchez-janssen19} for which color information on the host galaxy is available\footnote{In two cases, the galaxy data is presented in separate papers: \cite{ferrarese06b} contains the host galaxy information for the NSC sample of \cite{cote06}, while \cite{eigenthaler18} describe the host galaxy sample of \cite{ordenes-briceno18}.} to create the color-mass diagram in the left hand panel. Galaxies with and without identified NSCs are indicated by filled and open symbols, respectively. Note that the colors and stellar masses of the host galaxies are homogenized to a common photometric system. 

In contrast to earlier studies, we here classify NSC host galaxies based on their color rather than their Hubble type, in order to enable inference of NSC properties in modern large galaxy samples such as those from the Sloan Digital Sky Survey \citep[e.g.,][]{blanton05} and future galaxy samples with e.g.~the Large Synoptic Survey Telescope. We use a color cut to divide the galaxy sample into `blue cloud' and `red sequence' (corresponding to late- and early-types, respectively), indicated by the dividing line shown in Fig.~\ref{fig:demographics}. The amount and uniformity of data on the red sequence is exceptional, with large numbers of galaxies across the full mass range. On the other hand, there is a clear lack of low-mass blue cloud galaxies. This implies that the galaxy samples for which NSC searches have been conducted are significantly biased: at lower masses, the dominant population of galaxies in the universe is actually blue and of late Hubble-type \citep[][]{blanton05}. We note that the Hubble type classifications follow closely to the color cut shown here, with just a few galaxies from the early type samples falling in the blue cloud and vice versa.  In the rest of this review galaxies are divided based on the Hubble type and not based on the color-mass relation shown here due to a lack of available colors and masses for many NSC sample host galaxies.

The right panel of Fig.~\ref{fig:demographics} shows the NSC occupation as a function of galaxy mass, again separated into blue/late-type, and red/early-type galaxies.  For the red sequence, this figure is nearly identical to the one shown in \citet{sanchez-janssen19}, although the inclusion of the \citet{lauer05} data makes the downturn at the highest masses less pronounced. Remarkably, the blue cloud NSC occupation fraction depends on mass in a way that is very similar to the early types with a peak at $\sim10^9\,M_\odot$ as well.  However, note that (1) at low masses the consistency of early and late types seen here is apparently at odds with initial results from the MATLAS survey where early types have a higher occupation \citep{habas19}, and (2) at higher masses, the occupation fraction in late type galaxies appears to remain high with no clear drop in the highest mass galaxies.  
This is consistent with the fact that both the Milky Way and Andromeda host prominent NSCs despite having quite high stellar masses.  The sample of \citet{carollo98b} and \citet{carollo02} (which are not included in these plots due to a lack of homogeneous galaxy color information) show a nucleation fraction of at least 60\%, and given the high masses of these galaxies, this provides further evidence that the highest mass spirals are typically nucleated. 

The structural properties of the host galaxies may also affect nucleation. Two recent works have interesting implications for this topic: \citet{lim18} find lower-surface brightness galaxies to be less frequently nucleated than their higher-surface brightness counterparts of the same mass. Moreover, \cite{vandenbergh86}, \cite{ferguson89}, and \cite{lisker07} all find that low-mass early-type galaxies with NSCs are rounder 
than non-nucleated galaxies of the same galaxy mass. Similarly, \cite{denbrok15} show that in the Coma cluster, rounder galaxies harbor more massive NSCs than more flattened galaxies of the same stellar mass. \citet{sanchez-janssen19b} find that these results hold in multiple environments, and suggest that nucleation happens preferentially in galaxies that were formed more rapidly, and thus have a less flattened stellar mass distribution.

\subsection{Trends with host galaxy environment}

Recently, \citet{sanchez-janssen19} compared the NSC occupation fraction in lower mass early-type hosts ($M_\star \lesssim 10^9$\,\msol) across different galaxy environments. Using the data from \citet{denbrok14}, they found that the NSC occupation fraction is significantly higher in the Coma cluster than in Virgo, with the lowest occupation fractions found in the local group.  This conclusion appears to be at odds with a previous study of 28 field ellipticals by \citet{baldassare14}, who find an NSC occupation fraction similar to that in a mass-matched sample of galaxies in Virgo.  We note that a key difference between these two studies is the mass range of galaxies for which the comparison was conducted -- the \citet{baldassare14} work is focused on high-mass galaxies with $M_\star > 3 \times 10^9$\,\msol , while this regime was not probed by the study of \citet{sanchez-janssen19}.  

A related observation is the radial distribution of nucleated galaxies in cluster environments, a measurement that has only been done for spheroidal galaxies. Early studies of nucleated early-type dwarfs in Virgo ($M_\star \lesssim 10^8\,M_\odot$) suggested that they were more centrally concentrated within the cluster than non-nucleated galaxies \citep{ferguson89}.
Put differently, the NSC occupation fraction declines at lower galaxy number densities. This result was confirmed in subsequent work by \citet{lisker07}, and for low-surface brightness galaxies in Coma by \cite{lim18}. Overall, there is strong evidence that in denser environments, the NSC occupation fraction is enhanced for low-mass early-type galaxies ($M_\star \lesssim 10^9$\,\msol ). However, the same may not be true for higher host galaxy masses.  We note that all of these studies focus on early-type galaxies in cluster environments.  The \citet{georgiev09} and \citet{georgiev14} samples of late-type galaxies are based on archival data of galaxies, and mostly are in the field, yet their occupation is very high in the mass range between $M_\star \sim 10^8$ and $10^{10}\,M_\odot$.  This would seem to argue against the environmental dependence extending across all Hubble types, and perhaps being just restricted to nucleation in early-type galaxies.  Careful studies comparing satellites and central galaxies in a range of environments are clearly needed to resolve this issue.

\section{Properties of nuclear star clusters}\label{sec:properties}
Perhaps unsurprisingly given their unique location in the deep potential well of a dark-matter halo, NSCs have extreme properties: they are the brightest, most massive, and densest stellar clusters known. In this chapter, we will review the observational evidence for this statement, as well as the data for other NSC properties such as their stellar composition and kinematics. Finally, we briefly summarize the properties of the NSCs of the Milky Way and M31, in order to compare these unique objects to the population of NSCs. We tabulate typical properties of NSCs in different types and masses of galaxies, as well as the Milky Way NSC properties in Table~\ref{tab:properties}.  
\begin{table}
\caption[]{NSC properties\label{tab:properties} }
\begin{tabular}[t]{c c c c c c} \hline \hline 

(1) & (2) & (3) & (4) & (5) & (6) \\ 
Galaxy type & size & ellipticity & mass & mass fraction & surface mass density \\
& $r_\mathrm{eff}$ [pc]  & $\epsilon = 1-b/a $ & log(M$_{\rm{NSC}}$) &  M$_{\rm{NSC}}$/M$_{\star}$ & $\Sigma_{e}$ [M$_{\odot}/\mathrm{pc^2}$] \\
\hline \hline
\hline
Early types & & & & & \\
$<10^9$\msol & $4.4^{+6.5}_{-1.7}$  & $0.16^{+0.06}_{-0.08}$ & $5.8^{+0.6}_{-0.6}$ & $0.020$ & $10^{4.1} $  \\  
$>10^9$\msol & $4.4^{+17.9}_{-1.8}$  & $0.17^{+0.20}_{-0.08}$ & $7.0^{+0.8}_{-0.6}$ & $0.004$ & $10^{4.6} $  \\
\hline
Late types  & & & & & \\
$<10^9$\msol & $3.1^{+2.4}_{-2.0}$   & $0.37^{+0.14}_{-0.22}$  & $6.2^{+0.4}_{-0.5}$ & $0.007$ & $10^{3.6}$  \\
$>10^9$\msol & $4.8^{+7.5}_{-2.5}$   & $0.20^{+0.17}_{-0.09}$  & $6.8^{+0.8}_{-0.8}$ & $0.001$ & $10^{4.7}$  \\
\hline
Milky Way & $4.2\pm 0.4 $  & $0.29\pm 0.02$ & 7.5 & $0.0005$ & $10^{5.3}$ (within 4.2\,pc)  \\
&  &  &   &    &   $10^{6.4}$ (within 0.5\,pc) \\
\hline \hline
& & & & & \\
\end{tabular}

\small
{
\textsc{Notes} For all galaxy samples, NSC sizes and surface mass density values are derived from the data shown in Fig.~\ref{fig:mass_radius}, ellipticities from Fig.~\ref{fig:ellipticity}, and the masses and mass fractions from Fig.~\ref{fig:mass_scaling}.  We note that the samples of early and late-type galaxies differ in their stellar mass distribution, especially at low masses.  Values given are medians for the sample, while error bars indicate the 16th \& 84th percentiles. For more details on the Milky Way estimates, see Section~\ref{subsec:MWNSC}.  
}
\end{table}

\subsection{Sizes and morphologies}\label{subsec:sizes}
The size of an NSC, and in fact any star cluster, is not easily defined, because it depends on the analytic model used to describe the distribution of stars within the cluster. Typically, the metric used is the effective radius $r_{\rm eff}$ within which half the cluster light is contained. This measurement is preferred because it is fairly robust against the details of the model parameters used to fit the NSC profile \citep{larsen99}.

Because all but the closest NSCs are barely resolved even with HST, their effective radii are usually measured via PSF-fitting techniques that describe the observed NSC shape as a convolution of the instrumental point spread function (PSF) with an analytical function used to describe Galactic GCs or galaxy light profiles, e.g. the widely used King models \citep{king62} or S\'ersic profiles \citep{sersic68,graham05}. This approach has been applied both to one-dimensional surface brightness profiles \citep[e.g.,][]{boker02,cote06} and two-dimensional images of the NSC \citep[e.g.,][]{boker04,turner12,georgiev14,spengler17}. The latter approach has the advantage that one can simultaneously quantify the ellipticity of (well-resolved) NSCs.

\begin{figure}[h]
    \centering
    \includegraphics[width=10cm]{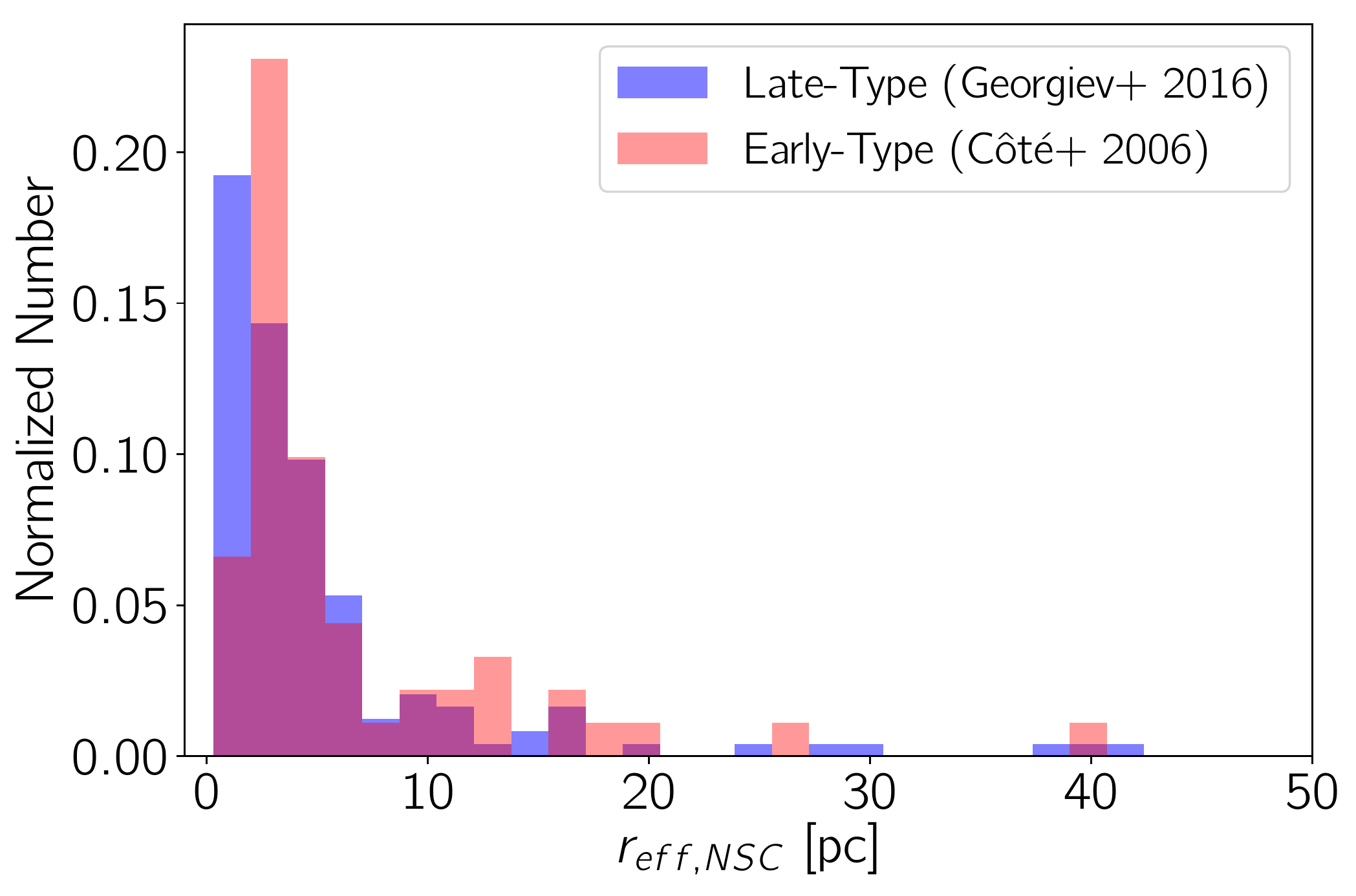}
    \caption{The effective radii of most NSCs are below 10~pc, but a tail of larger NSCs exists in both early and late-type galaxies. Data for early-type galaxies are taken from the ACSVCS survey \citep{cote06}, and for late-type spiral galaxies from \citet{georgiev16}.  Note that \citet{cote06} resolve effective radii only above $\sim$2~pc, but all data have been plotted. For the \citet{georgiev16} sample, we exclude objects with only upper limits for $r_{\rm eff}$, although this has little impact on the size distribution (because of the wide range of distances in this sample). }
    \label{fig:re_comparison}
\end{figure}

The results from the various studies are summarized in Fig.~\ref{fig:re_comparison}, which illustrates that the median size of NSCs ($r_{\mathrm{eff}}=3.3_{-1.9}^{+7.0}$\,pc) is comparable to that of globular clusters \citep{harris96,mclaughlin05}, although the distribution has a noticeable tail towards larger sizes. 

\begin{figure}[h]
    \centering
    \includegraphics[width=0.75\textwidth]{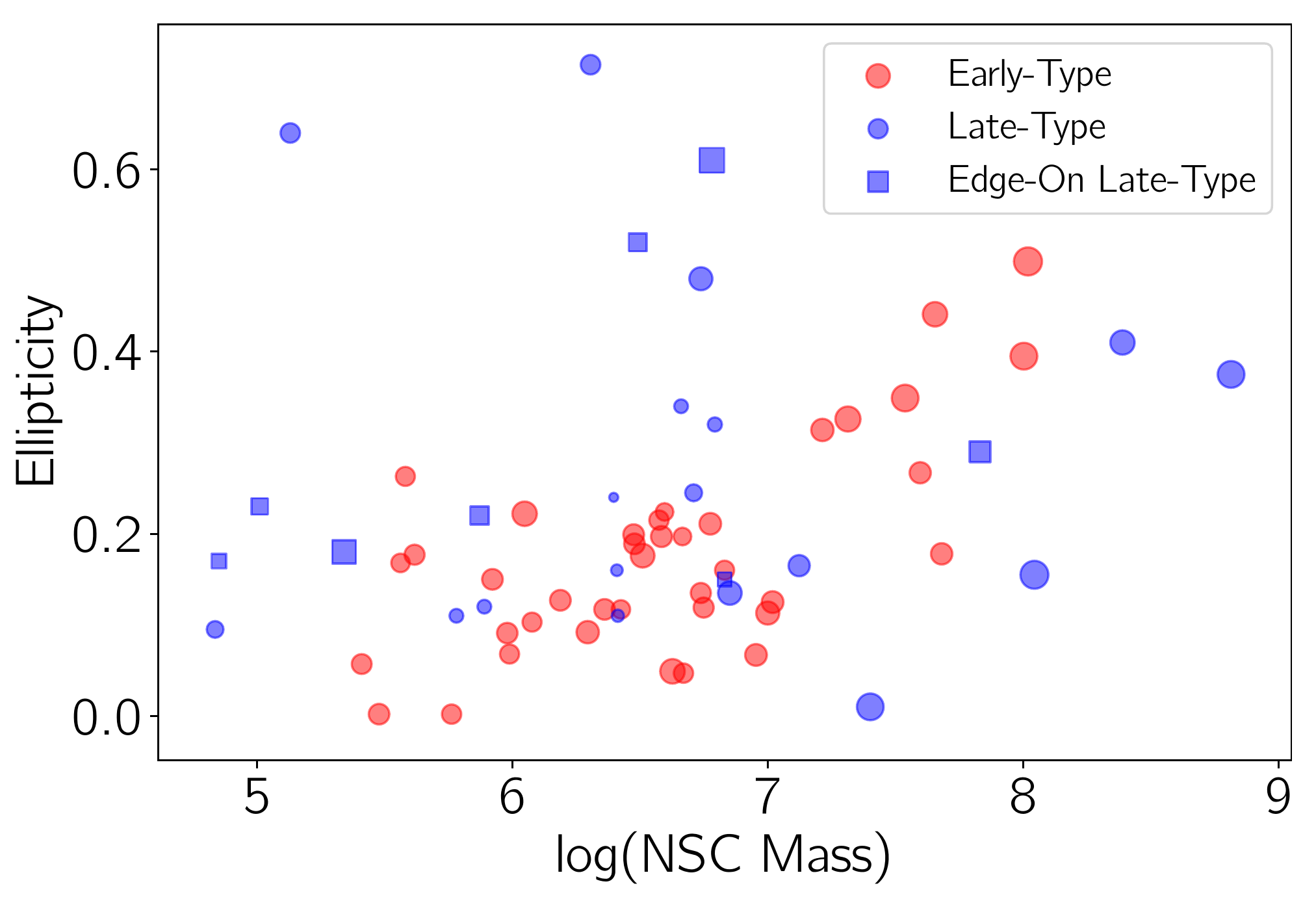}
    \caption{The highest mass NSCs tend to be more flattened (i.e., have higher ellipticity) than lower-mass NSCs.  This is especially true in the early-type galaxy sample of \citet{spengler17}, while late-type galaxies can have very flattened NSCs even in low mass galaxies.  The late-type galaxies plotted contain reliable measurements from \citet{georgiev14} (circles), while squares are edge-on galaxies including the Milky Way \citep{seth06,schoedel14a}.  All symbols are sized based on their effective radius; the range in the effective radii is 0.4 to 44~pc.  
    \label{fig:ellipticity}}
\end{figure}

Figure~\ref{fig:ellipticity} demonstrates that many NSCs appear non-spherical, with ellipticities as high as 0.6 (ellipticity: $\epsilon = 1-b/a $, a: major axis, b: minor axis) . Reliably measuring the axis ratio $b/a$ of barely resolved sources is difficult, especially in nuclei of spiral galaxies which are often affected by patchy dust extinction. Nevertheless, observations of edge-on spirals by \cite{seth06} have clearly identified elongated, disk-like structures in NSCs that are well-aligned with the disk of their host galaxies. The results of \cite{spengler17} demonstrate that many nuclei in spheroidal galaxies also show a pronounced ellipticity, and that the most elongated ones are also well aligned with their host bodies. Moreover, the amount of ellipticity appears to scale with NSC and host galaxy mass. We will come back to these results in Sect.~\ref{sec:formation} because they add important constraints on competing formation scenarios for NSCs.

\subsection{Luminosities and stellar masses}\label{subsec:lum}
One of the most directly measurable properties of NSCs is their luminosity. However, aperture photometry cannot usually be applied, as the underlying host galaxy structure is often complex and steeply rising towards the center, and thus cannot easily be removed. The preferred method to measure the magnitudes of resolved NSCs therefore is to integrate a model parameterization of their surface brightness distribution. This approach has been used both for spiral galaxies \citep{carollo02,boker02,georgiev14} as well as for early-type samples \citep{cote06,turner12}. 

\begin{figure}[h]
    \centering
    \includegraphics[width=0.75\textwidth]{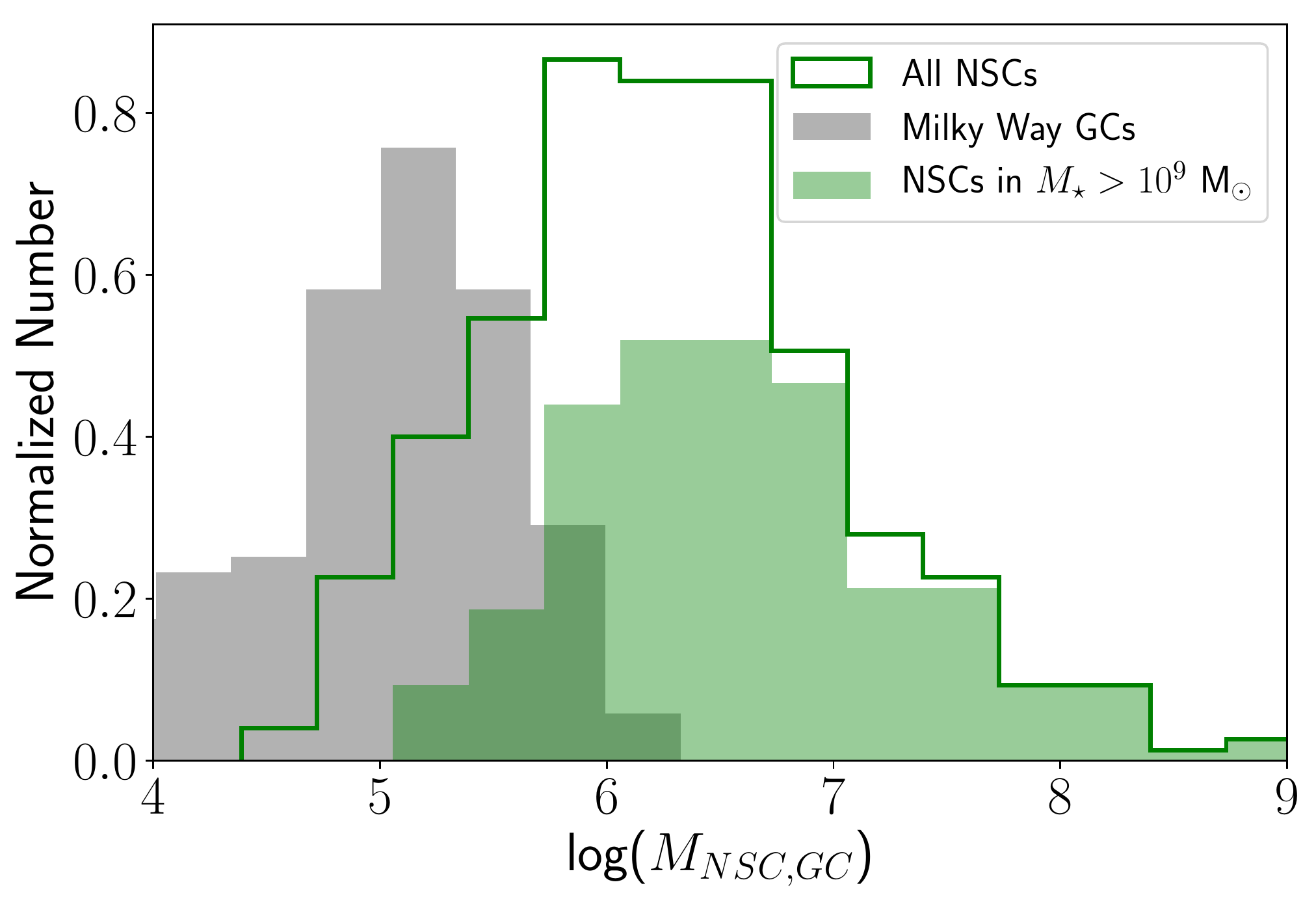}
    \caption{Mass distribution of NSCs relative to that of globular clusters, showing that NSCs in massive galaxies are typically an order of magnitude more massive. Data for NSCs (open and filled green histograms) are estimated via color--$M/L$ relations or SED fitting to determine their $M/L$, and using the \citet{bruzual03} models with a \citet{chabrier03} IMF.  The sample of NSCs includes both early-type galaxies from the Virgo cluster \citep{spengler17,sanchez-janssen19} and late-type galaxies from \citep{georgiev16}.  The masses of NSCs are very dependent on the host galaxy sample: the filled green histogram only includes host galaxies with masses above 10$^9$~M$_\odot$, while the open green histogram also includes less massive hosts.  The globular cluster masses are taken from the updated Harris catalog \citep{harris96} assuming an $M/L_V$ of 2, typical for GCs of all metallicities \citep[e.g.,][]{voggel19}.  }
    \label{fig:mass_histogram}
\end{figure}

The total stellar mass of an NSC is an even more fundamental quantity which, however, is not straightforward to measure. Generally speaking, the mass of a stellar system can be derived via three different observational approaches:
i) photometric, ii) spectroscopic, and iii) dynamical.

Most of the mass measurements for NSCs come from photometric studies, where luminosities are converted into mass using relations between the color and the mass-to-light ratio of a stellar population, which are selected based on (fits to) the spectral energy distribution (SED) of the stellar population  \citep{georgiev16,spengler17,sanchez-janssen19}. To make this method robust, multi-band photometry is required, with at least one colour measurement to constrain the stellar population. Unfortunately, there is no common set of filters used for all the observations, which adds further complexity.

Stellar population synthesis methods can also be used to estimate the mass-to-light ratio \citep{rossa06,seth10,kacharov18}. These mass-to-light ratios based on full spectrum fitting are more reliable than photometric measurements, especially if those are based on just a single color.  Dynamical mass measurements are considered the most accurate way to estimate masses. To measure the mass, the mass-to-light ratio of the NSC is determined by comparing the measured (typically integrated) velocity dispersion to a dynamical model for the stellar potential based on the luminosity profile of the NSC.  Since the integrated velocity dispersions of NSCs are typically low ($<20-30$km/s), the spectral resolution of the observations needs to be sufficiently high \citep[R$\gtrsim$5000, e.g.][]{walcher05,kormendy10}.  Recently, \citet{nguyen19} showed that mass-to-light ratios based on stellar population synthesis vary by less than 10\% with dynamical estimates based on kinematic maps of four NSCs, suggesting that stellar population synthesis also provide quite reliable masses.  

Our current knowledge of the NSC mass distribution is summarized in Fig.~\ref{fig:mass_histogram} which demonstrates that NSCs are, on average, much more massive than globular clusters, especially in hosts with masses above $10^9$\,\msol~-- we consider the scaling of NSC and galaxy mass further in Sect.~\ref{subsec:scaling}. The high masses of NSCs combined with their similar sizes to GCs (see Sect.~\ref{subsec:sizes}) already suggests that the stellar densities of NSCs are extremely high, a fact that we will discuss next.  

\subsection{Size-mass relation and stellar densities}\label{subsec:sizemass}

In Fig.~\ref{fig:mass_radius}, we compare the mass-density and size-mass relations of NSCs to those of other stellar systems. Similar plots have previously been published by various authors \citep[e.g.,][]{hopkins10,misgeld11,norris14}. Our version has a more complete sample of NSCs including recently published data, and distinguishes the most reliable NSC dynamical and stellar population synthesis mass measurements \citep{erwin12,nguyen18}. We draw three conclusions from Fig.\,\ref{fig:mass_radius}: 

\begin{figure}[h]
    \centering
    \includegraphics[width=0.4\textwidth]{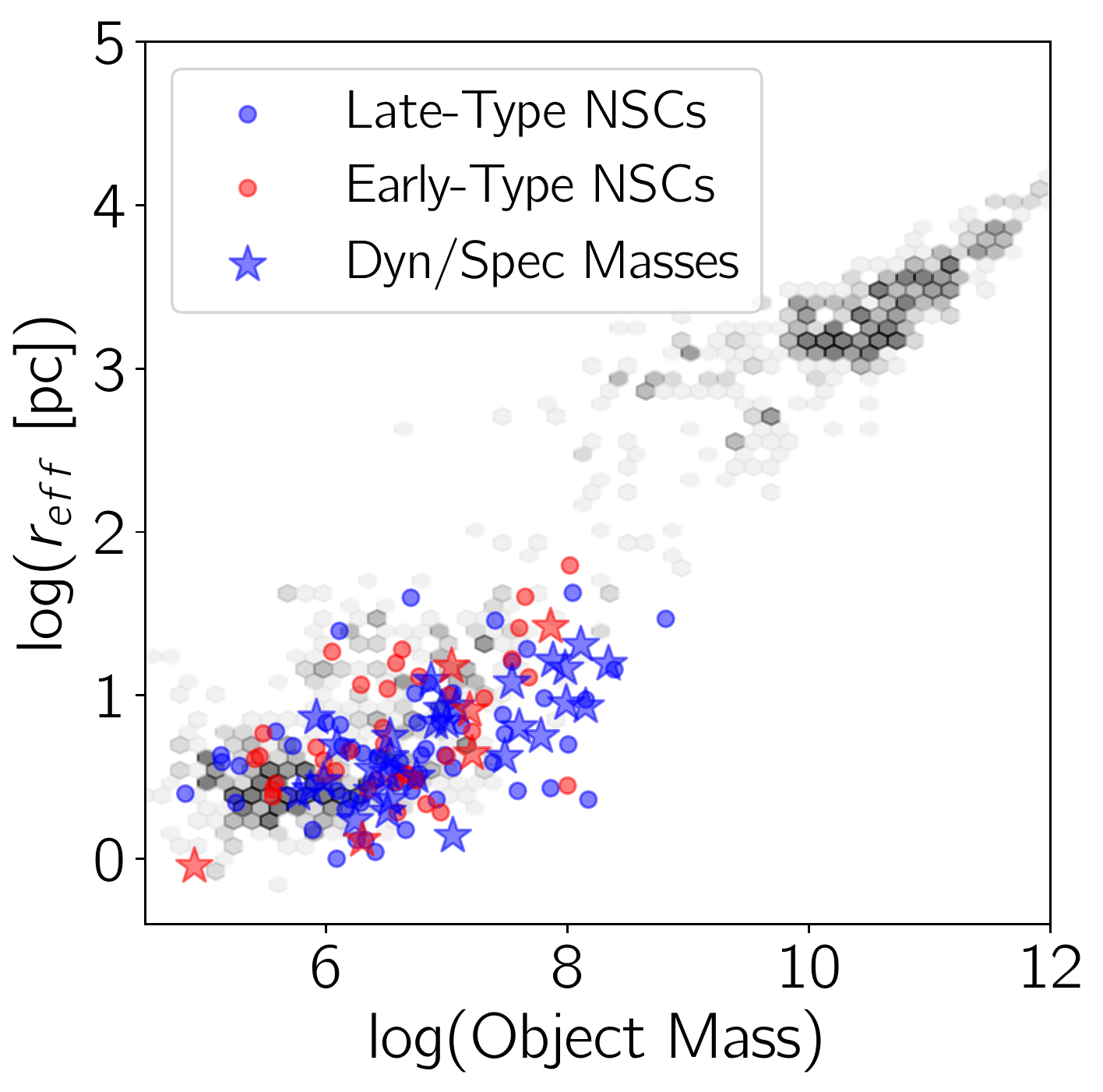}
    \includegraphics[width=0.4\textwidth]{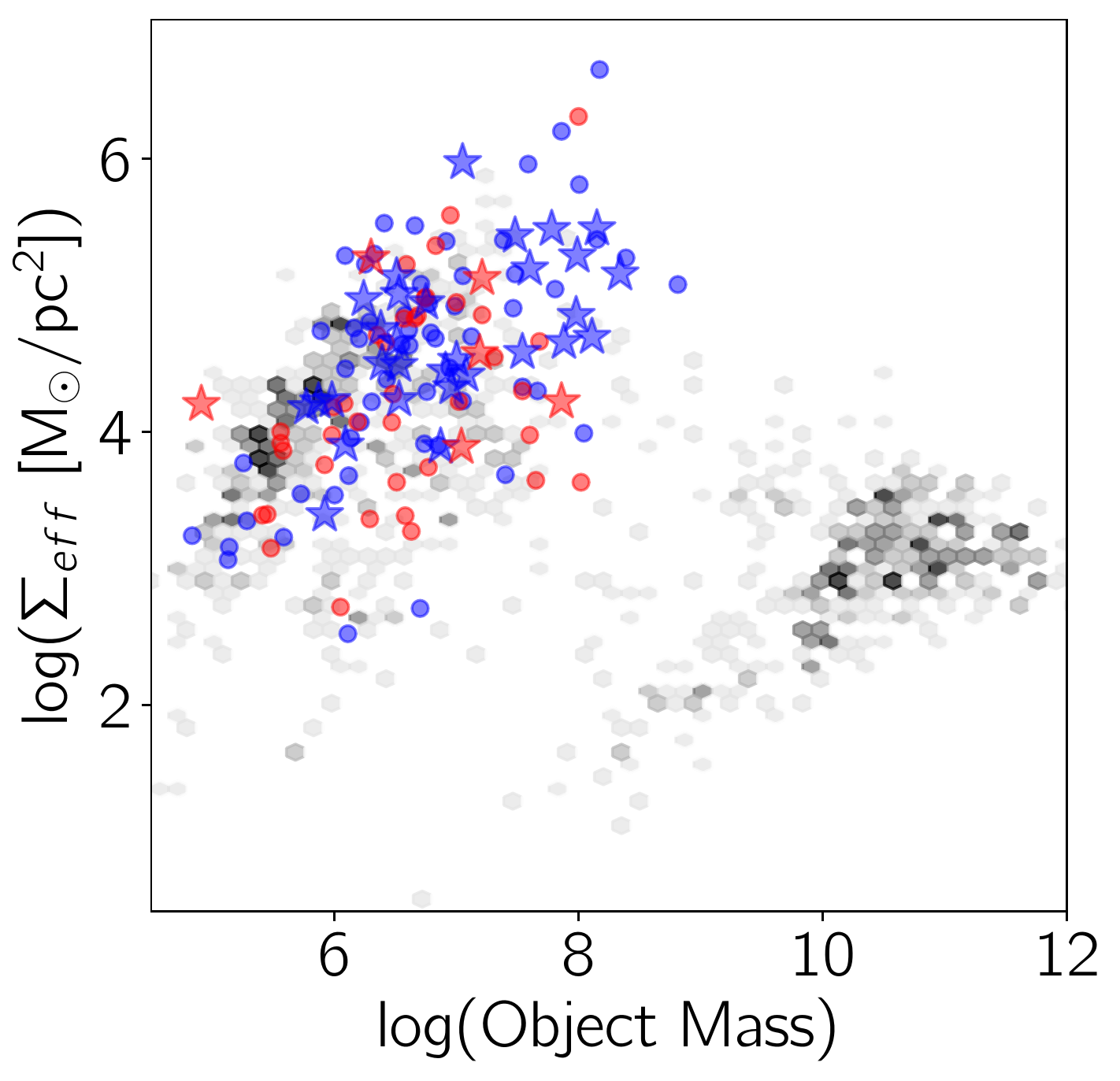} \\
    \includegraphics[width=0.4\textwidth]{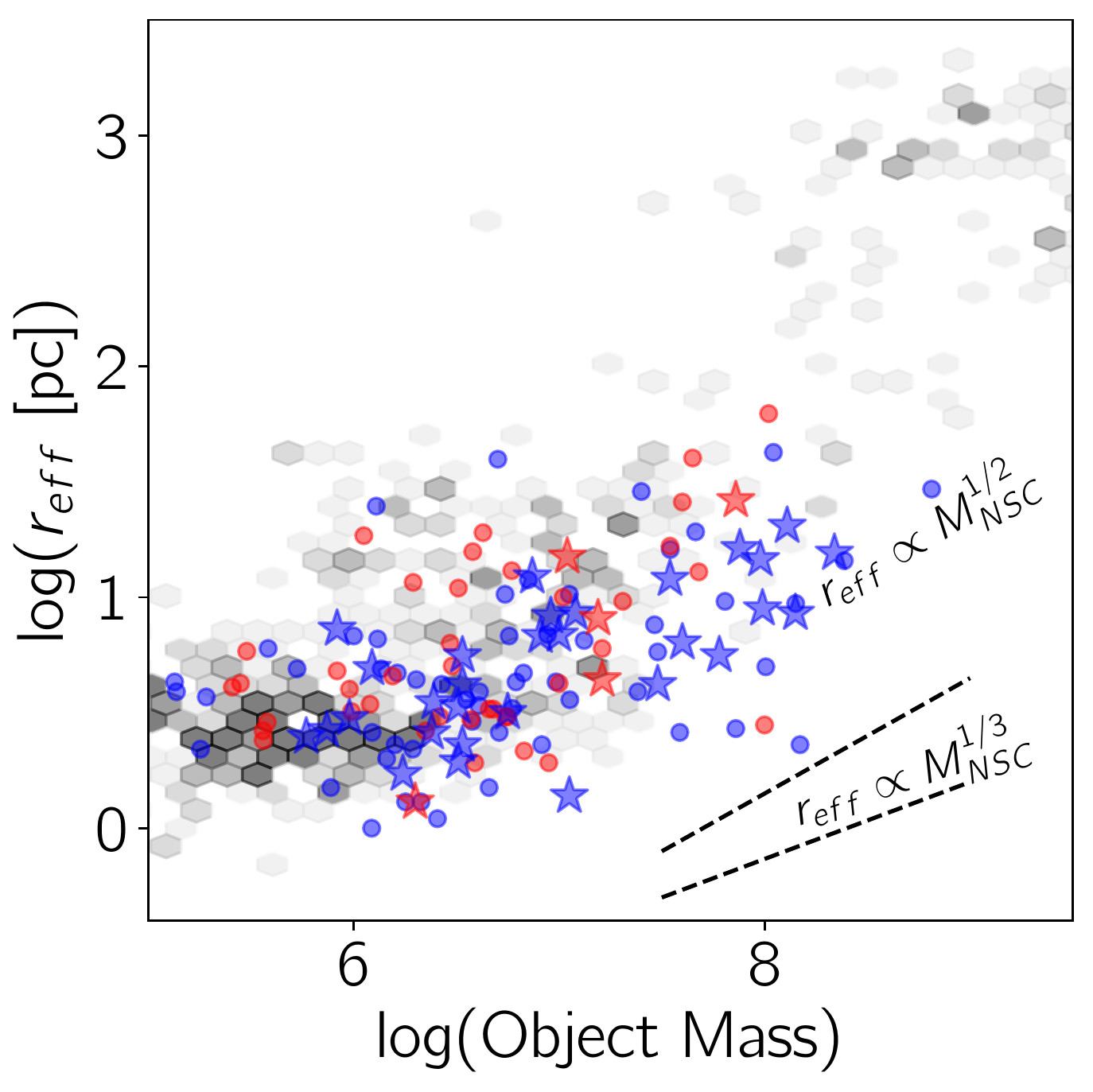}
    \includegraphics[width=0.4\textwidth]{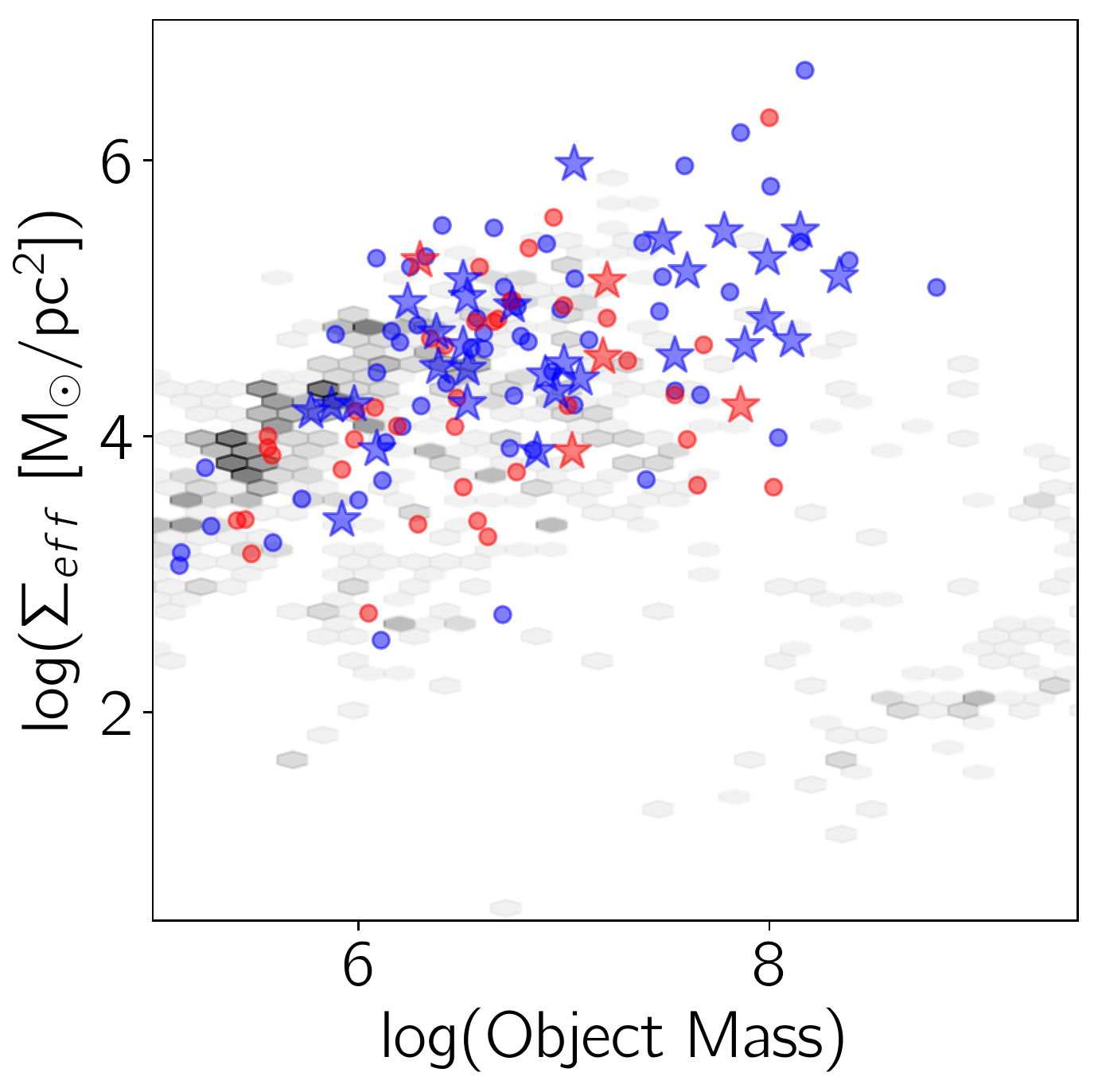}
    \caption{Mass--radius and mass--surface density relations for NSCs (colored points), compared to other hot dynamical systems (GCs, bulges, elliptical galaxies; grayscale) from \cite{norris14}.  
    The blue asterisks denote measurements compiled by \citet{erwin12}, which stem from dynamical measurements or spectral synthesis fits, taken mostly from \citet{walcher05} and \citet{rossa06}. Other NSC data is taken from \citet{georgiev16} for late types, and the ACSVCS for early types \citep{cote06,spengler17}. Other hot stellar systems plotted in gray include globular clusters, ultracompact dwarfs, compact ellipticals, and normal early-type galaxies. For these, the masses plotted denote the total stellar mass of the compact object/galaxy. The top two panels plot the data over the full mass range of hot stellar systems, while the bottom two panels 'zoom in' to just the region of parameter space covered by NSCs. The dashed lines in the lower left panel are a line of constant surface density (top line) and constant mass density (bottom line). Above a few million solar masses, NSCs show a positive mass-radius relation, which results in a flattening in surface densities with increasing mass.   }
    \label{fig:mass_radius}
\end{figure}

\begin{enumerate}
    \item For NSCs with masses below a few $10^6$\,\msol , the size of the NSC does not depend on their mass, while for higher masses (above $\sim 10^7$\,\msol), the size appears to increase proportionally to the square root of the mass, thus preserving a constant surface density. 
    \item At a given NSC stellar mass, NSCs in early-type galaxies are on average larger than those in late-type galaxies. A factor of two difference was previously noted by \citet{georgiev16}, but the data shown here suggests this offset may not extend down to the lowest mass NSCs.
    \item The most massive NSCs are the densest known stellar systems, and can reach mass surface densities of $\sim 10^6$\,\msol /$\rm pc^2$ or more.  This is true even for reliable (i.e., spectroscopically derived) NSC masses, which suggests that the $\sim$10$^5$\,\msol /$\rm pc^2$ upper limit in surface density suggested by \citet{hopkins10c} to be due to stellar feedback may need to be revised upward.  These dense massive clusters have masses derived primarily from the stellar population fits of \citet{rossa06}.  
\end{enumerate}

While surface mass densities are available in a wide range of NSCs, estimates of the volume densities densities of NSCs have been derived in only a handful of systems. 
\citet{lauer98} derived deprojected density profiles in M31, M32 and M33, and found the central densities (on scales of $\sim$0.1 pc) are $\sim 10^6$\,\msol /$\rm pc^3$ for M31 and M33 and $\sim 10^7$\,\msol /$\rm pc^3$ for M32, and that the volume density profile of M32 goes as $\rho(r) \propto r^{-3/2}$. \citet{schoedel18} derived the three dimensional stellar density for the Milky Way NSC on scales of 0.01 pc to $2.6 \pm 0.3 \times 10^7$\,\msol /$\rm pc^3$.
A recent study by \cite{pechetti19} derived volume density profiles in a number of the nearest NSCs and finds a clear correlation between the galaxy mass and the  density and slope of their NSC profiles, with higher mass galaxies having denser NSCs and shallower density profiles.  The NSC mass profiles range from $\rho(r) \propto r^{-1}$ to $r^{-3}$ with the densities at $r=5$\,pc ranging from  10$^2$ to $\sim$10$^4$\,\msol /$\rm pc^3$. 

\subsection{Stellar ages and metallicities}\label{subsec:stelpop}
Measuring individual stellar ages and metallicities within an NSC is the most direct way to derive its formation history. Unfortunately, this is an extremely difficult undertaking, because individual stars can only be resolved in a few very nearby NSCs. For most NSCs in external galaxies, we must rely on the analysis of the integrated light of the entire cluster after careful subtraction of the surrounding galaxy light.  Population fitting techniques are usually applied to this integrated light data to determine the combination of stellar ages and metallicities that best describes the observed spectral energy distribution. Even with high quality spectra or multi-band images, there are many challenges and limitations associated with stellar population modeling, which have been extensively discussed recently by \cite{conroy13}. Generally speaking, the youngest population of stars in an NSC can be fairly reliably determined because they dominate the light. On the other hand, the age and mass of the oldest populations (which provides the strongest constraints on when the NSC formed) are extremely challenging to derive. We focus first on late-type galaxies, then early-types, followed by a short section on the nucleus of the Sgr dwarf spheroidal galaxy.

\vspace{0.1cm}

\begin{figure}[h]
    \centering
   \includegraphics[width=0.9\textwidth]{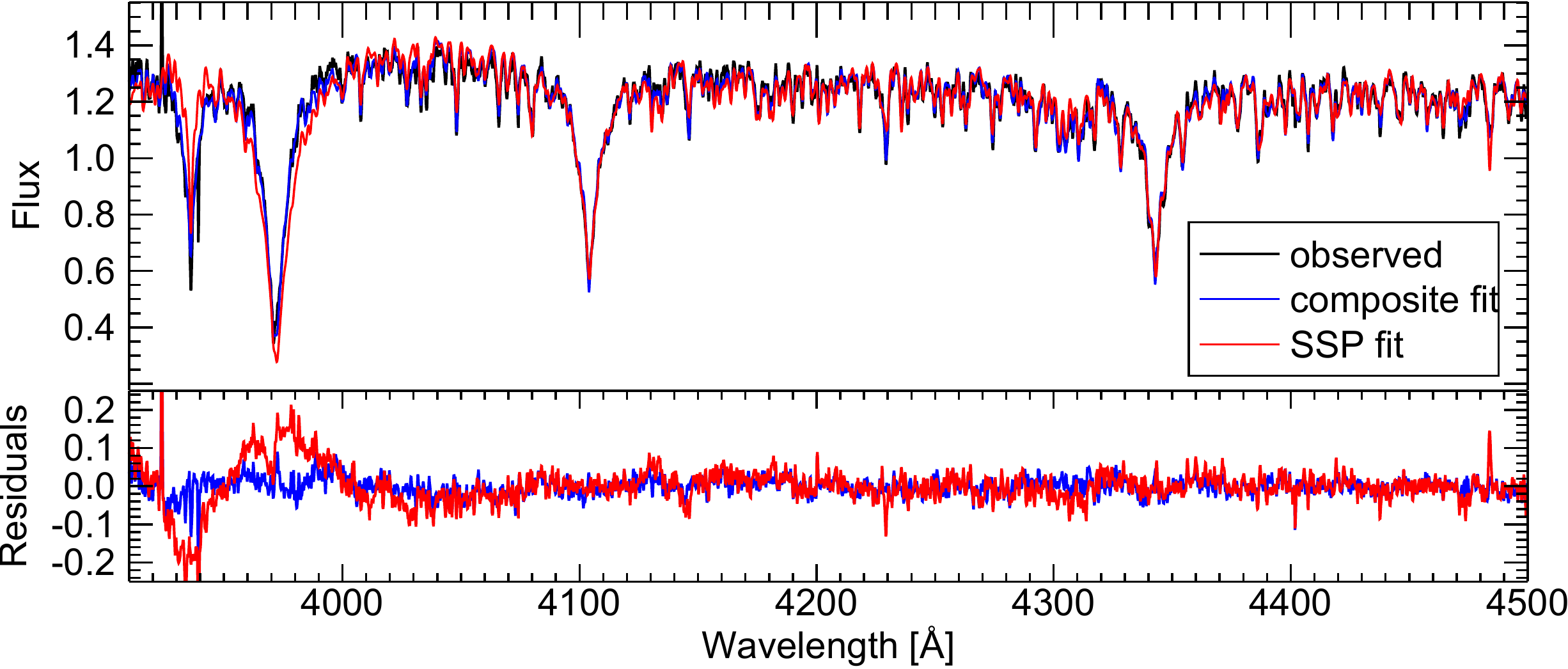}
   \caption{Stellar population fits for NGC\,247 \citep[from][]{kacharov18}. The fit residuals are significantly improved when including multiple populations of stars.  In this case, the best fit stellar population model fit (blue line) has $\sim$50\% of the light and 90\% of the mass coming from populations with ages $>$1~Gyr and subsolar metallicity, while significant light contributions are also seen from stars of $\sim$10 and $\sim$100 Myr, both with slightly super-solar metallicities. The best-fit single stellar population has an age of 0.22 Gyr, and [Fe/H]=+0.26.}
    \label{fig:ssp_comp}
\end{figure}

\noindent {\em Late-Type Galaxies:} Spectroscopic and multi-band photometric studies of late-type galaxies, for both individual objects \citep[e.g.,][]{boker97,boker01,seth06} and larger samples of galaxies \citep{walcher05,rossa06,carson15,kacharov18}, all agree that most late-type NSCs contain a mix of stellar populations, with the mass dominated by old stars with ages of more than a few Gyr.  In all cases, an extended star formation history provides significantly better fits to the NSC spectra than single stellar populations. This is demonstrated in Fig.\,\ref{fig:ssp_comp} for the case of NGC\,247. All NSCs in the latest-type spirals appear to contain stars younger than 100~Myr \citep{walcher05,kacharov18}.  In earlier-type spirals, \citet{rossa06} find that half their NSCs have a significant population of stars younger than one Gyr. The Milky Way NSC appears to be a typical example, containing both very young stars and a dominant old population \cite[][see also Sect.~\ref{subsec:MWNSC}]{blum03,pfuhl11}. A young, $\sim$200~Myr old stellar population is also seen at the center of the M31 NSC \citep{bender05}. These young stars are typically centrally concentrated within the NSC \citep{georgiev14,carson15}, although occasionally they will be found in a larger disk or ring structure \citep{seth06}.  As discussed further in Sect.~\ref{sec:formation}, the ubiquitous presence of a young population is strong evidence that NSCs in late-type hosts experience periodic in-situ star formation triggered by infalling gas. 

\vspace{0.1cm}

\noindent {\em Early-Type Galaxies:} The situation is less clear in early-type hosts. The nearest examples seem to have substantial young populations, with ages less than a Gyr \citep{monaco09,kacharov18}.  However, most early-type NSC hosts are found in dense galaxy clusters where typically, star formation has ceased due to the lack of molecular gas. Therefore, most stellar population studies of NSCs in early-type galaxies assume stellar ages above 1\,Gyr, and use single stellar population models to fit their spectra and spectral energy distributions \citep[e.g.,][]{koleva09,paudel11,spengler17}. 
These studies find median spectroscopic ages of NSCs of $\sim$3~Gyr \citep{paudel11,spengler17}. Furthermore, NSCs in early-type galaxies are typically younger than their surrounding galaxy \citep{koleva09,chilingarian09,paudel11}.  We caution against interpreting these age results too strongly, as the SSP assumptions made by these studies are likely incorrect (see section on M54 below).  Furthermore, the age and metallicity estimates are degenerate, with the age being significantly more uncertain \citep[e.g.,][]{worthey94,spengler17}.

\begin{figure}[h]
    \centering
   \includegraphics[width=0.45\textwidth]{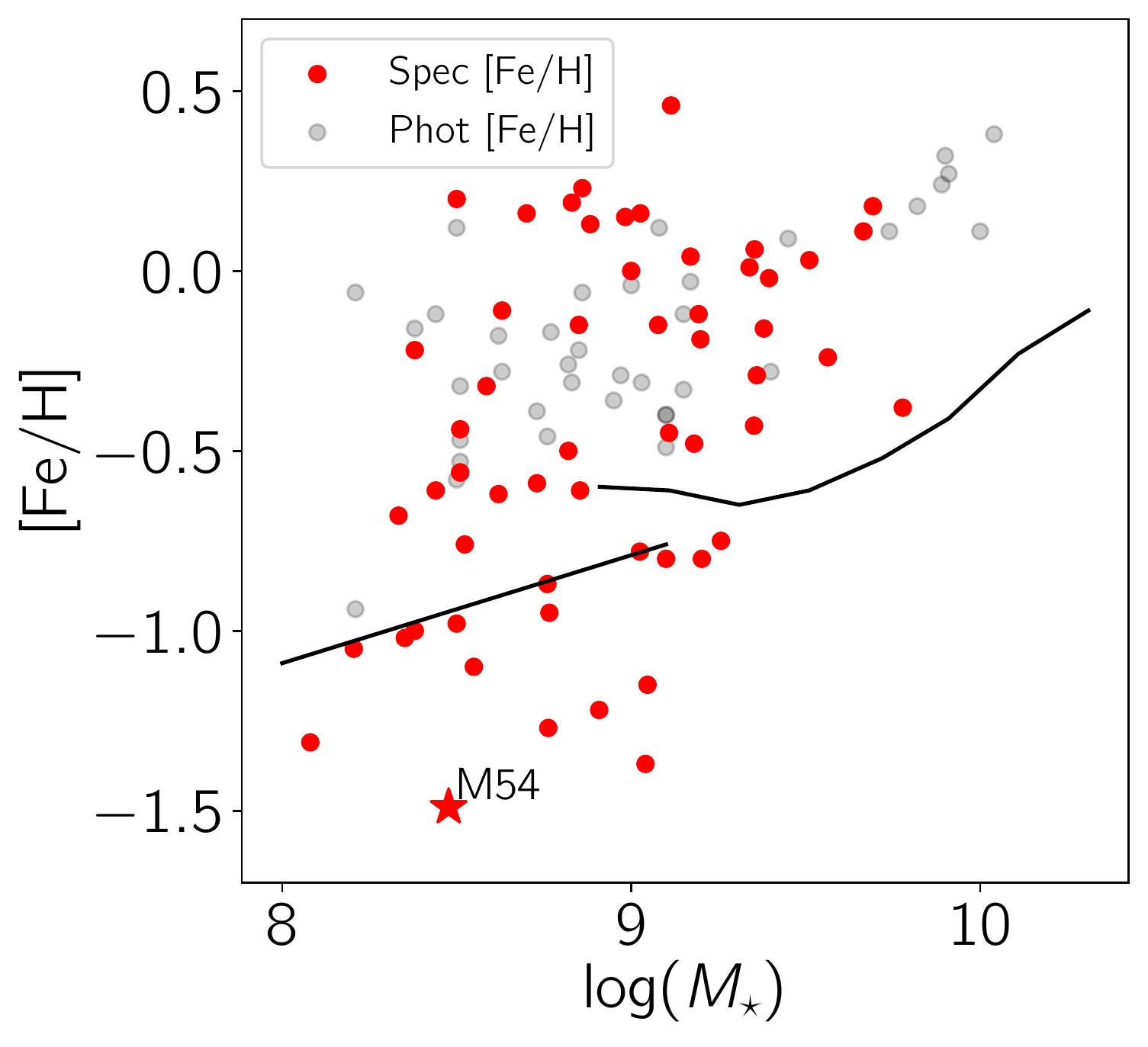}
   \includegraphics[width=0.45\textwidth]{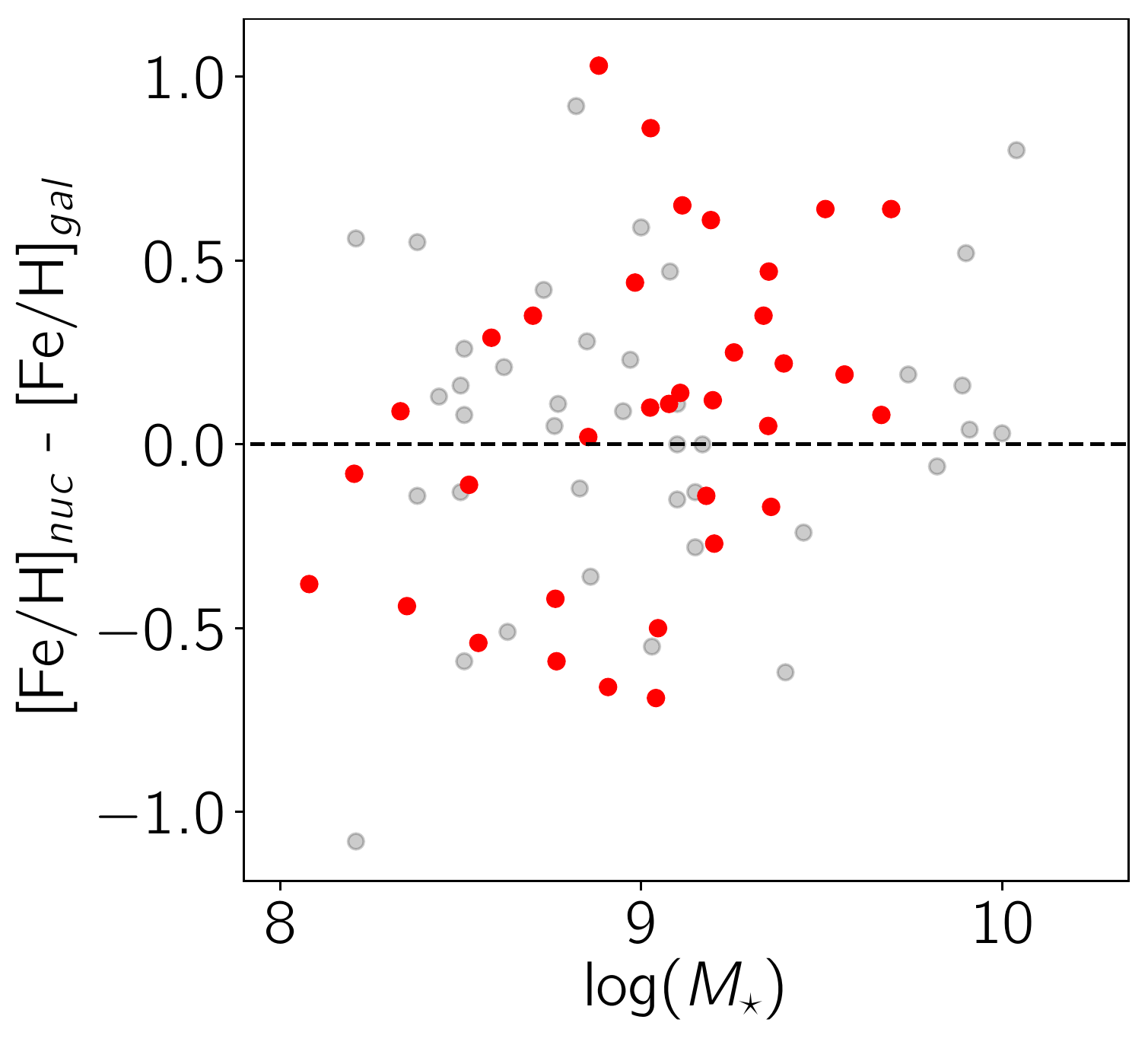}
    \caption{Metallicity estimates for NSCs in early-type galaxies. {\em Left:} NSCs in higher-mass galaxies are all metal-rich, while a range of metallicities exist at lower host masses. The spectroscopic data (red symbols) include results from four studies \citep{koleva09,paudel11, spengler17, kacharov18}. The photometric metallicity estimates (grey symbols) are from \citet{spengler17}. The estimate for the nucleus of the Sgr dwarf (labeled M\,54) is that of its dominant metal-poor population \citep[e.g.,][]{mucciarelli17}. M\,54 is the only NSC plotted here for which individual stars can be resolved and their abundances determined.  The two solid lines are the mass-metallicity relations of \citet{kirby13} at the low mass end, and \citet{gallazzi05} at the high mass end. {\em Right:} the difference in metallicities between the NSC and their host galaxies.  Data are from \citet{koleva09,paudel11, kacharov18}.  Difference values above zero indicate that the NSC is more metal-rich than its host, which predominantly occurs in high-mass galaxies. 
    }
    \label{fig:mass_metallicity}
\end{figure}

In Fig.~\ref{fig:mass_metallicity} we show metallicities for a sample of NSCs in early-type galaxies. The NSCs with spectroscopically measured metallicities \citep[red symbols;][]{koleva09,paudel11,spengler17} show a fairly clear transition at galaxy stellar masses of $\sim10^9\,M_\odot$.  Above this galaxy mass, NSCs are uniformly metal-rich, and fall above the median mass-metallicity relationship, as might be expected due to metallicity gradients within the galaxies \citep[e.g.,][]{koleva11}.  For galaxies with $M_\star < 10^9\,M_\odot$, a much wider range of NSC metallicities is found, with roughly half of the measurements falling on or below the median galaxy mass-metallicity relationship.  This same trend is seen when directly comparing spectroscopic estimates of NSCs and their hosts for a smaller sample of galaxies in the right panel of Fig.~\ref{fig:mass_metallicity}.  This trend with galaxy mass is not as evident in the photometric measurements alone, but we regard the spectroscopic results as more reliable, given the near independence of the Lick index-based metallicities on age \cite[e.g.,][]{schiavon07}.  

\vspace{0.1cm}

\noindent {\em M54 -- a unique example:}  The only NSC for which we can obtain optical spectroscopy of individual member stars is the nucleus of the Sgr dwarf spheroidal galaxy (a.k.a.\ the globular cluster M\,54). This makes it a key object for our understanding of NSCs, especially for low-mass early-type galaxies ($M_\star \sim 3\times10^8\,M_\odot$).  The dominant population in this NSC is old and metal-poor (${\rm [Fe/H]}=-1.5$), with a small number of much younger ($\lesssim$2 Gyr) and more metal-rich (${\rm [Fe/H]}=-0.5$) stars \citep{monaco05,siegel07,mucciarelli17,alfaro-cuello19}. The old metal-poor population has light-element abundance variations similar to those found in globular clusters \citep{mucciarelli17}, and recent work by \citet{alfaro-cuello19} suggests that it also has a significant spread in both age and metallicity.  We include the dominant metal-poor component of M54 in the left-hand panel of Fig.~\ref{fig:mass_metallicity}, and see that it fits with the general trend of low metallicites in similar mass galaxies.  

\subsection{Kinematics of NSCs}\label{subsec:kin}
To measure the kinematic structure of NSCs, one needs to resolve their morphology into many resolution elements. The sensitivity and high spatial resolution required to do this only became available with the advent of adaptive optics facilities on ground-based 8-m class telescopes. Earlier studies without the aid of adaptive optics were only able to get integrated velocity dispersion measurements \citep[e.g.,][]{boker99,walcher05,barth09}.

As discussed in Sect.~\ref{subsec:ground}, coupling the advantages of adaptive optics with integral-field spectrographs enabled spatially resolved studies of NSCs in a number of nearby galaxies \citep{seth08b, seth10}. These studies showed that most NSCs rotate in the same sense as the underlying host galaxy. This is true for both spiral and spheroidal galaxies, although in spheroidals, the amount of NSC rotation is typically lower and their kinematics can be quite complex: some show little or no rotation \citep[e.g. in NGC\,205;][]{nguyen18}, while others have high angular momentum \citep[e.g. FCC~47][]{fahrion19}, or even counter-rotating stars \citep{seth10,lyubenova13}. 
\begin{figure}[h]
    \centering
    \includegraphics[width=0.95\textwidth]{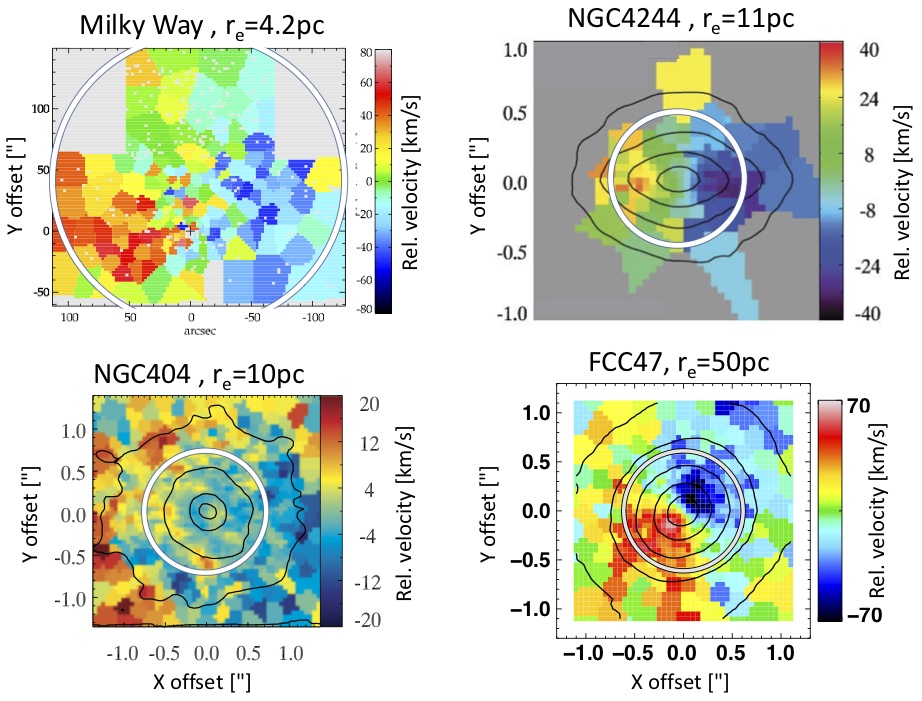}
    \caption{Examples for rotating NSCs: top: spiral galaxies Milky Way \citep[left, ][]{feldmeier14} and NGC\,4244 \cite[right, ][]{seth08b}. Bottom: early-type galaxies NGC\,404 \citep[left, ][]{seth10} and FCC\,47 \citep[right, ][]{lyubenova19}. In all panels, the effective radius of the NSC is marked with a white circle. }
    \label{fig:nsc_rotation}
\end{figure}
The ratio $v/\sigma$ of rotational velocity $v$ and velocity dispersion $\sigma$ provides a sense of whether the NSC dynamics are dominated by rotation or random motion.
Typical values of $(v/\sigma)_{\mathrm{r_{eff}}}$ (measured at the NSC effective radius) for NSCs in early-type galaxies are in the range of $0-0.5$ \citep{seth10,nguyen18,lyubenova19}. In comparison, the NSC of the Milky Way has $(v/\sigma)_{\mathrm{r_{eff}}}= 0.6$ \citep{feldmeier14}. Another edge-on spiral galaxy, NGC\,4244, has an even higher value of $(v/\sigma)_{\mathrm{r_{eff}}}\sim 1$ \citep{seth08a}.  A compilation of available kinematic data is shown in Fig.~\ref{fig:v_over_sigma}, which compares these rough NSC measurements with globular cluster measurements from \citet{bianchini13} and \citet{kamann18}.  While many NSCs fall in the same region of the diagram occupied by globular clusters, some are significantly more flattened and rapidly rotating.  We caution that detailed modeling is required to infer the orbital structure of NSCs; for instance, three-integral models of the flattened and rapidly rotating NSC in NGC\,4244 suggest that the object has a mildly negative vertical anisotropy \citep{delorenzi13}, while models of the Milky Way NSC suggest tangential anisotropy at radii $<$2~pc, with nearly isotropic orbits at larger radii \citep{feldmeier-krause17b}.

\begin{figure}[h]
    \centering
    \includegraphics[width=0.45\textwidth]{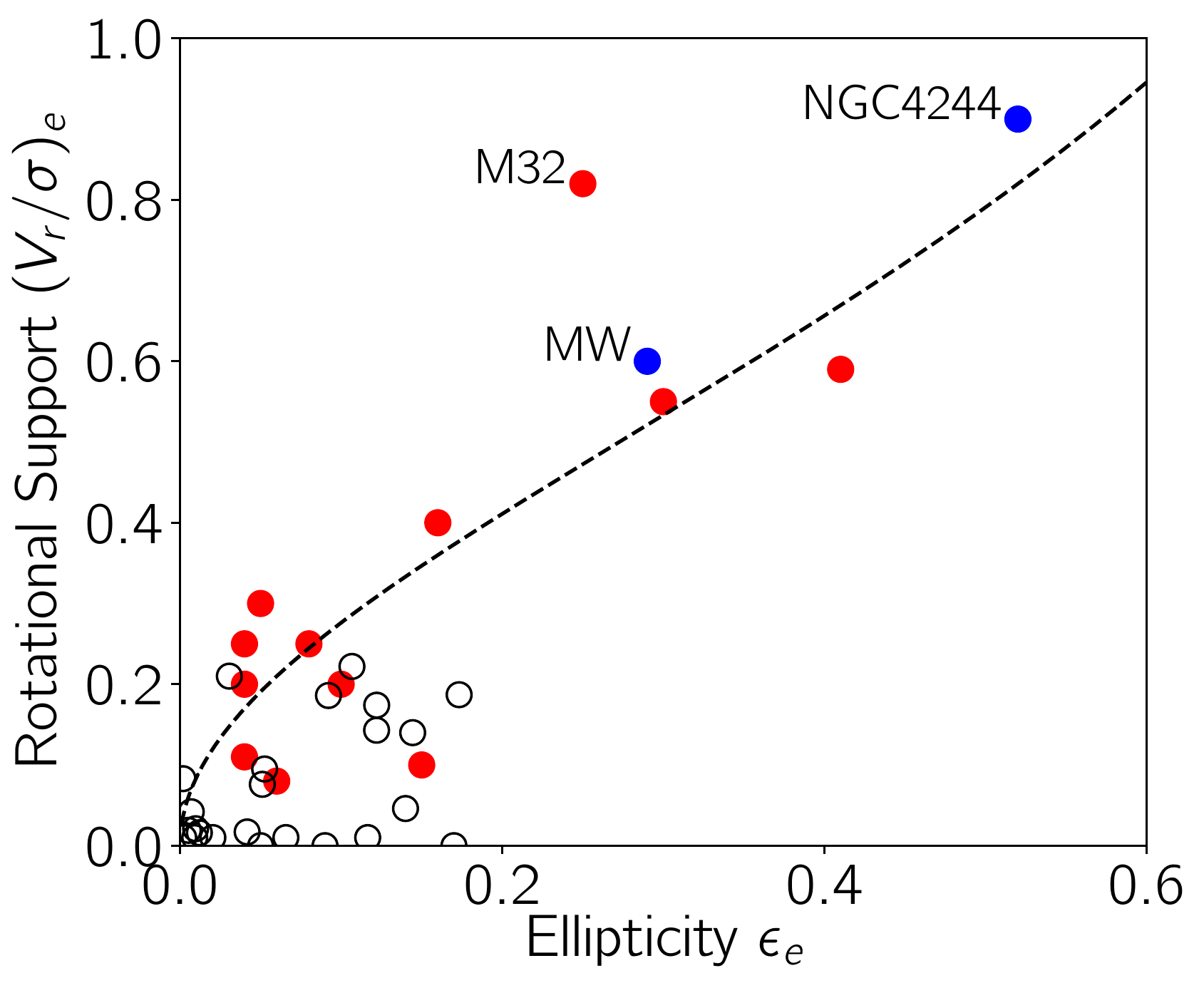}
    \caption{Many NSCs are rapidly rotating.  This plot shows the strength of rotation $v/\sigma_{\mathrm{r_{\rm eff}}}$ vs.~the ellipticity $\epsilon$ both estimated at the half-light radius for NSCs \citep[colored by Hubble type of their host galaxy; from][]{seth08b,seth10, feldmeier14,nguyen18,lyubenova19} and GCs \citep[open circles from][]{bianchini13,kamann18}.  The dashed line shows the expected trend for an edge-on isotropic, oblate rotator \citep{cappellari16}; many of the more rapidly rotating NSCs fall close to this trend. Both the Milky Way and NGC4244 are indeed seen edge-on; lower inclinations move objects to the left of the line (and then down). Note that the plotted NSCs sit at the center of the potential well of a galaxy, and should not necessarily be compared to entire galaxies.}
    \label{fig:v_over_sigma}
\end{figure}

We conclude that while some NSCs (mostly in spiral galaxies) can have significant levels of rotation, in general they are pressure-supported stellar systems. 

\subsection{Properties of the Milky Way nuclear star cluster}\label{subsec:MWNSC}
The NSC at the heart of the Milky Way warrants special attention because its proximity \citep[$d = 8.1 \pm 0.1$\,kpc,][]{abuter19,do19} offers a unique opportunity to study physical processes on scales that are impossible to resolve in external NSCs (at $d = 8.1$\,kpc, $1\as = 0.04$\,pc). In this section, we therefore provide a brief summary of the properties of the Milky Way NSC. For a more comprehensive overview, we refer the reader to the recent reviews by \cite{genzel10} and \cite{schoedel14b}. 

\begin{itemize}\itemsep5pt
\item{ {\bf Morphology, luminosity and mass:} }
The Milky Way NSC was first detected by \cite{becklin68} and later put into context of the larger Galactic structure by \cite{launhardt02}. \cite{becklin68} measured the FWHM of the Milky Way NSC to be $\sim 180''$ in K-band (i.e., $7.2$pc for $d = 8.1$kpc), with a shape elongated along the Galactic plane. Based on wide-field observations with the IRAS and COBE satellites, \cite{launhardt02} later described the Milky Way NSC as spherically symmetric and embedded in a larger disk structure, i.e., a nuclear stellar disk with a scale length of about 120\,pc. However, due to the limited spatial resolution of their data, they did not further characterise the NSC itself. More recently, \cite{schoedel14a} produced  mid-infrared images of the Milky Way NSC from multi-band Spitzer data that were largely corrected for extinction. Contrary to the assumption of \cite{launhardt02}, the Milky Way NSC appears to be intrinsically elliptical and flattened along the Galactic plane, with an ellipticity of $\epsilon = 0.29\pm 0.02$ \citep[i.e., an axis ratio $q=b/a=0.71\pm0.02$;][]{schoedel14a}.

Published values for the effective radius of the Milky Way NSC fall within the range of $r_{\mathrm{eff}} \sim 110''\pm11''$ (Spitzer 4.5$\mu$m) to $178''\pm 51''$ (K-band), i.e., between $4.2\pm0.4$pc and $7.2\pm 2.0$ pc \citep{schoedel14a,fritz16}. The total luminosity of the Milky Way NSC is $L_{\rm 4.5\mu\,m}= 4.1\pm0.4\times 10^7 L_{\odot}$ \cite{schoedel14a}.
At NIR wavelengths, \cite{fritz16} obtain M$_{\rm Ks}=-16.0\pm0.5$ within the effective radius, which corresponds to a total L$_{\rm Ks}=5.2\pm 3.0\times 10^7 L_{\odot}$, consistent with the earlier measurement of \cite{launhardt02}. 

 A number of recent studies have derived the total stellar mass of the Milky Way NSC. Using both photometric and dynamical methods, the results fall within the range between $~\sim 2.1\pm0.7 \times 10^7$\msol\ and $4.2\pm 1.1 \times 10^7$\msol\ \citep{schoedel14a,feldmeier14,chatzopoulos15,fritz16,feldmeier-krause17b}. Remarkably, this range is entirely consistent with the earliest mass estimates of the Milky Way NSC \citep{becklin68,launhardt02}.

\item{ {\bf Stellar populations and star formation history:} }
The majority of the resolved stars in the Milky Way NSC 
are old ($> 5$ Gyr) and evolved (spectral types K and M) giant and supergiant stars, as well as helium-burning stars on the horizontal branch and the `red clump', which have temperatures of $\sim 3000 - 5000$\,K, and low to intermediate masses 
\citep[\mstar $\sim 0.5 - 4$\msol ,][]{blum03,genzel10,feldmeier-krause17}. This is consistent with $\sim 80\%$ of the stellar mass having formed more than 5\,Gyr ago, with an initially high star formation rate 10\,Gyr ago that dropped to a minimum about 1-2\,Gyr ago, and then increased again over the past few hundred million years \citep{blum03, pfuhl11,nogueras-lara19b}. 

In addition to the old stars, there is a population of about 200 massive and young Wolf-Rayet and O- and B-type stars that are confined to the central 0.5\,pc \citep{allen90,krabbe91,ghez03,bartko10,pfuhl11,feldmeier-krause15}. The ages of these young stars are very well constrained within the range 3--8\,Myr \citep{paumard06,lu13}.
The total mass in young stars is between 14,000--37,000\msol \citep[measured via the K-band luminosity function,][]{lu13}; the measurement of photometric masses of the O/B stars gives a consistent lower limit $\geq 12 000$\msol \citep{feldmeier-krause15}. 
These stars have very likely formed \textit{in situ}, because the time it would take to bring them to the center is in conflict with their very young age. The lack of young stars outside of the central 0.5\,pc is further evidence against the infall of young star clusters \citep[e.g.,][]{feldmeier-krause15}.
In summary, the stars in the NSC are on average younger and more metal rich than the stars in the Galactic bulge \citep{ness13,feldmeier-krause15,feldmeier-krause17,nogueras-lara18,schultheis19}.
 
\item{ {\bf Mass density:} } 
The average surface mass density of the Milky Way NSC is $\sim 2 \times 10^5$\msol/pc$^2$ within the effective radius, and $\sim 2.5 \times 10^6$\msol/pc$^2$ within the central $\sim 0.5$\,pc \citep{genzel10,feldmeier14,schoedel14a,schoedel14b}. \citet{schoedel18} derived the three dimensional stellar mass density for the Milky Way NSC on scales of 0.01 pc to $2.6 \pm 0.3 \times 10^7$\,\msol /$\rm pc^3$.

\item{ {\bf Kinematic structure:}  } A variety of observational techniques have revealed that the Milky Way NSC is rotating in the same direction as the overall Galactic disk \citep{mcginn89,lindqvist92,genzel96,feldmeier14}. 
The line-of-sight velocity of the stars at the NSC effective radius 
is about $50\pm3$km/s, while the stellar velocity dispersion at the same radius is slightly higher \citep[$60\pm5$km/s]{mcginn89,feldmeier14}. The resulting value of $v/\sigma_{\mathrm{r_{eff}}} \sim0.8$ at the effective radius actually represents a maximum, with values between 0.2 and 0.8 for smaller radii. 

Using a long-slit drift scanning technique, \cite{feldmeier14} constructed contiguous velocity and velocity dispersion maps of the central $8\mathrm{pc}\times4\mathrm{pc}$. These maps show that the major axis of rotation is offset with respect to the Galactic plane by about $9^{\circ}$ (see their Fig.~8). This finding was confirmed by \cite{fritz16}. In addition, the line-of-sight velocity map shows a perpendicular rotating substructure, which may be the result of star cluster infall \citep{feldmeier14,tsatsi17}. 

\item {\bf Supermassive black hole:} At the center of the Milky Way NSC resides a supermassive black hole, SgrA*, with a mass of 
$4.04 \pm 0.06\times10^6$\msol \citep{abuter18,do19}.
Given the NSC mass of $\sim 3 \times10^7$\msol , this implies that in the case of the Milky Way, the mass fraction of the black hole relative to the nuclear star cluster is $\sim 13\%$. 
\end{itemize}

In summary, the Milky Way appears to contain a rather typical example for an NSC, because many of its properties are entirely consistent with the general sample of NSCs in external galaxies (see also Table~\ref{tab:properties}). 

\subsection{Properties of the M\,31 nuclear star cluster}\label{subsec:M31NSC}

Besides the Milky Way NSC, the nucleus of M\,31 also deserves special attention because it is one of the nearest extragalactic NSCs \citep[distance d=785 pc;][]{mcconnachie05}, and in fact was the first NSC to be  discovered, as described in Section~\ref{sec:history}.
As will become evident, its properties are quite different from those of the Milky Way NSC. For example, it represents a rare example of a nucleus in which the central black hole outweighs the NSC.

HST observations have made it clear that the M31 NSC has a complicated morphology \citep{lauer93,king95,lauer98}.  The surface brightness of the NSC rises above the M\,31 bulge at a radius of $\sim$5'' ($\sim$20~pc) \citep{kormendy99,peng02}.  The outermost component of the NSC is nearly spherical ($q=0.97$), with an effective radius of r$_{\rm eff}$=3.2'' ($\sim$12~pc) and a mass of $2.8\times 10^7$\,\msol \citep{peng02}. Embedded within this are two sources (dubbed P1 and P2) which are both offset from the central black hole and thought to be part of a flattened eccentric disk (discussed in more detail below).    This disk shows strong rotation \citep[with an amplitude up to $\sim$250~km/s][]{lockhart18}, has an effective radius of $<$1'' ($<$4~pc), and a total mass of $2.1 \times 10^7$\msol, resulting in a total NSC mass of $\sim 5 \times 10^7$\msol \citep{peng02}. Spectroscopic fitting shows that the stars in the central $\sim$5'' are old (7--13 Gyr) and extremely metal-rich (${\rm [Z/H]}=0.3$--$0.5$), i.e., even more metal-rich than the stars in the surrounding bulge \citep{saglia10}.  The three dimensional stellar mass density of the NSC on scales of 0.1 pc is $\sim 2 \times 10^6$\,\msol /$\rm pc^3$ \citep{lauer98}.

The NSC of M31 hosts a SMBH of $\sim1.1$--$2.3 \times10^8$ \msol \citep{peiris03,bender05}. Immediately surrounding the central black hole is a UV-bright source with an A-type spectrum known as P3 \citep{bender05}.  This source is blue, not because of AGN light but rather because it is dominated by hot stars. The spectrum and high resolution images are well described by a $\sim$10$^4$~\msol~100-200 Myr old stellar population \citep{bender05,lauer12}.  These stars are rotating in a co-planar way with the older eccentric disk \citep{bender05}.  This young population also appears to be associated with a lack of NIR $K$-band emission \citep{lockhart18}.  

The morphological structure of the central region of M\,31's NSC is best explained by an eccentric disk model, composed of stars traveling on nearly Keplerian orbits around a black hole \citep{tremaine95}. This model reproduces most of the features seen in HST photometry, in particular the bright off-center source P1 which is the apocenter region of the disk. Recent integral field kinematics are also well-fit by this model \citep{lockhart18}, and those data suggest a slow precession rate for the disk.  
The eccentric disk model also explains the fact that the velocity dispersion peak is offset by $\sim 0.2''$ from the UV peak, assumed to mark the location of the supermassive black hole \citep{bacon01}.  Moreover, the eccentric stellar disk model can explain the formation of young stars at the center of M31's NSC \citep{chang07}. The disk creates a non-axisymmetric perturbation to the potential and drives gas into the inner parsec around the SMBH. 
\cite{chang07} show that stellar mass loss from P1 and P2 would be sufficient to create a gravitationally unstable gaseous disk of $\sim 10^5$\msol\, every 0.1-1 Gyr, consistent with the young age of P3 ($\sim$ 200\,Myr). 

Although the mass of the MW and M\,31 NSCs are quite comparable, the ratio of this mass to the black hole mass is dramatically different, with the M\,31 BH being $\sim3\times$ the mass of the NSC, while in the Milky Way it makes up only $<$15\%.  The dominance of the BH mass accounts for M31's eccentric disk structure \citep{tremaine19}, providing us with a useful nearby case study in this regime which is more typical for higher-mass galaxies (see Section~\ref{sec:nsc-mbh}). 

\section{Nuclear star clusters and their host galaxies}\label{sec:hosts}

\begin{figure}
    \centering
    \includegraphics[width=0.45\textwidth]{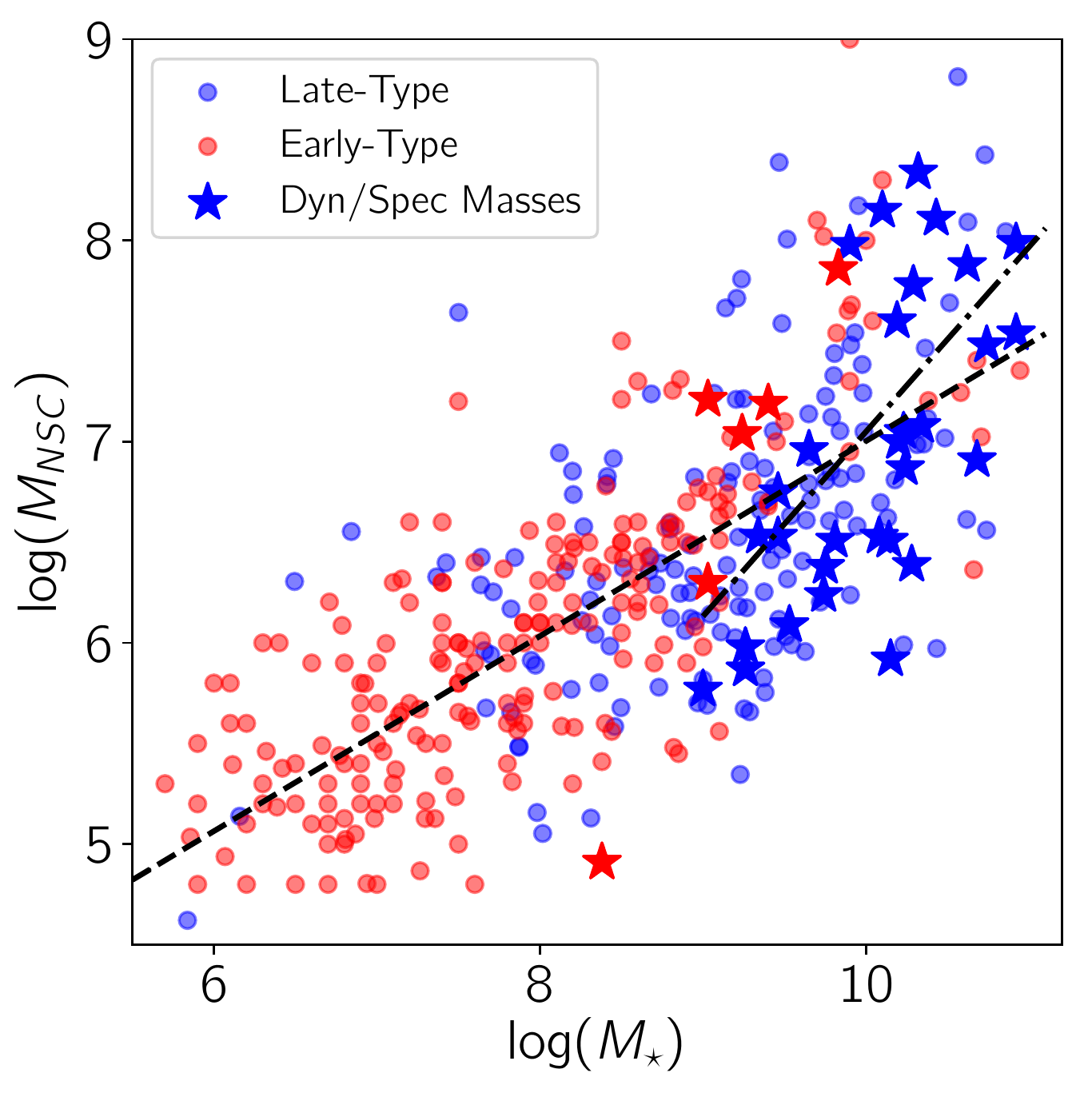}
    \includegraphics[width=0.48\textwidth]{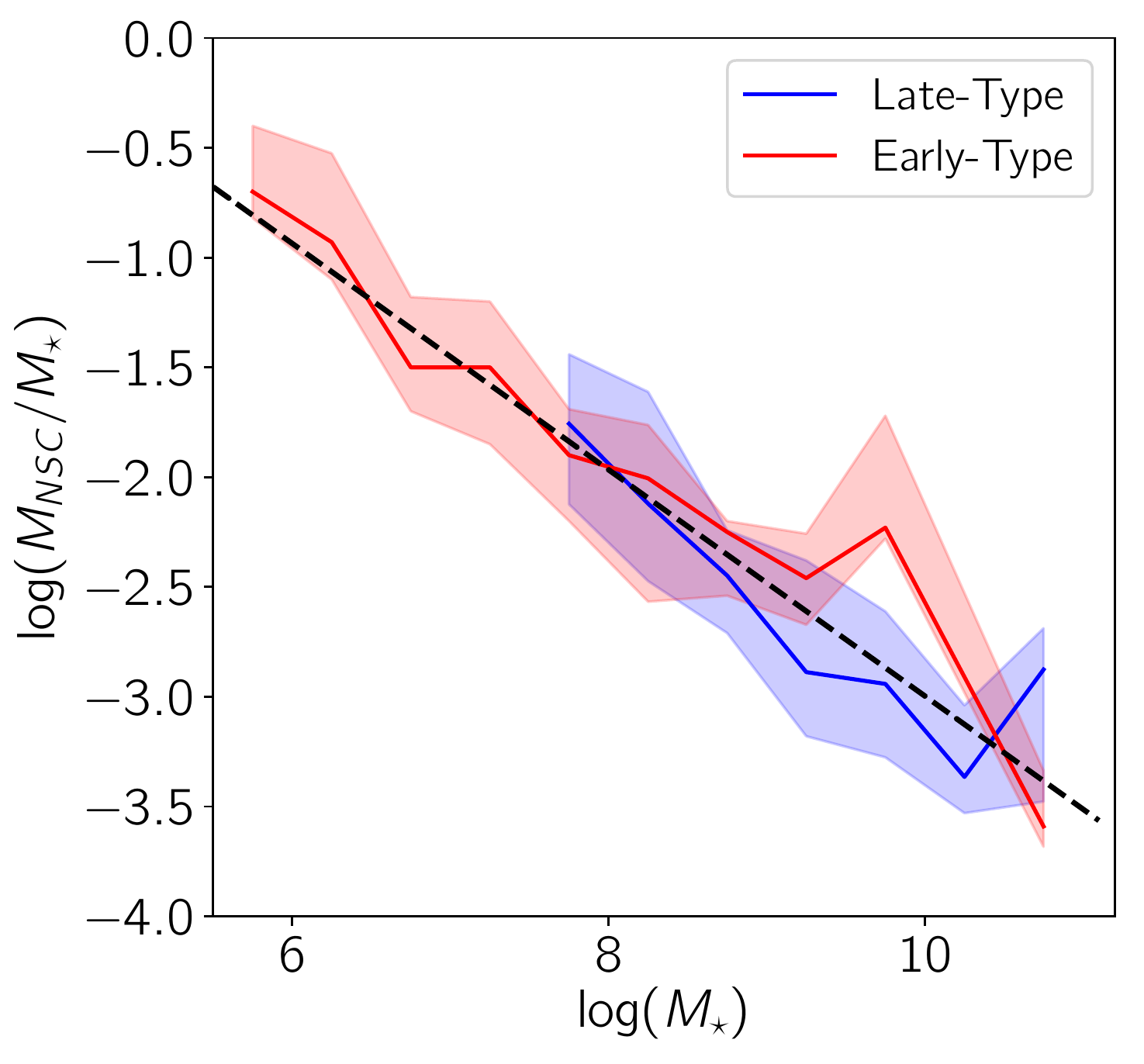}
    \caption{The masses of nuclear star clusters correlate with galaxy masses, but higher-mass galaxies have a lower fraction of their mass in their NSC.  {\em Left --} Galaxy and NSC masses for galaxies in which both quantities are available.  The compilation of dynamical and spectroscopically modeled NSC masses from \citet{erwin12} is shown with stars, while all other masses are derived from colors using stellar population models with a Chabrier or Kroupa IMF \citep{georgiev16,spengler17,ordenes-briceno18,sanchez-janssen19}.  Galaxies have been divided by their Hubble types into early and late types. {\em Right --}  The mass fraction of galaxies in NSCs as a function of galaxy mass, using the data from the left panel.  The line indicates the median galaxy within each mass bin, while shaded regions show the 25th and 75th percentiles of the distribution. }
    \label{fig:mass_scaling}
\end{figure}

In this section, we consider how NSCs are related to the host galaxies they live in. We first discuss the overall scaling of NSC mass with galaxy properties, and then investigate how other NSC properties vary with galaxy mass and type.

\subsection{Scaling relations of NSCs}\label{subsec:scaling}

The most straightforward quantities to compare between galaxies and NSCs are their luminosities and stellar masses, and there is a substantial body of literature that has examined these correlations.  The NSC luminosity was first compared to the host galaxy/bulge luminosity for small samples of galaxies by \citet{balcells03} and \citet{graham03}. Inspired by this, as well as the prominent black hole scaling relations, subsequent papers examined NSC--galaxy scaling relations using both luminosity and mass \citep{ferrarese06,wehner06,rossa06,balcells07,seth08b,erwin12,scott13,denbrok14,sanchez-janssen19}.  

While it was at one point claimed that the NSC and BH scaling relations may be comparable \citep{ferrarese06,wehner06}, it has since become clear that the NSC scaling relations are distinct from the BH scaling relations \citep{leigh12,erwin12,scott13}.  

In the left panel of Fig.~\ref{fig:mass_scaling}, we show a compilation of NSC and galaxy masses. The NSC sample is the same as in Fig.~\ref{fig:demographics}, with the addition of NSCs with masses measured using stellar population model fits or dynamical estimates  \citep[plotted as stars;][]{erwin12, nguyen18}.  
The best-fit log-linear relation to the full sample of 407 NSC and host galaxy mass estimates is:
\begin{equation}
    \log\, M_{\rm NSC} = 0.48\, \log\, \left( \frac{M_\star}{10^9 {\rm M}_\odot} \right) + 6.51
\end{equation}
Bootstrapping errors on both parameters of this fit are $\sim$0.04 dex, while the scatter around the fit is $\sim$0.6 dex.  While the errors on estimated M/L ratios for the individual measurements are large and heterogeneous, they should be below $\sim$0.3 dex \citep{roediger15}, which implies that the scatter is primarily intrinsic.  

This NSC--stellar mass scaling relation is consistent with previous results \citep[][but also see \citeauthor{georgiev16} \citeyear{georgiev16} who find a steeper relation]{balcells03,scott13,denbrok14,sanchez-janssen19}, and suggests that roughly $M_{\rm NSC} \propto M_\star^{1/2}$.  This sub-linear trend implies that NSCs in lower-mass galaxies contain a higher fraction of the galaxy mass than in higher-mass galaxies.  At $10^9\,M_\odot$, this mass is about 0.3\% of the total galaxy mass, in agreement with the findings from \citet{cote06}.  The NSC-galaxy mass fraction is shown in the right panel of Fig.~\ref{fig:mass_scaling}, which shows clearly that the mass scaling is very similar for both early- and late-type galaxies.  We will discuss this result in context of formation models in Sect.~\ref{sec:formation}.   

There are hints in both panels that the high mass end of NSCs may behave somewhat differently, with a possible steepening of the NSC mass scaling slope.  For early types, this steepening has previously been noted by \citet{denbrok14} and by \citet{sanchez-janssen19}. At the same time, \citet{scott13} exclude these more massive objects as apparent nuclear stellar disks with larger effective radii rather than NSCs.  Despite the fact that, given our inclusive defintion of NSCs, these objects remain in our sample, we nonetheless maintain a shallow slope even at higher masses when including all objects in the fit.  On the other hand, if we restrict ourselves to the smaller subsample of objects with more accurate NSC mass measurements (plotted as stars), we do indeed find a steeper slope, as also found by \citet{georgiev16}. More specifically, fitting just these galaxies which primarily contain massive NSCs in late-type galaxies, we get a nearly linear relationship:
\begin{equation}
    \log \,M_{\rm NSC} = 0.92\, \log\, \left( \frac{M_\star}{10^9 {\rm M}_\odot} \right) + 6.13
\end{equation}
with bootstrapping uncertainties of $\sim$0.18 dex on both quantities.  This line is shown as a dot-dashed line in the left panel of Fig.~\ref{fig:mass_scaling}.  Hints of this linear mass scaling relationship are also seen in the full sample of galaxies in the right panel, where the early-type galaxies have a bump above the best-fit relation above $10^9\,M_\odot$, while the late-type galaxies show a modest flattening at the highest mass end.  We note that the lower mass fraction in the highest-mass early-type bin in the right panel is due to our inclusion of galaxies from the \citet{lauer05} sample, which have not been included in previous scaling relation work \citep[i.e.,][]{scott13,sanchez-janssen19}.  Surprisingly, these objects seem to follow the trend found for low-mass early-types, rather than continue the upturn found by \citet{sanchez-janssen19}.  

\subsection{Other correlations between NSCs and their hosts}

Throughout this review, we have already touched on several links between NSCs and their hosts (apart from their mass scaling relations discussed above) which we now summarize.  In Section~\ref{sec:demographics} we showed that there is a clear dependence of the nucleation fraction on galaxy stellar mass.  At low masses, this dependence seems to be independent of galaxy type, rising steadily up to  $\sim10^9\,M_\odot$ where the occupation fraction peaks.  
Moreover, the mass scaling relation discussed in the previous subsection also seems to depend on galaxy stellar mass but not on galaxy type.

The stellar populations of NSCs (Section~\ref{subsec:stelpop}) also are linked to their host galaxy masses, at least in early-type galaxies.  Specifically, at low masses, NSC metallicities have a wide range and are often lower than those of their hosts (Fig.~\ref{fig:mass_metallicity}), while at high masses, they are always higher.  A comparison of the integrated colors of early-type NSCs vs.\ their hosts shows a trend consistent with this change in metallicity: \citet{turner12} and \citet{sanchez-janssen19} find that their lowest mass galaxies have NSCs that are typically bluer than their hosts (suggesting the NSCs are more metal-poor than their hosts), while many higher mass galaxies have NSCs that are redder (consistent with higher NSC metallicities).  The metallicities of older stars in late-type galaxies are very difficult to constrain \citep{kacharov18}, and existing late-type spectroscopic measurements are all for galaxies with $M_\star > 10^9\,M_\odot$. Therefore, it remains unknown if a similar trend towards low-metallicity NSCs exists in late-type galaxies, although the colors of NSCs in these galaxies do seem to be typical of metal-poor GCs \citep{georgiev09}.

While the occupation fraction and mass scaling of NSCs in early-type and late-type galaxies are remarkably similar, there are some NSC properties that seem to differ between the two galaxy types.  Most obviously, spectroscopic studies suggest that the ages and metallicities of stars in NSCs are correlated with those of their host galaxies \citep[][]{rossa06,kacharov18}, and young $<$100~Myr populations are ubiquitous only in late-type spirals \citep{walcher06}.  The size-mass relation also appears to differ, with NSCs in early-type galaxies being larger at a given NSC mass (and galaxy stellar mass) than late-type NSCs \citep{georgiev16}.  This observation may be related to the young stars in these NSCs being centrally concentrated \citep{carson15}. Another possible consequence of the presence of young stars is the large ellipticity of some young clusters (Fig.~\ref{fig:ellipticity}), which may be due to disky distributions of young stars \citep{seth06,carson15}.  

We will return to these observations to help constrain formation models of NSCs in the next section.

\section{Scenarios for the formation and growth of nuclear star clusters}\label{sec:formation}
In this section, we will attempt to address the questions raised in the Introduction, namely  \emph{when} and \emph{how} NSCs form, and whether the answer depends on the properties of their host galaxy. We emphasize that the mechanisms for the initial NSC formation and their subsequent growth do not necessarily have to be the same.  

\subsection{Initial formation of NSCs}

A high-level question we would like to address first is whether NSCs initially form before, together with, or after their hosts. Since they lie deepest in the potential well of galaxies where gas densities are likely to be highest, seed NSCs could be among the earliest structures to form in a galaxy. In this scenario, one would expect that at least some NSC stars are both older and more metal-poor than the stars in the surrounding galaxy. 

In principle, it is easier to observationally constrain the metallicity of individual stars, rather than their age. Individual metal-poor stars have been identified at the center of the two nearest NSCs: the Sgr dSph NSC (M54) contains a dominant old metal-poor population \citep{mucciarelli17,alfaro-cuello19},  while in the Milky Way NSC, a small number of stars with [Fe/H]$< -1$ are found \citep{do15,ryde16,rich17}. In fact, the Milky Way is the only case where we can ascertain that such a population exists alongside a dominant population of younger and more metal-rich stars.  

In the general case of more distant NSCs, however, one has to rely on integrated light measurements, which makes it nearly impossible to constrain the metal-poor population \citep{seth10,kacharov18}.  The exception is when the metal-poor population dominates the light of the NSC, as it does in some low-mass early-type galaxies (see Fig.\,\ref{fig:mass_metallicity} and Sect.~\ref{subsec:stelpop}).
The fact that these NSCs are dominated by stars that are more metal poor than those of the stellar body of their host galaxy implies that NSC stars must have either formed in-situ \emph{before} the stars that surround them, or they formed elsewhere from less enriched gas, and then inspiraled to the nucleus (see next subsection for details). Such early formation of dense clusters in dark matter halos has a rich theoretical history, focused primarily on the goal of forming globular cluster-like objects \citep[see recent review by][]{renaud18}. One particularly interesting scenario that may be relevant for NSCs is the proposal by  \citet{cen01} that during reionization, shocks could compress gas in the inner regions of dark matter halos, thus forming clusters with masses as high as $10^6\,M_\odot$.  

Ongoing efforts to model globular cluster formation in cosmological simulations \citep[e.g.,][]{li19,ma19} have the potential to address the important question of how quickly galaxies acquire their NSCs, especially if they include the effects of dynamical friction. However, given the current lack of strong observational constraints and theoretical work on the initial formation of NSCs, we now turn our focus to understanding their subsequent growth.

\subsection{Globular cluster infall and merging}\label{subsec:inspiral}
Soon after the breakthrough observation of the M31 NSC by \cite{light74}, the first formation scenario for NSCs was offered by \cite{tremaine75}.  They suggested that the compact nuclear structures in M\,31 and similar galaxies will inevitably be formed when globular clusters are 'dragged' into the nucleus by dynamical friction created as they orbit through the stellar body of their host galaxy.  The strength of this dynamical friction depends linearly on the mass of the cluster, and thus the most massive clusters are the most likely to inspiral into and form the NSC.  The strength of the dynamical friction also depends inversely on the velocity ($\propto 1/v^2$) of the stars in the host galaxy.  
This scenario has received much attention over the years, both observationally and theoretically. 

On the observational side, it has been pointed out that the radial profiles of GC systems in early-type galaxies seem to indicate a deficit of massive GCs in the inner parts of many galaxies \citep{lotz01,capuzzo09}, potentially indicating that dynamical friction has driven them into the nucleus.  Additional evidence comes from the recent observation that in low-mass early-types, the nucleation fraction closely tracks the fraction of galaxies that host (non-nuclear) globular clusters \citep{sanchez-janssen19}.  In addition, previous work has shown that the specific frequency of globular clusters (the number of globular clusters per unit galaxy luminosity) is substantially higher in nucleated early-type dwarfs than in their non-nucleated counterparts \citep{miller07,lim18}.

An additional interesting finding is that in early-type galaxies, the total luminosity/mass in globular clusters and in NSCs seems to be nearly equal. This was first noted by \citet{cote06} who found that the typical NSC-to-galaxy luminosity ratio in their sample of Virgo early-type galaxies (0.3\%) was nearly identical to the GC system-to-galaxy mass ratio calculated by \citet{mclaughlin99}.  Taking this a step further, \citet{denbrok14} showed that the typical GC system luminosity derived from previous work is within a factor of $\sim$2 of the NSC luminosity for their large sample of Coma galaxies with masses between roughly $10^7$ and $10^9\,M_\odot$.  The persistence of this mass fraction similarity between GCs and NSCs over a wide range of galaxy masses is something existing infall models have yet to explain.  

The presence of metal-poor stars in low-mass early-type galaxy nuclei discussed in Sections~\ref{subsec:stelpop}, Fig.~\ref{fig:mass_metallicity}, and in the previous subsection may also support the dynamical friction scenario, as it provides a natural way to bring more metal-poor stellar populations to the center of the galaxy.  The massive star clusters that can inspiral to NSCs appear to be more efficiently produced at low metallicities \citep{pfeffer18}.  Therefore, NSCs more metal-poor than their surrounding hosts are likely an expectation of dynamical friction-driven infall models.  However, we caution that the surviving globular cluster population can be quite different from the population of clusters that formed and fell into the NSC (or were subsequently destroyed) \citep[e.g.,][]{lamers17}.  We also note here that although we use the term globular clusters in this discussion, the infalling clusters can also be younger massive clusters \citep[e.g.,][]{agarwal11}.  

A large number of simulations and semi-analytic models have quantified the efficiency of dynamical friction over a wide range of starting conditions for the GC systems and their host galaxies. The general picture that has emerged is the following. Provided that early star formation in the host galaxy produces a sufficient number of high-mass (\mstar$\gtrsim 10^5$\msol ) clusters, dynamical friction indeed provides a plausible mechanism to form an NSC on time scales much shorter than a Hubble time \citep[e.g.,][]{capuzzo93,oh00,lotz01,agarwal11,neumayer11}.  

Detailed N-body simulations of the dynamical friction process have shown consistency with some properties of observed NSCs.  For instance, \citet{hartmann11} found they could match the density and shape of some local, well-resolved NSCs in late-type galaxies with merging star cluster simulations, but that matching the kinematics of the clusters likely requires roughly half the NSC mass to stem from accreted gas and subsequent in-situ star formation. In a similar vein, \citet{antonini12} successfully reproduced the density profile of the Milky Way NSC from the accretion of globular clusters, but also concluded that the observed luminosity function requires younger stars and therefore suggested that only about half the NSC mass can originate from globular cluster inspirals. The rotation of NSCs was investigated by \citet{tsatsi17}, who found that their models could match the observed flattening \citep{schoedel14a}, and rotation signature \citep{feldmeier14} of the Milky Way NSC (Figs.~\ref{fig:nsc_rotation} and \ref{fig:v_over_sigma}).  

Another class of simulations are semi-analytic models that follow the evolution of GC systems with dynamical friction to track the growth of their NSCs. These models are able to produce some of the properties of both the present-day NSCs and the `left over' GC systems \citep{antonini13,gnedin14}.  \citet{antonini12} and \citet{antonini13} find a size-luminosity (or mass) relation of $r_{\rm eff} \propto M_{\rm NSC}^{0.5}$ for a small number of mergers, similar to the one derived by \citet{cote06} for early-type NSCs, but a little steeper than the size-mass relationship of $r_{\rm eff} \propto M_{\rm NSC}^{0.35}$ found by \citet{georgiev16}.   Fig.~\ref{fig:mass_radius} shows the size-mass relations for NSCs, with the upper dashed line indicating the power-law $r_{\rm eff} \propto M_{\rm NSC}^{0.5}$.  At the low mass end, it is clear that the observations indicate a shallower relation than this, while at the high-mass end, the relation seems to be a good match to the data. The simulations of  \citet{gnedin14} find that the fraction of galaxy mass contained within the NSC scales as $M_{\rm NSC}/M_\star \propto M_\star^{-0.5}$ or, alternatively, $M_{\rm NSC} \propto M_\star^{0.5}$.  In Sect.~\ref{subsec:scaling} and Fig.~\ref{fig:mass_scaling}, we find a very good match to this mass scaling prediction for the full sample of both early- and late-type galaxies. We note, however, that while the power-law slope matches, the normalization we find is about an order of magnitude lower than the 0.25\% of mass suggested by \citet{gnedin14} for galaxies with $M_\star = 10^{11}\,M_\odot$.  

One additional interesting body of work on dynamical friction-induced cluster inspiral relates to the Fornax dSph galaxy, which has five bright globular clusters, but is not nucleated. The galaxy is dark matter dominated, and \citet{goerdt06} find that if the galaxy possessed a NFW halo (with a central dark matter cusp), the globular clusters should sink to the center over a timescale between 1.5 and 5\,Gyr.  They interpret this as evidence for a flattened (cored) dark matter halo, which would lengthen the timescale for cluster inspiral. This issue remains an area of active research \citep[e.g.,][]{arca-sedda17, boldrini19, meadows19}, and highlights the possibility that the properties of NSCs and star cluster populations in the lowest mass galaxies may be able to contribute to our understanding of dark matter halos.  This is especially true if these studies can be expanded from a single galaxy to large samples of low-mass NSCs and GCs that span a wide range in galaxy mass.  

To summarize, the infall of massive stellar clusters into the galaxy nucleus is an expected (and perhaps unavoidable) process that has likely contributed a significant amount of stellar mass to present-day NSCs. However, the cluster infall scenario alone cannot easily explain the presence of young stars in many NSCs (see Sect.~\ref{subsec:stelpop}), which is why additional mechanisms are needed to fully describe NSC evolution.

\subsection{In-situ star formation}\label{subsec:insitu}
To explain the presence of young stellar populations in many NSCs in late-type hosts, \textit{in situ} star formation is usually invoked, which occurs when gas reaches the central few pc and triggers an intense burst of star formation. This mechanism is potentially repetitive, driven by the mechanical feedback from supernovae and strong stellar winds which `blows out' the gas and temporarily halts star formation until the turbulent energy has been dissipated, and the gas re-assembles in the nucleus on timescales of $\sim$10$^8$~years \citep{loose82}.  One complication in separating this scenario from the cluster inspiral scenario, is that young stars can also inspiral to the NSC if they form sufficiently close to the center \citep[e.g.,][]{gerhard01,agarwal11,neumayer11}.

In this section, we re-analyze the evidence for young stellar populations in NSCs from Section~\ref{subsec:stelpop}, with a focus on arguments for \textit{in situ} star formation.  The Milky Way NSC has a population of very young stars with ages between 3--8 Myr at radii $<$0.5~pc (see Section~\ref{subsec:MWNSC}).  The presence of such young stellar ages argues in favor of \textit{in situ} formation rather than the inspiral of a young cluster, because the latter should leave detectable remnants at larger radii which are not found \citep{paumard06,lu09,feldmeier-krause15}. A similar concentration of young stars is seen in many other nearby NSCs including M\,31  \citep{bender05,georgiev14,carson15}, as well as in some of the closest early-type galaxies \citep{nguyen17,nguyen19}.  These younger populations are often flattened and rotating, as expected if they formed from gas accreted into the nucleus \citep[][see also Sect.~\ref{subsec:kin}]{seth06,seth08b,carson15,nguyen19}.  
Integrated spectroscopy of NSCs in $>10^9\,M_\odot$ late-type spirals shows that they all contain stellar populations younger than $10^8$ years \citep{walcher06,seth06,kacharov18}.  The mass of these young populations is of order $10^5\,M_\odot$, sufficient to explain the total mass of the NSC if they had formed through similar formation episodes repeated every $\sim$100~Myr over a Hubble time \citep{walcher06,seth06}. The presence of emission lines also indicate the presence of very young stars ($<$10\,Myr) in some NSCs \citep{walcher06,seth08a}.  Overall, there is strong evidence that \textit{in situ} star formation occurs frequently in NSCs, and is likely the dominant growth mechanism for NSCs in late-type galaxies.

The starbursting blue compact dwarf Henize 2-10 is a special case, which we mention to demonstrate that young cluster accretion can occur in some cases. In this galaxy, no NSC is currently visible, despite evidence for a central BH \citep[][although see \citeauthor{hebbar19}\,\citeyear{hebbar19}]{reines11,reines16}.  However, there are several young and massive star clusters within the central $\sim$100\,pc that due to dynamical friction, will likely inspiral and merge to form an NSC on timescales of a few hundred Myr \citep{nguyen14,arca-sedda15}.  However, based on the evidence presented above, this case appears to be somewhat atypical, with truly centralized star formation being the more common mode of building NSCs. This triggers the question of how to transport the necessary amounts of gas from larger radii to the central few pc.

It has long been accepted that interactions between gas-rich galaxies, a.k.a.\ `wet mergers', will inevitably lead to a central gas concentration in the merger remnant \citep{mihos94}. The ensuing localized star formation will leave behind ``a sharp break in the surface density profile at a few percent of the effective radius'' which matches well the extent of NSCs in early-type hosts \citep{cote07}. More recent work by \citet{hopkins10,hopkins10b} simulates mergers at high resolution to determine the amount of star formation in the nucleus (and the black hole accretion), and finds that they can form objects similar to the asymmetric disk in M31.  However, the fact that NSCs are also found in most `pure' disk galaxies which have never experienced any substantive interactions argues that even without mergers, gas can reach the nuclear regions of galaxies. In fact, there are various secular mechanisms that cause gas to accumulate in the galaxy nucleus:
\begin{itemize}
    \item {\bf bar-driven gas infall:} gas responding to a non-axisymmetric potential (e.g.  produced by a stellar bar) will move inwards, and will ultimately reach the central few pc \citep[e.g.,][]{shlosman90}. This can occur both via regular flows in so-called `nuclear spirals' \citep[e.g.,][]{maciejewski04,kim17} or through angular momentum loss in star formation rings caused by dynamical resonances \citep[e.g.,][]{hunt08}.  One clear example of bar-driven gas infall is in the late-type spiral NGC\,6946, where \cite{schinnerer06,schinnerer07} find $1.6\times10^7$\msol\ of molecular gas within a radius of 30\,pc.  Star formation in this gas will likely lead to the formation (or growth) of a massive NSC. This suggests that NSC formation may still be ongoing in the local universe \citep{emsellem15}.
    
    \item {\bf dissipative nucleation:} a similar process occuring at high-redshift that involves clumpy star formation in nuclear spirals, and the ensuing infall and merging of such clumps, has been proposed by \cite{bekki06} and \cite{bekki07} to explain the nucleation of spheroidal galaxies that are now devoid of gas. They suggest that the low nucleation rate in the least massive galaxies (see Fig.\,\ref{fig:demographics}) is a natural consequence of the fact that feedback from stellar winds and supernovae will more easily expel the gas before it reaches the nucleus.
        
    \item {\bf tidal compression:} \citet{emsellem08} show that for galaxies without a pre-existing NSC and a reasonably flat density profile (i.e., with Sersic index $n\lesssim 3.5$), the tidal forces acting on the gas become compressive within about 0.1\% of the galaxy's effective radius, which typically corresponds to a few pc. They point out that NSC formation via this mechanism would naturally explain the observed scaling between the mass of the NSC and that of the host spheroid.  
    
    \item {\bf magneto-rotational instability:} \cite{milosavljevic04} suggests that a differentially rotating disk of neutral gas subjected to a weak magnetic field will develop the so-called magneto-rotational instability, which causes radial gas transport towards the nucleus. Using conditions typical for those in late-type spiral galaxies such as M\,33, they conclude that the expected gas inflow rates are sufficient to produce a $10^6$\msol\ NSC within about a Gyr.
    
\end{itemize}

Overall, it is clear that gas infall into the central few pc and the ensuing nuclear star formation are a natural and expected process, which almost certainly contributes to the growth of NSCs in gas-rich galaxies.

Numerically simulating \textit{in situ} NSC star formation in a galactic or even cosmological context is challenging, although some initial work has been done \citep{hopkins10,hopkins10b,guillard16,brown18}. In a more specific context, \citet{aharon15} have tried to simulate whether the Milky Way NSC can be built from successive generations of star formation, and show that they can potentially explain the varying density profiles of different stellar populations. Using a different approach, \citet{guillard16} perform galaxy-scale simulations to investigate a `hybrid' wet-merger formation scenario, which combines cluster infall with \textit{in situ} star formation in a gas-rich dwarf. If NSCs form at high redshift, they would inevitably be embedded in a gas-rich system. They conclude that star formation triggered by infalling gas-rich clusters dominates the mass of the resulting NSC, and considerably increases its density.  The NSCs built in this way would fall near the high-mass end of the size-mass relation.  

The work of \citet{brown18} is the only study to focus on \textit{in situ} star formation using cosmological simulations. While these simulations run only to $z \sim 1.5$, they examine the effect of varying star formation efficiencies on the size, flattening and rotation of the resulting cluster. They also examine the metallicity and abundances of stars formed, with the goal comparing it to the `stripped NSC' candidate $\omega$~Cen (discussed further in Section~\ref{sec:stripped}).  The extended star formation histories of these NSCs result in metallicity spreads of comparable to that in observed NSCs and the most massive globular clusters.

To summarize, there is strong evidence for \textit{in situ} star formation happening in NSCs, and theoretical work shows that there are many ways for gas to reach the inner few parsec, both through merger activity and secular processes.  However, setting up theoretical work suitable to compare to observations remains challenging.  Hopefully, with increasing computational power and resolution, NSC formation can be more directly studied in galaxy-scale and cosmological simulations.  

\subsection{NSC Formation: Different galaxies, different mechanisms}

The evidence reviewed above suggests that NSCs grow through both globular cluster inspiral and \textit{in situ} formation.  In this section, we tackle the question of which mechanism dominates the growth of NSCs in different types of hosts.

Apart from the aforementioned work of \citet{hartmann11} and \citet{antonini12} who argue that both formation mechanisms are required to replicate observations of specific NSCs, the only theoretical effort to tackle this problem has been the semi-analytic model of \citet{antonini15}.  This model includes prescriptions for both cluster infall and \textit{in situ} formation, as well as a model for how central black holes influence NSC growth.  They successfully predict the mass and high occupation of NSCs in galaxies at $\sim10^9\,M_\odot$, and also the decline in occupation fractions at the highest masses due to the influence of black holes on the NSCs.  Their model suggests that over most galaxy masses, $\sim$40\% of the stars form from \textit{in situ} formation for both early and late types except for a higher fraction in the very highest mass early types ($M_\star \gtrsim 10^{11}$~M$_\odot$).  Due to discreteness issues, their model fails to predict the drop in occupation fraction towards lower galaxies, (see Fig.~\ref{fig:demographics}).  In addition, the model predicts a broadening and  flattening in the galaxy mass -- NSC mass correlation at high galaxy masses due to the ejection of NSC stars in massive black hole binary mergers.  While a broadening does seem to occur at high masses (Fig.~\ref{fig:mass_scaling}),  if anything the mass scaling relation appears to steepen rather than flattening (see discussion in Section~\ref{subsec:scaling}).  

\vspace{0.2cm}

\noindent {\em Evidence for a transition at $M_\star \sim 10^9\,M_\odot$:}  We have presented several pieces of evidence in this review that there appears to be a change in the growth mechanism of NSCs at galaxy masses of $\sim10^9\,M_\odot$.  We propose that at lower masses, NSCs grow primarily by globular cluster accretion, while at high masses \textit{in situ} formation becomes dominant.  Based on Fig.~\ref{fig:mass_scaling} and Eq.~1, the NSC mass that corresponds to the transition at $M_\star \sim 10^9\,M_\odot$ is $M_{\rm NSC} \sim 3 \times 10^6\,M_\odot$.  This transition is likely stochastic -- i.e., some lower mass galaxies may experience significant growth through \textit{in situ} star formation.  This suggestion has been initially discussed by \citet{turner12} and \citet{sanchez-janssen19}, but the evidence collected here supports it more strongly.  The evidence for the scenario is strongest for early-type galaxies, but is consistent with the evidence in late-type galaxies as well.  We enumerate the evidence in support of the scenario below, and then consider the challenges.
\begin{enumerate}
\item The first piece of evidence comes from the spectroscopic metallicities of early-type galaxies shown in Fig.~\ref{fig:mass_metallicity}.  It is clear in this figure that below $10^9\,M_\odot$, about half of NSCs are more metal-poor than their surrounding hosts.  This provides strong evidence for the importance of globular cluster accretion in this mass regime.  At higher masses, most NSCs are more metal-rich than their NSCs, suggesting \textit{in situ} star formation from gas enriched within the galaxy \citep{brown18}.
\item There is strong evidence for \textit{in situ} star formation in late-type galaxies above $10^9\,M_\odot$ (Section~\ref{subsec:insitu}), however there is a lack of comparable data on lower-mass late-type galaxies.  If these lower mass galaxies also transition to having NSCs dominated by globular cluster inspiral, we would expect these NSCs to be old and metal-poor; this possibility is hinted at in the colors of NSCs in \citet{georgiev09}.
\item The occupation fraction of NSCs (Fig.~\ref{fig:demographics}) increases at low masses as $M_\star^{1/4}$, exactly tracking the occupation fraction of non-nuclear globular clusters in early-type galaxies up until $\sim10^9\,M_\odot$, where NSCs are nearly ubiquitous in all galaxies \citep{sanchez-janssen19}.  Note that the NSC occupation fraction of early-type galaxies quickly declines above this mass, probably due to disruption or prevention from central black holes \citep{bekki10,antonini15}.
\item The NSCs in the highest mass early-type galaxies are significantly flattened \citep[Fig.~\ref{fig:ellipticity};][]{spengler17}, similar to the disks of young stars seen in late-type NSCs \citep{seth06} suggesting that their NSCs are grown (in the past) through \textit{in situ} star formation.
\item At low masses, the scaling of NSC mass with galaxy mass is exactly in line with the predictions of globular cluster inspiral \citep[Section~\ref{subsec:inspiral};][]{gnedin14,sanchez-janssen19}.  The mass function of NSCs shows some sign of steepening at the highest mass end (Section~\ref{subsec:scaling}), as might be expected if a new source of growth (i.e., \textit{situ}) becomes important. 
\item There appears to be a break in the NSC size-mass relation (Fig.~\ref{fig:mass_radius}) at $M_{\rm NSC} \sim 3 \times 10^6\,M_\odot$, as might be expected if the growth mechanism was changing at this NSC mass.  
\end{enumerate}

While there is no strong evidence contradicting our proposal, there are several issues that may not be consistent with it.  First of all, related to the last point raised, in Sect.~\ref{subsec:inspiral}, we noted that the slope of the mass-radius at the low mass-end is basically flat, and thus does not agree with the predictions from cluster merging simulations despite the other evidence that cluster merging dominates NSC growth at these masses. Second, it provides no clear guide as to why the NSC occupation fraction (at least for lower-mass NSCs) appears to vary with environment (Section~\ref{sec:demographics}).  Third, there is a lack of data in low-mass late-type galaxies that could support a transition in the dominant formation mechanism; the evidence all comes from the early types.  We return to this lack of data and what further evidence could be collected to test this scenario in Section~\ref{sec:summary}.  

This proposed change of NSC growth mechanism at $\sim10^9\,M_\odot$ also brings up the question of what is causing this change.  If this change really occurs, some mechanism must prevent lower mass galaxies from effectively channeling or retaining gas in their nuclei.  This could therefore be related to a change in the ability to form bars or to stellar feedback preventing formation in the nucleus. 
As discussed further in the next section, we have strong evidence for central black holes only for galaxies above $\sim10^9\,M_\odot$, while below this mass there are almost no clear detections.  So it is possible that the change in NSC growth mechanisms at  $\sim10^9\,M_\odot$ is related to the presence of massive black holes.  However, it is also possible that black hole growth is driven by the same gas reaching the center that forms the NSC through \textit{in situ} formation.  

\section{Nuclear star clusters and massive black holes}\label{sec:nsc-mbh}

Nuclear star clusters dominate the centers of galaxies with stellar masses in the range $\sim10^8$--$10^{10}$\msol\, (see Fig.~\ref{fig:demographics}). Above this mass range, one invariably finds supermassive black holes (SMBHs) occupying the centers of massive galaxies \citep[see recent review by][]{kormendy13}. Interestingly, the masses of both SMBHs and NSCs follow a tight relation with the masses of their host galaxies \citep{wehner06,rossa06,ferrarese06}. This has led to speculations that NSCs take over the role of SMBHs in low mass galaxies, leading to the term of `central massive object' (CMO) to combine both incarnations of a seemingly similar product of galaxy evolution \citep{ferrarese06,wehner06,cote06}. 

However, there are a number of cases where nuclear star clusters and massive black holes are found to co-exist \citep{filippenko03,seth08a,graham09,neumayer12,nguyen19}. The best-studied example is our own galaxy, where a SMBH with a mass of $4 \times 10^6\,M_\odot$ resides in an NSC with $\sim 3 \times 10^7\,M_\odot$ \citep[][see also Sect.~\ref{subsec:MWNSC}]{genzel10,schoedel14b,feldmeier-krause17b}. A few more examples are well studied, and accurate masses for both the NSC and the BH are available. 
These findings indicate that NSCs are not an alternative incarnation of SMBHs, and refute the notion that NSCs replace massive black holes as the central massive object in low mass galaxies. Instead, they suggest that the build-up of NSCs and the growth of massive black holes are closely related. 

\subsection{Evidence for massive black holes within NSCs}

Most NSCs are readily observed in images of nearby galaxies (see Fig.~\ref{fig:NSC_demo}). The detection of supermassive black holes, on the other hand, is generally much harder because it involves indirect detection techniques such as the characterisation of radiation from the black hole accretion disk, or kinematic measurements and dynamical modelling of the motion of gas or stars in the immediate vicinity of the black hole. This task is challenging even for the nearest galactic nuclei, and becomes ever more difficult with increasing distance and decreasing black hole mass.

In the closest example, the Milky Way, it is interesting to note that the NSC had been known \citep{becklin68} before the SMBH was discovered \citep{balick74}. The combined mass of both entities was measured using gas \citep{lacy80,lacy82} and stellar kinematics \citep{rieke88,mcginn89}. However, it took more than three decades of hard work and vast technological improvement in observing facilities and instrumentation to achieve the high spatial resolution and wide-field coverage necessary to disentangle the relative masses of the SMBH and the NSC \citep[see][for a recent review]{genzel10,schoedel14b}. 

Other nearby examples for black hole detections within NSCs are M\,31 \citep{bacon94,bender05} and M\,32 \citep{verolme02,nguyen18}. In contrast, M\,33 hosts a well-studied NSC \citep{kormendy95,kormendy10}, but shows no sign of a massive back hole \citep{gebhardt01}. The absence of a black hole in M\,33 has led to the assumption that low-mass spiral galaxies without a bulge do not host massive black holes. This notion was turned upside down with the detection of an active black hole in the NSC of a bulgeless galaxy with a very similar stellar mass to M33, NGC\,4395 \citep{filippenko03}. The mass of the black hole in NGC\,4395 has been consistently measured to be $\sim 4\times 10^5$\,\msol\ using reverberation mapping as well as gas dynamical modelling \citep{peterson05,denbrok15}, which for a long time constituted the lowest-mass central black hole detection.\footnote{Note that a recent study by \cite{woo19} claims an even lower mass of $\sim10^4$\,\msol\ for the massive black hole in NGC\,4395.}

The advancement of high spatial resolution observing techniques such as adaptive optics \citep{davies12} has been fundamental in pushing the detection of massive black holes down to even lower masses. Low-mass black holes ($< 10^6$\msol) dominate the dynamics only within a very small volume (typically within $<1$pc), and one needs to spatially resolve this region to detect the effect of the black hole on its surroundings. In principle, the high stellar densities within NSCs are conducive to the dynamical detection of low-mass black holes, as they ensure an abundance of stellar orbits very close to the (putative) black hole, which in turn helps to reveal its dynamical signature (i.e., enhanced rotational signal and rise in the velocity dispersion). To separate the gravitational effect of the NSC from that of the black hole, accurate high resolution mass models for the NSC are essential. 
Combining stellar population analysis with dynamical modelling has thereby enabled black hole mass detections below $\sim 10^5$\msol\ in NSCs \citep{nguyen19}.  Recent work on a volume limited sample of five $10^9$ to $10^{10}\,M_\odot$ early-type galaxies has shown that all have evidence for central black holes \citep{nguyen17,nguyen18,nguyen19}, providing strong evidence that at this mass range where NSC occupation peaks, many co-exist with massive black holes.  

In addition to dynamical constraints, the accretion signal in active galactic nuclei reveals the presence of central massive black holes in NSCs \citep{filippenko03,seth08a}. Tailored surveys in the X-rays \citep{gallo10}, as well as studies of large samples in the optical \citep{reines13,baldassare18} and infrared \citep{satyapal08,satyapal09} have revealed a significant number of black holes in low-mass galaxies.  An analysis of the X-ray detections in early-type galaxies suggests a high fraction of galaxies with black holes in galaxies with masses as low as $\sim10^9\,M_\odot$ \citep{miller15} in agreement with the dynamical results discussed above.  In addition AGN studies show similar detection fractions in early and late type nucleated galaxies \citep{foord17}.  Accretion observations have also revealed evidence for BHs below $10^5\,M_\odot$ \citep{baldassare15,chilingarian18}.  Tidal disruption events have also started to uncover populations of lower-mass galaxies; half of hosts have stellar masses below $10^{10}\,M_\odot$ \citep{law-smith17}, and 75\%  have dispersions below 100~km/s \citep{wevers17}.  Comparison of the galaxy demographics for observed tidal disruption events has the promise of unveiling the occupation fraction of black holes in low mass galaxies \citep[e.g.,][]{stone16,vanvelzen18}. Currently, there is little evidence for central massive black holes in galaxies below $M_\star \sim 10^9\,M_\odot$, although a handful of candidates do exist \citep[e.g.,][]{maksym14,wevers19,reines19}.

Black holes of even lower masses (crossing into the realm of intermediate-mass BHs) have been claimed to exist at the centers of the likely stripped NSCs G1 \citep{gebhardt02}, $\omega$Cen \citep{noyola10}, and M\,54 \citep{ibata09,baumgardt17}. These dynamical detections are still debated \citep[e.g.,][]{vandermarel10,miller-jones12,tremou18,baumgardt19}, and further work at even higher spatial resolution is needed to reliably confirm or reject intermediate-mass black holes in these objects.  The recent review by \citet{greene19} provides an in-depth discussion of the reliability of globular cluster black hole measurements, as well as a table of low-mass black hole candidates and their overlap with NSCs.  

\subsection{Co-existence of NSCs and massive black holes}
As outlined in the last section, NSCs and massive black holes are found to coexist in many galaxies, especially in the host galaxy mass range $10^9-10^{10}$\msol. Towards higher host galaxy masses, NSCs are less common, and their demise in galaxies of masses above $\sim$10$^{11}$\msol\, (see Fig.~\ref{fig:demographics}) might be related to the presence and the dynamical interaction with the massive black hole \citep{antonini15}. 

\begin{figure}[h]
    \centering
    \includegraphics[width=0.45\textwidth]{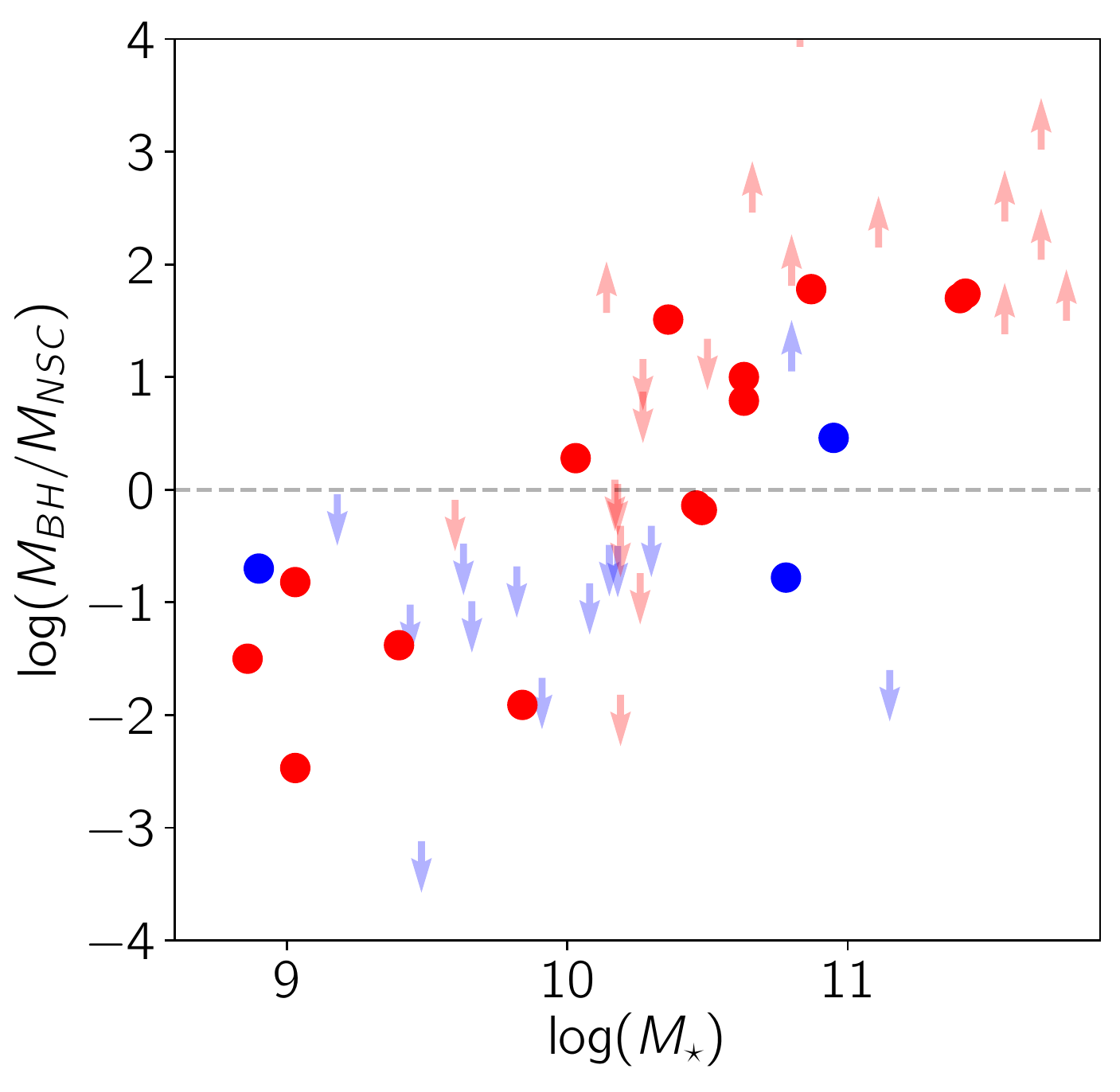}
    \includegraphics[width=0.45\textwidth]{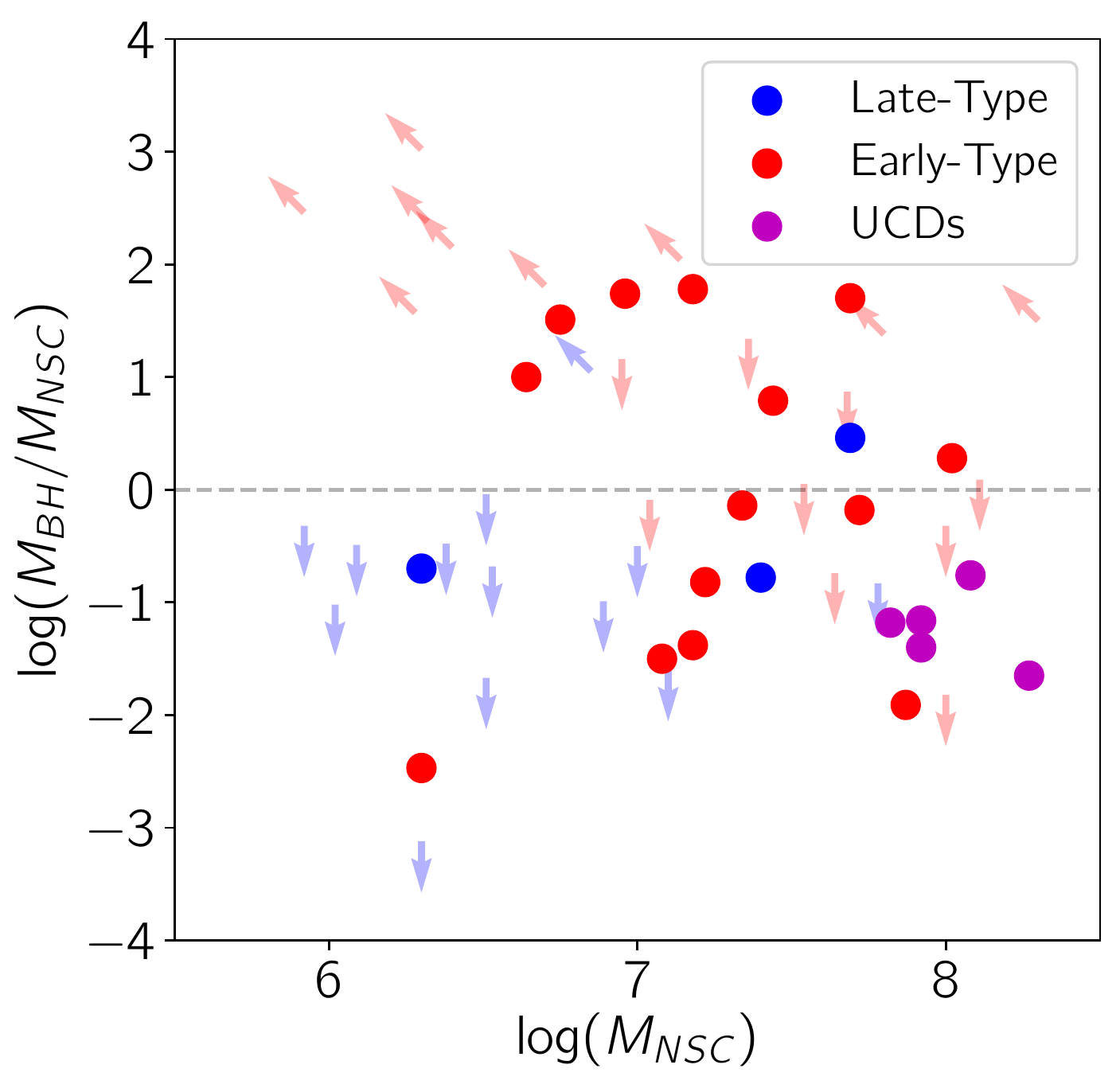}
    \caption{The ratio of the masses of the central massive black hole and the NSC plotted against their host galaxy stellar mass (left) and against the NSC dynamical mass (right).  The data plotted is compiled in Table~\ref{tab:bhnsc}.  The dashed line indicates equal mass NSCs and black holes in galaxies, while objects above the line have more massive black holes than NSCs.  Coloring breaks the galaxies into early-type (red), late-type (blue), and UCDs/stripped galaxy nuclei (purple; plotted only in the right hand plot. Upper \& lower limits result from objects with upper limits on black hole masses and NSC masses respectively.  This plot builds on compilations presented in several previous works \citep{seth08a,graham09,neumayer12,georgiev16,nguyen18}. }
    \label{fig:mbh-mnsc}
\end{figure}

If the growth of the massive black hole and the NSC are indeed interrelated, their mass ratio might be an interesting metric to study. Several authors have investigated this quantity \citep{graham09,seth08a,neumayer12,georgiev16,nguyen18}, and we present an extensive compilation in Fig.~\ref{fig:mbh-mnsc} and Table~\ref{tab:bhnsc}, which suggests that the mass fraction $M_{\rm BH}/M_{\rm NSC}$ increases with host galaxy mass. However, the scatter in this ratio at a given galaxy mass is still up to three orders of magnitude. This likely reflects the fact that the growth history of both NSC and BH are very stochastic. 

The host galaxy mass at which BHs start to dominate (the ratio $M_{\rm BH}/M_{\rm NSC}$ becomes $>1$), is a few times $10^{10}$\msol, corresponding to a host galaxy velocity dispersion of $>100$km/s. 
\cite{nayakshin09} find that the changeover from NSCs to SMBHs occurs when the dynamical time is about equal to the Salpeter time, i.e., the time over which the black hole mass doubles. In other words, low-dispersion galaxies with dynamical times shorter than the Salpeter time cannot grow their central massive black holes quickly enough to affect the gas infall, and this gas accumulates at the center and will form NSCs. These galaxies might still host massive black holes, but they would be underweight with respect to their host galaxy mass.
An alternative approach to explaining the dividing line between NSC- and SMBH-dominated galactic nuclei was given by \cite{stone17} who show that the low-mass end of the SMBH mass function can be explained by collisions of stellar mass black holes and subsequent runaway tidal capture and tidal disruption events in NSCs. We will discuss the formation channel of massive black holes in the next section.

\subsection{Massive black hole formation and growth within NSCs}

The formation of SMBHs remains a mystery, and is therefore a topic of vast current interest.  Several recent reviews summarise the different proposed formation channels \citep{volonteri10,volonteri12,mezcua17,greene19,inayoshi19}. Here, we focus on one of the proposed channels, namely the formation and runaway collisions of black holes within dense (nuclear) star clusters. The idea for the formation of an (intermediate mass) black hole within a dense stellar cluster has been discussed for a long time \citep{begelman78,quinlan87,ebisuzaki01,miller02,portegies-zwart02,portegies-zwart04,gurkan04,freitag06a,freitag06b}.
The most massive and dense NSCs have high enough velocity dispersions that binary heating cannot prevent core collapse, and this can result in the formation of massive black holes \citep{miller12}.  However, lower mass NSCs (typically in lower mass galaxies) fall below this threshold.

\cite{antonini19} point out that in a star cluster with a sufficiently large escape velocity, black holes that are produced by black hole mergers can be retained, merge again and grow to larger masses. 
This means that runaway collisions are favoured in the most massive clusters, as they can retain black holes against the gravitational recoil kicks produced during the merger of two black holes \citep{holley-bockelmann08}.
While NSCs are the densest stellar systems known in the local universe, even denser clusters may have existed at high redshift, with central subclusters driven to even higher densities through gas dissipational processes \citep{davies11,leigh14}. These would be even more susceptible to the early formation of massive black holes.

The subsequent growth of the massive black hole can be sustained through different channels. 
As mentioned in Sect.~\ref{subsec:stelpop} and \ref{subsec:insitu}, many NSCs experience periodic bursts of star formation \citep{walcher06,carson15,kacharov18}. It is likely that the same gas that is feeding star formation also feeds the black hole. In that sense, the presence of an NSC may, in fact, enhance the growth rate of the central black hole \citep{naiman15}. This notion is supported by the simulations of \cite{hopkins10b} who find that nuclear star formation is more tightly coupled to the growth of central black holes than the global star formation rate. 

As discussed in the previous subsection, another growth channel for the central massive black hole is via the tidal disruption and capture of stars coming close to the central black hole \citep[e.g.,][]{strubbe09}.  This is especially true for lower mass black holes ($<2 \times 10^6$~M$_\odot$), which may acquire the majority of mass through tidal capture \citep{milosavljevic06}.  However, in the lowest mass galaxies (below $\sigma \sim 35$~km/s), NSCs are not dense enough to fuel black hole growth by runaway tidal capture of NSC stars \citep{miller12, stone17}.  This may be related to the lack of massive central black holes in the lowest mass galaxies.  

\begin{table}
\caption[]{Data and References for Figure 13}
\begin{tabular}[t]{l c c c c c c c} \hline \hline 
(1) & (2) & (3) & (4) & (5) & (6) & (7) & (8) \\ 
Galaxy & Type & log($M_{\rm BH}$) & Ref. & log($M_{\rm NSC})$ & Ref. & log($M_{\rm gal})$  & Ref.  \\
\hline \hline
\multicolumn{8}{c}{Objects with NSCs and BH masses} \\
\hline \hline
Milky Way & late & 6.62 & 1,2 & 7.40 & 34 & 10.78 & 42 \\
M\,31 & late & 8.15 & 3 & 7.69 & 35 & 10.95 & 43 \\
NGC\,205 & early & 3.83 & 4 & 6.30 & 5 & 9.03 & 5 \\
M\,32 & early & 6.40 & 5 & 7.22 & 5 & 9.03 & 5 \\
NGC\,404 & early & 5.58 & 6 & 7.08 & 36 & 8.86 & 44 \\
NGC\,5102 & early & 5.96 & 4 &  7.87 & 5 & 9.84 & 5 \\
NGC\,5206 & early & 5.80 & 4 &  7.18 & 5 &  9.40 & 5 \\
NGC\,4395 & late & 5.60 & 7,8 & 6.30 & 8 & 8.90 & 45 \\
M\,60-UCD1 & UCD & 7.32 & 9 &  8.08 & 9 & - & - \\
M\,59cO & UCD & 6.76 & 10 &  7.92 &  10 & - & - \\
VUCD3 & UCD & 6.64 & 10 &  7.82 &  10 & - & - \\
M\,59-UCD3 & UCD & 6.62 & 11 &  8.27 &  11 & - & - \\
FUCD3 & UCD & 6.52 & 12 &  7.92 &  12 & - & - \\
NGC\,4578 & early & 7.54 & 13 &  7.72 & 37 &  10.48 & 46 \\
NGC\,1023 & early & 7.64 & 14 &  6.64 & 38 &  10.63 & 45 \\
NGC\,3115 & early & 8.96 & 15 &  7.18 & 38 &  10.87 & 38 \\
NGC\,3384 & early & 7.20 & 16 &  7.34 & 38 &  10.46 & 45 \\
NGC\,4026 & early & 8.26 & 17 &  6.75 & 38 &  10.36 & 18 \\
NGC\,4697 & early & 8.23 & 16 &  7.44 & 38 &  10.63 & 45 \\
IC\,1459 & early & 9.39 & 18 &  7.69 & 39 &  11.40 & 18 \\
NGC\,4552 & early & 8.70 & 16 &  6.96 & 39 &  11.42 & 18 \\
NGC\,4558/VCC1146 & early & 8.30 & 19 &  8.02 & 40 &  10.03 & 46 \\
\hline \hline
\multicolumn{8}{c}{Objects with NSC masses and BH upper limits} \\
\hline \hline

M\,33 & late & $<3.18$ & 20 & 6.30 & 38 & 9.48 & 47  \\
NGC\,4474/VCC\,1242 & early & $<6.18$ & 13 & 8.0 & 40 & 10.19 & 46   \\
NGC\,4551/VCC\,1630 & early & $<6.90$ & 13 & 7.64 & 40 &  10.26 & 46   \\
NGC\,300 & late & $<5.00$ & 21 & 6.02 & 41 & 9.44 & 5   \\
NGC\,428 & late & $<4.84$ & 21 & 6.51 & 41 & 9.91 & 5   \\
NGC\,1042 & late & $<6.47$ & 21 & 6.51 & 41 & 9.18 & 5 \\
NGC\,1493 & late & $<5.90$ & 21 & 6.38 & 41 & 9.63 & 5   \\
NGC\,2139 & late & $<5.60$ & 21 & 5.92 & 41 & 10.30 & 5  \\
NGC\,3423 & late & $<5.85$ & 21 & 6.53 & 41 & 9.82 & 5   \\
NGC\,7418 & late & $<6.95$ & 21 & 7.78 & 41 & 10.08 & 5   \\
NGC\,7424 & late & $<5.60$ & 21 & 6.09 & 41 & 10.15 & 5   \\
NGC\,7793 & late & $<5.90$ & 21 & 6.89 & 41 & 9.66 & 5   \\
VCC\,1254 & early & $<6.95$ & 22 & 7.04 & 38 & 9.60 & 38   \\
NGC\,3621 & late & $<6.50$ & 23 & 7.00 & 23 & 10.18 & 48  \\
IC\,342 & late & $<5.50$ & 24 & 7.1 & 24 & 11.15 & 49 \\
NGC\,2778 & early & $<8.70$ & 19 & 7.36 & 39 & 10.50 & 46   \\
NGC\,4379/VCC\,784 & early &  $<8.55$ & 19 & 7.68 & 40 & 10.27 & 46   \\
NGC\,4387/VCC\,828 & early & $<7.59$ & 19 & 7.54 & 40 & 10.18 & 46   \\
NGC\,4474/VCC\,1242 & early & $<7.68$ & 19 & 8.00 & 40 & 10.19 & 46   \\
NGC\,4612/VCC\,1883 & early & $<8.11$ & 19 & 6.95 & 40 & 10.27 & 46   \\
NGC\,4623/VCC\,1913 & early & $<8.20$ & 19 & 8.11 & 37 & 10.17 & 46    \\

\hline \hline
\multicolumn{8}{c}{Objects with BH masses and NSC upper limits} \\
\hline \hline

NGC\,4486 & early & 9.80 & 25 & $<8.30$ & 21 & 11.78 & 21  \\
NGC\,4374 & early & 9.18 & 26 & $<7.80$ & 21 & 11.56 & 21   \\
NGC\,1332 & early & 9.18 & 27 & $<7.14$ & 21 & 11.69 & 5  \\
NGC\,3031 & late & 7.90 & 28 & $<6.85$ & 21 & 10.80 & 50  \\
NGC\,4261 & early & 8.70 & 29 & $<6.32$ & 21 & 11.56 & 21   \\
NGC\,4649 & early & 9.32 & 16 & $<6.30$ & 21 & 11.69 & 21   \\
NGC\,3998 & early & 8.38 & 30 & $<5.92$ & 21 & 10.66 & 51   \\
NGC\,2787 & early & 7.85 & 31 & $<6.28$ & 21 & 10.14 & 51  \\
NGC\,3379 & early & 8.08 & 32 & $<4.15$ & 21 & 10.83 & 21  \\
NGC\,4342 & early & 8.55 & 33 & $<6.40$ & 21 & 11.11 & 21   \\
NGC\,4291 & early & 8.51 & 16 & $<6.70$ & 21 & 10.80 & 21  \\

\hline \hline
\end{tabular}

\small
{
\textsc{Notes} - (1),(2): Galaxy name and morphological type. (3),(4): logarithm of black hole mass \& assoc. literature reference. (5),(6): logarithm of NSC mass \& reference. (7),(8): logarithm of galaxy stellar mass \& reference.  

\textsc{References} - (1) \cite{abuter19}, (2) \cite{do19}, (3) \cite{bender05}, (4) \cite{nguyen19}, (5) \cite{nguyen18}, (6) Davis et al. in prep., (7) \cite{peterson05}, (8) \cite{denbrok15}, (9) \cite{seth14}, (10) \cite{ahn17}, (11) \cite{ahn18}, (12) \cite{afanasiev18}, (13) \cite{krajnovic18}, (14) \cite{bower01}, (15) \cite{emsellem99}, (16) \cite{gebhardt03}, (17) \cite{gueltekin09}, (18) \cite{saglia16}, (19) \cite{pechetti17}, (20) \cite{gebhardt01}, (21) \cite{neumayer12}, (22) \cite{geha02}, (23) \cite{barth09}, (24) \cite{boker99}, (25) \cite{gebhardt11}, (26) \cite{bower98}, (27) \cite{rusli11}, (28) \cite{devereux03}, (29) \cite{ferrarese96}, (30) \cite{defrancesco06}, (31) \cite{sarzi01}, (32) \cite{shapiro06}, (33) \cite{cretton99}, (34) \cite{schoedel14a}, (35) \cite{peng02}, (36) \cite{nguyen17}, (37) \cite{cote06}, (38) \cite{graham09}, (39) \cite{lauer05}, (40) \cite{spengler17}, (41) \cite{walcher05}, (42) \cite{bland-hawthorn16}, (43) \cite{williams17}, (44) \cite{seth10}, (45) \cite{reines15}, (46) \cite{cappellari13}, (47) \cite{mcconnachie12}, (48) derived from \cite{jarrett03}, (49) \cite{karachentsev13}, (50) \cite{querejeta15}, (51) derived from \cite{jarrett00}

\label{tab:bhnsc} 
}
\end{table}


\section{Stripped down: Connecting nuclear star clusters, globular clusters, and ultra-compact dwarf galaxies}\label{sec:stripped}

Given the hierarchical nature of galaxy formation, smaller galaxies often fall into larger galaxies, and are tidally disrupted in the process.  The gravitationally self-bound NSCs in these galaxies are likely to survive this disruption process \citep{bassino94,bekki03,pfeffer13} and will continue their lives orbiting in the halo of the larger galaxy.  This process is currently ongoing in the Sgr dSph galaxy, whose NSC was first discovered as the globular cluster M\,54.  In this section we discuss the evidence for stripped NSCs, both outside our Milky Way and hidden amongst the Milky Way's globular cluster system.  These stripped NSCs can be important for: (1) understanding the formation and evolution of NSCs, (2) preserving a record of a galaxy's accretion history, and (3) obtaining a census of massive black holes.  

\subsection{Extragalactic evidence for stripped NSCs: Ultracompact dwarfs}

Outside the Milky Way, evidence for the survival of nuclei after tidal disruption is found in the existence of ultracompact dwarfs (UCDs), a class of objects that constitutes the brightest compact stellar systems \citep[][]{hilker99,drinkwater00}.  In the size-mass plane, no clear boundary exists between ``ordinary'' massive globular clusters and UCDs, while NSCs and UCDs appear in overlapping regions \citep[see Fig.~\ref{fig:mass_radius};][]{norris14}. Identifying which UCDs are stripped NSCs is therefore a complicated task, a difficulty that is enhanced by our current lack of understanding of the globular cluster formation process. Stripped NSCs have been identified through the presence of (1) an extended SFH, (2) a massive black hole, or (3) a significant halo or stream of tidal material.  None of these signatures are expected for ordinary globular clusters\footnote{Throughout this section we will refer to ordinary globular clusters as massive clusters that were not once NSCs}. However, not all stripped NSCs are expected to have these signatures, and obtaining a census of all stripped NSCs therefore is challenging. In what follows, we summarize the observational evidence for (some) UCDs being stripped NSCs.  

The extended SFH of many NSCs has been discussed in Section~\ref{subsec:stelpop}.  For tidally stripped systems, in addition to the extended SFH in M\,54, a clearly extended SFH has been found in NGC4546-UCD1 \citep{norris15}.  In several other UCDs, a younger age or higher metallicity than typical GCs is inferred from single stellar population modeling, suggesting that they also are stripped NSCs \citep{janz16}.  

As discussed in Sect.~\ref{sec:nsc-mbh}, BHs with masses that are a significant fraction of their NSC mass are common in massive NSCs.  In UCDs, detection of these black holes can provide strong evidence for stripping.  \citet{mieske13} proposed that the inflated dynamical mass estimates seen in many UCDs could be explained with black holes of order 10\% of the total UCD mass.  Adaptive optics observations have now found evidence for BHs with mass fractions of 2-18\% in all five UCDs searched with luminosities above 10$^7$~L$_\odot$ \citep{seth14,ahn17,ahn18,afanasiev18}. By applying the well-known scaling relations between BH and host galaxy, the BH masses can provide additional insight into the UCD progenitor galaxies.  On the other hand, in less luminous systems, BHs have been more elusive \citep[e.g.,][]{gebhardt05,voggel18}, despite the dynamical BH detections in some lower-luminosity NSCs \citep{nguyen18,nguyen19}.  
With the evidence of inflated dynamical mass estimates being an indicator of BHs, the fraction of UCDs with dynamical evidence for the presence of a massive BH from integrated spectroscopy has recently been reanalyzed, showing a rise from $\sim$20\% at $10^6\,L_\odot$ to $\sim$70\% at the highest luminosities \citep{voggel19}. 

Simulations by \citet{pfeffer13} suggest that NSC properties are mostly unchanged during tidal stripping, but that a remnant halo (or stream) of stars from the stripped galaxy may surround these NSCs for an extended period of time. Such halos have indeed been detected in a number of cases, ranging from sources embedded in tidal streams \citep{jennings15,voggel16} to extended halos and multiple component fits being required to model the surface brightness profiles of UCDs \citep[e.g.,][]{martini04,evstigneeva08,chilingarian08, strader13, ahn17}.

\subsection{Abundance of stripped NSCs}

Overall, there are multiple lines of evidence suggesting that many massive UCDs are in fact stripped NSCs.  Two related questions to this are (1) what fraction of the UCD population are stripped NSCs, and (2) what is the relative number of ``present-day'' NSCs (i.e., that still occupy their host nucleus) to the number of 'former' NSCs from stripped galaxies. Tackling these questions from the theoretical side using semi-analytic methods based on the Millenium-II simulations, \citet{pfeffer14,pfeffer16} suggest that stripped NSCs can account for all UCDs observed in the Virgo and Fornax cluster above a mass of $\sim$2$\times$10$^7$~M$_\odot$, while only a small fraction of lower mass UCDs are likely to be stripped NSCs.  However, accurately tracking the fate of infalling subhalos within N-body simulations has shown to be challenging \citep[e.g.,][]{vandenbosch18}, suggesting there is significant uncertainty in predicting which satellites are disrupted and which are not.   

From the observational side, \citet{ferrarese16} show that, assuming all UCDs in Virgo are stripped NSCs, the number of stripped NSCs outnumbers the number of present-day NSCs by a factor of $\sim$3.  Taking into account only those UCDs with dynamical evidence for massive black holes, \citet{voggel19} estimates the ratio of stripped NSCs to present-day NSCs in nearby galaxy clusters is $\sim$1. Furthermore, they show that this ratio depends on the luminosity of the NSCs/UCDs (and therefore the stellar masses of the host galaxies), in the sense that stripped NSCs are more abundant at lower luminosity.  Given that some stripped NSCs may lack massive black holes, the true ratio of stripped NSCs to present-day NSCs in clusters likely falls between one and three.

The ratio of stripped to present-day NSCs may also be dependent on the galaxy environment, with the Milky Way providing a useful testbed for predictions in lower mass halos.  \citet{pfeffer14} suggest that 1--3 stripped NSCs may be hidden amongst the globular cluster population.  Using a different approach to simulate the formation of massive clusters based on a suite of ``zoomed-in'' SPH simulations, \citet{kruijssen19} suggest $\sim$6 stripped NSCs in the Milky Way.  We examine the observational evidence for stripped NSCs in the Milky Way in the next subsection.   

\subsection{Stripped NSCs amongst Milky Way globular clusters}

The brightest, most massive globular clusters in the Milky Way, $\omega$\,Cen has long been suspected to be a former NSC \citep[e.g.,][]{bekki03b}, due to its large spread in metallicity \citep[e.g.,][]{norris96,johnson10}, and possibly also an age spread of several Gyr \citep{hilker04,villanova14}.  However, this age spread is difficult to interpret, due to the fact that the abundance spreads seen in many stars can also affect their age estimates \citep[e.g.,][]{marino12}.  However, $\omega$\,Cen's status as a stripped NSC has recently been significantly strengthened through the identification of material that may be associated with its progenitor galaxy \citep{majewski12, ibata19}.  As discussed in Section~\ref{subsec:stelpop}, M54, the NSC of the Sgr dwarf galaxy, shares many similarities with $\omega$\,Cen, suggesting there are at least two very strong candidates for (partially) stripped NSCs amongst the Milky Way's globular clusters.

Moreover, the complex stellar population of $\omega$\,Cen and M54 are becoming less and less unique as the Milky Way's GC system is being studied in more detail, with metallicity spreads now found in at least nine clusters \citep{marino15,dacosta16}.  An overlapping set of clusters show color-magnitude and color-color diagrams similar to that of $\omega$~Cen, also suggesting metallicity spreads \citep[Type II clusters in][]{milone17}.  If these globular clusters are indeed all stripped NSCs, it would suggest $\sim$10 stripped NSCs, exceeding the model predictions of \citet{pfeffer14} and \citet{kruijssen19}.   

Despite the wealth of information available for Milky Way globular clusters, identifying these objects as stripped NSCs remains challenging.  This is at least in part due to the range of likely  formation mechanisms for NSCs.  It is feasible to produce a stripped NSC with no spread in age or metallicity (e.g. a single cluster inspiraling to the nucleus or single star formation event), a spread in metallicity, but not age (e.g. dynamical friction inspiral of two ancient globular clusters with different metallicities), or a spread in both (e.g. inspiral of globular cluster(s) followed by \textit{in situ} star formation events).  Note that only in the latter case do the stellar populations of a stripped NSC provide clear evidence of its origin, and thus an improved understanding of age spreads in globular clusters (including their complex chemical variations) seems like the next step required to obtain at least a lower limit on the number of stripped NSCs in the Milky Way.

\section{The unique astrophysics of very dense stellar systems}\label{sec:physics}
 
The extreme stellar densities of NSCs ($>10^6\,M_\odot/{\rm pc}^3$; Sect.~\ref{subsec:sizemass}), and their coincidence with massive black holes (Section~\ref{sec:nsc-mbh}) make NSCs an interesting laboratory.  In the following section, we briefly highlight some of the work done on the unique dynamics of NSCs, as well as the consequences of these dynamics for unusual stellar populations, tidal disruption events, and compact binary systems that may emit gravitational waves.

\subsection{Stellar dynamical evolution of NSCs}
 
Due to their higher masses and larger sizes, the dynamical evolution of NSCs is slower than the evolution of typical globular clusters.  Half-mass relaxation times are many gigayears, with the larger, more massive NSCs like the one in the Milky Way having $>$10~Gyr relaxation time at the half mass radius \citep[e.g.,][]{merritt09,panamarev19}.  The unique location of NSCs at the centers of galaxies means that in addition to their internal dynamical evolution, they can be dynamically heated by both their surrounding galaxy and the massive black holes at their centers.  
\cite{merritt09} suggests two regimes exist: (1) lower mass galaxies have compact enough clusters to core-collapse despite the energy input from the surrounding galaxies, while (2) in higher mass galaxies little dynamical evolution seems to be occuring due to their long relaxation times.  Dynamical heating from binaries can also be important in NSCs that lack massive black holes \citep[e.g.,][]{antonini19}.  

The presence of a central massive black hole within an NSC should lead to the formation of a cusp surrounding the massive black hole with a density slope of $\rho \propto r^{-\gamma}$ with $\gamma = 7/4$ in single-mass models \citep{bahcall76}.  The presence or absence of this Bahcall-Wolf cusp in NSCs, especially the Milky Way NSC, is debated in both the theoretical \citep[e.g.,][]{merritt10,vasiliev17} and observational literature \citep[e.g.,][]{buchholz09,schoedel14a,gallego-cano18}.  The current theoretical literature \citep{vasiliev17,baumgardt18,panamarev19} suggests that the mass-segregation of stellar-mass black holes towards the center leads to a steep cusp ($\gamma > 1.5$) in their distribution, while the stellar distribution remains somewhat flatter. This seems to be consistent with more recent observations that include fainter stars and unresolved light at the center \citep{gallego-cano18,schoedel18}.

\subsection{Nuclear star formation and stellar collisions}

As discussed in Section~\ref{sec:properties}, the NSCs in the Milky Way, M\,31, and many other galaxies seem to have concentrated young stellar populations at their centers. The presence of young stars is challenging to explain, as infalling young star clusters would disrupt at larger radii than they are observed \citep[e.g.,][]{gerhard01,fujii09}.  This suggests that \textit{in situ} star formation is taking place, despite the disruptive tidal shear forces present at these small radii \citep[e.g.,][]{levin03,mapelli12}.   Another possibility for the formation of the apparently young ($\sim$200 Myr) stars in M31 is that they are collision-induced blue stragglers that form due to the high stellar density \citep{demarque07,leigh16}.

In addition to the presence of a young stellar component very close to the SMBH, another puzzling phenomenon is the apparent absence of red giant branch (RGB) stars near the center of the Milky Way \citep[e.g.,][]{do09}.  This has been hypothesized to be a result of stellar collisions \citep{davies98,dale09}, although it is unclear whether the stellar/black hole density is high enough to effectively destroy the stars \citep{leigh16}.  Other possible explanations are that the lack of RGB stars was caused by the destruction of these stars as they passed through a star-forming disk such as the one that created the young stars \citep{amaro-seoane14,kieffer16,amaro-seoane19}, or that the central region of the Milky Way NSC has not yet had enough time to form a cusp \citep{merritt10,antonini14}.  

\subsection{Tidal disruption events and SMBH growth}
Tidal disruption events (TDEs) are a natural consequence of having a dense NSC surrounding a massive black hole.  A TDE occurs when a star approaches the SMBH to within less than its tidal radius\footnote{approximately given by $r_t= (M/m)^{1/3}\times r$ , where $M$ is the mass of the SMBH, m is the mass of the star, and r is the stellar radius.}. The star is then pulled apart by the tidal forces imparted by the BH; part of its material is ejected away, while part is accreted onto the SMBH \citep[e.g.,][]{rees88}.

The observation of TDEs is a rapidly growing field with detections at optical, UV and X-ray wavelengths \citep[see][for a recent compilation of events]{auchettl17}.  These TDEs are occuring in relatively low-mass galaxies, with nearly half of them having stellar masses between $10^{9}$ and $10^{10}\,M_\odot$ \citep{wevers19}, a mass range in which NSCs are ubiquitous.  An important aspect of understanding the observed targets are predictions of TDE rates in galaxies, which strongly depend on the density of stars within the sphere of influence of the black hole \citep[e.g.,][]{stone16}.  Therefore, the density profiles of NSCs are important for understanding the demographics of TDEs and the underlying black hole population in galaxies.  The formation mechanism of NSCs may also affect TDE rates, as merging of clusters or infall of gas could enhance TDE rates \citep{aharon16, lezhnin16, arca-sedda17}.  Furthermore, as noted in Section~\ref{sec:nsc-mbh}, the NSC can be an important source of mass growth for black holes over time \citep{stone17}.  A better understanding of NSC formation and density profiles is needed to quantify the TDE-induced BH growth throughout cosmic history, and to  constrain the expected TDE detection rates for future time resolved survey telescopes such as the Large Synoptic Survey Telescope (LSST).
\subsection{NSCs as gravitational-wave emitters}

The recent detection of gravitational wave (GW) emission from merging stellar mass black holes \citep{abbott16,abbott19} has highlighted the role of galactic nuclei as potential sources for GW emission. Given their high stellar densities and escape velocities, NSCs are prime candidates for hosting dynamically created compact object mergers that can result in detectable LIGO sources.  Stellar mass BH binary mergers appear to be more efficiently produced in more massive clusters and in younger clusters, thus favoring production in NSCs \citep[e.g.,][]{askar18}.  The high mass of NSCs means there is a higher expected retention of stellar mass BHs after supernova kicks \citep{miller09}.  Furthermore, stellar mass BHs also can experience successive mergers in NSCs, since a merged BH can be retained despite gravitational wave recoil during the merger \citep[e.g.,][]{antonini19}.  In NSCs lacking a central massive BH, stellar mass BHs will mass segregate to the center of the cluster and then can form binaries and harden through three-body interactions  \citep[e.g.,][]{antonini16,choksi19}.  The presence of a massive BH also enables stellar mass BH binaries to harden through Kozai--Lidov interactions with the massive BH \citep[e.g.,][]{antonini12b,zhang19,bub19}, or via gas torques and/or hardening encounters with single stars in AGN disks \citep{mckernan18,leigh18}. The presence of gas can even facilitate the formation of compact object mergers via dissipative effects \citep{secunda19}.
In addition to stellar mass BH binary mergers, NSCs can also be a possible site of triple mergers of BHs and neutron stars \citep{fragione19}.

Lower-frequency GW events will be detected in the future with LISA, and many of the events expected to be detected will be in NSCs \citep{amaro-seoane17}.  For instance, extreme mass-ratio inspirals (EMRIs) occur when a compact stellar remnant (BH or neutron star) falls into a massive BH \citep[e.g.,][]{hopman06,merritt11}.  Similar to the TDE rates, the formation scenarios of NSCs and their subsequent dynamical evolution can significantly impact EMRI rates \citep{aharon16b}.  Current rate estimates suggest that LISA should be able to detect multiple EMRIs at any given time \citep{emami19}.  NSCs also likely play a role in the ``final-parsec'' problem that needs to be overcome for massive BH binaries to merge \citep[e.g.,][]{vasiliev15,khan19}.  Overall, it is clear that the density profiles, morphology, and kinematics of NSCs play an important role in many types of GW events.

\section{Summary, open issues, and outlook}\label{sec:summary}

The hundreds of papers we have discussed here have elevated our understanding of NSCs from merely bright points of light at the centers of nearby galaxies to the extremely dense and complex stellar systems we now know them to be.  

Still, in the process of writing this review we have found a number of topics where our present knowledge is lacking. Perhaps the most pressing need is for theoretical work on NSC formation.  This includes simulations of NSC formation in the cosmological context, especially in the early universe.  The simulations targeting the formation of the first stars and black holes \citep[e.g.,][]{wise19}, and the formation of globular clusters in the early Universe \citep[e.g.,][]{li17,ma19} should provide us with valuable insight on the early epochs of NSC formation, and whether seed clusters or central black holes play a significant role in promoting the formation of NSCs.  
Also, theoretical work is urgently needed to understand the data that suggests a transition in the dominant growth mechanism at galaxy masses of $\sim10^9\,M_\odot$.  Are there mechanisms that can prevent gas from reaching the center of galaxies in low mass galaxies, or is there a barrier to globular cluster accretion in higher mass galaxies?  

Observationally, our knowledge of NSCs in lower-mass late-type galaxies is woefully lacking.  The impressive results from large samples of NSCs in low-mass early types \citep{denbrok14,ordenes-briceno18,sanchez-janssen19} clearly highlight the importance for obtaining comparable (i.e., large and homogenously selected) samples also for lower-mass late-type galaxies.  This data should also help test the environmental dependence of NSC formation.  Furthermore, understanding the stellar populations of these NSCs, and determining if they more closely resemble those in low-mass early types or higher-mass late-types would also help us understand NSC formation more generally.  The availability of robust mass measurements (see Fig.~\ref{fig:mass_radius}) is limited primarily to NSCs in higher mass galaxies. More NSC mass measurements at both the high- and low-mass extreme would be of clear interest for understanding NSC scaling relations, enabling us to understand whether they steepen at the high mass end.  Also, density measurements in galaxies across the mass spectrum are a key input for understanding tidal disruption event rates, which depend strongly on the NSC density profiles.  Finally, we note the lack of data on the resolved kinematics of NSCs (Fig.~\ref{fig:v_over_sigma}), especially for late-type galaxies, which would provide important insights into the formation history of their NSCs.  

Future facilities promise to greatly improve our understanding of NSCs.  With the launch of JWST, a better census of NSCs in higher mass galaxies will be possible through high resolution mid-infrared observations of dusty galaxy cores.  Also, the identification of faint AGN enabled by JWST will aid in understanding the frequency of black hole / NSC co-existence in lower mass galaxies \citep[e.g.,][]{dumont19}.  LSST will also play an important role in characterizing the population of black holes in low mass galaxies through statistics on tidal disruption events, and variability of AGN \citep[e.g.,][]{baldassare18}. Finally, NSCs will likely be prime targets for adaptive optics observations on the 30-meter class telescopes \citep{gullieuszik14}, enabling characterization of their internal kinematics, stellar population variations, and central massive black holes.  

\begin{acknowledgements}
We thank Iskren Georgiev, Dieu Nguyen, Renuka Pechetti, Ruben Sanchez-Janssen, and Chelsea Spengler for sharing their data, and Nikolay Kacharov for providing Fig.~\ref{fig:ssp_comp}. We are grateful to Fabio Antonini, Mark den Brok, Nelson Caldwell, Eric Emsellem, Iskren Georgiev, Oleg Gnedin, Jenny Greene, Nikolay Kacharov, Alessandra Mastrobuono-Battisti, Mark Norris, Ruben Sanchez-Janssen, Rainer Sch\"odel,  Hans Walter-Rix, and Monica Valluri for insightful comments.
\end{acknowledgements}

\bibliographystyle{spbasic}
\bibliography{NSC_review.bib}

\begin{thebibliography}{400}
\providecommand{\natexlab}[1]{#1}
\providecommand{\url}[1]{{#1}}
\providecommand{\urlprefix}{URL }
\expandafter\ifx\csname urlstyle\endcsname\relax
  \providecommand{\doi}[1]{DOI~\discretionary{}{}{}#1}\else
  \providecommand{\doi}{DOI~\discretionary{}{}{}\begingroup
  \urlstyle{rm}\Url}\fi
\providecommand{\eprint}[2][]{\url{#2}}

\bibitem[{{Abbott} et~al.(2016){Abbott}, {Abbott}, {Abbott}, {Abernathy},
  {Acernese}, {Ackley}, {Adams}, {Adams}, {Addesso}, {Adhikari}, and
  et~al.}]{abbott16}
{Abbott} BP, {Abbott} R, {Abbott} TD, {Abernathy} MR, {Acernese} F, {Ackley} K,
  {Adams} C, {Adams} T, {Addesso} P, {Adhikari} RX, et~al (2016) {Observation
  of Gravitational Waves from a Binary Black Hole Merger}. Physical Review
  Letters 116(6):061102, \doi{10.1103/PhysRevLett.116.061102},
  \eprint{1602.03837}

\bibitem[{{Abbott} et~al.(2019){Abbott}, {Abbott}, {Abbott}, {Abraham},
  {Acernese}, {Ackley}, {Adams}, {Adhikari}, {Adya}, {Affeldt}, {Agathos},
  {Agatsuma}, {Aggarwal}, {Aguiar}, {Aiello}, {Ain}, {Ajith}, {Allen},
  {Allocca}, {Aloy}, {Altin}, {Amato}, {Ananyeva}, {Anderson}, {Anderson},
  {Angelova}, {LIGO Scientific Collaboration}, and {Virgo
  Collaboration}}]{abbott19}
{Abbott} BP, {Abbott} R, {Abbott} TD, {Abraham} S, {Acernese} F, {Ackley} K,
  {Adams} C, {Adhikari} RX, {Adya} VB, {Affeldt} C, {Agathos} M, {Agatsuma} K,
  {Aggarwal} N, {Aguiar} OD, {Aiello} L, {Ain} A, {Ajith} P, {Allen} G,
  {Allocca} A, {Aloy} MA, {Altin} PA, {Amato} A, {Ananyeva} A, {Anderson} SB,
  {Anderson} WG, {Angelova} SV, {LIGO Scientific Collaboration}, {Virgo
  Collaboration} (2019) {Binary Black Hole Population Properties Inferred from
  the First and Second Observing Runs of Advanced LIGO and Advanced Virgo}.
  \apjl 882(2):L24, \doi{10.3847/2041-8213/ab3800}, \eprint{1811.12940}

\bibitem[{{Afanasiev} et~al.(2018){Afanasiev}, {Chilingarian}, {Mieske},
  {Voggel}, {Picotti}, {Hilker}, {Seth}, {Neumayer}, {Frank}, {Romanowsky},
  {Hau}, {Baumgardt}, {Ahn}, {Strader}, {den Brok}, {McDermid}, {Spitler},
  {Brodie}, and {Walsh}}]{afanasiev18}
{Afanasiev} AV, {Chilingarian} IV, {Mieske} S, {Voggel} KT, {Picotti} A,
  {Hilker} M, {Seth} A, {Neumayer} N, {Frank} M, {Romanowsky} AJ, {Hau} G,
  {Baumgardt} H, {Ahn} C, {Strader} J, {den Brok} M, {McDermid} R, {Spitler} L,
  {Brodie} J, {Walsh} JL (2018) {A 3.5 million Solar masses black hole in the
  centre of the ultracompact dwarf galaxy fornax UCD3}. \mnras
  477(4):4856--4865, \doi{10.1093/mnras/sty913}, \eprint{1804.02938}

\bibitem[{{Agarwal} and {Milosavljevi{\'c}}(2011)}]{agarwal11}
{Agarwal} M, {Milosavljevi{\'c}} M (2011) {Nuclear Star Clusters from Clustered
  Star Formation}. \apj 729(1):35, \doi{10.1088/0004-637X/729/1/35},
  \eprint{1008.2986}

\bibitem[{{Aharon} and {Perets}(2015)}]{aharon15}
{Aharon} D, {Perets} HB (2015) {Formation and Evolution of Nuclear Star
  Clusters with In Situ Star Formation: Nuclear Cores and Age Segregation}.
  \apj 799(2):185, \doi{10.1088/0004-637X/799/2/185}, \eprint{1409.5121}

\bibitem[{{Aharon} and {Perets}(2016)}]{aharon16b}
{Aharon} D, {Perets} HB (2016) {The Impact of Mass Segregation and Star
  Formation on the Rates of Gravitational-wave Sources from Extreme Mass Ratio
  Inspirals}. \apjl 830(1):L1, \doi{10.3847/2041-8205/830/1/L1},
  \eprint{1609.01715}

\bibitem[{{Aharon} et~al.(2016){Aharon}, {Mastrobuono Battisti}, and
  {Perets}}]{aharon16}
{Aharon} D, {Mastrobuono Battisti} A, {Perets} HB (2016) {The History of Tidal
  Disruption Events in Galactic Nuclei}. \apj 823:137,
  \doi{10.3847/0004-637X/823/2/137}, \eprint{1507.08287}

\bibitem[{{Ahn} et~al.(2017){Ahn}, {Seth}, {den Brok}, {Strader}, {Baumgardt},
  {van den Bosch}, {Chilingarian}, {Frank}, {Hilker}, {McDermid}, {Mieske},
  {Romanowsky}, {Spitler}, {Brodie}, {Neumayer}, and {Walsh}}]{ahn17}
{Ahn} CP, {Seth} AC, {den Brok} M, {Strader} J, {Baumgardt} H, {van den Bosch}
  R, {Chilingarian} I, {Frank} M, {Hilker} M, {McDermid} R, {Mieske} S,
  {Romanowsky} AJ, {Spitler} L, {Brodie} J, {Neumayer} N, {Walsh} JL (2017)
  {Detection of Supermassive Black Holes in Two Virgo Ultracompact Dwarf
  Galaxies}. \apj 839(2):72, \doi{10.3847/1538-4357/aa6972},
  \eprint{1703.09221}

\bibitem[{{Ahn} et~al.(2018){Ahn}, {Seth}, {Cappellari}, {Krajnovi{\'c}},
  {Strader}, {Voggel}, {Walsh}, {Bahramian}, {Baumgardt}, {Brodie},
  {Chilingarian}, {Chomiuk}, {den Brok}, {Frank}, {Hilker}, {McDermid},
  {Mieske}, {Neumayer}, {Nguyen}, {Pechetti}, {Romanowsky}, and
  {Spitler}}]{ahn18}
{Ahn} CP, {Seth} AC, {Cappellari} M, {Krajnovi{\'c}} D, {Strader} J, {Voggel}
  KT, {Walsh} JL, {Bahramian} A, {Baumgardt} H, {Brodie} J, {Chilingarian} I,
  {Chomiuk} L, {den Brok} M, {Frank} M, {Hilker} M, {McDermid} RM, {Mieske} S,
  {Neumayer} N, {Nguyen} DD, {Pechetti} R, {Romanowsky} AJ, {Spitler} L (2018)
  {The Black Hole in the Most Massive Ultracompact Dwarf Galaxy M59-UCD3}. \apj
  858(2):102, \doi{10.3847/1538-4357/aabc57}, \eprint{1804.02399}

\bibitem[{{Alfaro-Cuello} et~al.(2019){Alfaro-Cuello}, {Kacharov}, {Neumayer},
  {Luetzgendorf}, {Seth}, {Boeker}, {Kamann}, {Leaman}, {van de Ven},
  {Bianchini}, {Watkins}, and {Lyubenova}}]{alfaro-cuello19}
{Alfaro-Cuello} M, {Kacharov} N, {Neumayer} N, {Luetzgendorf} N, {Seth} AC,
  {Boeker} T, {Kamann} S, {Leaman} R, {van de Ven} G, {Bianchini} P, {Watkins}
  LL, {Lyubenova} M (2019) {A deep view into the nucleus of the Sagittarius
  Dwarf Spheroidal Galaxy with MUSE. I. Data and stellar population
  characterization}. arXiv e-prints arXiv:1909.10529, \eprint{1909.10529}

\bibitem[{{Allen} et~al.(1990){Allen}, {Hyland}, and {Hillier}}]{allen90}
{Allen} DA, {Hyland} AR, {Hillier} DJ (1990) {The source of luminosity at the
  Galactic Centre.} \mnras 244:706

\bibitem[{{Amaro-Seoane} and {Chen}(2014)}]{amaro-seoane14}
{Amaro-Seoane} P, {Chen} X (2014) {The Fragmenting Past of the Disk at the
  Galactic Center: The Culprit for the Missing Red Giants}. \apjl 781(1):L18,
  \doi{10.1088/2041-8205/781/1/L18}, \eprint{1310.0458}

\bibitem[{{Amaro-Seoane} et~al.(2017){Amaro-Seoane}, {Audley}, {Babak},
  {Baker}, {Barausse}, {Bender}, {Berti}, {Binetruy}, and
  {Born}}]{amaro-seoane17}
{Amaro-Seoane} P, {Audley} H, {Babak} S, {Baker} J, {Barausse} E, {Bender} P,
  {Berti} E, {Binetruy} P, {Born} M (2017) {Laser Interferometer Space
  Antenna}. arXiv e-prints arXiv:1702.00786, \eprint{1702.00786}

\bibitem[{{Amaro-Seoane} et~al.(2019){Amaro-Seoane}, {Chen}, {Sch{\"o}del}, and
  {Casanellas}}]{amaro-seoane19}
{Amaro-Seoane} P, {Chen} X, {Sch{\"o}del} R, {Casanellas} J (2019) {Making
  Bright Giants Invisible At The Galactic Centre}. arXiv e-prints
  arXiv:1910.04774, \eprint{1910.04774}

\bibitem[{{Antonini}(2013)}]{antonini13}
{Antonini} F (2013) {Origin and Growth of Nuclear Star Clusters around Massive
  Black Holes}. \apj 763(1):62, \doi{10.1088/0004-637X/763/1/62},
  \eprint{1207.6589}

\bibitem[{{Antonini}(2014)}]{antonini14}
{Antonini} F (2014) {On the Distribution of Stellar Remnants around Massive
  Black Holes: Slow Mass Segregation, Star Cluster Inspirals, and Correlated
  Orbits}. \apj 794(2):106, \doi{10.1088/0004-637X/794/2/106},
  \eprint{1402.4865}

\bibitem[{{Antonini} and {Perets}(2012)}]{antonini12b}
{Antonini} F, {Perets} HB (2012) {Secular Evolution of Compact Binaries near
  Massive Black Holes: Gravitational Wave Sources and Other Exotica}. \apj
  757(1):27, \doi{10.1088/0004-637X/757/1/27}, \eprint{1203.2938}

\bibitem[{{Antonini} and {Rasio}(2016)}]{antonini16}
{Antonini} F, {Rasio} FA (2016) {Merging Black Hole Binaries in Galactic
  Nuclei: Implications for Advanced-LIGO Detections}. \apj 831:187,
  \doi{10.3847/0004-637X/831/2/187}, \eprint{1606.04889}

\bibitem[{{Antonini} et~al.(2012){Antonini}, {Capuzzo-Dolcetta},
  {Mastrobuono-Battisti}, and {Merritt}}]{antonini12}
{Antonini} F, {Capuzzo-Dolcetta} R, {Mastrobuono-Battisti} A, {Merritt} D
  (2012) {Dissipationless Formation and Evolution of the Milky Way Nuclear Star
  Cluster}. \apj 750(2):111, \doi{10.1088/0004-637X/750/2/111},
  \eprint{1110.5937}

\bibitem[{{Antonini} et~al.(2015){Antonini}, {Barausse}, and
  {Silk}}]{antonini15}
{Antonini} F, {Barausse} E, {Silk} J (2015) {The Coevolution of Nuclear Star
  Clusters, Massive Black Holes, and Their Host Galaxies}. \apj 812:72,
  \doi{10.1088/0004-637X/812/1/72}, \eprint{1506.02050}

\bibitem[{{Antonini} et~al.(2019){Antonini}, {Gieles}, and
  {Gualandris}}]{antonini19}
{Antonini} F, {Gieles} M, {Gualandris} A (2019) {Black hole growth through
  hierarchical black hole mergers in dense star clusters: implications for
  gravitational wave detections}. \mnras 486(4):5008--5021,
  \doi{10.1093/mnras/stz1149}, \eprint{1811.03640}

\bibitem[{{Arca-Sedda} and {Capuzzo-Dolcetta}(2017)}]{arca-sedda17}
{Arca-Sedda} M, {Capuzzo-Dolcetta} R (2017) {Lack of nuclear clusters in dwarf
  spheroidal galaxies: implications for massive black holes formation and the
  cusp/core problem}. \mnras 464(3):3060--3070, \doi{10.1093/mnras/stw2483},
  \eprint{1611.01088}

\bibitem[{{Arca-Sedda} et~al.(2015){Arca-Sedda}, {Capuzzo-Dolcetta},
  {Antonini}, and {Seth}}]{arca-sedda15}
{Arca-Sedda} M, {Capuzzo-Dolcetta} R, {Antonini} F, {Seth} A (2015) {Henize
  2-10: The Ongoing Formation of a Nuclear Star Cluster around a Massive Black
  Hole}. \apj 806(2):220, \doi{10.1088/0004-637X/806/2/220},
  \eprint{1501.04567}

\bibitem[{{Askar} et~al.(2018){Askar}, {Arca Sedda}, and {Giersz}}]{askar18}
{Askar} A, {Arca Sedda} M, {Giersz} M (2018) {MOCCA-SURVEY Database I: Galactic
  globular clusters harbouring a black hole subsystem}. \mnras
  478(2):1844--1854, \doi{10.1093/mnras/sty1186}, \eprint{1802.05284}

\bibitem[{{Auchettl} et~al.(2017){Auchettl}, {Guillochon}, and
  {Ramirez-Ruiz}}]{auchettl17}
{Auchettl} K, {Guillochon} J, {Ramirez-Ruiz} E (2017) {New Physical Insights
  about Tidal Disruption Events from a Comprehensive Observational Inventory at
  X-Ray Wavelengths}. \apj 838(2):149, \doi{10.3847/1538-4357/aa633b},
  \eprint{1611.02291}

\bibitem[{{Bacon} et~al.(1994){Bacon}, {Emsellem}, {Monnet}, and
  {Nieto}}]{bacon94}
{Bacon} R, {Emsellem} E, {Monnet} G, {Nieto} JL (1994) {Sub-arcsecond 2D
  photometry and spectroscopy of the nucleus of M 31 : the supermassive black
  hole revisited.} \aap 281:691--717, \eprint{astro-ph/9309008}

\bibitem[{{Bacon} et~al.(2001){Bacon}, {Emsellem}, {Combes}, {Copin}, {Monnet},
  and {Martin}}]{bacon01}
{Bacon} R, {Emsellem} E, {Combes} F, {Copin} Y, {Monnet} G, {Martin} P (2001)
  {The M 31 double nucleus probed with OASIS. A natural vec m = 1 mode?} \aap
  371:409--428, \doi{10.1051/0004-6361:20010317}, \eprint{astro-ph/0010567}

\bibitem[{{Bahcall} and {Wolf}(1976)}]{bahcall76}
{Bahcall} JN, {Wolf} RA (1976) {Star distribution around a massive black hole
  in a globular cluster.} \apj 209:214--232, \doi{10.1086/154711}

\bibitem[{{Bailey}(1980)}]{bailey80}
{Bailey} ME (1980) {On the fate of stellar mass loss in galactic nuclei}.
  \mnras 191:195--206, \doi{10.1093/mnras/191.2.195}

\bibitem[{{Balcells} et~al.(2003){Balcells}, {Graham},
  {Dom{\'\i}nguez-Palmero}, and {Peletier}}]{balcells03}
{Balcells} M, {Graham} AW, {Dom{\'\i}nguez-Palmero} L, {Peletier} RF (2003)
  {Galactic Bulges from Hubble Space Telescope Near-Infrared Camera
  Multi-Object Spectrometer Observations: The Lack of r$^{1/4}$ Bulges}. \apj
  582(2):L79--L82, \doi{10.1086/367783}, \eprint{astro-ph/0212184}

\bibitem[{{Balcells} et~al.(2007){Balcells}, {Graham}, and
  {Peletier}}]{balcells07}
{Balcells} M, {Graham} AW, {Peletier} RF (2007) {Galactic Bulges from Hubble
  Space Telescope NICMOS Observations: Central Galaxian Objects, and Nuclear
  Profile Slopes}. \apj 665:1084--1103, \doi{10.1086/519752},
  \eprint{arXiv:astro-ph/0404379}

\bibitem[{{Baldassare} et~al.(2014){Baldassare}, {Gallo}, {Miller}, {Plotkin},
  {Treu}, {Valluri}, and {Woo}}]{baldassare14}
{Baldassare} VF, {Gallo} E, {Miller} BP, {Plotkin} RM, {Treu} T, {Valluri} M,
  {Woo} JH (2014) {AMUSE-Field. II. Nucleation of Early-type Galaxies in the
  Field versus Cluster Environment}. \apj 791(2):133,
  \doi{10.1088/0004-637X/791/2/133}, \eprint{1406.6697}

\bibitem[{{Baldassare} et~al.(2015){Baldassare}, {Reines}, {Gallo}, and
  {Greene}}]{baldassare15}
{Baldassare} VF, {Reines} AE, {Gallo} E, {Greene} JE (2015) {A $\sim 50,000
  M_{\odot}$ Solar Mass Black Hole in the Nucleus of RGG 118}. \apjl
  809(1):L14, \doi{10.1088/2041-8205/809/1/L14}, \eprint{1506.07531}

\bibitem[{{Baldassare} et~al.(2018){Baldassare}, {Geha}, and
  {Greene}}]{baldassare18}
{Baldassare} VF, {Geha} M, {Greene} J (2018) {Identifying AGNs in Low-mass
  Galaxies via Long-term Optical Variability}. \apj 868(2):152,
  \doi{10.3847/1538-4357/aae6cf}, \eprint{1808.09578}

\bibitem[{{Balick} and {Brown}(1974)}]{balick74}
{Balick} B, {Brown} RL (1974) {Intense sub-arcsecond structure in the galactic
  center.} \apj 194:265--270, \doi{10.1086/153242}

\bibitem[{{Barth} et~al.(2009){Barth}, {Strigari}, {Bentz}, {Greene}, and
  {Ho}}]{barth09}
{Barth} AJ, {Strigari} LE, {Bentz} MC, {Greene} JE, {Ho} LC (2009) {Dynamical
  Constraints on the Masses of the Nuclear Star Cluster and Black Hole in the
  Late-Type Spiral Galaxy NGC 3621}. \apj 690(1):1031--1044,
  \doi{10.1088/0004-637X/690/1/1031}, \eprint{0809.1066}

\bibitem[{{Bartko} et~al.(2010){Bartko}, {Martins}, {Trippe}, {Fritz},
  {Genzel}, {Ott}, {Eisenhauer}, {Gillessen}, {Paumard}, and
  {Alexander}}]{bartko10}
{Bartko} H, {Martins} F, {Trippe} S, {Fritz} TK, {Genzel} R, {Ott} T,
  {Eisenhauer} F, {Gillessen} S, {Paumard} T, {Alexander} T (2010) {An
  Extremely Top-Heavy Initial Mass Function in the Galactic Center Stellar
  Disks}. \apj 708(1):834--840, \doi{10.1088/0004-637X/708/1/834},
  \eprint{0908.2177}

\bibitem[{{Bassino} et~al.(1994){Bassino}, {Muzzio}, and {Rabolli}}]{bassino94}
{Bassino} LP, {Muzzio} JC, {Rabolli} M (1994) {Are Globular Clusters the Nuclei
  of Cannibalized Dwarf Galaxies?} \apj 431:634, \doi{10.1086/174514}

\bibitem[{{Baumgardt}(2017)}]{baumgardt17}
{Baumgardt} H (2017) {N -body modelling of globular clusters: masses,
  mass-to-light ratios and intermediate-mass black holes}. \mnras
  464(2):2174--2202, \doi{10.1093/mnras/stw2488}, \eprint{1609.08794}

\bibitem[{{Baumgardt} et~al.(2018){Baumgardt}, {Amaro-Seoane}, and
  {Sch{\"o}del}}]{baumgardt18}
{Baumgardt} H, {Amaro-Seoane} P, {Sch{\"o}del} R (2018) {The distribution of
  stars around the Milky Way's central black hole. III. Comparison with
  simulations}. \aap 609:A28, \doi{10.1051/0004-6361/201730462},
  \eprint{1701.03818}

\bibitem[{{Baumgardt} et~al.(2019){Baumgardt}, {He}, {Sweet}, {Drinkwater},
  {Sollima}, {Hurley}, {Usher}, {Kamann}, {Dalgleish}, {Dreizler}, and
  {Husser}}]{baumgardt19}
{Baumgardt} H, {He} C, {Sweet} SM, {Drinkwater} M, {Sollima} A, {Hurley} J,
  {Usher} C, {Kamann} S, {Dalgleish} H, {Dreizler} S, {Husser} TO (2019) {No
  evidence for intermediate-mass black holes in the globular clusters
  {\ensuremath{\omega}} Cen and NGC 6624}. \mnras 488(4):5340--5351,
  \doi{10.1093/mnras/stz2060}, \eprint{1907.10845}

\bibitem[{{Becklin} and {Neugebauer}(1968)}]{becklin68}
{Becklin} EE, {Neugebauer} G (1968) {Infrared Observations of the Galactic
  Center}. \apj 151:145, \doi{10.1086/149425}

\bibitem[{{Becklin} and {Neugebauer}(1975)}]{becklin75}
{Becklin} EE, {Neugebauer} G (1975) {High-resolution maps of the galactic
  center at 2.2 and 10 microns.} \apj 200:L71--L74, \doi{10.1086/181899}

\bibitem[{{Begelman} and {Rees}(1978)}]{begelman78}
{Begelman} MC, {Rees} MJ (1978) {The fate of dense stellar systems}. \mnras
  185:847--860, \doi{10.1093/mnras/185.4.847}

\bibitem[{{Bekki}(2007)}]{bekki07}
{Bekki} K (2007) {The Formation of Stellar Galactic Nuclei through Dissipative
  Gas Dynamics}. \pasa 24(2):77--94, \doi{10.1071/AS07008}

\bibitem[{{Bekki} and {Freeman}(2003)}]{bekki03b}
{Bekki} K, {Freeman} KC (2003) {Formation of {\ensuremath{\omega}} Centauri
  from an ancient nucleated dwarf galaxy in the young Galactic disc}. \mnras
  346(2):L11--L15, \doi{10.1046/j.1365-2966.2003.07275.x},
  \eprint{astro-ph/0310348}

\bibitem[{{Bekki} and {Graham}(2010)}]{bekki10}
{Bekki} K, {Graham} AW (2010) {On the Transition from Nuclear-cluster- to
  Black-hole-dominated Galaxy Cores}. \apjl 714(2):L313--L317,
  \doi{10.1088/2041-8205/714/2/L313}, \eprint{1004.3627}

\bibitem[{{Bekki} et~al.(2003){Bekki}, {Couch}, {Drinkwater}, and
  {Shioya}}]{bekki03}
{Bekki} K, {Couch} WJ, {Drinkwater} MJ, {Shioya} Y (2003) {Galaxy threshing and
  the origin of ultra-compact dwarf galaxies in the Fornax cluster}. \mnras
  344(2):399--411, \doi{10.1046/j.1365-8711.2003.06916.x},
  \eprint{astro-ph/0308243}

\bibitem[{{Bekki} et~al.(2006){Bekki}, {Couch}, and {Shioya}}]{bekki06}
{Bekki} K, {Couch} WJ, {Shioya} Y (2006) {Dissipative Transformation of
  Nonnucleated Dwarf Galaxies into Nucleated Systems}. \apj 642(2):L133--L136,
  \doi{10.1086/504588}, \eprint{astro-ph/0604340}

\bibitem[{{Bender} et~al.(2005){Bender}, {Kormendy}, {Bower}, {Green},
  {Thomas}, {Danks}, {Gull}, {Hutchings}, {Joseph}, {Kaiser}, {Lauer},
  {Nelson}, {Richstone}, {Weistrop}, and {Woodgate}}]{bender05}
{Bender} R, {Kormendy} J, {Bower} G, {Green} R, {Thomas} J, {Danks} AC, {Gull}
  T, {Hutchings} JB, {Joseph} CL, {Kaiser} ME, {Lauer} TR, {Nelson} CH,
  {Richstone} D, {Weistrop} D, {Woodgate} B (2005) {HST STIS Spectroscopy of
  the Triple Nucleus of M31: Two Nested Disks in Keplerian Rotation around a
  Supermassive Black Hole}. \apj 631(1):280--300, \doi{10.1086/432434},
  \eprint{astro-ph/0509839}

\bibitem[{{Bianchini} et~al.(2013){Bianchini}, {Varri}, {Bertin}, and
  {Zocchi}}]{bianchini13}
{Bianchini} P, {Varri} AL, {Bertin} G, {Zocchi} A (2013) {Rotating Globular
  Clusters}. \apj 772(1):67, \doi{10.1088/0004-637X/772/1/67},
  \eprint{1305.6025}

\bibitem[{{Binggeli} et~al.(1985){Binggeli}, {Sandage}, and
  {Tammann}}]{binggeli85}
{Binggeli} B, {Sandage} A, {Tammann} GA (1985) {Studies of the Virgo Cluster.
  II - A catalog of 2096 galaxies in the Virgo Cluster area.} \aj
  90:1681--1759, \doi{10.1086/113874}

\bibitem[{{Bland-Hawthorn} and {Gerhard}(2016)}]{bland-hawthorn16}
{Bland-Hawthorn} J, {Gerhard} O (2016) {The Galaxy in Context: Structural,
  Kinematic, and Integrated Properties}. \araa 54:529--596,
  \doi{10.1146/annurev-astro-081915-023441}, \eprint{1602.07702}

\bibitem[{{Bland-Hawthorn} et~al.(2005){Bland-Hawthorn}, {Vlaji{\'c}},
  {Freeman}, and {Draine}}]{bland-hawthorn05}
{Bland-Hawthorn} J, {Vlaji{\'c}} M, {Freeman} KC, {Draine} BT (2005) {NGC 300:
  An Extremely Faint, Outer Stellar Disk Observed to 10 Scale Lengths}. The
  Astrophysical Journal 629(1):239--249, \doi{10.1086/430512},
  \eprint{astro-ph/0503488}

\bibitem[{{Blanton} et~al.(2005){Blanton}, {Lupton}, {Schlegel}, {Strauss},
  {Brinkmann}, {Fukugita}, and {Loveday}}]{blanton05}
{Blanton} MR, {Lupton} RH, {Schlegel} DJ, {Strauss} MA, {Brinkmann} J,
  {Fukugita} M, {Loveday} J (2005) {The Properties and Luminosity Function of
  Extremely Low Luminosity Galaxies}. \apj 631(1):208--230,
  \doi{10.1086/431416}, \eprint{astro-ph/0410164}

\bibitem[{{Blum} et~al.(2003){Blum}, {Ram{\'\i}rez}, {Sellgren}, and
  {Olsen}}]{blum03}
{Blum} RD, {Ram{\'\i}rez} SV, {Sellgren} K, {Olsen} K (2003) {Really Cool Stars
  and the Star Formation History at the Galactic Center}. \apj 597(1):323--346,
  \doi{10.1086/378380}, \eprint{astro-ph/0307291}

\bibitem[{{B{\"o}ker} et~al.(1997){B{\"o}ker}, {F{\"o}rster-Schreiber}, and
  {Genzel}}]{boker97}
{B{\"o}ker} T, {F{\"o}rster-Schreiber} NM, {Genzel} R (1997) {Near-Infrared
  Imaging Spectroscopy of IC 342: Evolution of a Bar-Driven Central Starburst}.
  \aj 114:1883, \doi{10.1086/118612}

\bibitem[{{B{\"o}ker} et~al.(1999){B{\"o}ker}, {van der Marel}, and
  {Vacca}}]{boker99}
{B{\"o}ker} T, {van der Marel} RP, {Vacca} WD (1999) {CO Band Head Spectroscopy
  of IC 342: Mass and Age of the Nuclear Star Cluster}. \aj 118(2):831--842,
  \doi{10.1086/300985}, \eprint{astro-ph/9903457}

\bibitem[{{B{\"o}ker} et~al.(2001){B{\"o}ker}, {van der Marel}, {Mazzuca},
  {Rix}, {Rudnick}, {Ho}, and {Shields}}]{boker01}
{B{\"o}ker} T, {van der Marel} RP, {Mazzuca} L, {Rix} HW, {Rudnick} G, {Ho} LC,
  {Shields} JC (2001) {A Young Stellar Cluster in the Nucleus of NGC 4449}. \aj
  121(3):1473--1481, \doi{10.1086/319415}, \eprint{astro-ph/0010542}

\bibitem[{{B{\"o}ker} et~al.(2002){B{\"o}ker}, {Laine}, {van der Marel},
  {Sarzi}, {Rix}, {Ho}, and {Shields}}]{boker02}
{B{\"o}ker} T, {Laine} S, {van der Marel} RP, {Sarzi} M, {Rix} HW, {Ho} LC,
  {Shields} JC (2002) {A Hubble Space Telescope Census of Nuclear Star Clusters
  in Late-Type Spiral Galaxies. I. Observations and Image Analysis}. \aj
  123:1389--1410, \doi{10.1086/339025}

\bibitem[{{B{\"o}ker} et~al.(2004){B{\"o}ker}, {Sarzi}, {McLaughlin}, {van der
  Marel}, {Rix}, {Ho}, and {Shields}}]{boker04}
{B{\"o}ker} T, {Sarzi} M, {McLaughlin} DE, {van der Marel} RP, {Rix} HW, {Ho}
  LC, {Shields} JC (2004) {A Hubble Space Telescope Census of Nuclear Star
  Clusters in Late-Type Spiral Galaxies. II. Cluster Sizes and Structural
  Parameter Correlations}. \aj 127:105--118, \doi{10.1086/380231},
  \eprint{astro-ph/0309761}

\bibitem[{{Boldrini} et~al.(2019){Boldrini}, {Mohayaee}, and
  {Silk}}]{boldrini19}
{Boldrini} P, {Mohayaee} R, {Silk} J (2019) {Fornax globular cluster
  distributions: implications for the cusp-core problem}. \mnras
  485(2):2546--2557, \doi{10.1093/mnras/stz573}, \eprint{1903.00354}

\bibitem[{{Bothun} and {Mould}(1988)}]{bothun88}
{Bothun} GD, {Mould} JR (1988) {Medium-Resolution Spectroscopy of Nucleated
  Dwarf Elliptical Galaxies}. \apj 324:123, \doi{10.1086/165885}

\bibitem[{{Bothun} et~al.(1985){Bothun}, {Mould}, {Wirth}, and
  {Caldwell}}]{bothun85}
{Bothun} GD, {Mould} JR, {Wirth} A, {Caldwell} N (1985) {An investigation of
  dwarf galaxies in the Virgo cluster}. \aj 90:697--707, \doi{10.1086/113778}

\bibitem[{{Bower} et~al.(1998){Bower}, {Green}, {Danks}, {Gull}, {Heap},
  {Hutchings}, {Joseph}, {Kaiser}, {Kimble}, {Kraemer}, {Weistrop}, {Woodgate},
  {Lindler}, {Hill}, {Malumuth}, {Baum}, {Sarajedini}, {Heckman}, {Wilson}, and
  {Richstone}}]{bower98}
{Bower} GA, {Green} RF, {Danks} A, {Gull} T, {Heap} S, {Hutchings} J, {Joseph}
  C, {Kaiser} ME, {Kimble} R, {Kraemer} S, {Weistrop} D, {Woodgate} B,
  {Lindler} D, {Hill} RS, {Malumuth} EM, {Baum} S, {Sarajedini} V, {Heckman}
  TM, {Wilson} AS, {Richstone} DO (1998) {Kinematics of the Nuclear Ionized Gas
  in the Radio Galaxy M84 (NGC 4374)}. \apjl 492(2):L111--L114,
  \doi{10.1086/311109}, \eprint{astro-ph/9710264}

\bibitem[{{Bower} et~al.(2001){Bower}, {Green}, {Bender}, {Gebhardt}, {Lauer},
  {Magorrian}, {Richstone}, {Danks}, {Gull}, {Hutchings}, {Joseph}, {Kaiser},
  {Weistrop}, {Woodgate}, {Nelson}, and {Malumuth}}]{bower01}
{Bower} GA, {Green} RF, {Bender} R, {Gebhardt} K, {Lauer} TR, {Magorrian} J,
  {Richstone} DO, {Danks} A, {Gull} T, {Hutchings} J, {Joseph} C, {Kaiser} ME,
  {Weistrop} D, {Woodgate} B, {Nelson} C, {Malumuth} EM (2001) {Evidence of a
  Supermassive Black Hole in the Galaxy NGC 1023 from the Nuclear Stellar
  Dynamics}. \apj 550(1):75--86, \doi{10.1086/319730},
  \eprint{astro-ph/0011204}

\bibitem[{{Brown} et~al.(2018){Brown}, {Gnedin}, and {Li}}]{brown18}
{Brown} G, {Gnedin} OY, {Li} H (2018) {Nuclear Star Clusters in Cosmological
  Simulations}. \apj 864(1):94, \doi{10.3847/1538-4357/aad595},
  \eprint{1804.09819}

\bibitem[{{Bruzual} and {Charlot}(2003)}]{bruzual03}
{Bruzual} G, {Charlot} S (2003) {Stellar population synthesis at the resolution
  of 2003}. \mnras 344(4):1000--1028, \doi{10.1046/j.1365-8711.2003.06897.x},
  \eprint{astro-ph/0309134}

\bibitem[{{Bub} and {Petrovich}(2019)}]{bub19}
{Bub} MW, {Petrovich} C (2019) {Compact-Object Mergers in the Galactic Center:
  Evolution in Triaxial Clusters}. arXiv e-prints arXiv:1910.02079,
  \eprint{1910.02079}

\bibitem[{{Buchholz} et~al.(2009){Buchholz}, {Sch{\"o}del}, and
  {Eckart}}]{buchholz09}
{Buchholz} RM, {Sch{\"o}del} R, {Eckart} A (2009) {Composition of the galactic
  center star cluster. Population analysis from adaptive optics narrow band
  spectral energy distributions}. \aap 499(2):483--501,
  \doi{10.1051/0004-6361/200811497}, \eprint{0903.2135}

\bibitem[{{Bullock} and {Boylan-Kolchin}(2017)}]{bullock17}
{Bullock} JS, {Boylan-Kolchin} M (2017) {Small-Scale Challenges to the
  {\ensuremath{\Lambda}}CDM Paradigm}. \araa 55(1):343--387,
  \doi{10.1146/annurev-astro-091916-055313}, \eprint{1707.04256}

\bibitem[{{Caldwell}(1983)}]{caldwell83}
{Caldwell} N (1983) {Structure and stellar content of dwarf elliptical
  galaxies.} \aj 88:804--812, \doi{10.1086/113367}

\bibitem[{{Caldwell} and {Bothun}(1987)}]{caldwell87}
{Caldwell} N, {Bothun} GD (1987) {Dwarf Elliptical Galaxies in the Fornax
  Cluster. II. Their Structure and Stellar Populations}. \aj 94:1126,
  \doi{10.1086/114550}

\bibitem[{{Caldwell} et~al.(2017){Caldwell}, {Strader}, {Sand}, {Willman}, and
  {Seth}}]{caldwell17}
{Caldwell} N, {Strader} J, {Sand} DJ, {Willman} B, {Seth} AC (2017) {The Faint
  Globular Cluster in the Dwarf Galaxy Andromeda I}. \pasa 34:e039,
  \doi{10.1017/pasa.2017.35}, \eprint{1708.03596}

\bibitem[{{Cappellari}(2016)}]{cappellari16}
{Cappellari} M (2016) {Structure and Kinematics of Early-Type Galaxies from
  Integral Field Spectroscopy}. \araa 54:597--665,
  \doi{10.1146/annurev-astro-082214-122432}, \eprint{1602.04267}

\bibitem[{{Cappellari} et~al.(2013){Cappellari}, {McDermid}, {Alatalo},
  {Blitz}, {Bois}, {Bournaud}, {Bureau}, {Crocker}, {Davies}, {Davis}, {de
  Zeeuw}, {Duc}, {Emsellem}, {Khochfar}, {Krajnovi{\'c}}, {Kuntschner},
  {Morganti}, {Naab}, {Oosterloo}, {Sarzi}, {Scott}, {Serra}, {Weijmans}, and
  {Young}}]{cappellari13}
{Cappellari} M, {McDermid} RM, {Alatalo} K, {Blitz} L, {Bois} M, {Bournaud} F,
  {Bureau} M, {Crocker} AF, {Davies} RL, {Davis} TA, {de Zeeuw} PT, {Duc} PA,
  {Emsellem} E, {Khochfar} S, {Krajnovi{\'c}} D, {Kuntschner} H, {Morganti} R,
  {Naab} T, {Oosterloo} T, {Sarzi} M, {Scott} N, {Serra} P, {Weijmans} AM,
  {Young} LM (2013) {The ATLAS$^{3D}$ project - XX. Mass-size and
  mass-{\ensuremath{\sigma}} distributions of early-type galaxies: bulge
  fraction drives kinematics, mass-to-light ratio, molecular gas fraction and
  stellar initial mass function}. \mnras 432(3):1862--1893,
  \doi{10.1093/mnras/stt644}, \eprint{1208.3523}

\bibitem[{{Capuzzo-Dolcetta}(1993)}]{capuzzo93}
{Capuzzo-Dolcetta} R (1993) {The Evolution of the Globular Cluster System in a
  Triaxial Galaxy: Can a Galactic Nucleus Form by Globular Cluster Capture?}
  \apj 415:616, \doi{10.1086/173189}, \eprint{astro-ph/9301006}

\bibitem[{{Capuzzo-Dolcetta} and {Mastrobuono-Battisti}(2009)}]{capuzzo09}
{Capuzzo-Dolcetta} R, {Mastrobuono-Battisti} A (2009) {Globular cluster system
  erosion in elliptical galaxies}. \aap 507(1):183--193,
  \doi{10.1051/0004-6361/200912255}, \eprint{0904.0526}

\bibitem[{{Carlsten} et~al.(2019){Carlsten}, {Greco}, {Beaton}, and
  {Greene}}]{carlsten19}
{Carlsten} SG, {Greco} JP, {Beaton} RL, {Greene} JE (2019) {Wide-Field Survey
  of Dwarf Satellite Systems Around 10 Hosts in the Local Volume}. arXiv
  e-prints arXiv:1909.07389, \eprint{1909.07389}

\bibitem[{{Carollo} and {Stiavelli}(1998)}]{carollo98a}
{Carollo} CM, {Stiavelli} M (1998) {Spiral Galaxies with WFPC2. III. Nuclear
  Cusp Slopes}. \aj 115:2306--2319, \doi{10.1086/300373},
  \eprint{astro-ph/9804010}

\bibitem[{{Carollo} et~al.(1997){Carollo}, {Stiavelli}, {de Zeeuw}, and
  {Mack}}]{carollo97}
{Carollo} CM, {Stiavelli} M, {de Zeeuw} PT, {Mack} J (1997) {Spiral Galaxies
  with WFPC2.I.Nuclear Morphology, Bulges, Star Clusters, and Surface
  Brightness Profiles}. \aj 114:2366--+, \doi{10.1086/118654}

\bibitem[{{Carollo} et~al.(1998){Carollo}, {Stiavelli}, and
  {Mack}}]{carollo98b}
{Carollo} CM, {Stiavelli} M, {Mack} J (1998) {Spiral Galaxies with WFPC2. II.
  The Nuclear Properties of 40 Objects}. \aj 116:68--84, \doi{10.1086/300407},
  \eprint{astro-ph/9804007}

\bibitem[{{Carollo} et~al.(2002){Carollo}, {Stiavelli}, {Seigar}, {de Zeeuw},
  and {Dejonghe}}]{carollo02}
{Carollo} CM, {Stiavelli} M, {Seigar} M, {de Zeeuw} PT, {Dejonghe} H (2002)
  {Spiral Galaxies with HST/NICMOS. I. Nuclear Morphologies, Color Maps, and
  Distinct Nuclei}. \aj 123:159--183, \doi{10.1086/324725},
  \eprint{astro-ph/0110281}

\bibitem[{{Carson} et~al.(2015){Carson}, {Barth}, {Seth}, {den Brok},
  {Cappellari}, {Greene}, {Ho}, and {Neumayer}}]{carson15}
{Carson} DJ, {Barth} AJ, {Seth} AC, {den Brok} M, {Cappellari} M, {Greene} JE,
  {Ho} LC, {Neumayer} N (2015) {The Structure of Nuclear Star Clusters in
  Nearby Late-type Spiral Galaxies from Hubble Space Telescope Wide Field
  Camera 3 Imaging}. \aj 149:170, \doi{10.1088/0004-6256/149/5/170},
  \eprint{1501.05586}

\bibitem[{{Cen}(2001)}]{cen01}
{Cen} R (2001) {Synchronized Formation of Subgalactic Systems at Cosmological
  Reionization: Origin of Halo Globular Clusters}. \apj 560(2):592--598,
  \doi{10.1086/323071}, \eprint{astro-ph/0101197}

\bibitem[{{Chabrier}(2003)}]{chabrier03}
{Chabrier} G (2003) {Galactic Stellar and Substellar Initial Mass Function}.
  \pasp 115(809):763--795, \doi{10.1086/376392}, \eprint{astro-ph/0304382}

\bibitem[{{Chang} et~al.(2007){Chang}, {Murray-Clay}, {Chiang}, and
  {Quataert}}]{chang07}
{Chang} P, {Murray-Clay} R, {Chiang} E, {Quataert} E (2007) {The Origin of the
  Young Stars in the Nucleus of M31}. \apj 668(1):236--244,
  \doi{10.1086/521018}, \eprint{0704.3831}

\bibitem[{{Chatzopoulos} et~al.(2015){Chatzopoulos}, {Fritz}, {Gerhard},
  {Gillessen}, {Wegg}, {Genzel}, and {Pfuhl}}]{chatzopoulos15}
{Chatzopoulos} S, {Fritz} TK, {Gerhard} O, {Gillessen} S, {Wegg} C, {Genzel} R,
  {Pfuhl} O (2015) {The old nuclear star cluster in the Milky Way: dynamics,
  mass, statistical parallax, and black hole mass}. \mnras 447(1):948--968,
  \doi{10.1093/mnras/stu2452}, \eprint{1403.5266}

\bibitem[{{Chilingarian}(2009)}]{chilingarian09}
{Chilingarian} IV (2009) {Evolution of dwarf early-type galaxies - I. Spatially
  resolved stellar populations and internal kinematics of Virgo cluster dE/dS0
  galaxies}. \mnras 394(3):1229--1248, \doi{10.1111/j.1365-2966.2009.14450.x},
  \eprint{0812.3272}

\bibitem[{{Chilingarian} and {Mamon}(2008)}]{chilingarian08}
{Chilingarian} IV, {Mamon} GA (2008) {SDSSJ124155.33+114003.7 - a missing link
  between compact elliptical and ultracompact dwarf galaxies}. \mnras
  385(1):L83--L87, \doi{10.1111/j.1745-3933.2008.00438.x}, \eprint{0712.2724}

\bibitem[{{Chilingarian} et~al.(2018){Chilingarian}, {Katkov}, {Zolotukhin},
  {Grishin}, {Beletsky}, {Boutsia}, and {Osip}}]{chilingarian18}
{Chilingarian} IV, {Katkov} IY, {Zolotukhin} IY, {Grishin} KA, {Beletsky} Y,
  {Boutsia} K, {Osip} DJ (2018) {A Population of Bona Fide Intermediate-mass
  Black Holes Identified as Low-luminosity Active Galactic Nuclei}. \apj
  863(1):1, \doi{10.3847/1538-4357/aad184}, \eprint{1805.01467}

\bibitem[{{Choksi} et~al.(2019){Choksi}, {Volonteri}, {Colpi}, {Gnedin}, and
  {Li}}]{choksi19}
{Choksi} N, {Volonteri} M, {Colpi} M, {Gnedin} OY, {Li} H (2019) {The Star
  Clusters That Make Black Hole Binaries across Cosmic Time}. \apj 873(1):100,
  \doi{10.3847/1538-4357/aaffde}, \eprint{1809.01164}

\bibitem[{{Conroy}(2013)}]{conroy13}
{Conroy} C (2013) {Modeling the Panchromatic Spectral Energy Distributions of
  Galaxies}. \araa 51(1):393--455, \doi{10.1146/annurev-astro-082812-141017},
  \eprint{1301.7095}

\bibitem[{{C{\^o}t{\'e}} et~al.(2006){C{\^o}t{\'e}}, {Piatek}, {Ferrarese},
  {Jord{\'a}n}, {Merritt}, {Peng}, {Ha{\c s}egan}, {Blakeslee}, {Mei}, {West},
  {Milosavljevi{\'c}}, and {Tonry}}]{cote06}
{C{\^o}t{\'e}} P, {Piatek} S, {Ferrarese} L, {Jord{\'a}n} A, {Merritt} D,
  {Peng} EW, {Ha{\c s}egan} M, {Blakeslee} JP, {Mei} S, {West} MJ,
  {Milosavljevi{\'c}} M, {Tonry} JL (2006) {The ACS Virgo Cluster Survey. VIII.
  The Nuclei of Early-Type Galaxies}. \apjs 165:57--94, \doi{10.1086/504042}

\bibitem[{{C{\^o}t{\'e}} et~al.(2007){C{\^o}t{\'e}}, {Ferrarese}, {Jord{\'a}n},
  {Blakeslee}, {Chen}, {Infante}, {Merritt}, {Mei}, {Peng}, {Tonry}, {West},
  and {West}}]{cote07}
{C{\^o}t{\'e}} P, {Ferrarese} L, {Jord{\'a}n} A, {Blakeslee} JP, {Chen} CW,
  {Infante} L, {Merritt} D, {Mei} S, {Peng} EW, {Tonry} JL, {West} AA, {West}
  MJ (2007) {The ACS Fornax Cluster Survey. II. The Central Brightness Profiles
  of Early-Type Galaxies: A Characteristic Radius on Nuclear Scales and the
  Transition from Central Luminosity Deficit to Excess}. \apj 671:1456--1465,
  \doi{10.1086/522822}, \eprint{0711.1358}

\bibitem[{{Cretton} and {van den Bosch}(1999)}]{cretton99}
{Cretton} N, {van den Bosch} FC (1999) {Evidence for a Massive Black Hole in
  the S0 Galaxy NGC 4342}. \apj 514(2):704--724, \doi{10.1086/306971},
  \eprint{astro-ph/9805324}

\bibitem[{{Crnojevi{\'c}} et~al.(2016){Crnojevi{\'c}}, {Sand}, {Zaritsky},
  {Spekkens}, {Willman}, and {Hargis}}]{crnojevic16}
{Crnojevi{\'c}} D, {Sand} DJ, {Zaritsky} D, {Spekkens} K, {Willman} B, {Hargis}
  JR (2016) {Deep Imaging of Eridanus II and Its Lone Star Cluster}. \apj
  824(1):L14, \doi{10.3847/2041-8205/824/1/L14}, \eprint{1604.08590}

\bibitem[{{Da Costa}(2016)}]{dacosta16}
{Da Costa} GS (2016) {Are the globular clusters with significant internal
  [Fe/H] spreads all former dwarf galaxy nuclei?} In: {Bragaglia} A,
  {Arnaboldi} M, {Rejkuba} M, {Romano} D (eds) The General Assembly of Galaxy
  Halos: Structure, Origin and Evolution, IAU Symposium, vol 317, pp 110--115,
  \doi{10.1017/S174392131500678X}, \eprint{1510.00873}

\bibitem[{{Dale} et~al.(2009){Dale}, {Davies}, {Church}, and
  {Freitag}}]{dale09}
{Dale} JE, {Davies} MB, {Church} RP, {Freitag} M (2009) {Red giant stellar
  collisions in the Galactic Centre}. \mnras 393:1016--1033,
  \doi{10.1111/j.1365-2966.2008.14254.x}, \eprint{0811.3111}

\bibitem[{{Davies} et~al.(1998){Davies}, {Blackwell}, {Bailey}, and
  {Sigurdsson}}]{davies98}
{Davies} MB, {Blackwell} R, {Bailey} VC, {Sigurdsson} S (1998) {The destructive
  effects of binary encounters on red giants in the Galactic Centre}. \mnras
  301(3):745--753, \doi{10.1046/j.1365-8711.1998.02027.x}

\bibitem[{{Davies} et~al.(2011){Davies}, {Miller}, and {Bellovary}}]{davies11}
{Davies} MB, {Miller} MC, {Bellovary} JM (2011) {Supermassive Black Hole
  Formation Via Gas Accretion in Nuclear Stellar Clusters}. \apjl 740(2):L42,
  \doi{10.1088/2041-8205/740/2/L42}, \eprint{1106.5943}

\bibitem[{{Davies} and {Kasper}(2012)}]{davies12}
{Davies} R, {Kasper} M (2012) {Adaptive Optics for Astronomy}. \araa
  50:305--351, \doi{10.1146/annurev-astro-081811-125447}, \eprint{1201.5741}

\bibitem[{{de Francesco} et~al.(2006){de Francesco}, {Capetti}, and
  {Marconi}}]{defrancesco06}
{de Francesco} G, {Capetti} A, {Marconi} A (2006) {Measuring supermassive black
  holes with gas kinematics: the active S0 galaxy <ASTROBJ>NGC 3998</ASTROBJ>}.
  \aap 460(2):439--448, \doi{10.1051/0004-6361:20065826},
  \eprint{astro-ph/0609603}

\bibitem[{{De Lorenzi} et~al.(2013){De Lorenzi}, {Hartmann}, {Debattista},
  {Seth}, and {Gerhard}}]{delorenzi13}
{De Lorenzi} F, {Hartmann} M, {Debattista} VP, {Seth} AC, {Gerhard} O (2013)
  {Three-integral multicomponent dynamical models and simulations of the
  nuclear star cluster in NGC 4244}. \mnras 429:2974--2985,
  \doi{10.1093/mnras/sts545}, \eprint{1208.2161}

\bibitem[{{Demarque} and {Virani}(2007)}]{demarque07}
{Demarque} P, {Virani} S (2007) {The hot stars in orbit around the M 31 central
  supermassive black hole: are they young or old?} \aap 461(2):651--656,
  \doi{10.1051/0004-6361:20065921}, \eprint{astro-ph/0603326}

\bibitem[{{den Brok} et~al.(2014){den Brok}, {Peletier}, {Seth}, {Balcells},
  {Dominguez}, {Graham}, {Carter}, {Erwin}, {Ferguson}, {Goudfrooij},
  {Guzm{\'a}n}, {Hoyos}, {Jogee}, {Lucey}, {Phillipps}, {Puzia}, {Valentijn},
  {Kleijn}, and {Weinzirl}}]{denbrok14}
{den Brok} M, {Peletier} RF, {Seth} A, {Balcells} M, {Dominguez} L, {Graham}
  AW, {Carter} D, {Erwin} P, {Ferguson} HC, {Goudfrooij} P, {Guzm{\'a}n} R,
  {Hoyos} C, {Jogee} S, {Lucey} J, {Phillipps} S, {Puzia} T, {Valentijn} E,
  {Kleijn} GV, {Weinzirl} T (2014) {The HST/ACS Coma Cluster Survey - X.
  Nuclear star clusters in low-mass early-type galaxies: scaling relations}.
  \mnras 445:2385--2403, \doi{10.1093/mnras/stu1906}, \eprint{1409.4766}

\bibitem[{{den Brok} et~al.(2015){den Brok}, {Seth}, {Barth}, {Carson},
  {Neumayer}, {Cappellari}, {Debattista}, {Ho}, {Hood}, and
  {McDermid}}]{denbrok15}
{den Brok} M, {Seth} AC, {Barth} AJ, {Carson} DJ, {Neumayer} N, {Cappellari} M,
  {Debattista} VP, {Ho} LC, {Hood} CE, {McDermid} RM (2015) {Measuring the Mass
  of the Central Black Hole in the Bulgeless Galaxy NGC 4395 from Gas Dynamical
  Modeling}. \apj 809(1):101, \doi{10.1088/0004-637X/809/1/101},
  \eprint{1507.04358}

\bibitem[{{Devereux} et~al.(2003){Devereux}, {Ford}, {Tsvetanov}, and
  {Jacoby}}]{devereux03}
{Devereux} N, {Ford} H, {Tsvetanov} Z, {Jacoby} G (2003) {STIS Spectroscopy of
  the Central 10 Parsecs of M81: Evidence for a Massive Black Hole}. \aj
  125(3):1226--1235, \doi{10.1086/367595}

\bibitem[{{Do} et~al.(2009){Do}, {Ghez}, {Morris}, {Lu}, {Matthews}, {Yelda},
  and {Larkin}}]{do09}
{Do} T, {Ghez} AM, {Morris} MR, {Lu} JR, {Matthews} K, {Yelda} S, {Larkin} J
  (2009) {High Angular Resolution Integral-Field Spectroscopy of the Galaxy's
  Nuclear Cluster: A Missing Stellar Cusp?} \apj 703(2):1323--1337,
  \doi{10.1088/0004-637X/703/2/1323}, \eprint{0908.0311}

\bibitem[{{Do} et~al.(2015){Do}, {Kerzendorf}, {Winsor}, {St{\o}stad},
  {Morris}, {Lu}, and {Ghez}}]{do15}
{Do} T, {Kerzendorf} W, {Winsor} N, {St{\o}stad} M, {Morris} MR, {Lu} JR,
  {Ghez} AM (2015) {Discovery of Low-metallicity Stars in the Central Parsec of
  the Milky Way}. \apj 809(2):143, \doi{10.1088/0004-637X/809/2/143},
  \eprint{1506.07891}

\bibitem[{{Do} et~al.(2019){Do}, {Hees}, {Ghez}, {Martinez}, {Chu}, {Jia},
  {Sakai}, {Lu}, {Gautam}, {O'Neil}, {Becklin}, {Morris}, {Matthews},
  {Nishiyama}, {Campbell}, {Chappell}, {Chen}, {Ciurlo}, {Dehghanfar},
  {Gallego-Cano}, {Kerzendorf}, {Lyke}, {Naoz}, {Saida}, {Sch{\"o}del},
  {Takahashi}, {Takamori}, {Witzel}, and {Wizinowich}}]{do19}
{Do} T, {Hees} A, {Ghez} A, {Martinez} GD, {Chu} DS, {Jia} S, {Sakai} S, {Lu}
  JR, {Gautam} AK, {O'Neil} KK, {Becklin} EE, {Morris} MR, {Matthews} K,
  {Nishiyama} S, {Campbell} R, {Chappell} S, {Chen} Z, {Ciurlo} A, {Dehghanfar}
  A, {Gallego-Cano} E, {Kerzendorf} WE, {Lyke} JE, {Naoz} S, {Saida} H,
  {Sch{\"o}del} R, {Takahashi} M, {Takamori} Y, {Witzel} G, {Wizinowich} P
  (2019) {Relativistic redshift of the star S0-2 orbiting the Galactic Center
  supermassive black hole}. Science 365(6454):664--668,
  \doi{10.1126/science.aav8137}, \eprint{1907.10731}

\bibitem[{{Drinkwater} et~al.(2000){Drinkwater}, {Jones}, {Gregg}, and
  {Phillipps}}]{drinkwater00}
{Drinkwater} MJ, {Jones} JB, {Gregg} MD, {Phillipps} S (2000) {Compact Stellar
  Systems in the Fornax Cluster: Super-massive Star Clusters or Extremely
  Compact Dwarf Galaxies?} \pasa 17(3):227--233, \doi{10.1071/AS00034},
  \eprint{astro-ph/0002003}

\bibitem[{{Dumont} et~al.(2019){Dumont}, {Seth}, {Strader}, {Greene},
  {Burtscher}, and {Neumayer}}]{dumont19}
{Dumont} A, {Seth} A, {Strader} J, {Greene} JE, {Burtscher} L, {Neumayer} N
  (2019) {Surprisingly Strong K-band Emission Found in Low Luminosity Active
  Galactic Nuclei}. arXiv e-prints arXiv:1909.07323, \eprint{1909.07323}

\bibitem[{{Ebisuzaki} et~al.(2001){Ebisuzaki}, {Makino}, {Tsuru}, {Funato},
  {Portegies Zwart}, {Hut}, {McMillan}, {Matsushita}, {Matsumoto}, and
  {Kawabe}}]{ebisuzaki01}
{Ebisuzaki} T, {Makino} J, {Tsuru} TG, {Funato} Y, {Portegies Zwart} S, {Hut}
  P, {McMillan} S, {Matsushita} S, {Matsumoto} H, {Kawabe} R (2001) {Missing
  Link Found? The ``Runaway'' Path to Supermassive Black Holes}. \apjl
  562(1):L19--L22, \doi{10.1086/338118}, \eprint{astro-ph/0106252}

\bibitem[{{Eigenthaler} et~al.(2018){Eigenthaler}, {Puzia}, {Taylor},
  {Ordenes-Brice{\~n}o}, {Mu{\~n}oz}, {Ribbeck}, {Alamo-Mart{\'\i}nez},
  {Zhang}, {{\'A}ngel}, and {Capaccioli}}]{eigenthaler18}
{Eigenthaler} P, {Puzia} TH, {Taylor} MA, {Ordenes-Brice{\~n}o} Y, {Mu{\~n}oz}
  RP, {Ribbeck} KX, {Alamo-Mart{\'\i}nez} KA, {Zhang} H, {{\'A}ngel} S,
  {Capaccioli} M (2018) {The Next Generation Fornax Survey (NGFS). II. The
  Central Dwarf Galaxy Population}. \apj 855(2):142,
  \doi{10.3847/1538-4357/aaab60}, \eprint{1801.02633}

\bibitem[{{Emami} and {Loeb}(2019)}]{emami19}
{Emami} R, {Loeb} A (2019) {Gravitational Waves from Stellar Mass Black Holes
  Around SgrA*}. arXiv e-prints arXiv:1903.02579, \eprint{1903.02579}

\bibitem[{{Emsellem} and {van de Ven}(2008)}]{emsellem08}
{Emsellem} E, {van de Ven} G (2008) {Formation of Central Massive Objects via
  Tidal Compression}. \apj 674(2):653--659, \doi{10.1086/524720},
  \eprint{0710.3161}

\bibitem[{{Emsellem} et~al.(1999){Emsellem}, {Dejonghe}, and
  {Bacon}}]{emsellem99}
{Emsellem} E, {Dejonghe} H, {Bacon} R (1999) {Dynamical models of NGC 3115}.
  \mnras 303(3):495--514, \doi{10.1046/j.1365-8711.1999.02210.x},
  \eprint{astro-ph/9810306}

\bibitem[{{Emsellem} et~al.(2015){Emsellem}, {Renaud}, {Bournaud}, {Elmegreen},
  {Combes}, and {Gabor}}]{emsellem15}
{Emsellem} E, {Renaud} F, {Bournaud} F, {Elmegreen} B, {Combes} F, {Gabor} JM
  (2015) {The interplay between a galactic bar and a supermassive black hole:
  nuclear fuelling in a subparsec resolution galaxy simulation}. \mnras
  446(3):2468--2482, \doi{10.1093/mnras/stu2209}, \eprint{1410.6479}

\bibitem[{{Erwin} and {Gadotti}(2012)}]{erwin12}
{Erwin} P, {Gadotti} DA (2012) {Do Nuclear Star Clusters and Supermassive Black
  Holes Follow the Same Host-Galaxy Correlations?} Advances in Astronomy
  2012:946368, \doi{10.1155/2012/946368}, \eprint{1112.2740}

\bibitem[{{Evstigneeva} et~al.(2008){Evstigneeva}, {Drinkwater}, {Peng},
  {Hilker}, {De Propris}, {Jones}, {Phillipps}, {Gregg}, and
  {Karick}}]{evstigneeva08}
{Evstigneeva} EA, {Drinkwater} MJ, {Peng} CY, {Hilker} M, {De Propris} R,
  {Jones} JB, {Phillipps} S, {Gregg} MD, {Karick} AM (2008) {Structural
  Properties of Ultra-Compact Dwarf Galaxies in the Fornax and Virgo Clusters}.
  \aj 136(1):461--478, \doi{10.1088/0004-6256/136/1/461}, \eprint{0804.4353}

\bibitem[{{Fahrion} et~al.(2019){Fahrion}, {Lyubenova}, {van de Ven}, {Leaman},
  {Hilker}, {Mart{\'\i}n-Navarro}, {Zhu}, {Alfaro-Cuello}, {Coccato},
  {Corsini}, {Falc{\'o}n-Barroso}, {Iodice}, {McDermid}, {Sarzi}, and {de
  Zeeuw}}]{fahrion19}
{Fahrion} K, {Lyubenova} M, {van de Ven} G, {Leaman} R, {Hilker} M,
  {Mart{\'\i}n-Navarro} I, {Zhu} L, {Alfaro-Cuello} M, {Coccato} L, {Corsini}
  EM, {Falc{\'o}n-Barroso} J, {Iodice} E, {McDermid} RM, {Sarzi} M, {de Zeeuw}
  T (2019) {Constraining nuclear star cluster formation using MUSE-AO
  observations of the early-type galaxy FCC 47}. \aap 628:A92,
  \doi{10.1051/0004-6361/201935832}, \eprint{1907.01007}

\bibitem[{{Feldmeier} et~al.(2014){Feldmeier}, {Neumayer}, {Seth},
  {Sch{\"o}del}, {L{\"u}tzgendorf}, {de Zeeuw}, {Kissler-Patig}, {Nishiyama},
  and {Walcher}}]{feldmeier14}
{Feldmeier} A, {Neumayer} N, {Seth} A, {Sch{\"o}del} R, {L{\"u}tzgendorf} N,
  {de Zeeuw} PT, {Kissler-Patig} M, {Nishiyama} S, {Walcher} CJ (2014) {Large
  scale kinematics and dynamical modelling of the Milky Way nuclear star
  cluster}. \aap 570:A2, \doi{10.1051/0004-6361/201423777}, \eprint{1406.2849}

\bibitem[{{Feldmeier-Krause} et~al.(2015){Feldmeier-Krause}, {Neumayer},
  {Sch{\"o}del}, {Seth}, {Hilker}, {de Zeeuw}, {Kuntschner}, {Walcher},
  {L{\"u}tzgendorf}, and {Kissler-Patig}}]{feldmeier-krause15}
{Feldmeier-Krause} A, {Neumayer} N, {Sch{\"o}del} R, {Seth} A, {Hilker} M, {de
  Zeeuw} PT, {Kuntschner} H, {Walcher} CJ, {L{\"u}tzgendorf} N, {Kissler-Patig}
  M (2015) {KMOS view of the Galactic centre. I. Young stars are centrally
  concentrated}. \aap 584:A2, \doi{10.1051/0004-6361/201526336},
  \eprint{1509.04707}

\bibitem[{{Feldmeier-Krause} et~al.(2017{\natexlab{a}}){Feldmeier-Krause},
  {Kerzendorf}, {Neumayer}, {Sch{\"o}del}, {Nogueras-Lara}, {Do}, {de Zeeuw},
  and {Kuntschner}}]{feldmeier-krause17}
{Feldmeier-Krause} A, {Kerzendorf} W, {Neumayer} N, {Sch{\"o}del} R,
  {Nogueras-Lara} F, {Do} T, {de Zeeuw} PT, {Kuntschner} H (2017{\natexlab{a}})
  {KMOS view of the Galactic Centre - II. Metallicity distribution of late-type
  stars}. \mnras 464(1):194--209, \doi{10.1093/mnras/stw2339},
  \eprint{1610.01623}

\bibitem[{{Feldmeier-Krause} et~al.(2017{\natexlab{b}}){Feldmeier-Krause},
  {Zhu}, {Neumayer}, {van de Ven}, {de Zeeuw}, and
  {Sch{\"o}del}}]{feldmeier-krause17b}
{Feldmeier-Krause} A, {Zhu} L, {Neumayer} N, {van de Ven} G, {de Zeeuw} PT,
  {Sch{\"o}del} R (2017{\natexlab{b}}) {Triaxial orbit-based modelling of the
  Milky Way Nuclear Star Cluster}. \mnras 466(4):4040--4052,
  \doi{10.1093/mnras/stw3377}, \eprint{1701.01583}

\bibitem[{{Ferguson} and {Sandage}(1989)}]{ferguson89}
{Ferguson} HC, {Sandage} A (1989) {The Spatial Distributions and Intrinsic
  Shapes of Dwarf Elliptical Galaxies in the Virgo and Fornax Clusters}. \apj
  346:L53, \doi{10.1086/185577}

\bibitem[{{Ferrarese} et~al.(1996){Ferrarese}, {Ford}, and
  {Jaffe}}]{ferrarese96}
{Ferrarese} L, {Ford} HC, {Jaffe} W (1996) {Evidence for a Massive Black Hole
  in the Active Galaxy NGC 4261 from Hubble Space Telescope Images and
  Spectra}. \apj 470:444, \doi{10.1086/177876}

\bibitem[{{Ferrarese} et~al.(2006{\natexlab{a}}){Ferrarese}, {C{\^o}t{\'e}},
  {Dalla Bont{\`a}}, {Peng}, {Merritt}, {Jord{\'a}n}, {Blakeslee}, {Ha{\c
  s}egan}, {Mei}, {Piatek}, {Tonry}, and {West}}]{ferrarese06}
{Ferrarese} L, {C{\^o}t{\'e}} P, {Dalla Bont{\`a}} E, {Peng} EW, {Merritt} D,
  {Jord{\'a}n} A, {Blakeslee} JP, {Ha{\c s}egan} M, {Mei} S, {Piatek} S,
  {Tonry} JL, {West} MJ (2006{\natexlab{a}}) {A Fundamental Relation between
  Compact Stellar Nuclei, Supermassive Black Holes, and Their Host Galaxies}.
  \apjl 644:L21--L24, \doi{10.1086/505388}, \eprint{astro-ph/0603840}

\bibitem[{{Ferrarese} et~al.(2006{\natexlab{b}}){Ferrarese}, {C{\^o}t{\'e}},
  {Jord{\'a}n}, {Peng}, {Blakeslee}, {Piatek}, {Mei}, {Merritt},
  {Milosavljevi{\'c}}, and {Tonry}}]{ferrarese06b}
{Ferrarese} L, {C{\^o}t{\'e}} P, {Jord{\'a}n} A, {Peng} EW, {Blakeslee} JP,
  {Piatek} S, {Mei} S, {Merritt} D, {Milosavljevi{\'c}} M, {Tonry} JL
  (2006{\natexlab{b}}) {The ACS Virgo Cluster Survey. VI. Isophotal Analysis
  and the Structure of Early-Type Galaxies}. \apjs 164(2):334--434,
  \doi{10.1086/501350}, \eprint{astro-ph/0602297}

\bibitem[{{Ferrarese} et~al.(2016){Ferrarese}, {C{\^o}t{\'e}},
  {S{\'a}nchez-Janssen}, {Roediger}, {McConnachie}, {Durrell}, {MacArthur},
  {Blakeslee}, {Duc}, {Boissier}, {Boselli}, {Courteau}, {Cuillandre},
  {Emsellem}, {Gwyn}, {Guhathakurta}, {Jord{\'a}n}, {Lan{\c c}on}, {Liu},
  {Mei}, {Mihos}, {Navarro}, {Peng}, {Puzia}, {Taylor}, {Toloba}, and
  {Zhang}}]{ferrarese16}
{Ferrarese} L, {C{\^o}t{\'e}} P, {S{\'a}nchez-Janssen} R, {Roediger} J,
  {McConnachie} AW, {Durrell} PR, {MacArthur} LA, {Blakeslee} JP, {Duc} PA,
  {Boissier} S, {Boselli} A, {Courteau} S, {Cuillandre} JC, {Emsellem} E,
  {Gwyn} SDJ, {Guhathakurta} P, {Jord{\'a}n} A, {Lan{\c c}on} A, {Liu} C, {Mei}
  S, {Mihos} JC, {Navarro} JF, {Peng} EW, {Puzia} TH, {Taylor} JE, {Toloba} E,
  {Zhang} H (2016) {The Next Generation Virgo Cluster Survey (NGVS). XIII. The
  Luminosity and Mass Function of Galaxies in the Core of the Virgo Cluster and
  the Contribution from Disrupted Satellites}. \apj 824:10,
  \doi{10.3847/0004-637X/824/1/10}, \eprint{1604.06462}

\bibitem[{{Filippenko} and {Ho}(2003)}]{filippenko03}
{Filippenko} AV, {Ho} LC (2003) {A Low-Mass Central Black Hole in the Bulgeless
  Seyfert 1 Galaxy NGC 4395}. \apjl 588(1):L13--L16, \doi{10.1086/375361},
  \eprint{astro-ph/0303429}

\bibitem[{{Foord} et~al.(2017){Foord}, {Gallo}, {Hodges-Kluck}, {Miller},
  {Baldassare}, {G{\"u}ltekin}, and {Gnedin}}]{foord17}
{Foord} A, {Gallo} E, {Hodges-Kluck} E, {Miller} BP, {Baldassare} VF,
  {G{\"u}ltekin} K, {Gnedin} OY (2017) {AGN Activity in Nucleated Galaxies as
  Measured by Chandra}. \apj 841(1):51, \doi{10.3847/1538-4357/aa6d63},
  \eprint{1704.03882}

\bibitem[{{Fragione} et~al.(2019){Fragione}, {Leigh}, and {Perna}}]{fragione19}
{Fragione} G, {Leigh} NWC, {Perna} R (2019) {Black hole and neutron star
  mergers in Galactic Nuclei: the role of triples}. \mnras p 1759,
  \doi{10.1093/mnras/stz1803}, \eprint{1903.09160}

\bibitem[{{Freitag} et~al.(2006{\natexlab{a}}){Freitag}, {G{\"u}rkan}, and
  {Rasio}}]{freitag06b}
{Freitag} M, {G{\"u}rkan} MA, {Rasio} FA (2006{\natexlab{a}}) {Runaway
  collisions in young star clusters - II. Numerical results}. \mnras
  368(1):141--161, \doi{10.1111/j.1365-2966.2006.10096.x},
  \eprint{astro-ph/0503130}

\bibitem[{{Freitag} et~al.(2006{\natexlab{b}}){Freitag}, {Rasio}, and
  {Baumgardt}}]{freitag06a}
{Freitag} M, {Rasio} FA, {Baumgardt} H (2006{\natexlab{b}}) {Runaway collisions
  in young star clusters - I. Methods and tests}. \mnras 368(1):121--140,
  \doi{10.1111/j.1365-2966.2006.10095.x}, \eprint{astro-ph/0503129}

\bibitem[{{Fritz} et~al.(2011){Fritz}, {Gillessen}, {Dodds-Eden}, {Lutz},
  {Genzel}, {Raab}, {Ott}, {Pfuhl}, {Eisenhauer}, and {Yusef-Zadeh}}]{fritz11}
{Fritz} TK, {Gillessen} S, {Dodds-Eden} K, {Lutz} D, {Genzel} R, {Raab} W,
  {Ott} T, {Pfuhl} O, {Eisenhauer} F, {Yusef-Zadeh} F (2011) {Line Derived
  Infrared Extinction toward the Galactic Center}. \apj 737(2):73,
  \doi{10.1088/0004-637X/737/2/73}, \eprint{1105.2822}

\bibitem[{{Fritz} et~al.(2016){Fritz}, {Chatzopoulos}, {Gerhard}, {Gillessen},
  {Genzel}, {Pfuhl}, {Tacchella}, {Eisenhauer}, and {Ott}}]{fritz16}
{Fritz} TK, {Chatzopoulos} S, {Gerhard} O, {Gillessen} S, {Genzel} R, {Pfuhl}
  O, {Tacchella} S, {Eisenhauer} F, {Ott} T (2016) {The Nuclear Cluster of the
  Milky Way: Total Mass and Luminosity}. \apj 821(1):44,
  \doi{10.3847/0004-637X/821/1/44}, \eprint{1406.7568}

\bibitem[{{Fujii} et~al.(2009){Fujii}, {Iwasawa}, {Funato}, and
  {Makino}}]{fujii09}
{Fujii} M, {Iwasawa} M, {Funato} Y, {Makino} J (2009) {Trojan Stars in the
  Galactic Center}. \apj 695(2):1421--1429, \doi{10.1088/0004-637X/695/2/1421},
  \eprint{0807.2818}

\bibitem[{{Gallagher} et~al.(1982){Gallagher}, {Goad}, and
  {Mould}}]{gallagher82}
{Gallagher} JS, {Goad} JW, {Mould} J (1982) {Structure of the M33 nucleus}.
  \apj 263:101--107, \doi{10.1086/160484}

\bibitem[{{Gallazzi} et~al.(2005){Gallazzi}, {Charlot}, {Brinchmann}, {White},
  and {Tremonti}}]{gallazzi05}
{Gallazzi} A, {Charlot} S, {Brinchmann} J, {White} SDM, {Tremonti} CA (2005)
  {The ages and metallicities of galaxies in the local universe}. \mnras
  362(1):41--58, \doi{10.1111/j.1365-2966.2005.09321.x},
  \eprint{astro-ph/0506539}

\bibitem[{{Gallego-Cano} et~al.(2018){Gallego-Cano}, {Sch{\"o}del}, {Dong},
  {Nogueras-Lara}, {Gallego-Calvente}, {Amaro-Seoane}, and
  {Baumgardt}}]{gallego-cano18}
{Gallego-Cano} E, {Sch{\"o}del} R, {Dong} H, {Nogueras-Lara} F,
  {Gallego-Calvente} AT, {Amaro-Seoane} P, {Baumgardt} H (2018) {The
  distribution of stars around the Milky Way's central black hole. I. Deep star
  counts}. \aap 609:A26, \doi{10.1051/0004-6361/201730451}, \eprint{1701.03816}

\bibitem[{{Gallo} et~al.(2010){Gallo}, {Treu}, {Marshall}, {Woo}, {Leipski},
  and {Antonucci}}]{gallo10}
{Gallo} E, {Treu} T, {Marshall} PJ, {Woo} JH, {Leipski} C, {Antonucci} R (2010)
  {AMUSE-Virgo. II. Down-sizing in Black Hole Accretion}. \apj 714(1):25--36,
  \doi{10.1088/0004-637X/714/1/25}, \eprint{1002.3619}

\bibitem[{{Gebhardt} et~al.(2001){Gebhardt}, {Lauer}, {Kormendy}, {Pinkney},
  {Bower}, {Green}, {Gull}, {Hutchings}, {Kaiser}, and {Nelson}}]{gebhardt01}
{Gebhardt} K, {Lauer} TR, {Kormendy} J, {Pinkney} J, {Bower} GA, {Green} R,
  {Gull} T, {Hutchings} JB, {Kaiser} ME, {Nelson} CH (2001) {M33: A Galaxy with
  No Supermassive Black Hole}. \aj 122(5):2469--2476, \doi{10.1086/323481},
  \eprint{astro-ph/0107135}

\bibitem[{{Gebhardt} et~al.(2002){Gebhardt}, {Rich}, and {Ho}}]{gebhardt02}
{Gebhardt} K, {Rich} RM, {Ho} LC (2002) {A 20,000 M$_{solar}$ Black Hole in the
  Stellar Cluster G1}. \apjl 578(1):L41--L45, \doi{10.1086/342980},
  \eprint{astro-ph/0209313}

\bibitem[{{Gebhardt} et~al.(2003){Gebhardt}, {Richstone}, {Tremaine}, {Lauer},
  {Bender}, {Bower}, {Dressler}, {Faber}, {Filippenko}, {Green}, {Grillmair},
  {Ho}, {Kormendy}, {Magorrian}, and {Pinkney}}]{gebhardt03}
{Gebhardt} K, {Richstone} D, {Tremaine} S, {Lauer} TR, {Bender} R, {Bower} G,
  {Dressler} A, {Faber} SM, {Filippenko} AV, {Green} R, {Grillmair} C, {Ho} LC,
  {Kormendy} J, {Magorrian} J, {Pinkney} J (2003) {Axisymmetric Dynamical
  Models of the Central Regions of Galaxies}. \apj 583(1):92--115,
  \doi{10.1086/345081}, \eprint{astro-ph/0209483}

\bibitem[{{Gebhardt} et~al.(2005){Gebhardt}, {Rich}, and {Ho}}]{gebhardt05}
{Gebhardt} K, {Rich} RM, {Ho} LC (2005) {An Intermediate-Mass Black Hole in the
  Globular Cluster G1: Improved Significance from New Keck and Hubble Space
  Telescope Observations}. \apj 634(2):1093--1102, \doi{10.1086/497023},
  \eprint{astro-ph/0508251}

\bibitem[{{Gebhardt} et~al.(2011){Gebhardt}, {Adams}, {Richstone}, {Lauer},
  {Faber}, {G{\"u}ltekin}, {Murphy}, and {Tremaine}}]{gebhardt11}
{Gebhardt} K, {Adams} J, {Richstone} D, {Lauer} TR, {Faber} SM, {G{\"u}ltekin}
  K, {Murphy} J, {Tremaine} S (2011) {The Black Hole Mass in M87 from
  Gemini/NIFS Adaptive Optics Observations}. \apj 729(2):119,
  \doi{10.1088/0004-637X/729/2/119}, \eprint{1101.1954}

\bibitem[{{Geha} et~al.(2002){Geha}, {Guhathakurta}, and {van der
  Marel}}]{geha02}
{Geha} M, {Guhathakurta} P, {van der Marel} RP (2002) {Internal Dynamics,
  Structure, and Formation of Dwarf Elliptical Galaxies. I. A Keck/Hubble Space
  Telescope Study of Six Virgo Cluster Dwarf Galaxies}. \aj 124(6):3073--3087,
  \doi{10.1086/344764}, \eprint{astro-ph/0206153}

\bibitem[{{Genzel} et~al.(1996){Genzel}, {Thatte}, {Krabbe}, {Kroker}, and
  {Tacconi-Garman}}]{genzel96}
{Genzel} R, {Thatte} N, {Krabbe} A, {Kroker} H, {Tacconi-Garman} LE (1996) {The
  Dark Mass Concentration in the Central Parsec of the Milky Way}. \apj
  472:153, \doi{10.1086/178051}

\bibitem[{{Genzel} et~al.(2010){Genzel}, {Eisenhauer}, and
  {Gillessen}}]{genzel10}
{Genzel} R, {Eisenhauer} F, {Gillessen} S (2010) {The Galactic Center massive
  black hole and nuclear star cluster}. Reviews of Modern Physics
  82:3121--3195, \doi{10.1103/RevModPhys.82.3121}, \eprint{1006.0064}

\bibitem[{{Georgiev} and {B{\"o}ker}(2014)}]{georgiev14}
{Georgiev} IY, {B{\"o}ker} T (2014) {Nuclear star clusters in 228 spiral
  galaxies in the HST/WFPC2 archive: catalogue and comparison to other stellar
  systems}. \mnras 441:3570--3590, \doi{10.1093/mnras/stu797},
  \eprint{1404.5956}

\bibitem[{{Georgiev} et~al.(2009){Georgiev}, {Hilker}, {Puzia}, {Goudfrooij},
  and {Baumgardt}}]{georgiev09}
{Georgiev} IY, {Hilker} M, {Puzia} TH, {Goudfrooij} P, {Baumgardt} H (2009)
  {Globular cluster systems in nearby dwarf galaxies - II. Nuclear star
  clusters and their relation to massive Galactic globular clusters}. \mnras
  396(2):1075--1085, \doi{10.1111/j.1365-2966.2009.14776.x}, \eprint{0903.2857}

\bibitem[{{Georgiev} et~al.(2016){Georgiev}, {B{\"o}ker}, {Leigh},
  {L{\"u}tzgendorf}, and {Neumayer}}]{georgiev16}
{Georgiev} IY, {B{\"o}ker} T, {Leigh} N, {L{\"u}tzgendorf} N, {Neumayer} N
  (2016) {Masses and scaling relations for nuclear star clusters, and their
  co-existence with central black holes}. \mnras 457:2122--2138,
  \doi{10.1093/mnras/stw093}, \eprint{1601.02613}

\bibitem[{{Gerhard}(2001)}]{gerhard01}
{Gerhard} O (2001) {The Galactic Center HE I Stars: Remains of a Dissolved
  Young Cluster?} \apjl 546(1):L39--L42, \doi{10.1086/318054},
  \eprint{astro-ph/0005096}

\bibitem[{{Ghez} et~al.(2003){Ghez}, {Duch{\^e}ne}, {Matthews}, {Hornstein},
  {Tanner}, {Larkin}, {Morris}, {Becklin}, {Salim}, and {Kremenek}}]{ghez03}
{Ghez} AM, {Duch{\^e}ne} G, {Matthews} K, {Hornstein} SD, {Tanner} A, {Larkin}
  J, {Morris} M, {Becklin} EE, {Salim} S, {Kremenek} T (2003) {The First
  Measurement of Spectral Lines in a Short-Period Star Bound to the Galaxy's
  Central Black Hole: A Paradox of Youth}. \apj 586(2):L127--L131,
  \doi{10.1086/374804}, \eprint{astro-ph/0302299}

\bibitem[{{Gnedin} et~al.(2014){Gnedin}, {Ostriker}, and {Tremaine}}]{gnedin14}
{Gnedin} OY, {Ostriker} JP, {Tremaine} S (2014) {Co-evolution of Galactic
  Nuclei and Globular Cluster Systems}. \apj 785(1):71,
  \doi{10.1088/0004-637X/785/1/71}, \eprint{1308.0021}

\bibitem[{{Goerdt} et~al.(2006){Goerdt}, {Moore}, {Read}, {Stadel}, and
  {Zemp}}]{goerdt06}
{Goerdt} T, {Moore} B, {Read} JI, {Stadel} J, {Zemp} M (2006) {Does the Fornax
  dwarf spheroidal have a central cusp or core?} \mnras 368(3):1073--1077,
  \doi{10.1111/j.1365-2966.2006.10182.x}, \eprint{astro-ph/0601404}

\bibitem[{{Graham} and {Driver}(2005)}]{graham05}
{Graham} AW, {Driver} SP (2005) {A Concise Reference to (Projected) S{\'e}rsic
  R$^{1/n}$ Quantities, Including Concentration, Profile Slopes, Petrosian
  Indices, and Kron Magnitudes}. \pasa 22(2):118--127, \doi{10.1071/AS05001},
  \eprint{astro-ph/0503176}

\bibitem[{{Graham} and {Guzm{\'a}n}(2003)}]{graham03}
{Graham} AW, {Guzm{\'a}n} R (2003) {HST Photometry of Dwarf Elliptical Galaxies
  in Coma, and an Explanation for the Alleged Structural Dichotomy between
  Dwarf and Bright Elliptical Galaxies}. \aj 125(6):2936--2950,
  \doi{10.1086/374992}, \eprint{astro-ph/0303391}

\bibitem[{{Graham} and {Spitler}(2009)}]{graham09}
{Graham} AW, {Spitler} LR (2009) {Quantifying the coexistence of massive black
  holes and dense nuclear star clusters}. \mnras 397(4):2148--2162,
  \doi{10.1111/j.1365-2966.2009.15118.x}, \eprint{0907.5250}

\bibitem[{{Gravity Collaboration} et~al.(2018){Gravity Collaboration},
  {Abuter}, {Amorim}, {Anugu}, {Baub{\"o}ck}, {Benisty}, {Berger}, {Blind},
  {Bonnet}, {Brandner}, {Buron}, {Collin}, {Chapron}, {Cl{\'e}net}, {Coud{\'e}
  Du Foresto}, {de Zeeuw}, {Deen}, {Delplancke-Str{\"o}bele}, {Dembet},
  {Dexter}, {Duvert}, {Eckart}, {Eisenhauer}, {Finger}, {F{\"o}rster
  Schreiber}, {F{\'e}dou}, {Garcia}, {Garcia Lopez}, {Gao}, {Gendron},
  {Genzel}, {Gillessen}, {Gordo}, {Habibi}, {Haubois}, {Haug}, {Hau{\ss}mann},
  {Henning}, {Hippler}, {Horrobin}, {Hubert}, {Hubin}, {Jimenez Rosales},
  {Jochum}, {Jocou}, {Kaufer}, {Kellner}, {Kendrew}, {Kervella}, {Kok},
  {Kulas}, {Lacour}, {Lapeyr{\`e}re}, {Lazareff}, {Le Bouquin}, {L{\'e}na},
  {Lippa}, {Lenzen}, {M{\'e}rand}, {M{\"u}ler}, {Neumann}, {Ott}, {Palanca},
  {Paumard}, {Pasquini}, {Perraut}, {Perrin}, {Pfuhl}, {Plewa}, {Rabien},
  {Ram{\'\i}rez}, {Ramos}, {Rau}, {Rodr{\'\i}guez-Coira}, {Rohloff}, {Rousset},
  {Sanchez-Bermudez}, {Scheithauer}, {Sch{\"o}ller}, {Schuler}, {Spyromilio},
  {Straub}, {Straubmeier}, {Sturm}, {Tacconi}, {Tristram}, {Vincent}, {von
  Fellenberg}, {Wank}, {Waisberg}, {Widmann}, {Wieprecht}, {Wiest},
  {Wiezorrek}, {Woillez}, {Yazici}, {Ziegler}, and {Zins}}]{abuter18}
{Gravity Collaboration}, {Abuter} R, {Amorim} A, {Anugu} N, {Baub{\"o}ck} M,
  {Benisty} M, {Berger} JP, {Blind} N, {Bonnet} H, {Brandner} W, {Buron} A,
  {Collin} C, {Chapron} F, {Cl{\'e}net} Y, {Coud{\'e} Du Foresto} V, {de Zeeuw}
  PT, {Deen} C, {Delplancke-Str{\"o}bele} F, {Dembet} R, {Dexter} J, {Duvert}
  G, {Eckart} A, {Eisenhauer} F, {Finger} G, {F{\"o}rster Schreiber} NM,
  {F{\'e}dou} P, {Garcia} P, {Garcia Lopez} R, {Gao} F, {Gendron} E, {Genzel}
  R, {Gillessen} S, {Gordo} P, {Habibi} M, {Haubois} X, {Haug} M,
  {Hau{\ss}mann} F, {Henning} T, {Hippler} S, {Horrobin} M, {Hubert} Z, {Hubin}
  N, {Jimenez Rosales} A, {Jochum} L, {Jocou} K, {Kaufer} A, {Kellner} S,
  {Kendrew} S, {Kervella} P, {Kok} Y, {Kulas} M, {Lacour} S, {Lapeyr{\`e}re} V,
  {Lazareff} B, {Le Bouquin} JB, {L{\'e}na} P, {Lippa} M, {Lenzen} R,
  {M{\'e}rand} A, {M{\"u}ler} E, {Neumann} U, {Ott} T, {Palanca} L, {Paumard}
  T, {Pasquini} L, {Perraut} K, {Perrin} G, {Pfuhl} O, {Plewa} PM, {Rabien} S,
  {Ram{\'\i}rez} A, {Ramos} J, {Rau} C, {Rodr{\'\i}guez-Coira} G, {Rohloff} RR,
  {Rousset} G, {Sanchez-Bermudez} J, {Scheithauer} S, {Sch{\"o}ller} M,
  {Schuler} N, {Spyromilio} J, {Straub} O, {Straubmeier} C, {Sturm} E,
  {Tacconi} LJ, {Tristram} KRW, {Vincent} F, {von Fellenberg} S, {Wank} I,
  {Waisberg} I, {Widmann} F, {Wieprecht} E, {Wiest} M, {Wiezorrek} E, {Woillez}
  J, {Yazici} S, {Ziegler} D, {Zins} G (2018) {Detection of the gravitational
  redshift in the orbit of the star S2 near the Galactic centre massive black
  hole}. \aap 615:L15, \doi{10.1051/0004-6361/201833718}, \eprint{1807.09409}

\bibitem[{{Gravity Collaboration} et~al.(2019){Gravity Collaboration},
  {Abuter}, {Amorim}, {Baub{\"o}ck}, {Berger}, {Bonnet}, {Brand ner},
  {Cl{\'e}net}, {Coud{\'e} Du Foresto}, and {de Zeeuw}}]{abuter19}
{Gravity Collaboration}, {Abuter} R, {Amorim} A, {Baub{\"o}ck} M, {Berger} JP,
  {Bonnet} H, {Brand ner} W, {Cl{\'e}net} Y, {Coud{\'e} Du Foresto} V, {de
  Zeeuw} PT (2019) {A geometric distance measurement to the Galactic center
  black hole with 0.3\% uncertainty}. \aap 625:L10,
  \doi{10.1051/0004-6361/201935656}

\bibitem[{{Greene} et~al.(2019){Greene}, {Strader}, and {Ho}}]{greene19}
{Greene} JE, {Strader} J, {Ho} LC (2019) {Intermediate Mass Black Holes}. \araa
  58(1):xxxx, \doi{XXX}, \eprint{XXX}

\bibitem[{{Guillard} et~al.(2016){Guillard}, {Emsellem}, and
  {Renaud}}]{guillard16}
{Guillard} N, {Emsellem} E, {Renaud} F (2016) {New insights on the formation of
  nuclear star clusters}. \mnras 461:3620--3629, \doi{10.1093/mnras/stw1570},
  \eprint{1606.09537}

\bibitem[{{Gullieuszik} et~al.(2014){Gullieuszik}, {Greggio}, {Falomo},
  {Schreiber}, and {Uslenghi}}]{gullieuszik14}
{Gullieuszik} M, {Greggio} L, {Falomo} R, {Schreiber} L, {Uslenghi} M (2014)
  {Probing the nuclear star cluster of galaxies with extremely large
  telescopes}. \aap 568:A89, \doi{10.1051/0004-6361/201424279},
  \eprint{1406.7818}

\bibitem[{{G{\"u}ltekin} et~al.(2009){G{\"u}ltekin}, {Richstone}, {Gebhardt},
  {Lauer}, {Pinkney}, {Aller}, {Bender}, {Dressler}, {Faber}, {Filippenko},
  {Green}, {Ho}, {Kormendy}, and {Siopis}}]{gueltekin09}
{G{\"u}ltekin} K, {Richstone} DO, {Gebhardt} K, {Lauer} TR, {Pinkney} J,
  {Aller} MC, {Bender} R, {Dressler} A, {Faber} SM, {Filippenko} AV, {Green} R,
  {Ho} LC, {Kormendy} J, {Siopis} C (2009) {A Quintet of Black Hole Mass
  Determinations}. \apj 695(2):1577--1590, \doi{10.1088/0004-637X/695/2/1577},
  \eprint{0901.4162}

\bibitem[{{G{\"u}rkan} et~al.(2004){G{\"u}rkan}, {Freitag}, and
  {Rasio}}]{gurkan04}
{G{\"u}rkan} MA, {Freitag} M, {Rasio} FA (2004) {Formation of Massive Black
  Holes in Dense Star Clusters. I. Mass Segregation and Core Collapse}. \apj
  604(2):632--652, \doi{10.1086/381968}, \eprint{astro-ph/0308449}

\bibitem[{{Habas} et~al.(2019){Habas}, {Marleau}, {Duc}, {Durrell}, {Paudel},
  {Poulain}, {S{\'a}nchez-Janssen}, {Sreejith}, {Ramasawmy}, {Stemock},
  {Leach}, {Cuillandre}, {Gwyn}, {Agnello}, {B{\'\i}lek}, {Fensch},
  {M{\"u}ller}, {Peng}, and {van der Burg}}]{habas19}
{Habas} R, {Marleau} FR, {Duc} PA, {Durrell} PR, {Paudel} S, {Poulain} M,
  {S{\'a}nchez-Janssen} R, {Sreejith} S, {Ramasawmy} J, {Stemock} B, {Leach} C,
  {Cuillandre} JC, {Gwyn} S, {Agnello} A, {B{\'\i}lek} M, {Fensch} J,
  {M{\"u}ller} O, {Peng} EW, {van der Burg} RFJ (2019) {Newly discovered dwarf
  galaxies in the MATLAS low density fields}. \mnras p 2656,
  \doi{10.1093/mnras/stz3045}

\bibitem[{{Harris}(1996)}]{harris96}
{Harris} WE (1996) {A Catalog of Parameters for Globular Clusters in the Milky
  Way}. \aj 112:1487, \doi{10.1086/118116}

\bibitem[{{Hartmann} et~al.(2011){Hartmann}, {Debattista}, {Seth},
  {Cappellari}, and {Quinn}}]{hartmann11}
{Hartmann} M, {Debattista} VP, {Seth} A, {Cappellari} M, {Quinn} TR (2011)
  {Constraining the role of star cluster mergers in nuclear cluster formation:
  simulations confront integral-field data}. \mnras 418:2697--2714,
  \doi{10.1111/j.1365-2966.2011.19659.x}, \eprint{1103.5464}

\bibitem[{{Hebbar} et~al.(2019){Hebbar}, {Heinke}, {Sivakoff}, and
  {Shaw}}]{hebbar19}
{Hebbar} PR, {Heinke} CO, {Sivakoff} GR, {Shaw} AW (2019) {X-ray spectroscopy
  of the candidate AGNs in Henize 2-10 and NGC 4178: likely supernova
  remnants}. \mnras 485(4):5604--5615, \doi{10.1093/mnras/stz553},
  \eprint{1902.08293}

\bibitem[{{Hilker} et~al.(1999){Hilker}, {Infante}, {Vieira}, {Kissler-Patig},
  and {Richtler}}]{hilker99}
{Hilker} M, {Infante} L, {Vieira} G, {Kissler-Patig} M, {Richtler} T (1999)
  {The central region of the Fornax cluster. II. Spectroscopy and radial
  velocities of member and background galaxies}. \aaps 134:75--86,
  \doi{10.1051/aas:1999434}, \eprint{astro-ph/9807144}

\bibitem[{{Hilker} et~al.(2004){Hilker}, {Kayser}, {Richtler}, and
  {Willemsen}}]{hilker04}
{Hilker} M, {Kayser} A, {Richtler} T, {Willemsen} P (2004) {The extended star
  formation history of {\ensuremath{\omega}} Centauri}. \aap 422:L9--L12,
  \doi{10.1051/0004-6361:20040188}, \eprint{astro-ph/0406017}

\bibitem[{{Holley-Bockelmann} et~al.(2008){Holley-Bockelmann}, {G{\"u}ltekin},
  {Shoemaker}, and {Yunes}}]{holley-bockelmann08}
{Holley-Bockelmann} K, {G{\"u}ltekin} K, {Shoemaker} D, {Yunes} N (2008)
  {Gravitational Wave Recoil and the Retention of Intermediate-Mass Black
  Holes}. \apj 686(2):829--837, \doi{10.1086/591218}, \eprint{0707.1334}

\bibitem[{{Hopkins} and {Quataert}(2010{\natexlab{a}})}]{hopkins10b}
{Hopkins} PF, {Quataert} E (2010{\natexlab{a}}) {How do massive black holes get
  their gas?} \mnras 407(3):1529--1564, \doi{10.1111/j.1365-2966.2010.17064.x},
  \eprint{0912.3257}

\bibitem[{{Hopkins} and {Quataert}(2010{\natexlab{b}})}]{hopkins10}
{Hopkins} PF, {Quataert} E (2010{\natexlab{b}}) {The nuclear stellar disc in
  Andromeda: a fossil from the era of black hole growth}. \mnras 405:L41--L45,
  \doi{10.1111/j.1745-3933.2010.00855.x}, \eprint{1002.1079}

\bibitem[{{Hopkins} et~al.(2009){Hopkins}, {Cox}, {Dutta}, {Hernquist},
  {Kormendy}, and {Lauer}}]{hopkins09}
{Hopkins} PF, {Cox} TJ, {Dutta} SN, {Hernquist} L, {Kormendy} J, {Lauer} TR
  (2009) {Dissipation and Extra Light in Galactic Nuclei. II. ``Cusp''
  Ellipticals}. \apjs 181(1):135--182, \doi{10.1088/0067-0049/181/1/135},
  \eprint{0805.3533}

\bibitem[{{Hopkins} et~al.(2010){Hopkins}, {Murray}, {Quataert}, and
  {Thompson}}]{hopkins10c}
{Hopkins} PF, {Murray} N, {Quataert} E, {Thompson} TA (2010) {A maximum stellar
  surface density in dense stellar systems}. \mnras 401(1):L19--L23,
  \doi{10.1111/j.1745-3933.2009.00777.x}, \eprint{0908.4088}

\bibitem[{{Hopman} and {Alexander}(2006)}]{hopman06}
{Hopman} C, {Alexander} T (2006) {The Effect of Mass Segregation on
  Gravitational Wave Sources near Massive Black Holes}. \apjl
  645(2):L133--L136, \doi{10.1086/506273}, \eprint{astro-ph/0603324}

\bibitem[{{Hughes} et~al.(2005){Hughes}, {Axon}, {Atkinson}, {Alonso-Herrero},
  {Scarlata}, {Marconi}, {Batcheldor}, {Binney}, {Capetti}, {Carollo},
  {Dressel}, {Gerssen}, {Macchetto}, {Maciejewski}, {Merrifield}, {Ruiz},
  {Sparks}, {Stiavelli}, and {Tsvetanov}}]{hughes05}
{Hughes} MA, {Axon} D, {Atkinson} J, {Alonso-Herrero} A, {Scarlata} C,
  {Marconi} A, {Batcheldor} D, {Binney} J, {Capetti} A, {Carollo} CM, {Dressel}
  L, {Gerssen} J, {Macchetto} D, {Maciejewski} W, {Merrifield} M, {Ruiz} M,
  {Sparks} W, {Stiavelli} M, {Tsvetanov} Z (2005) {Nuclear Properties of Nearby
  Spiral Galaxies from Hubble Space Telescope NICMOS Imaging and STIS
  Spectroscopy}. \aj 130:73--83, \doi{10.1086/430531},
  \eprint{astro-ph/0503693}

\bibitem[{{Hunt} et~al.(2008){Hunt}, {Combes}, {Garc{\'\i}a-Burillo},
  {Schinnerer}, {Krips}, {Baker}, {Boone}, {Eckart}, {L{\'e}on}, {Neri}, and
  {Tacconi}}]{hunt08}
{Hunt} LK, {Combes} F, {Garc{\'\i}a-Burillo} S, {Schinnerer} E, {Krips} M,
  {Baker} AJ, {Boone} F, {Eckart} A, {L{\'e}on} S, {Neri} R, {Tacconi} LJ
  (2008) {Molecular Gas in NUclei of GAlaxies (NUGA). IX. The decoupled bars
  and gas inflow in NGC 2782}. \aap 482(1):133--150,
  \doi{10.1051/0004-6361:20078874}, \eprint{0802.2775}

\bibitem[{{Ibata} et~al.(2009){Ibata}, {Bellazzini}, {Chapman}, {Dalessand ro},
  {Ferraro}, {Irwin}, {Lanzoni}, {Lewis}, {Mackey}, {Miocchi}, and
  {Varghese}}]{ibata09}
{Ibata} R, {Bellazzini} M, {Chapman} SC, {Dalessand ro} E, {Ferraro} F, {Irwin}
  M, {Lanzoni} B, {Lewis} GF, {Mackey} AD, {Miocchi} P, {Varghese} A (2009)
  {Density and Kinematic Cusps in M54 at the Heart of the Sagittarius Dwarf
  Galaxy: Evidence for A {}10$^{4}$ M $_{sun}$ Black Hole?} \apjl
  699(2):L169--L173, \doi{10.1088/0004-637X/699/2/L169}, \eprint{0906.4894}

\bibitem[{{Ibata} et~al.(2019){Ibata}, {Bellazzini}, {Malhan}, {Martin}, and
  {Bianchini}}]{ibata19}
{Ibata} RA, {Bellazzini} M, {Malhan} K, {Martin} N, {Bianchini} P (2019)
  {Identification of the long stellar stream of the prototypical massive
  globular cluster {\ensuremath{\omega}} Centauri}. Nature Astronomy
  3:667--672, \doi{10.1038/s41550-019-0751-x}, \eprint{1902.09544}

\bibitem[{{Inayoshi} et~al.(2019){Inayoshi}, {Visbal}, and
  {Haiman}}]{inayoshi19}
{Inayoshi} K, {Visbal} E, {Haiman} Z (2019) {The Assembly of the First Massive
  Black Holes}. arXiv e-prints arXiv:1911.05791, \eprint{1911.05791}

\bibitem[{{Janz} et~al.(2016){Janz}, {Norris}, {Forbes}, {Huxor}, {Romanowsky},
  {Frank}, {Escudero}, {Faifer}, {Forte}, {Kannappan}, {Maraston}, {Brodie},
  {Strader}, and {Thompson}}]{janz16}
{Janz} J, {Norris} MA, {Forbes} DA, {Huxor} A, {Romanowsky} AJ, {Frank} MJ,
  {Escudero} CG, {Faifer} FR, {Forte} JC, {Kannappan} SJ, {Maraston} C,
  {Brodie} JP, {Strader} J, {Thompson} BR (2016) {The AIMSS Project - III. The
  stellar populations of compact stellar systems}. \mnras 456(1):617--632,
  \doi{10.1093/mnras/stv2636}, \eprint{1511.03264}

\bibitem[{{Jarrett} et~al.(2000){Jarrett}, {Chester}, {Cutri}, {Schneider},
  {Skrutskie}, and {Huchra}}]{jarrett00}
{Jarrett} TH, {Chester} T, {Cutri} R, {Schneider} S, {Skrutskie} M, {Huchra} JP
  (2000) {2MASS Extended Source Catalog: Overview and Algorithms}. \aj
  119(5):2498--2531, \doi{10.1086/301330}, \eprint{astro-ph/0004318}

\bibitem[{{Jarrett} et~al.(2003){Jarrett}, {Chester}, {Cutri}, {Schneider}, and
  {Huchra}}]{jarrett03}
{Jarrett} TH, {Chester} T, {Cutri} R, {Schneider} SE, {Huchra} JP (2003) {The
  2MASS Large Galaxy Atlas}. \aj 125(2):525--554, \doi{10.1086/345794}

\bibitem[{{Jennings} et~al.(2015){Jennings}, {Romanowsky}, {Brodie}, {Janz},
  {Norris}, {Forbes}, {Martinez-Delgado}, {Fagioli}, and {Penny}}]{jennings15}
{Jennings} ZG, {Romanowsky} AJ, {Brodie} JP, {Janz} J, {Norris} MA, {Forbes}
  DA, {Martinez-Delgado} D, {Fagioli} M, {Penny} SJ (2015) {NGC 3628-UCD1: A
  Possible {\ensuremath{\omega}} Cen Analog Embedded in a Stellar Stream}.
  \apjl 812(1):L10, \doi{10.1088/2041-8205/812/1/L10}, \eprint{1509.04710}

\bibitem[{{Johnson} and {Pilachowski}(2010)}]{johnson10}
{Johnson} CI, {Pilachowski} CA (2010) {Chemical Abundances for 855 Giants in
  the Globular Cluster Omega Centauri (NGC 5139)}. \apj 722(2):1373--1410,
  \doi{10.1088/0004-637X/722/2/1373}, \eprint{1008.2232}

\bibitem[{{Jord{\'a}n} et~al.(2009){Jord{\'a}n}, {Peng}, {Blakeslee},
  {C{\^o}t{\'e}}, {Eyheramendy}, {Ferrarese}, {Mei}, {Tonry}, and
  {West}}]{jordan09}
{Jord{\'a}n} A, {Peng} EW, {Blakeslee} JP, {C{\^o}t{\'e}} P, {Eyheramendy} S,
  {Ferrarese} L, {Mei} S, {Tonry} JL, {West} MJ (2009) {The ACS Virgo Cluster
  Survey XVI. Selection Procedure and Catalogs of Globular Cluster Candidates}.
  \apjs 180(1):54--66, \doi{10.1088/0067-0049/180/1/54}

\bibitem[{{Kacharov} et~al.(2018){Kacharov}, {Neumayer}, {Seth}, {Cappellari},
  {McDermid}, {Walcher}, and {B{\"o}ker}}]{kacharov18}
{Kacharov} N, {Neumayer} N, {Seth} AC, {Cappellari} M, {McDermid} R, {Walcher}
  CJ, {B{\"o}ker} T (2018) {Stellar populations and star formation histories of
  the nuclear star clusters in six nearby galaxies}. \mnras 480(2):1973--1998,
  \doi{10.1093/mnras/sty1985}, \eprint{1807.08765}

\bibitem[{{Kamann} et~al.(2018){Kamann}, {Husser}, {Dreizler}, {Emsellem},
  {Weilbacher}, {Martens}, {Bacon}, {den Brok}, {Giesers}, {Krajnovi{\'c}},
  {Roth}, {Wendt}, and {Wisotzki}}]{kamann18}
{Kamann} S, {Husser} TO, {Dreizler} S, {Emsellem} E, {Weilbacher} PM, {Martens}
  S, {Bacon} R, {den Brok} M, {Giesers} B, {Krajnovi{\'c}} D, {Roth} MM,
  {Wendt} M, {Wisotzki} L (2018) {A stellar census in globular clusters with
  MUSE: The contribution of rotation to cluster dynamics studied with 200 000
  stars}. \mnras 473(4):5591--5616, \doi{10.1093/mnras/stx2719},
  \eprint{1710.07257}

\bibitem[{{Karachentsev} et~al.(2013){Karachentsev}, {Makarov}, and
  {Kaisina}}]{karachentsev13}
{Karachentsev} ID, {Makarov} DI, {Kaisina} EI (2013) {Updated Nearby Galaxy
  Catalog}. \aj 145(4):101, \doi{10.1088/0004-6256/145/4/101},
  \eprint{1303.5328}

\bibitem[{{Khan} et~al.(2019){Khan}, {Awais Mirza}, and
  {Holley-Bockelmann}}]{khan19}
{Khan} FM, {Awais Mirza} M, {Holley-Bockelmann} K (2019) {Inward Bound: The
  incredible journey of massive black holes as they pair and merge; I. The
  effect of mass ratio in flattened rotating galactic nuclei}. arXiv e-prints
  arXiv:1911.07946, \eprint{1911.07946}

\bibitem[{{Kieffer} and {Bogdanovi{\'c}}(2016)}]{kieffer16}
{Kieffer} TF, {Bogdanovi{\'c}} T (2016) {Can Star-Disk Collisions Explain the
  Missing Red Giants Problem in the Galactic Center?} \apj 823(2):155,
  \doi{10.3847/0004-637X/823/2/155}, \eprint{1602.03527}

\bibitem[{{Kim} et~al.(2004){Kim}, {Sung}, {Park}, and {Sung}}]{kim04}
{Kim} SC, {Sung} H, {Park} HS, {Sung} EC (2004) {UBVI Surface Photometry of the
  Spiral Galaxy NGC 300 in the Sculptor Group}. Chinese Journal of Astronomy
  and Astrophysics 4:299--310, \doi{10.1088/1009-9271/4/4/299},
  \eprint{astro-ph/0404036}

\bibitem[{{Kim} and {Elmegreen}(2017)}]{kim17}
{Kim} WT, {Elmegreen} BG (2017) {Nuclear Spiral Shocks and Induced Gas Inflows
  in Weak Oval Potentials}. \apjl 841(1):L4, \doi{10.3847/2041-8213/aa70a1},
  \eprint{1705.00863}

\bibitem[{{King}(1962)}]{king62}
{King} I (1962) {The structure of star clusters. I. an empirical density law}.
  \aj 67:471, \doi{10.1086/108756}

\bibitem[{{King} et~al.(1995){King}, {Stanford}, and {Crane}}]{king95}
{King} IR, {Stanford} SA, {Crane} P (1995) {Far-UV Properties of the Nuclear
  Region of M31}. \aj 109:164, \doi{10.1086/117264}

\bibitem[{{Kirby} et~al.(2013){Kirby}, {Cohen}, {Guhathakurta}, {Cheng},
  {Bullock}, and {Gallazzi}}]{kirby13}
{Kirby} EN, {Cohen} JG, {Guhathakurta} P, {Cheng} L, {Bullock} JS, {Gallazzi} A
  (2013) {The Universal Stellar Mass-Stellar Metallicity Relation for Dwarf
  Galaxies}. \apj 779(2):102, \doi{10.1088/0004-637X/779/2/102},
  \eprint{1310.0814}

\bibitem[{{Koleva} et~al.(2009){Koleva}, {de Rijcke}, {Prugniel}, {Zeilinger},
  and {Michielsen}}]{koleva09}
{Koleva} M, {de Rijcke} S, {Prugniel} P, {Zeilinger} WW, {Michielsen} D (2009)
  {Formation and evolution of dwarf elliptical galaxies - II. Spatially
  resolved star formation histories}. \mnras 396(4):2133--2151,
  \doi{10.1111/j.1365-2966.2009.14820.x}, \eprint{0903.4393}

\bibitem[{{Koleva} et~al.(2011){Koleva}, {Prugniel}, {De Rijcke}, and
  {Zeilinger}}]{koleva11}
{Koleva} M, {Prugniel} P, {De Rijcke} S, {Zeilinger} WW (2011) {Age and
  metallicity gradients in early-type galaxies: a dwarf-to-giant sequence}.
  \mnras 417(3):1643--1671, \doi{10.1111/j.1365-2966.2011.19057.x},
  \eprint{1105.4809}

\bibitem[{{Kormendy}(1985)}]{kormendy85}
{Kormendy} J (1985) {Brightness profiles of the cores of bulges and elliptical
  galaxies}. \apjl 292:L9--L13, \doi{10.1086/184463}

\bibitem[{{Kormendy} and {Bender}(1999)}]{kormendy99}
{Kormendy} J, {Bender} R (1999) {The Double Nucleus and Central Black Hole of
  M31}. \apj 522(2):772--792, \doi{10.1086/307665}

\bibitem[{{Kormendy} and {Djorgovski}(1989)}]{kormendy89}
{Kormendy} J, {Djorgovski} S (1989) {Surface photometry and the structure of
  elliptical galaxies}. \araa 27:235--277,
  \doi{10.1146/annurev.aa.27.090189.001315}

\bibitem[{{Kormendy} and {Ho}(2013)}]{kormendy13}
{Kormendy} J, {Ho} LC (2013) {Coevolution (Or Not) of Supermassive Black Holes
  and Host Galaxies}. \araa 51(1):511--653,
  \doi{10.1146/annurev-astro-082708-101811}, \eprint{1304.7762}

\bibitem[{{Kormendy} and {McClure}(1993)}]{kormendy95}
{Kormendy} J, {McClure} RD (1993) {The Nucleus of M33}. \aj 105:1793,
  \doi{10.1086/116555}

\bibitem[{{Kormendy} et~al.(2009){Kormendy}, {Fisher}, {Cornell}, and
  {Bender}}]{kormendy09}
{Kormendy} J, {Fisher} DB, {Cornell} ME, {Bender} R (2009) {Structure and
  Formation of Elliptical and Spheroidal Galaxies}. \apjs 182(1):216--309,
  \doi{10.1088/0067-0049/182/1/216}, \eprint{0810.1681}

\bibitem[{{Kormendy} et~al.(2010){Kormendy}, {Drory}, {Bender}, and
  {Cornell}}]{kormendy10}
{Kormendy} J, {Drory} N, {Bender} R, {Cornell} ME (2010) {Bulgeless Giant
  Galaxies Challenge Our Picture of Galaxy Formation by Hierarchical
  Clustering}. \apj 723(1):54--80, \doi{10.1088/0004-637X/723/1/54},
  \eprint{1009.3015}

\bibitem[{{Krabbe} et~al.(1991){Krabbe}, {Genzel}, {Drapatz}, and
  {Rotaciuc}}]{krabbe91}
{Krabbe} A, {Genzel} R, {Drapatz} S, {Rotaciuc} V (1991) {A Cluster of He i
  Emission-Line Stars in the Galactic Center}. \apj 382:L19,
  \doi{10.1086/186204}

\bibitem[{{Krajnovi{\'c}} et~al.(2018){Krajnovi{\'c}}, {Cappellari},
  {McDermid}, {Thater}, {Nyland}, {de Zeeuw}, {Falc{\'o}n-Barroso}, {Khochfar},
  {Kuntschner}, {Sarzi}, and {Young}}]{krajnovic18}
{Krajnovi{\'c}} D, {Cappellari} M, {McDermid} RM, {Thater} S, {Nyland} K, {de
  Zeeuw} PT, {Falc{\'o}n-Barroso} J, {Khochfar} S, {Kuntschner} H, {Sarzi} M,
  {Young} LM (2018) {A quartet of black holes and a missing duo: probing the
  low end of the M$_{BH}$-{\ensuremath{\sigma}} relation with the adaptive
  optics assisted integral-field spectroscopy}. \mnras 477(3):3030--3064,
  \doi{10.1093/mnras/sty778}, \eprint{1803.08055}

\bibitem[{{Kruijssen} et~al.(2019){Kruijssen}, {Pfeffer}, {Reina-Campos},
  {Crain}, and {Bastian}}]{kruijssen19}
{Kruijssen} JMD, {Pfeffer} JL, {Reina-Campos} M, {Crain} RA, {Bastian} N (2019)
  {The formation and assembly history of the Milky Way revealed by its globular
  cluster population}. \mnras 486(3):3180--3202, \doi{10.1093/mnras/sty1609},
  \eprint{1806.05680}

\bibitem[{{Krumholz} et~al.(2019){Krumholz}, {McKee}, and {Bland
  -Hawthorn}}]{krumholz18}
{Krumholz} MR, {McKee} CF, {Bland -Hawthorn} J (2019) {Star Clusters Across
  Cosmic Time}. \araa 57:227--303, \doi{10.1146/annurev-astro-091918-104430},
  \eprint{1812.01615}

\bibitem[{{Lacy} et~al.(1980){Lacy}, {Townes}, {Geballe}, and
  {Hollenbach}}]{lacy80}
{Lacy} JH, {Townes} CH, {Geballe} TR, {Hollenbach} DJ (1980) {Observations of
  the motion and distribution of the ionized gas in the central parsec of the
  Galaxy. II.} \apj 241:132--146, \doi{10.1086/158324}

\bibitem[{{Lacy} et~al.(1982){Lacy}, {Townes}, and {Hollenbach}}]{lacy82}
{Lacy} JH, {Townes} CH, {Hollenbach} DJ (1982) {The nature of the central
  parsec of the Galaxy}. \apj 262:120--134, \doi{10.1086/160402}

\bibitem[{{Lamers} et~al.(2017){Lamers}, {Kruijssen}, {Bastian}, {Rejkuba},
  {Hilker}, and {Kissler-Patig}}]{lamers17}
{Lamers} HJGLM, {Kruijssen} JMD, {Bastian} N, {Rejkuba} M, {Hilker} M,
  {Kissler-Patig} M (2017) {The difference in metallicity distribution
  functions of halo stars and globular clusters as a function of galaxy type. A
  tracer of globular cluster formation and evolution}. \aap 606:A85,
  \doi{10.1051/0004-6361/201731062}, \eprint{1706.00939}

\bibitem[{{Larsen}(1999)}]{larsen99}
{Larsen} SS (1999) {Young massive star clusters in nearby galaxies. II.
  Software tools, data reductions and cluster sizes}. \aaps 139:393--415,
  \doi{10.1051/aas:1999509}, \eprint{astro-ph/9907163}

\bibitem[{{Lauer} et~al.(1993){Lauer}, {Faber}, {Groth}, {Shaya}, {Campbell},
  {Code}, {Currie}, {Baum}, {Ewald}, {Hester}, {Holtzman}, {Kristian}, {Light},
  {Ligynds}, {O'Neil}, and {Westphal}}]{lauer93}
{Lauer} TR, {Faber} SM, {Groth} EJ, {Shaya} EJ, {Campbell} B, {Code} A,
  {Currie} DG, {Baum} WA, {Ewald} SP, {Hester} JJ, {Holtzman} JA, {Kristian} J,
  {Light} RM, {Ligynds} CR, {O'Neil} J E~J, {Westphal} JA (1993) {Planetary
  Camera Observations of the Double Nucleus of M31}. \aj 106:1436,
  \doi{10.1086/116737}

\bibitem[{{Lauer} et~al.(1998){Lauer}, {Faber}, {Ajhar}, {Grillmair}, and
  {Scowen}}]{lauer98}
{Lauer} TR, {Faber} SM, {Ajhar} EA, {Grillmair} CJ, {Scowen} PA (1998) {M32 +/-
  1}. \aj 116(5):2263--2286, \doi{10.1086/300617}, \eprint{astro-ph/9806277}

\bibitem[{{Lauer} et~al.(2005){Lauer}, {Faber}, {Gebhardt}, {Richstone},
  {Tremaine}, {Ajhar}, {Aller}, {Bender}, {Dressler}, {Filippenko}, {Green},
  {Grillmair}, {Ho}, {Kormendy}, {Magorrian}, {Pinkney}, and
  {Siopis}}]{lauer05}
{Lauer} TR, {Faber} SM, {Gebhardt} K, {Richstone} D, {Tremaine} S, {Ajhar} EA,
  {Aller} MC, {Bender} R, {Dressler} A, {Filippenko} AV, {Green} R, {Grillmair}
  CJ, {Ho} LC, {Kormendy} J, {Magorrian} J, {Pinkney} J, {Siopis} C (2005) {The
  Centers of Early-Type Galaxies with Hubble Space Telescope. V. New WFPC2
  Photometry}. \aj 129(5):2138--2185, \doi{10.1086/429565},
  \eprint{astro-ph/0412040}

\bibitem[{{Lauer} et~al.(2012){Lauer}, {Bender}, {Kormendy}, {Rosenfield}, and
  {Green}}]{lauer12}
{Lauer} TR, {Bender} R, {Kormendy} J, {Rosenfield} P, {Green} RF (2012) {The
  Cluster of Blue Stars Surrounding the M31 Nuclear Black Hole}. \apj
  745(2):121, \doi{10.1088/0004-637X/745/2/121}, \eprint{1112.1419}

\bibitem[{{Launhardt} et~al.(2002){Launhardt}, {Zylka}, and
  {Mezger}}]{launhardt02}
{Launhardt} R, {Zylka} R, {Mezger} PG (2002) {The nuclear bulge of the Galaxy.
  III. Large-scale physical characteristics of stars and interstellar matter}.
  \aap 384:112--139, \doi{10.1051/0004-6361:20020017},
  \eprint{astro-ph/0201294}

\bibitem[{{Law-Smith} et~al.(2017){Law-Smith}, {MacLeod}, {Guillochon},
  {Macias}, and {Ramirez-Ruiz}}]{law-smith17}
{Law-Smith} J, {MacLeod} M, {Guillochon} J, {Macias} P, {Ramirez-Ruiz} E (2017)
  {Low-mass White Dwarfs with Hydrogen Envelopes as a Missing Link in the Tidal
  Disruption Menu}. \apj 841(2):132, \doi{10.3847/1538-4357/aa6ffb},
  \eprint{1701.08162}

\bibitem[{{Leigh} et~al.(2012){Leigh}, {B{\"o}ker}, and {Knigge}}]{leigh12}
{Leigh} N, {B{\"o}ker} T, {Knigge} C (2012) {Nuclear star clusters and the
  stellar spheroids of their host galaxies}. \mnras 424(3):2130--2138,
  \doi{10.1111/j.1365-2966.2012.21365.x}, \eprint{1205.5033}

\bibitem[{{Leigh} et~al.(2014){Leigh}, {Mastrobuono-Battisti}, {Perets}, and
  {B{\"o}ker}}]{leigh14}
{Leigh} NWC, {Mastrobuono-Battisti} A, {Perets} HB, {B{\"o}ker} T (2014)
  {Stellar dynamics in gas: the role of gas damping}. \mnras 441(2):919--932,
  \doi{10.1093/mnras/stu622}, \eprint{1404.0379}

\bibitem[{{Leigh} et~al.(2016){Leigh}, {Antonini}, {Stone}, {Shara}, and
  {Merritt}}]{leigh16}
{Leigh} NWC, {Antonini} F, {Stone} NC, {Shara} MM, {Merritt} D (2016) {On the
  origins of enigmatic stellar populations in Local Group galactic nuclei}.
  \mnras 463(2):1605--1623, \doi{10.1093/mnras/stw2018}, \eprint{1608.02944}

\bibitem[{{Leigh} et~al.(2018){Leigh}, {Geller}, {McKernan}, {Ford}, {Mac Low},
  {Bellovary}, {Haiman}, {Lyra}, {Samsing}, {O'Dowd}, {Kocsis}, and
  {Endlich}}]{leigh18}
{Leigh} NWC, {Geller} AM, {McKernan} B, {Ford} KES, {Mac Low} MM, {Bellovary}
  J, {Haiman} Z, {Lyra} W, {Samsing} J, {O'Dowd} M, {Kocsis} B, {Endlich} S
  (2018) {On the rate of black hole binary mergers in galactic nuclei due to
  dynamical hardening}. \mnras 474(4):5672--5683, \doi{10.1093/mnras/stx3134},
  \eprint{1711.10494}

\bibitem[{{Levin} and {Beloborodov}(2003)}]{levin03}
{Levin} Y, {Beloborodov} AM (2003) {Stellar Disk in the Galactic Center: A
  Remnant of a Dense Accretion Disk?} \apjl 590(1):L33--L36,
  \doi{10.1086/376675}, \eprint{astro-ph/0303436}

\bibitem[{{Lezhnin} and {Vasiliev}(2016)}]{lezhnin16}
{Lezhnin} K, {Vasiliev} E (2016) {Tidal Disruption Rates in Non-spherical
  Galactic Nuclei Formed by Galaxy Mergers}. \apj 831(1):84,
  \doi{10.3847/0004-637X/831/1/84}, \eprint{1609.00009}

\bibitem[{{Li} and {Gnedin}(2019)}]{li19}
{Li} H, {Gnedin} OY (2019) {Star cluster formation in cosmological simulations
  - III. Dynamical and chemical evolution}. \mnras 486(3):4030--4043,
  \doi{10.1093/mnras/stz1114}, \eprint{1810.11036}

\bibitem[{{Li} et~al.(2017){Li}, {Gnedin}, {Gnedin}, {Meng}, {Semenov}, and
  {Kravtsov}}]{li17}
{Li} H, {Gnedin} OY, {Gnedin} NY, {Meng} X, {Semenov} VA, {Kravtsov} AV (2017)
  {Star Cluster Formation in Cosmological Simulations. I. Properties of Young
  Clusters}. \apj 834(1):69, \doi{10.3847/1538-4357/834/1/69},
  \eprint{1608.03244}

\bibitem[{{Light} et~al.(1974){Light}, {Danielson}, and
  {Schwarzschild}}]{light74}
{Light} ES, {Danielson} RE, {Schwarzschild} M (1974) {The nucleus of M31}. \apj
  194:257--263, \doi{10.1086/153241}

\bibitem[{{Lim} et~al.(2018){Lim}, {Peng}, {C{\^o}t{\'e}}, {Sales}, {den Brok},
  {Blakeslee}, and {Guhathakurta}}]{lim18}
{Lim} S, {Peng} EW, {C{\^o}t{\'e}} P, {Sales} LV, {den Brok} M, {Blakeslee} JP,
  {Guhathakurta} P (2018) {The Globular Cluster Systems of Ultra-diffuse
  Galaxies in the Coma Cluster}. \apj 862(1):82,
  \doi{10.3847/1538-4357/aacb81}, \eprint{1806.05425}

\bibitem[{{Lindqvist} et~al.(1992){Lindqvist}, {Habing}, and
  {Winnberg}}]{lindqvist92}
{Lindqvist} M, {Habing} HJ, {Winnberg} A (1992) {OH/IR stars close to the
  galactic centre. II. Their spatial and kinematics properties and the mass
  distribution within 5-100 PC from the galactic centre.} \aap 259:118--127

\bibitem[{{Lisker} et~al.(2007){Lisker}, {Grebel}, {Binggeli}, and
  {Glatt}}]{lisker07}
{Lisker} T, {Grebel} EK, {Binggeli} B, {Glatt} K (2007) {Virgo Cluster
  Early-Type Dwarf Galaxies with the Sloan Digital Sky Survey. III.
  Subpopulations: Distributions, Shapes, Origins}. \apj 660(2):1186--1197,
  \doi{10.1086/513090}, \eprint{astro-ph/0701429}

\bibitem[{{Lockhart} et~al.(2018){Lockhart}, {Lu}, {Peiris}, {Rich}, {Bouchez},
  and {Ghez}}]{lockhart18}
{Lockhart} KE, {Lu} JR, {Peiris} HV, {Rich} RM, {Bouchez} A, {Ghez} AM (2018)
  {A Slowly Precessing Disk in the Nucleus of M31 as the Feeding Mechanism for
  a Central Starburst}. \apj 854(2):121, \doi{10.3847/1538-4357/aaaa71},
  \eprint{1710.01394}

\bibitem[{{Loose} et~al.(1982){Loose}, {Kruegel}, and {Tutukov}}]{loose82}
{Loose} HH, {Kruegel} E, {Tutukov} A (1982) {Bursts of star formation in the
  galactic centre}. \aap 105(2):342--350

\bibitem[{{Lotz} et~al.(2001){Lotz}, {Telford}, {Ferguson}, {Miller},
  {Stiavelli}, and {Mack}}]{lotz01}
{Lotz} JM, {Telford} R, {Ferguson} HC, {Miller} BW, {Stiavelli} M, {Mack} J
  (2001) {Dynamical Friction in DE Globular Cluster Systems}. \apj 552:572--581

\bibitem[{{Lu} et~al.(2009){Lu}, {Ghez}, {Hornstein}, {Morris}, {Becklin}, and
  {Matthews}}]{lu09}
{Lu} JR, {Ghez} AM, {Hornstein} SD, {Morris} MR, {Becklin} EE, {Matthews} K
  (2009) {A Disk of Young Stars at the Galactic Center as Determined by
  Individual Stellar Orbits}. \apj 690(2):1463--1487,
  \doi{10.1088/0004-637X/690/2/1463}, \eprint{0808.3818}

\bibitem[{{Lu} et~al.(2013){Lu}, {Do}, {Ghez}, {Morris}, {Yelda}, and
  {Matthews}}]{lu13}
{Lu} JR, {Do} T, {Ghez} AM, {Morris} MR, {Yelda} S, {Matthews} K (2013)
  {Stellar Populations in the Central 0.5 pc of the Galaxy. II. The Initial
  Mass Function}. \apj 764(2):155, \doi{10.1088/0004-637X/764/2/155},
  \eprint{1301.0540}

\bibitem[{{Lyubenova} and {Tsatsi}(2019)}]{lyubenova19}
{Lyubenova} M, {Tsatsi} A (2019) {Nuclear angular momentum of early-type
  galaxies hosting nuclear star clusters}. arXiv e-prints \eprint{1903.10918}

\bibitem[{{Lyubenova} et~al.(2013){Lyubenova}, {van den Bosch}, {C{\^o}t{\'e}},
  {Kuntschner}, {van de Ven}, {Ferrarese}, {Jord{\'a}n}, {Infante}, and
  {Peng}}]{lyubenova13}
{Lyubenova} M, {van den Bosch} RCE, {C{\^o}t{\'e}} P, {Kuntschner} H, {van de
  Ven} G, {Ferrarese} L, {Jord{\'a}n} A, {Infante} L, {Peng} EW (2013) {The
  complex nature of the nuclear star cluster in FCC 277}. \mnras
  431(4):3364--3372, \doi{10.1093/mnras/stt414}, \eprint{1303.1210}

\bibitem[{{Ma} et~al.(2019){Ma}, {Grudi{\'c}}, {Quataert}, {Hopkins},
  {Faucher-Gigu{\`e}re}, {Boylan-Kolchin}, {Wetzel}, {Kim}, {Murray}, and
  {Kere{\v{s}}}}]{ma19}
{Ma} X, {Grudi{\'c}} MY, {Quataert} E, {Hopkins} PF, {Faucher-Gigu{\`e}re} CA,
  {Boylan-Kolchin} M, {Wetzel} A, {Kim} Jh, {Murray} N, {Kere{\v{s}}} D (2019)
  {Self-consistent proto-globular cluster formation in cosmological simulations
  of high-redshift galaxies}. arXiv e-prints arXiv:1906.11261,
  \eprint{1906.11261}

\bibitem[{{Maciejewski}(2004)}]{maciejewski04}
{Maciejewski} W (2004) {Nuclear spirals in galaxies: gas response to an
  asymmetric potential - II. Hydrodynamical models}. \mnras 354(3):892--904,
  \doi{10.1111/j.1365-2966.2004.08254.x}, \eprint{astro-ph/0408100}

\bibitem[{{Majewski} et~al.(2012){Majewski}, {Nidever}, {Smith}, {Damke},
  {Kunkel}, {Patterson}, {Bizyaev}, and {Garc{\'\i}a P{\'e}rez}}]{majewski12}
{Majewski} SR, {Nidever} DL, {Smith} VV, {Damke} GJ, {Kunkel} WE, {Patterson}
  RJ, {Bizyaev} D, {Garc{\'\i}a P{\'e}rez} AE (2012) {Exploring Halo
  Substructure with Giant Stars: Substructure in the Local Halo as Seen in the
  Grid Giant Star Survey Including Extended Tidal Debris from
  {\ensuremath{\omega}}Centauri}. \apjl 747(2):L37,
  \doi{10.1088/2041-8205/747/2/L37}, \eprint{1202.1832}

\bibitem[{{Maksym} et~al.(2014){Maksym}, {Ulmer}, {Roth}, {Irwin}, {Dupke},
  {Ho}, {Keel}, and {Adami}}]{maksym14}
{Maksym} WP, {Ulmer} MP, {Roth} KC, {Irwin} JA, {Dupke} R, {Ho} LC, {Keel} WC,
  {Adami} C (2014) {Deep spectroscopy of the $M_{V} \sim -14.8$ host galaxy of
  a tidal disruption flare in A1795}. \mnras 444(1):866--873,
  \doi{10.1093/mnras/stu1485}, \eprint{1407.6737}

\bibitem[{{Mapelli} et~al.(2012){Mapelli}, {Hayfield}, {Mayer}, and
  {Wadsley}}]{mapelli12}
{Mapelli} M, {Hayfield} T, {Mayer} L, {Wadsley} J (2012) {In Situ Formation of
  SgrA* Stars Via Disk Fragmentation: Parent Cloud Properties and
  Thermodynamics}. \apj 749(2):168, \doi{10.1088/0004-637X/749/2/168},
  \eprint{1202.0555}

\bibitem[{{Marino} et~al.(2012){Marino}, {Milone}, {Piotto}, {Cassisi},
  {D'Antona}, {Anderson}, {Aparicio}, {Bedin}, {Renzini}, and
  {Villanova}}]{marino12}
{Marino} AF, {Milone} AP, {Piotto} G, {Cassisi} S, {D'Antona} F, {Anderson} J,
  {Aparicio} A, {Bedin} LR, {Renzini} A, {Villanova} S (2012) {The C+N+O
  Abundance of {\ensuremath{\omega}} Centauri Giant Stars: Implications for the
  Chemical-enrichment Scenario and the Relative Ages of Different Stellar
  Populations}. \apj 746(1):14, \doi{10.1088/0004-637X/746/1/14},
  \eprint{1111.1891}

\bibitem[{{Marino} et~al.(2015){Marino}, {Milone}, {Karakas}, {Casagrand e},
  {Yong}, {Shingles}, {Da Costa}, {Norris}, {Stetson}, {Lind}, {Asplund},
  {Collet}, {Jerjen}, {Sbordone}, {Aparicio}, and {Cassisi}}]{marino15}
{Marino} AF, {Milone} AP, {Karakas} AI, {Casagrand e} L, {Yong} D, {Shingles}
  L, {Da Costa} G, {Norris} JE, {Stetson} PB, {Lind} K, {Asplund} M, {Collet}
  R, {Jerjen} H, {Sbordone} L, {Aparicio} A, {Cassisi} S (2015) {Iron and
  s-elements abundance variations in NGC 5286: comparison with `anomalous'
  globular clusters and Milky Way satellites}. \mnras 450(1):815--845,
  \doi{10.1093/mnras/stv420}, \eprint{1502.07438}

\bibitem[{{Martini} and {Ho}(2004)}]{martini04}
{Martini} P, {Ho} LC (2004) {A Population of Massive Globular Clusters in NGC
  5128}. \apj 610(1):233--246, \doi{10.1086/421458}, \eprint{astro-ph/0404003}

\bibitem[{{Matthews} and {Gallagher}(1997)}]{matthews97}
{Matthews} LD, {Gallagher} JS III (1997) {B and V CCD Photometry of Southern,
  Extreme Late-Type Spiral Galaxies}. \aj 114:1899, \doi{10.1086/118613},
  \eprint{astro-ph/9709145}

\bibitem[{{Matthews} et~al.(1999){Matthews}, {Gallagher}, {Krist}, {Watson},
  {Burrows}, {Griffiths}, {Hester}, {Trauger}, {Ballester}, {Clarke}, {Crisp},
  {Evans}, {Hoessel}, {Holtzman}, {Mould}, {Scowen}, {Stapelfeldt}, and
  {Westphal}}]{matthews99}
{Matthews} LD, {Gallagher} JS III, {Krist} JE, {Watson} AM, {Burrows} CJ,
  {Griffiths} RE, {Hester} JJ, {Trauger} JT, {Ballester} GE, {Clarke} JT,
  {Crisp} D, {Evans} RW, {Hoessel} JG, {Holtzman} JA, {Mould} JR, {Scowen} PA,
  {Stapelfeldt} KR, {Westphal} JA (1999) {WFPC2 Observations of Compact Star
  Cluster Nuclei in Low-Luminosity Spiral Galaxies}. \aj 118:208--235,
  \doi{10.1086/300909}, \eprint{astro-ph/9904205}

\bibitem[{{McConnachie}(2012)}]{mcconnachie12}
{McConnachie} AW (2012) {The Observed Properties of Dwarf Galaxies in and
  around the Local Group}. \aj 144(1):4, \doi{10.1088/0004-6256/144/1/4},
  \eprint{1204.1562}

\bibitem[{{McConnachie} et~al.(2005){McConnachie}, {Irwin}, {Ferguson},
  {Ibata}, {Lewis}, and {Tanvir}}]{mcconnachie05}
{McConnachie} AW, {Irwin} MJ, {Ferguson} AMN, {Ibata} RA, {Lewis} GF, {Tanvir}
  N (2005) {Distances and metallicities for 17 Local Group galaxies}. \mnras
  356(3):979--997, \doi{10.1111/j.1365-2966.2004.08514.x},
  \eprint{astro-ph/0410489}

\bibitem[{{McGinn} et~al.(1989){McGinn}, {Sellgren}, {Becklin}, and
  {Hall}}]{mcginn89}
{McGinn} MT, {Sellgren} K, {Becklin} EE, {Hall} DNB (1989) {Stellar Kinematics
  in the Galactic Center}. \apj 338:824, \doi{10.1086/167239}

\bibitem[{{McKernan} et~al.(2018){McKernan}, {Ford}, {Bellovary}, {Leigh},
  {Haiman}, {Kocsis}, {Lyra}, {Mac Low}, {Metzger}, {O'Dowd}, {Endlich}, and
  {Rosen}}]{mckernan18}
{McKernan} B, {Ford} KES, {Bellovary} J, {Leigh} NWC, {Haiman} Z, {Kocsis} B,
  {Lyra} W, {Mac Low} MM, {Metzger} B, {O'Dowd} M, {Endlich} S, {Rosen} DJ
  (2018) {Constraining Stellar-mass Black Hole Mergers in AGN Disks Detectable
  with LIGO}. \apj 866(1):66, \doi{10.3847/1538-4357/aadae5},
  \eprint{1702.07818}

\bibitem[{{McLaughlin}(1999)}]{mclaughlin99}
{McLaughlin} DE (1999) {The Efficiency of Globular Cluster Formation}. \aj
  117(5):2398--2427, \doi{10.1086/300836}, \eprint{astro-ph/9901283}

\bibitem[{{McLaughlin} and {van der Marel}(2005)}]{mclaughlin05}
{McLaughlin} DE, {van der Marel} RP (2005) {Resolved Massive Star Clusters in
  the Milky Way and Its Satellites: Brightness Profiles and a Catalog of
  Fundamental Parameters}. \apjs 161(2):304--360, \doi{10.1086/497429},
  \eprint{astro-ph/0605132}

\bibitem[{{Meadows} et~al.(2019){Meadows}, {Navarro}, {Santos-Santos},
  {Ben{\'\i}tez-Llambay}, and {Frenk}}]{meadows19}
{Meadows} N, {Navarro} JF, {Santos-Santos} I, {Ben{\'\i}tez-Llambay} A, {Frenk}
  C (2019) {Cusp or core? Revisiting the globular cluster timing problem in
  Fornax}. \mnras p 2938, \doi{10.1093/mnras/stz3280}

\bibitem[{{Merritt}(2009)}]{merritt09}
{Merritt} D (2009) {Evolution of Nuclear Star Clusters}. \apj 694:959--970,
  \doi{10.1088/0004-637X/694/2/959}, \eprint{0802.3186}

\bibitem[{{Merritt}(2010)}]{merritt10}
{Merritt} D (2010) {The Distribution of Stars and Stellar Remnants at the
  Galactic Center}. \apj 718(2):739--761, \doi{10.1088/0004-637X/718/2/739},
  \eprint{0909.1318}

\bibitem[{{Merritt} et~al.(2011){Merritt}, {Alexander}, {Mikkola}, and
  {Will}}]{merritt11}
{Merritt} D, {Alexander} T, {Mikkola} S, {Will} CM (2011) {Stellar dynamics of
  extreme-mass-ratio inspirals}. \prd 84(4):044024,
  \doi{10.1103/PhysRevD.84.044024}, \eprint{1102.3180}

\bibitem[{{Mezcua}(2017)}]{mezcua17}
{Mezcua} M (2017) {Observational evidence for intermediate-mass black holes}.
  International Journal of Modern Physics D 26(11):1730021,
  \doi{10.1142/S021827181730021X}, \eprint{1705.09667}

\bibitem[{{Mieske} et~al.(2013){Mieske}, {Frank}, {Baumgardt},
  {L{\"u}tzgendorf}, {Neumayer}, and {Hilker}}]{mieske13}
{Mieske} S, {Frank} MJ, {Baumgardt} H, {L{\"u}tzgendorf} N, {Neumayer} N,
  {Hilker} M (2013) {On central black holes in ultra-compact dwarf galaxies}.
  \aap 558:A14, \doi{10.1051/0004-6361/201322167}, \eprint{1308.1398}

\bibitem[{{Mihos} and {Hernquist}(1994)}]{mihos94}
{Mihos} JC, {Hernquist} L (1994) {Dense Stellar Cores in Merger Remnants}. \apj
  437:L47, \doi{10.1086/187679}

\bibitem[{{Miller} et~al.(2015){Miller}, {Gallo}, {Greene}, {Kelly}, {Treu},
  {Woo}, and {Baldassare}}]{miller15}
{Miller} BP, {Gallo} E, {Greene} JE, {Kelly} BC, {Treu} T, {Woo} JH,
  {Baldassare} V (2015) {X-Ray Constraints on the Local Supermassive Black Hole
  Occupation Fraction}. \apj 799(1):98, \doi{10.1088/0004-637X/799/1/98},
  \eprint{1403.4246}

\bibitem[{{Miller} and {Lotz}(2007)}]{miller07}
{Miller} BW, {Lotz} JM (2007) {The Globular Cluster Luminosity Function and
  Specific Frequency in Dwarf Elliptical Galaxies}. \apj 670(2):1074--1089,
  \doi{10.1086/522323}, \eprint{0708.2511}

\bibitem[{{Miller} and {Davies}(2012)}]{miller12}
{Miller} MC, {Davies} MB (2012) {An Upper Limit to the Velocity Dispersion of
  Relaxed Stellar Systems without Massive Black Holes}. \apj 755(1):81,
  \doi{10.1088/0004-637X/755/1/81}, \eprint{1206.6167}

\bibitem[{{Miller} and {Hamilton}(2002)}]{miller02}
{Miller} MC, {Hamilton} DP (2002) {Production of intermediate-mass black holes
  in globular clusters}. \mnras 330(1):232--240,
  \doi{10.1046/j.1365-8711.2002.05112.x}, \eprint{astro-ph/0106188}

\bibitem[{{Miller} and {Lauburg}(2009)}]{miller09}
{Miller} MC, {Lauburg} VM (2009) {Mergers of Stellar-Mass Black Holes in
  Nuclear Star Clusters}. \apj 692(1):917--923,
  \doi{10.1088/0004-637X/692/1/917}, \eprint{0804.2783}

\bibitem[{{Miller-Jones} et~al.(2012){Miller-Jones}, {Wrobel}, {Sivakoff},
  {Heinke}, {Miller}, {Plotkin}, {Di Stefano}, {Greene}, {Ho}, {Joseph},
  {Kong}, and {Maccarone}}]{miller-jones12}
{Miller-Jones} JCA, {Wrobel} JM, {Sivakoff} GR, {Heinke} CO, {Miller} RE,
  {Plotkin} RM, {Di Stefano} R, {Greene} JE, {Ho} LC, {Joseph} TD, {Kong} AKH,
  {Maccarone} TJ (2012) {The Absence of Radio Emission from the Globular
  Cluster G1}. \apjl 755(1):L1, \doi{10.1088/2041-8205/755/1/L1},
  \eprint{1206.5729}

\bibitem[{{Milone} et~al.(2017){Milone}, {Piotto}, {Renzini}, {Marino},
  {Bedin}, {Vesperini}, {D'Antona}, {Nardiello}, {Anderson}, {King}, {Yong},
  {Bellini}, {Aparicio}, {Barbuy}, {Brown}, {Cassisi}, {Ortolani}, {Salaris},
  {Sarajedini}, and {van der Marel}}]{milone17}
{Milone} AP, {Piotto} G, {Renzini} A, {Marino} AF, {Bedin} LR, {Vesperini} E,
  {D'Antona} F, {Nardiello} D, {Anderson} J, {King} IR, {Yong} D, {Bellini} A,
  {Aparicio} A, {Barbuy} B, {Brown} TM, {Cassisi} S, {Ortolani} S, {Salaris} M,
  {Sarajedini} A, {van der Marel} RP (2017) {The Hubble Space Telescope UV
  Legacy Survey of Galactic globular clusters - IX. The Atlas of multiple
  stellar populations}. \mnras 464(3):3636--3656, \doi{10.1093/mnras/stw2531},
  \eprint{1610.00451}

\bibitem[{{Milosavljevi{\'c}}(2004)}]{milosavljevic04}
{Milosavljevi{\'c}} M (2004) {On the Origin of Nuclear Star Clusters in
  Late-Type Spiral Galaxies}. \apj 605(1):L13--L16, \doi{10.1086/420696},
  \eprint{astro-ph/0310574}

\bibitem[{{Milosavljevi{\'c}} and {Merritt}(2001)}]{milosavljevic01}
{Milosavljevi{\'c}} M, {Merritt} D (2001) {Formation of Galactic Nuclei}. \apj
  563(1):34--62, \doi{10.1086/323830}, \eprint{astro-ph/0103350}

\bibitem[{{Milosavljevi{\'c}} et~al.(2006){Milosavljevi{\'c}}, {Merritt}, and
  {Ho}}]{milosavljevic06}
{Milosavljevi{\'c}} M, {Merritt} D, {Ho} LC (2006) {Contribution of Stellar
  Tidal Disruptions to the X-Ray Luminosity Function of Active Galaxies}. \apj
  652(1):120--125, \doi{10.1086/508134}, \eprint{astro-ph/0602289}

\bibitem[{{Misgeld} and {Hilker}(2011)}]{misgeld11}
{Misgeld} I, {Hilker} M (2011) {Families of dynamically hot stellar systems
  over 10 orders of magnitude in mass}. \mnras 414(4):3699--3710,
  \doi{10.1111/j.1365-2966.2011.18669.x}, \eprint{1103.1628}

\bibitem[{{Monaco} et~al.(2005){Monaco}, {Bellazzini}, {Ferraro}, and
  {Pancino}}]{monaco05}
{Monaco} L, {Bellazzini} M, {Ferraro} FR, {Pancino} E (2005) {The central
  density cusp of the Sagittarius dwarf spheroidal galaxy}. \mnras
  356:1396--1402, \doi{10.1111/j.1365-2966.2004.08579.x},
  \eprint{arXiv:astro-ph/0411107}

\bibitem[{{Monaco} et~al.(2009){Monaco}, {Saviane}, {Perina}, {Bellazzini},
  {Buzzoni}, {Federici}, {Fusi Pecci}, and {Galleti}}]{monaco09}
{Monaco} L, {Saviane} I, {Perina} S, {Bellazzini} M, {Buzzoni} A, {Federici} L,
  {Fusi Pecci} F, {Galleti} S (2009) {The young stellar population at the
  center of NGC 205}. \aap 502(2):L9--L12, \doi{10.1051/0004-6361/200912412},
  \eprint{0907.0029}

\bibitem[{{Morelli} et~al.(2010){Morelli}, {Cesetti}, {Corsini}, {Pizzella},
  {Dalla Bont{\`a}}, {Sarzi}, and {Bertola}}]{morelli10}
{Morelli} L, {Cesetti} M, {Corsini} EM, {Pizzella} A, {Dalla Bont{\`a}} E,
  {Sarzi} M, {Bertola} F (2010) {Multiband photometric decomposition of nuclear
  stellar disks}. \aap 518:A32, \doi{10.1051/0004-6361/201014285},
  \eprint{1004.2190}

\bibitem[{{Mucciarelli} et~al.(2017){Mucciarelli}, {Bellazzini}, {Ibata},
  {Romano}, {Chapman}, and {Monaco}}]{mucciarelli17}
{Mucciarelli} A, {Bellazzini} M, {Ibata} R, {Romano} D, {Chapman} SC, {Monaco}
  L (2017) {Chemical abundances in the nucleus of the Sagittarius dwarf
  spheroidal galaxy}. \aap 605:A46, \doi{10.1051/0004-6361/201730707},
  \eprint{1705.03251}

\bibitem[{{Naiman} et~al.(2015){Naiman}, {Ramirez-Ruiz}, {Debuhr}, and
  {Ma}}]{naiman15}
{Naiman} JP, {Ramirez-Ruiz} E, {Debuhr} J, {Ma} CP (2015) {The Role of Nuclear
  Star Clusters in Enhancing Supermassive Black Hole Feeding Rates During
  Galaxy Mergers}. \apj 803(2):81, \doi{10.1088/0004-637X/803/2/81},
  \eprint{1410.7381}

\bibitem[{{Nayakshin} et~al.(2009){Nayakshin}, {Wilkinson}, and
  {King}}]{nayakshin09}
{Nayakshin} S, {Wilkinson} MI, {King} A (2009) {Competitive feedback in galaxy
  formation}. \mnras 398(1):L54--L57, \doi{10.1111/j.1745-3933.2009.00709.x},
  \eprint{0907.1002}

\bibitem[{{Ness} et~al.(2013){Ness}, {Freeman}, {Athanassoula},
  {Wylie-de-Boer}, {Bland-Hawthorn}, {Asplund}, {Lewis}, {Yong}, {Lane}, and
  {Kiss}}]{ness13}
{Ness} M, {Freeman} K, {Athanassoula} E, {Wylie-de-Boer} E, {Bland-Hawthorn} J,
  {Asplund} M, {Lewis} GF, {Yong} D, {Lane} RR, {Kiss} LL (2013) {ARGOS - III.
  Stellar populations in the Galactic bulge of the Milky Way}. \mnras
  430(2):836--857, \doi{10.1093/mnras/sts629}, \eprint{1212.1540}

\bibitem[{{Neumayer} and {Walcher}(2012)}]{neumayer12}
{Neumayer} N, {Walcher} CJ (2012) {Are Nuclear Star Clusters the Precursors of
  Massive Black Holes?} Advances in Astronomy 2012:709038,
  \doi{10.1155/2012/709038}, \eprint{1201.4950}

\bibitem[{{Neumayer} et~al.(2011){Neumayer}, {Walcher}, {Andersen},
  {S{\'a}nchez}, {B{\"o}ker}, and {Rix}}]{neumayer11}
{Neumayer} N, {Walcher} CJ, {Andersen} D, {S{\'a}nchez} SF, {B{\"o}ker} T,
  {Rix} HW (2011) {Two-dimensional H{$\alpha$} kinematics of bulgeless disc
  galaxies}. \mnras 413:1875--1888, \doi{10.1111/j.1365-2966.2011.18266.x},
  \eprint{1101.5154}

\bibitem[{{Nguyen} et~al.(2014){Nguyen}, {Seth}, {Reines}, {den Brok}, {Sand},
  and {McLeod}}]{nguyen14}
{Nguyen} DD, {Seth} AC, {Reines} AE, {den Brok} M, {Sand} D, {McLeod} B (2014)
  {Extended Structure and Fate of the Nucleus in Henize 2-10}. \apj 794(1):34,
  \doi{10.1088/0004-637X/794/1/34}, \eprint{1408.4446}

\bibitem[{{Nguyen} et~al.(2017){Nguyen}, {Seth}, {den Brok}, {Neumayer},
  {Cappellari}, {Barth}, {Caldwell}, {Williams}, and {Binder}}]{nguyen17}
{Nguyen} DD, {Seth} AC, {den Brok} M, {Neumayer} N, {Cappellari} M, {Barth} AJ,
  {Caldwell} N, {Williams} BF, {Binder} B (2017) {Improved Dynamical
  Constraints on the Mass of the Central Black Hole in NGC 404}. \apj
  836(2):237, \doi{10.3847/1538-4357/aa5cb4}, \eprint{1610.09385}

\bibitem[{{Nguyen} et~al.(2018){Nguyen}, {Seth}, {Neumayer}, {Kamann},
  {Voggel}, {Cappellari}, {Picotti}, {Nguyen}, {B{\"o}ker}, {Debattista},
  {Caldwell}, {McDermid}, {Bastian}, {Ahn}, and {Pechetti}}]{nguyen18}
{Nguyen} DD, {Seth} AC, {Neumayer} N, {Kamann} S, {Voggel} KT, {Cappellari} M,
  {Picotti} A, {Nguyen} PM, {B{\"o}ker} T, {Debattista} V, {Caldwell} N,
  {McDermid} R, {Bastian} N, {Ahn} CC, {Pechetti} R (2018) {Nearby Early-type
  Galactic Nuclei at High Resolution: Dynamical Black Hole and Nuclear Star
  Cluster Mass Measurements}. \apj 858:118, \doi{10.3847/1538-4357/aabe28},
  \eprint{1711.04314}

\bibitem[{{Nguyen} et~al.(2019){Nguyen}, {Seth}, {Neumayer}, {Iguchi},
  {Cappellari}, {Strader}, {Chomiuk}, {Tremou}, {Pacucci}, and
  {Nakanishi}}]{nguyen19}
{Nguyen} DD, {Seth} AC, {Neumayer} N, {Iguchi} S, {Cappellari} M, {Strader} J,
  {Chomiuk} L, {Tremou} E, {Pacucci} F, {Nakanishi} K (2019) {Improved
  Dynamical Constraints on the Masses of the Central Black Holes in Nearby
  Low-mass Early-type Galactic Nuclei and the First Black Hole Determination
  for NGC 205}. \apj 872(1):104, \doi{10.3847/1538-4357/aafe7a},
  \eprint{1901.05496}

\bibitem[{{Nogueras-Lara} et~al.(2018){Nogueras-Lara}, {Sch{\"o}del}, {Dong},
  {Najarro}, {Gallego-Calvente}, {Hilker}, {Gallego-Cano}, {Nishiyama},
  {Neumayer}, and {Feldmeier-Krause}}]{nogueras-lara18}
{Nogueras-Lara} F, {Sch{\"o}del} R, {Dong} H, {Najarro} F, {Gallego-Calvente}
  AT, {Hilker} M, {Gallego-Cano} E, {Nishiyama} S, {Neumayer} N,
  {Feldmeier-Krause} A (2018) {Star formation history and metallicity in the
  Galactic inner bulge revealed by the red giant branch bump}. \aap 620:A83,
  \doi{10.1051/0004-6361/201833518}, \eprint{1809.07627}

\bibitem[{{Nogueras-Lara} et~al.(2019{\natexlab{a}}){Nogueras-Lara},
  {Sch{\"o}del}, {Gallego-Calvente}, {Gallego-Cano}, {Shahzamanian}, {Dong},
  {Neumayer}, {Hilker}, {Najarro}, {Nishiyama}, {Feldmeier-Krause}, {Girard},
  and {Cassisi}}]{nogueras-lara19b}
{Nogueras-Lara} F, {Sch{\"o}del} R, {Gallego-Calvente} AT, {Gallego-Cano} E,
  {Shahzamanian} B, {Dong} H, {Neumayer} N, {Hilker} M, {Najarro} F,
  {Nishiyama} S, {Feldmeier-Krause} A, {Girard} JHV, {Cassisi} S
  (2019{\natexlab{a}}) {The nuclear disc of the Milky Way: Early formation,
  long quiescence, and starburst activity one billion years ago}. Nature
  Astronomy arXiv:1910.06968, \eprint{1910.06968}

\bibitem[{{Nogueras-Lara} et~al.(2019{\natexlab{b}}){Nogueras-Lara},
  {Sch{\"o}del}, {Najarro}, {Gallego-Calvente}, {Gallego-Cano}, {Shahzamanian},
  and {Neumayer}}]{nogueras-lara19}
{Nogueras-Lara} F, {Sch{\"o}del} R, {Najarro} F, {Gallego-Calvente} AT,
  {Gallego-Cano} E, {Shahzamanian} B, {Neumayer} N (2019{\natexlab{b}})
  {Variability of the near-infrared extinction curve towards the Galactic
  centre}. \aap 630:L3, \doi{10.1051/0004-6361/201936322}

\bibitem[{{Norris} et~al.(1996){Norris}, {Freeman}, and {Mighell}}]{norris96}
{Norris} JE, {Freeman} KC, {Mighell} KJ (1996) {The Giant Branch of omega
  Centauri. V. The Calcium Abundance Distribution}. \apj 462:241,
  \doi{10.1086/177145}

\bibitem[{{Norris} et~al.(2014){Norris}, {Kannappan}, {Forbes}, {Romanowsky},
  {Brodie}, {Faifer}, {Huxor}, {Maraston}, {Moffett}, and {Penny}}]{norris14}
{Norris} MA, {Kannappan} SJ, {Forbes} DA, {Romanowsky} AJ, {Brodie} JP,
  {Faifer} FR, {Huxor} A, {Maraston} C, {Moffett} AJ, {Penny} SJ (2014) {The
  AIMSS Project -- I. Bridging the star cluster-galaxy divide}. \mnras
  443(2):1151--1172, \doi{10.1093/mnras/stu1186}, \eprint{1406.6065}

\bibitem[{{Norris} et~al.(2015){Norris}, {Escudero}, {Faifer}, {Kannappan},
  {Forte}, and {van den Bosch}}]{norris15}
{Norris} MA, {Escudero} CG, {Faifer} FR, {Kannappan} SJ, {Forte} JC, {van den
  Bosch} RCE (2015) {An extended star formation history in an ultra-compact
  dwarf}. \mnras 451(4):3615--3626, \doi{10.1093/mnras/stv1221},
  \eprint{1506.00004}

\bibitem[{{Noyola} et~al.(2010){Noyola}, {Gebhardt}, {Kissler-Patig},
  {L{\"u}tzgendorf}, {Jalali}, {de Zeeuw}, and {Baumgardt}}]{noyola10}
{Noyola} E, {Gebhardt} K, {Kissler-Patig} M, {L{\"u}tzgendorf} N, {Jalali} B,
  {de Zeeuw} PT, {Baumgardt} H (2010) {Very Large Telescope Kinematics for
  Omega Centauri: Further Support for a Central Black Hole}. \apjl
  719(1):L60--L64, \doi{10.1088/2041-8205/719/1/L60}, \eprint{1007.4559}

\bibitem[{{Oh} and {Lin}(2000)}]{oh00}
{Oh} KS, {Lin} DNC (2000) {Nucleation of Dwarf Galaxies in the Virgo Cluster}.
  \apj 543(2):620--633, \doi{10.1086/317118}

\bibitem[{{Ordenes-Brice{\~n}o} et~al.(2018){Ordenes-Brice{\~n}o}, {Puzia},
  {Eigenthaler}, {Taylor}, {Mu{\~n}oz}, {Zhang}, {Alamo-Mart{\'{\i}}nez},
  {Ribbeck}, {Grebel}, {{\'A}ngel}, {C{\^o}t{\'e}}, {Ferrarese}, {Hilker},
  {Lan{\c c}on}, {Mieske}, {Miller}, {Rong}, and
  {S{\'a}nchez-Janssen}}]{ordenes-briceno18}
{Ordenes-Brice{\~n}o} Y, {Puzia} TH, {Eigenthaler} P, {Taylor} MA, {Mu{\~n}oz}
  RP, {Zhang} H, {Alamo-Mart{\'{\i}}nez} K, {Ribbeck} KX, {Grebel} EK,
  {{\'A}ngel} S, {C{\^o}t{\'e}} P, {Ferrarese} L, {Hilker} M, {Lan{\c c}on} A,
  {Mieske} S, {Miller} BW, {Rong} Y, {S{\'a}nchez-Janssen} R (2018) {The Next
  Generation Fornax Survey (NGFS). IV. Mass and Age Bimodality of Nuclear
  Clusters in the Fornax Core Region}. \apj 860:4,
  \doi{10.3847/1538-4357/aac1b8}, \eprint{1805.00491}

\bibitem[{{Panamarev} et~al.(2019){Panamarev}, {Just}, {Spurzem}, {Berczik},
  {Wang}, and {Arca Sedda}}]{panamarev19}
{Panamarev} T, {Just} A, {Spurzem} R, {Berczik} P, {Wang} L, {Arca Sedda} M
  (2019) {Direct N-body simulation of the Galactic centre}. \mnras
  484(3):3279--3290, \doi{10.1093/mnras/stz208}, \eprint{1805.02153}

\bibitem[{{Paudel} et~al.(2011){Paudel}, {Lisker}, and {Kuntschner}}]{paudel11}
{Paudel} S, {Lisker} T, {Kuntschner} H (2011) {Nuclei of early-type dwarf
  galaxies: insights from stellar populations}. \mnras 413(3):1764--1776,
  \doi{10.1111/j.1365-2966.2011.18256.x}, \eprint{1012.4092}

\bibitem[{{Paumard} et~al.(2006){Paumard}, {Genzel}, {Martins}, {Nayakshin},
  {Beloborodov}, {Levin}, {Trippe}, {Eisenhauer}, {Ott}, and
  {Gillessen}}]{paumard06}
{Paumard} T, {Genzel} R, {Martins} F, {Nayakshin} S, {Beloborodov} AM, {Levin}
  Y, {Trippe} S, {Eisenhauer} F, {Ott} T, {Gillessen} S (2006) {The Two Young
  Star Disks in the Central Parsec of the Galaxy: Properties, Dynamics, and
  Formation}. \apj 643(2):1011--1035, \doi{10.1086/503273},
  \eprint{astro-ph/0601268}

\bibitem[{{Pechetti} et~al.(2017){Pechetti}, {Seth}, {Cappellari}, {McDermid},
  {den Brok}, {Mieske}, and {Strader}}]{pechetti17}
{Pechetti} R, {Seth} A, {Cappellari} M, {McDermid} R, {den Brok} M, {Mieske} S,
  {Strader} J (2017) {Detection of Enhanced Central Mass-to-light Ratios in
  Low-mass Early-type Galaxies: Evidence for Black Holes?} \apj 850(1):15,
  \doi{10.3847/1538-4357/aa9021}, \eprint{1709.09172}

\bibitem[{{Pechetti} et~al.(2019){Pechetti}, {Seth}, {Neumayer}, {Georgiev},
  {Kacharov}, and {den Brok}}]{pechetti19}
{Pechetti} R, {Seth} A, {Neumayer} N, {Georgiev} I, {Kacharov} N, {den Brok} M
  (2019) {Luminosity Models and Density Profiles for Nuclear Star Clusters for
  a Nearby Volume-Limited Sample of 29 Galaxies}. arXiv e-prints
  arXiv:1911.09686, \eprint{1911.09686}

\bibitem[{{Peiris} and {Tremaine}(2003)}]{peiris03}
{Peiris} HV, {Tremaine} S (2003) {Eccentric-Disk Models for the Nucleus of
  M31}. \apj 599(1):237--257, \doi{10.1086/378638}, \eprint{astro-ph/0307412}

\bibitem[{{Peng}(2002)}]{peng02}
{Peng} CY (2002) {Assessing Formation Scenarios for the Double Nucleus of M31
  Using Two-Dimensional Image Decomposition}. \aj 124(1):294--309,
  \doi{10.1086/340958}, \eprint{astro-ph/0204184}

\bibitem[{{Peterson} et~al.(2005){Peterson}, {Bentz}, {Desroches},
  {Filippenko}, {Ho}, {Kaspi}, {Laor}, {Maoz}, {Moran}, {Pogge}, and
  {Quillen}}]{peterson05}
{Peterson} BM, {Bentz} MC, {Desroches} LB, {Filippenko} AV, {Ho} LC, {Kaspi} S,
  {Laor} A, {Maoz} D, {Moran} EC, {Pogge} RW, {Quillen} AC (2005)
  {Multiwavelength Monitoring of the Dwarf Seyfert 1 Galaxy NGC 4395. I. A
  Reverberation-based Measurement of the Black Hole Mass}. \apj
  632(2):799--808, \doi{10.1086/444494}, \eprint{astro-ph/0506665}

\bibitem[{{Pfeffer} and {Baumgardt}(2013)}]{pfeffer13}
{Pfeffer} J, {Baumgardt} H (2013) {Ultra-compact dwarf galaxy formation by
  tidal stripping of nucleated dwarf galaxies}. \mnras 433:1997--2005,
  \doi{10.1093/mnras/stt867}, \eprint{1305.3656}

\bibitem[{{Pfeffer} et~al.(2014){Pfeffer}, {Griffen}, {Baumgardt}, and
  {Hilker}}]{pfeffer14}
{Pfeffer} J, {Griffen} BF, {Baumgardt} H, {Hilker} M (2014) {Contribution of
  stripped nuclear clusters to globular cluster and ultracompact dwarf galaxy
  populations}. \mnras 444:3670--3683, \doi{10.1093/mnras/stu1705},
  \eprint{1408.4467}

\bibitem[{{Pfeffer} et~al.(2016){Pfeffer}, {Hilker}, {Baumgardt}, and
  {Griffen}}]{pfeffer16}
{Pfeffer} J, {Hilker} M, {Baumgardt} H, {Griffen} BF (2016) {Constraining
  ultracompact dwarf galaxy formation with galaxy clusters in the local
  universe}. \mnras 458(3):2492--2508, \doi{10.1093/mnras/stw498},
  \eprint{1603.00032}

\bibitem[{{Pfeffer} et~al.(2018){Pfeffer}, {Kruijssen}, {Crain}, and
  {Bastian}}]{pfeffer18}
{Pfeffer} J, {Kruijssen} JMD, {Crain} RA, {Bastian} N (2018) {The E-MOSAICS
  project: simulating the formation and co-evolution of galaxies and their star
  cluster populations}. \mnras 475(4):4309--4346, \doi{10.1093/mnras/stx3124},
  \eprint{1712.00019}

\bibitem[{{Pfuhl} et~al.(2011){Pfuhl}, {Fritz}, {Zilka}, {Maness},
  {Eisenhauer}, {Genzel}, {Gillessen}, {Ott}, {Dodds-Eden}, and
  {Sternberg}}]{pfuhl11}
{Pfuhl} O, {Fritz} TK, {Zilka} M, {Maness} H, {Eisenhauer} F, {Genzel} R,
  {Gillessen} S, {Ott} T, {Dodds-Eden} K, {Sternberg} A (2011) {The Star
  Formation History of the Milky Way's Nuclear Star Cluster}. \apj 741(2):108,
  \doi{10.1088/0004-637X/741/2/108}, \eprint{1110.1633}

\bibitem[{{Phillips} et~al.(1996){Phillips}, {Illingworth}, {MacKenty}, and
  {Franx}}]{phillips96}
{Phillips} AC, {Illingworth} GD, {MacKenty} JW, {Franx} M (1996) {Nuclei of
  Nearby Disk Galaxies.I.A Hubble Space Telescope Imaging Survey}. \aj
  111:1566, \doi{10.1086/117896}

\bibitem[{{Portegies Zwart} and {McMillan}(2002)}]{portegies-zwart02}
{Portegies Zwart} SF, {McMillan} SLW (2002) {The Runaway Growth of
  Intermediate-Mass Black Holes in Dense Star Clusters}. \apj 576(2):899--907,
  \doi{10.1086/341798}, \eprint{astro-ph/0201055}

\bibitem[{{Portegies Zwart} et~al.(2004){Portegies Zwart}, {Baumgardt}, {Hut},
  {Makino}, and {McMillan}}]{portegies-zwart04}
{Portegies Zwart} SF, {Baumgardt} H, {Hut} P, {Makino} J, {McMillan} SLW (2004)
  {Formation of massive black holes through runaway collisions in dense young
  star clusters}. \nat 428(6984):724--726, \doi{10.1038/nature02448},
  \eprint{astro-ph/0402622}

\bibitem[{{Querejeta} et~al.(2015){Querejeta}, {Meidt}, {Schinnerer},
  {Cisternas}, {Mu{\~n}oz-Mateos}, {Sheth}, {Knapen}, {van de Ven}, {Norris},
  {Peletier}, {Laurikainen}, {Salo}, {Holwerda}, {Athanassoula}, {Bosma},
  {Groves}, {Ho}, {Gadotti}, {Zaritsky}, {Regan}, {Hinz}, {Gil de Paz},
  {Menendez-Delmestre}, {Seibert}, {Mizusawa}, {Kim}, {Erroz-Ferrer}, {Laine},
  and {Comer{\'o}n}}]{querejeta15}
{Querejeta} M, {Meidt} SE, {Schinnerer} E, {Cisternas} M, {Mu{\~n}oz-Mateos}
  JC, {Sheth} K, {Knapen} J, {van de Ven} G, {Norris} MA, {Peletier} R,
  {Laurikainen} E, {Salo} H, {Holwerda} BW, {Athanassoula} E, {Bosma} A,
  {Groves} B, {Ho} LC, {Gadotti} DA, {Zaritsky} D, {Regan} M, {Hinz} J, {Gil de
  Paz} A, {Menendez-Delmestre} K, {Seibert} M, {Mizusawa} T, {Kim} T,
  {Erroz-Ferrer} S, {Laine} J, {Comer{\'o}n} S (2015) {The Spitzer Survey of
  Stellar Structure in Galaxies (S$^{4}$G): Precise Stellar Mass Distributions
  from Automated Dust Correction at 3.6 {\ensuremath{\mu}}m}. \apjs 219(1):5,
  \doi{10.1088/0067-0049/219/1/5}, \eprint{1410.0009}

\bibitem[{{Quinlan} and {Hernquist}(1997)}]{quinlan97}
{Quinlan} GD, {Hernquist} L (1997) {The dynamical evolution of massive black
  hole binaries {\textemdash} II. Self-consistent N-body integrations}. \na
  2(6):533--554, \doi{10.1016/S1384-1076(97)00039-0}, \eprint{astro-ph/9706298}

\bibitem[{{Quinlan} and {Shapiro}(1987)}]{quinlan87}
{Quinlan} GD, {Shapiro} SL (1987) {The Collapse of Dense Star Clusters to
  Supermassive Black Holes: Binaries and Gravitational Radiation}. \apj
  321:199, \doi{10.1086/165624}

\bibitem[{{Ravindranath} et~al.(2001){Ravindranath}, {Ho}, {Peng},
  {Filippenko}, and {Sargent}}]{ravindranath01}
{Ravindranath} S, {Ho} LC, {Peng} CY, {Filippenko} AV, {Sargent} WLW (2001)
  {Central Structural Parameters of Early-Type Galaxies as Viewed with Nicmos
  on the Hubble Space Telescope}. \aj 122(2):653--678, \doi{10.1086/321175},
  \eprint{astro-ph/0105390}

\bibitem[{{Reaves}(1983)}]{reaves83}
{Reaves} G (1983) {A catalog of dwarf galaxies in Virgo}. \apjs 53:375--395,
  \doi{10.1086/190895}

\bibitem[{{Rees}(1988)}]{rees88}
{Rees} MJ (1988) {Tidal disruption of stars by black holes of
  {}10$^{6}$-{}10$^{8}$ solar masses in nearby galaxies}. \nat
  333(6173):523--528, \doi{10.1038/333523a0}

\bibitem[{{Reines} et~al.(2019){Reines}, {Condon}, {Darling}, and
  {Greene}}]{reines19}
{Reines} A, {Condon} J, {Darling} J, {Greene} J (2019) {A New Sample of
  (Wandering) Massive Black Holes in Dwarf Galaxies from High Resolution Radio
  Observations}. arXiv e-prints arXiv:1909.04670, \eprint{1909.04670}

\bibitem[{{Reines} and {Volonteri}(2015)}]{reines15}
{Reines} AE, {Volonteri} M (2015) {Relations between Central Black Hole Mass
  and Total Galaxy Stellar Mass in the Local Universe}. \apj 813(2):82,
  \doi{10.1088/0004-637X/813/2/82}, \eprint{1508.06274}

\bibitem[{{Reines} et~al.(2011){Reines}, {Sivakoff}, {Johnson}, and
  {Brogan}}]{reines11}
{Reines} AE, {Sivakoff} GR, {Johnson} KE, {Brogan} CL (2011) {An actively
  accreting massive black hole in the dwarf starburst galaxy Henize2-10}. \nat
  470(7332):66--68, \doi{10.1038/nature09724}, \eprint{1101.1309}

\bibitem[{{Reines} et~al.(2013){Reines}, {Greene}, and {Geha}}]{reines13}
{Reines} AE, {Greene} JE, {Geha} M (2013) {Dwarf Galaxies with Optical
  Signatures of Active Massive Black Holes}. \apj 775(2):116,
  \doi{10.1088/0004-637X/775/2/116}, \eprint{1308.0328}

\bibitem[{{Reines} et~al.(2016){Reines}, {Reynolds}, {Miller}, {Sivakoff},
  {Greene}, {Hickox}, and {Johnson}}]{reines16}
{Reines} AE, {Reynolds} MT, {Miller} JM, {Sivakoff} GR, {Greene} JE, {Hickox}
  RC, {Johnson} KE (2016) {Deep Chandra Observations of the Compact Starburst
  Galaxy Henize 2-10: X-Rays from the Massive Black Hole}. \apjl 830(2):L35,
  \doi{10.3847/2041-8205/830/2/L35}, \eprint{1610.01598}

\bibitem[{{Renaud}(2018)}]{renaud18}
{Renaud} F (2018) {Star clusters in evolving galaxies}. \nar 81:1--38,
  \doi{10.1016/j.newar.2018.03.001}, \eprint{1801.04278}

\bibitem[{{Rich} et~al.(2017){Rich}, {Ryde}, {Thorsbro}, {Fritz}, {Schultheis},
  {Origlia}, and {J{\"o}nsson}}]{rich17}
{Rich} RM, {Ryde} N, {Thorsbro} B, {Fritz} TK, {Schultheis} M, {Origlia} L,
  {J{\"o}nsson} H (2017) {Detailed Abundances for the Old Population near the
  Galactic Center. I. Metallicity Distribution of the Nuclear Star Cluster}.
  \aj 154(6):239, \doi{10.3847/1538-3881/aa970a}, \eprint{1710.08477}

\bibitem[{{Rieke} and {Rieke}(1988)}]{rieke88}
{Rieke} GH, {Rieke} MJ (1988) {Stellar Velocities and the Mass Distribution in
  the Galactic Center}. \apjl 330:L33, \doi{10.1086/185199}

\bibitem[{{Roediger} and {Courteau}(2015)}]{roediger15}
{Roediger} JC, {Courteau} S (2015) {On the uncertainties of stellar mass
  estimates via colour measurements}. \mnras 452(3):3209--3225,
  \doi{10.1093/mnras/stv1499}, \eprint{1507.03016}

\bibitem[{{Romanishin} et~al.(1983){Romanishin}, {Strom}, and
  {Strom}}]{romanishin83}
{Romanishin} W, {Strom} KM, {Strom} SE (1983) {A study of low surface
  brightness spiral galaxies. II Optical surface photometry, infrared
  photometry, and H II region spectrophotometry}. \apjs 53:105--128,
  \doi{10.1086/190886}

\bibitem[{{Rossa} et~al.(2006){Rossa}, {van der Marel}, {B{\"o}ker}, {Gerssen},
  {Ho}, {Rix}, {Shields}, and {Walcher}}]{rossa06}
{Rossa} J, {van der Marel} RP, {B{\"o}ker} T, {Gerssen} J, {Ho} LC, {Rix} HW,
  {Shields} JC, {Walcher} CJ (2006) {Hubble Space Telescope STIS Spectra of
  Nuclear Star Clusters in Spiral Galaxies: Dependence of Age and Mass on
  Hubble Type}. \aj 132:1074--1099, \doi{10.1086/505968},
  \eprint{astro-ph/0604140}

\bibitem[{{Rusli} et~al.(2011){Rusli}, {Thomas}, {Erwin}, {Saglia}, {Nowak},
  and {Bender}}]{rusli11}
{Rusli} SP, {Thomas} J, {Erwin} P, {Saglia} RP, {Nowak} N, {Bender} R (2011)
  {The central black hole mass of the high-{\ensuremath{\sigma}} but
  low-bulge-luminosity lenticular galaxy NGC 1332}. \mnras 410(2):1223--1236,
  \doi{10.1111/j.1365-2966.2010.17610.x}, \eprint{1009.0515}

\bibitem[{{Ryde} et~al.(2016){Ryde}, {Fritz}, {Rich}, {Thorsbro}, {Schultheis},
  {Origlia}, and {Chatzopoulos}}]{ryde16}
{Ryde} N, {Fritz} TK, {Rich} RM, {Thorsbro} B, {Schultheis} M, {Origlia} L,
  {Chatzopoulos} S (2016) {Detailed Abundance Analysis of a Metal-poor Giant in
  the Galactic Center}. \apj 831(1):40, \doi{10.3847/0004-637X/831/1/40},
  \eprint{1608.07562}

\bibitem[{{Saglia} et~al.(2010){Saglia}, {Fabricius}, {Bender}, {Montalto},
  {Lee}, {Riffeser}, {Seitz}, {Morganti}, {Gerhard}, and {Hopp}}]{saglia10}
{Saglia} RP, {Fabricius} M, {Bender} R, {Montalto} M, {Lee} CH, {Riffeser} A,
  {Seitz} S, {Morganti} L, {Gerhard} O, {Hopp} U (2010) {The old and heavy
  bulge of M 31 . I. Kinematics and stellar populations}. \aap 509:A61,
  \doi{10.1051/0004-6361/200912805}, \eprint{0910.5590}

\bibitem[{{Saglia} et~al.(2016){Saglia}, {Opitsch}, {Erwin}, {Thomas},
  {Beifiori}, {Fabricius}, {Mazzalay}, {Nowak}, {Rusli}, and
  {Bender}}]{saglia16}
{Saglia} RP, {Opitsch} M, {Erwin} P, {Thomas} J, {Beifiori} A, {Fabricius} M,
  {Mazzalay} X, {Nowak} N, {Rusli} SP, {Bender} R (2016) {The SINFONI Black
  Hole Survey: The Black Hole Fundamental Plane Revisited and the Paths of
  (Co)evolution of Supermassive Black Holes and Bulges}. \apj 818(1):47,
  \doi{10.3847/0004-637X/818/1/47}, \eprint{1601.00974}

\bibitem[{{S{\'a}nchez-Janssen}
  et~al.(2019{\natexlab{a}}){S{\'a}nchez-Janssen}, {C{\^o}t{\'e}}, {Ferrarese},
  {Peng}, {Roediger}, {Blakeslee}, {Emsellem}, {Puzia}, {Spengler}, and
  {Taylor}}]{sanchez-janssen19}
{S{\'a}nchez-Janssen} R, {C{\^o}t{\'e}} P, {Ferrarese} L, {Peng} EW, {Roediger}
  J, {Blakeslee} JP, {Emsellem} E, {Puzia} TH, {Spengler} C, {Taylor} J
  (2019{\natexlab{a}}) {The Next Generation Virgo Cluster Survey. XXIII.
  Fundamentals of Nuclear Star Clusters over Seven Decades in Galaxy Mass}.
  \apj 878(1):18, \doi{10.3847/1538-4357/aaf4fd}, \eprint{1812.01019}

\bibitem[{{S{\'a}nchez-Janssen}
  et~al.(2019{\natexlab{b}}){S{\'a}nchez-Janssen}, {Puzia}, {Ferrarese},
  {C{\^o}t{\'e}}, {Eigenthaler}, {Miller}, {Ordenes-Brice{\~n}o}, {Peng},
  {Ribbeck}, and {Roediger}}]{sanchez-janssen19b}
{S{\'a}nchez-Janssen} R, {Puzia} TH, {Ferrarese} L, {C{\^o}t{\'e}} P,
  {Eigenthaler} P, {Miller} B, {Ordenes-Brice{\~n}o} Y, {Peng} EW, {Ribbeck}
  KX, {Roediger} J (2019{\natexlab{b}}) {How nucleation and luminosity shape
  faint dwarf galaxies}. \mnras 486(1):L1--L5, \doi{10.1093/mnrasl/slz008},
  \eprint{1901.04509}

\bibitem[{{Sandage}(1961)}]{sandage61}
{Sandage} A (1961) {The Hubble Atlas of Galaxies}. Carnegie Institution,
  Washington

\bibitem[{{Sandage} and {Bedke}(1994)}]{sandage94}
{Sandage} A, {Bedke} J (1994) {The Carnegie Atlas of Galaxies. Volumes I, II}.
  Carnegie Institution, Washington

\bibitem[{{Sarajedini} et~al.(1996){Sarajedini}, {Green}, {Griffiths}, and
  {Ratnatunga}}]{sarajedini96}
{Sarajedini} VL, {Green} RF, {Griffiths} RE, {Ratnatunga} K (1996) {Compact
  Nuclei in Moderately Redshifted Galaxies}. \apjl 471:L15,
  \doi{10.1086/310333}, \eprint{astro-ph/9608180}

\bibitem[{{Sarzi} et~al.(2001){Sarzi}, {Rix}, {Shields}, {Rudnick}, {Ho},
  {McIntosh}, {Filippenko}, and {Sargent}}]{sarzi01}
{Sarzi} M, {Rix} HW, {Shields} JC, {Rudnick} G, {Ho} LC, {McIntosh} DH,
  {Filippenko} AV, {Sargent} WLW (2001) {Supermassive Black Holes in Bulges}.
  \apj 550(1):65--74, \doi{10.1086/319724}, \eprint{astro-ph/0010240}

\bibitem[{{Satyapal} et~al.(2008){Satyapal}, {Vega}, {Dudik}, {Abel}, and
  {Heckman}}]{satyapal08}
{Satyapal} S, {Vega} D, {Dudik} RP, {Abel} NP, {Heckman} T (2008) {Spitzer
  Uncovers Active Galactic Nuclei Missed by Optical Surveys in Seven Late-Type
  Galaxies}. \apj 677(2):926--942, \doi{10.1086/529014}, \eprint{0801.2759}

\bibitem[{{Satyapal} et~al.(2009){Satyapal}, {B{\"o}ker}, {Mcalpine},
  {Gliozzi}, {Abel}, and {Heckman}}]{satyapal09}
{Satyapal} S, {B{\"o}ker} T, {Mcalpine} W, {Gliozzi} M, {Abel} NP, {Heckman} T
  (2009) {The Incidence of Active Galactic Nuclei in Pure Disk Galaxies: The
  Spitzer View}. \apj 704(1):439--452, \doi{10.1088/0004-637X/704/1/439},
  \eprint{0908.1820}

\bibitem[{{Scarlata} et~al.(2004){Scarlata}, {Stiavelli}, {Hughes}, {Axon},
  {Alonso-Herrero}, {Atkinson}, {Batcheldor}, {Binney}, {Capetti}, {Carollo},
  {Dressel}, {Gerssen}, {Macchetto}, {Maciejewski}, {Marconi}, {Merrifield},
  {Ruiz}, {Sparks}, {Tsvetanov}, and {van der Marel}}]{scarlata04}
{Scarlata} C, {Stiavelli} M, {Hughes} MA, {Axon} D, {Alonso-Herrero} A,
  {Atkinson} J, {Batcheldor} D, {Binney} J, {Capetti} A, {Carollo} CM,
  {Dressel} L, {Gerssen} J, {Macchetto} D, {Maciejewski} W, {Marconi} A,
  {Merrifield} M, {Ruiz} M, {Sparks} W, {Tsvetanov} Z, {van der Marel} RP
  (2004) {Nuclear Properties of a Sample of Nearby Spiral Galaxies from Hubble
  Space Telescope STIS Imaging}. \aj 128:1124--1137, \doi{10.1086/423036},
  \eprint{astro-ph/0408435}

\bibitem[{{Schiavon}(2007)}]{schiavon07}
{Schiavon} RP (2007) {Population Synthesis in the Blue. IV. Accurate Model
  Predictions for Lick Indices and UBV Colors in Single Stellar Populations}.
  \apjs 171(1):146--205, \doi{10.1086/511753}, \eprint{astro-ph/0611464}

\bibitem[{{Schinnerer} et~al.(2006){Schinnerer}, {B{\"o}ker}, {Emsellem}, and
  {Lisenfeld}}]{schinnerer06}
{Schinnerer} E, {B{\"o}ker} T, {Emsellem} E, {Lisenfeld} U (2006) {Molecular
  Gas Dynamics in NGC 6946: A Bar-driven Nuclear Starburst ``Caught in the
  Act''}. \apj 649(1):181--200, \doi{10.1086/506265}, \eprint{astro-ph/0605702}

\bibitem[{{Schinnerer} et~al.(2007){Schinnerer}, {B{\"o}ker}, {Emsellem}, and
  {Downes}}]{schinnerer07}
{Schinnerer} E, {B{\"o}ker} T, {Emsellem} E, {Downes} D (2007) {Bar-driven mass
  build-up within the central 50 pc of NGC 6946}. \aap 462(3):L27--L30,
  \doi{10.1051/0004-6361:20066711}, \eprint{astro-ph/0701599}

\bibitem[{{Sch{\"o}del} et~al.(2014{\natexlab{a}}){Sch{\"o}del}, {Feldmeier},
  {Kunneriath}, {Stolovy}, {Neumayer}, {Amaro-Seoane}, and
  {Nishiyama}}]{schoedel14a}
{Sch{\"o}del} R, {Feldmeier} A, {Kunneriath} D, {Stolovy} S, {Neumayer} N,
  {Amaro-Seoane} P, {Nishiyama} S (2014{\natexlab{a}}) {Surface brightness
  profile of the Milky Way's nuclear star cluster}. \aap 566:A47,
  \doi{10.1051/0004-6361/201423481}, \eprint{1403.6657}

\bibitem[{{Sch{\"o}del} et~al.(2014{\natexlab{b}}){Sch{\"o}del}, {Feldmeier},
  {Neumayer}, {Meyer}, and {Yelda}}]{schoedel14b}
{Sch{\"o}del} R, {Feldmeier} A, {Neumayer} N, {Meyer} L, {Yelda} S
  (2014{\natexlab{b}}) {The nuclear cluster of the Milky Way: our primary
  testbed for the interaction of a dense star cluster with a massive black
  hole}. Classical and Quantum Gravity 31(24):244007,
  \doi{10.1088/0264-9381/31/24/244007}, \eprint{1411.4504}

\bibitem[{{Sch{\"o}del} et~al.(2018){Sch{\"o}del}, {Gallego-Cano}, {Dong},
  {Nogueras-Lara}, {Gallego-Calvente}, {Amaro-Seoane}, and
  {Baumgardt}}]{schoedel18}
{Sch{\"o}del} R, {Gallego-Cano} E, {Dong} H, {Nogueras-Lara} F,
  {Gallego-Calvente} AT, {Amaro-Seoane} P, {Baumgardt} H (2018) {The
  distribution of stars around the Milky Way's central black hole. II. Diffuse
  light from sub-giants and dwarfs}. \aap 609:A27,
  \doi{10.1051/0004-6361/201730452}, \eprint{1701.03817}

\bibitem[{{Schultheis} et~al.(2019){Schultheis}, {Rich}, {Origlia}, {Ryde},
  {Nandakumar}, {Thorsbro}, and {Neumayer}}]{schultheis19}
{Schultheis} M, {Rich} RM, {Origlia} L, {Ryde} N, {Nandakumar} G, {Thorsbro} B,
  {Neumayer} N (2019) {The inner two degrees of the Milky Way. Evidence of a
  chemical difference between the Galactic Center and the surrounding inner
  bulge stellar populations}. Astronomy and Astrophysics 627:A152,
  \doi{10.1051/0004-6361/201935772}, \eprint{1906.07985}

\bibitem[{{Scott} and {Graham}(2013)}]{scott13}
{Scott} N, {Graham} AW (2013) {Updated Mass Scaling Relations for Nuclear Star
  Clusters and a Comparison to Supermassive Black Holes}. \apj 763:76,
  \doi{10.1088/0004-637X/763/2/76}, \eprint{1205.5338}

\bibitem[{{Secunda} et~al.(2019){Secunda}, {Bellovary}, {Mac Low}, {Ford},
  {McKernan}, {Leigh}, {Lyra}, and {S{\'a}ndor}}]{secunda19}
{Secunda} A, {Bellovary} J, {Mac Low} MM, {Ford} KES, {McKernan} B, {Leigh}
  NWC, {Lyra} W, {S{\'a}ndor} Z (2019) {Orbital Migration of Interacting
  Stellar Mass Black Holes in Disks around Supermassive Black Holes}. \apj
  878(2):85, \doi{10.3847/1538-4357/ab20ca}, \eprint{1807.02859}

\bibitem[{{Seigar} et~al.(2002){Seigar}, {Carollo}, {Stiavelli}, {de Zeeuw},
  and {Dejonghe}}]{seigar02}
{Seigar} M, {Carollo} CM, {Stiavelli} M, {de Zeeuw} PT, {Dejonghe} H (2002)
  {Spiral Galaxies with HST/NICMOS. II. Isophotal Fits and Nuclear Cusp
  Slopes}. \aj 123:184--194, \doi{10.1086/324730}, \eprint{astro-ph/0110282}

\bibitem[{{S\'ersic}(1968)}]{sersic68}
{S\'ersic} JL (1968) {Atlas de Galaxias Australes}. Observatorio Astronomico,
  Cordoba, Argentina

\bibitem[{{Seth} et~al.(2008{\natexlab{a}}){Seth}, {Ag{\"u}eros}, {Lee}, and
  {Basu-Zych}}]{seth08a}
{Seth} A, {Ag{\"u}eros} M, {Lee} D, {Basu-Zych} A (2008{\natexlab{a}}) {The
  Coincidence of Nuclear Star Clusters and Active Galactic Nuclei}. \apj
  678:116--130, \doi{10.1086/528955}, \eprint{0801.0439}

\bibitem[{{Seth} et~al.(2006){Seth}, {Dalcanton}, {Hodge}, and
  {Debattista}}]{seth06}
{Seth} AC, {Dalcanton} JJ, {Hodge} PW, {Debattista} VP (2006) {Clues to Nuclear
  Star Cluster Formation from Edge-on Spirals}. \aj 132:2539--2555,
  \doi{10.1086/508994}, \eprint{arXiv:astro-ph/0609302}

\bibitem[{{Seth} et~al.(2008{\natexlab{b}}){Seth}, {Blum}, {Bastian},
  {Caldwell}, and {Debattista}}]{seth08b}
{Seth} AC, {Blum} RD, {Bastian} N, {Caldwell} N, {Debattista} VP
  (2008{\natexlab{b}}) {The Rotating Nuclear Star Cluster in NGC 4244}. \apj
  687(2):997--1003, \doi{10.1086/591935}, \eprint{0807.3044}

\bibitem[{{Seth} et~al.(2010){Seth}, {Cappellari}, {Neumayer}, {Caldwell},
  {Bastian}, {Olsen}, {Blum}, {Debattista}, {McDermid}, {Puzia}, and
  {Stephens}}]{seth10}
{Seth} AC, {Cappellari} M, {Neumayer} N, {Caldwell} N, {Bastian} N, {Olsen} K,
  {Blum} RD, {Debattista} VP, {McDermid} R, {Puzia} T, {Stephens} A (2010) {The
  NGC 404 Nucleus: Star Cluster and Possible Intermediate-mass Black Hole}.
  \apj 714(1):713--731, \doi{10.1088/0004-637X/714/1/713}, \eprint{1003.0680}

\bibitem[{{Seth} et~al.(2014){Seth}, {van den Bosch}, {Mieske}, {Baumgardt},
  {Brok}, {Strader}, {Neumayer}, {Chilingarian}, {Hilker}, {McDermid},
  {Spitler}, {Brodie}, {Frank}, and {Walsh}}]{seth14}
{Seth} AC, {van den Bosch} R, {Mieske} S, {Baumgardt} H, {Brok} MD, {Strader}
  J, {Neumayer} N, {Chilingarian} I, {Hilker} M, {McDermid} R, {Spitler} L,
  {Brodie} J, {Frank} MJ, {Walsh} JL (2014) {A supermassive black hole in an
  ultra-compact dwarf galaxy}. \nat 513:398--400, \doi{10.1038/nature13762},
  \eprint{1409.4769}

\bibitem[{{Shapiro} et~al.(2006){Shapiro}, {Cappellari}, {de Zeeuw},
  {McDermid}, {Gebhardt}, {van den Bosch}, and {Statler}}]{shapiro06}
{Shapiro} KL, {Cappellari} M, {de Zeeuw} T, {McDermid} RM, {Gebhardt} K, {van
  den Bosch} RCE, {Statler} TS (2006) {The black hole in NGC 3379: a comparison
  of gas and stellar dynamical mass measurements with HST and integral-field
  data}. \mnras 370(2):559--579, \doi{10.1111/j.1365-2966.2006.10537.x},
  \eprint{astro-ph/0605479}

\bibitem[{{Shlosman} et~al.(1990){Shlosman}, {Begelman}, and
  {Frank}}]{shlosman90}
{Shlosman} I, {Begelman} MC, {Frank} J (1990) {The fuelling of active galactic
  nuclei}. \nat 345(6277):679--686, \doi{10.1038/345679a0}

\bibitem[{{Siegel} et~al.(2007){Siegel}, {Dotter}, {Majewski}, {Sarajedini},
  {Chaboyer}, {Nidever}, {Anderson}, {Mar{\'{\i}}n-Franch}, {Rosenberg},
  {Bedin}, {Aparicio}, {King}, {Piotto}, and {Reid}}]{siegel07}
{Siegel} MH, {Dotter} A, {Majewski} SR, {Sarajedini} A, {Chaboyer} B, {Nidever}
  DL, {Anderson} J, {Mar{\'{\i}}n-Franch} A, {Rosenberg} A, {Bedin} LR,
  {Aparicio} A, {King} I, {Piotto} G, {Reid} IN (2007) {The ACS Survey of
  Galactic Globular Clusters: M54 and Young Populations in the Sagittarius
  Dwarf Spheroidal Galaxy}. \apjl 667:L57--L60, \doi{10.1086/522003},
  \eprint{arXiv:0708.0027}

\bibitem[{{Silk} et~al.(1987){Silk}, {Wyse}, and {Shields}}]{silk87}
{Silk} J, {Wyse} RFG, {Shields} GA (1987) {On the Origin of Dwarf Galaxies}.
  \apj 322:L59, \doi{10.1086/185037}

\bibitem[{{Spengler} et~al.(2017){Spengler}, {C{\^o}t{\'e}}, {Roediger},
  {Ferrarese}, {S{\'a}nchez-Janssen}, {Toloba}, {Liu}, {Guhathakurta},
  {Cuillandre}, {Gwyn}, {Zirm}, {Mu{\~n}oz}, {Puzia}, {Lan{\c{c}}on}, {Peng},
  {Mei}, and {Powalka}}]{spengler17}
{Spengler} C, {C{\^o}t{\'e}} P, {Roediger} J, {Ferrarese} L,
  {S{\'a}nchez-Janssen} R, {Toloba} E, {Liu} Y, {Guhathakurta} P, {Cuillandre}
  JC, {Gwyn} S, {Zirm} A, {Mu{\~n}oz} R, {Puzia} T, {Lan{\c{c}}on} A, {Peng}
  EW, {Mei} S, {Powalka} M (2017) {Virgo Redux: The Masses and Stellar Content
  of Nuclei in Early-type Galaxies from Multiband Photometry and Spectroscopy}.
  \apj 849(1):55, \doi{10.3847/1538-4357/aa8a78}, \eprint{1709.00406}

\bibitem[{{Stone} and {Metzger}(2016)}]{stone16}
{Stone} NC, {Metzger} BD (2016) {Rates of stellar tidal disruption as probes of
  the supermassive black hole mass function}. \mnras 455:859--883,
  \doi{10.1093/mnras/stv2281}, \eprint{1410.7772}

\bibitem[{{Stone} et~al.(2017){Stone}, {K{\"u}pper}, and {Ostriker}}]{stone17}
{Stone} NC, {K{\"u}pper} AHW, {Ostriker} JP (2017) {Formation of massive black
  holes in galactic nuclei: runaway tidal encounters}. \mnras 467:4180--4199,
  \doi{10.1093/mnras/stx097}, \eprint{1606.01909}

\bibitem[{{Strader} et~al.(2013){Strader}, {Seth}, {Forbes}, {Fabbiano},
  {Romanowsky}, {Brodie}, {Conroy}, {Caldwell}, {Pota}, {Usher}, and
  {Arnold}}]{strader13}
{Strader} J, {Seth} AC, {Forbes} DA, {Fabbiano} G, {Romanowsky} AJ, {Brodie}
  JP, {Conroy} C, {Caldwell} N, {Pota} V, {Usher} C, {Arnold} JA (2013) {The
  Densest Galaxy}. \apjl 775(1):L6, \doi{10.1088/2041-8205/775/1/L6},
  \eprint{1307.7707}

\bibitem[{{Strubbe} and {Quataert}(2009)}]{strubbe09}
{Strubbe} LE, {Quataert} E (2009) {Optical flares from the tidal disruption of
  stars by massive black holes}. \mnras 400(4):2070--2084,
  \doi{10.1111/j.1365-2966.2009.15599.x}, \eprint{0905.3735}

\bibitem[{{Tremaine}(1995)}]{tremaine95}
{Tremaine} S (1995) {An Eccentric-Disk Model for the Nucleus of M31}. \aj
  110:628, \doi{10.1086/117548}, \eprint{astro-ph/9502065}

\bibitem[{{Tremaine}(2019)}]{tremaine19}
{Tremaine} S (2019) {Order-disorder phase transition in black-hole star
  clusters. II. A scale-free cluster}. \mnras p 2761,
  \doi{10.1093/mnras/stz3181}

\bibitem[{{Tremaine} et~al.(1975){Tremaine}, {Ostriker}, and
  {Spitzer}}]{tremaine75}
{Tremaine} SD, {Ostriker} JP, {Spitzer} L (1975) {The formation of the nuclei
  of galaxies. I - M31}. \apj 196:407--411

\bibitem[{{Tremou} et~al.(2018){Tremou}, {Strader}, {Chomiuk}, {Shishkovsky},
  {Maccarone}, {Miller-Jones}, {Tudor}, {Heinke}, {Sivakoff}, {Seth}, and
  {Noyola}}]{tremou18}
{Tremou} E, {Strader} J, {Chomiuk} L, {Shishkovsky} L, {Maccarone} TJ,
  {Miller-Jones} JCA, {Tudor} V, {Heinke} CO, {Sivakoff} GR, {Seth} AC,
  {Noyola} E (2018) {The MAVERIC Survey: Still No Evidence for Accreting
  Intermediate-mass Black Holes in Globular Clusters}. \apj 862(1):16,
  \doi{10.3847/1538-4357/aac9b9}, \eprint{1806.00259}

\bibitem[{{Tsatsi} et~al.(2017){Tsatsi}, {Mastrobuono-Battisti}, {van de Ven},
  {Perets}, {Bianchini}, and {Neumayer}}]{tsatsi17}
{Tsatsi} A, {Mastrobuono-Battisti} A, {van de Ven} G, {Perets} HB, {Bianchini}
  P, {Neumayer} N (2017) {On the rotation of nuclear star clusters formed by
  cluster inspirals}. \mnras 464(3):3720--3727, \doi{10.1093/mnras/stw2593},
  \eprint{1610.01162}

\bibitem[{{Turner} et~al.(2012){Turner}, {C{\^o}t{\'e}}, {Ferrarese},
  {Jord{\'a}n}, {Blakeslee}, {Mei}, {Peng}, and {West}}]{turner12}
{Turner} ML, {C{\^o}t{\'e}} P, {Ferrarese} L, {Jord{\'a}n} A, {Blakeslee} JP,
  {Mei} S, {Peng} EW, {West} MJ (2012) {The ACS Fornax Cluster Survey. VI. The
  Nuclei of Early-type Galaxies in the Fornax Cluster}. \apjs 203:5,
  \doi{10.1088/0067-0049/203/1/5}, \eprint{1208.0338}

\bibitem[{{Valluri} et~al.(2005){Valluri}, {Ferrarese}, {Merritt}, and
  {Joseph}}]{valluri05}
{Valluri} M, {Ferrarese} L, {Merritt} D, {Joseph} CL (2005) {The Low End of the
  Supermassive Black Hole Mass Function: Constraining the Mass of a Nuclear
  Black Hole in NGC 205 via Stellar Kinematics}. The Astrophysical Journal
  628(1):137--152, \doi{10.1086/430752}, \eprint{astro-ph/0502493}

\bibitem[{{van den Bergh}(1986)}]{vandenbergh86}
{van den Bergh} S (1986) {Nucleated spheroidal galaxies.} \aj 91:271--274,
  \doi{10.1086/114006}

\bibitem[{{van den Bosch} et~al.(2018){van den Bosch}, {Ogiya}, {Hahn}, and
  {Burkert}}]{vandenbosch18}
{van den Bosch} FC, {Ogiya} G, {Hahn} O, {Burkert} A (2018) {Disruption of dark
  matter substructure: fact or fiction?} \mnras 474(3):3043--3066,
  \doi{10.1093/mnras/stx2956}, \eprint{1711.05276}

\bibitem[{{van der Marel} and {Anderson}(2010)}]{vandermarel10}
{van der Marel} RP, {Anderson} J (2010) {New Limits on an Intermediate-Mass
  Black Hole in Omega Centauri. II. Dynamical Models}. \apj 710(2):1063--1088,
  \doi{10.1088/0004-637X/710/2/1063}, \eprint{0905.0638}

\bibitem[{{van Velzen}(2018)}]{vanvelzen18}
{van Velzen} S (2018) {On the Mass and Luminosity Functions of Tidal Disruption
  Flares: Rate Suppression due to Black Hole Event Horizons}. \apj 852(2):72,
  \doi{10.3847/1538-4357/aa998e}, \eprint{1707.03458}

\bibitem[{{Vasiliev}(2017)}]{vasiliev17}
{Vasiliev} E (2017) {A New Fokker-Planck Approach for the Relaxation-driven
  Evolution of Galactic Nuclei}. \apj 848(1):10,
  \doi{10.3847/1538-4357/aa8cc8}, \eprint{1709.04467}

\bibitem[{{Vasiliev} et~al.(2015){Vasiliev}, {Antonini}, and
  {Merritt}}]{vasiliev15}
{Vasiliev} E, {Antonini} F, {Merritt} D (2015) {The Final-parsec Problem in the
  Collisionless Limit}. \apj 810(1):49, \doi{10.1088/0004-637X/810/1/49},
  \eprint{1505.05480}

\bibitem[{{Verolme} et~al.(2002){Verolme}, {Cappellari}, {Copin}, {van der
  Marel}, {Bacon}, {Bureau}, {Davies}, {Miller}, and {de Zeeuw}}]{verolme02}
{Verolme} EK, {Cappellari} M, {Copin} Y, {van der Marel} RP, {Bacon} R,
  {Bureau} M, {Davies} RL, {Miller} BM, {de Zeeuw} PT (2002) {A SAURON study of
  M32: measuring the intrinsic flattening and the central black hole mass}.
  \mnras 335(3):517--525, \doi{10.1046/j.1365-8711.2002.05664.x},
  \eprint{astro-ph/0201086}

\bibitem[{{Villanova} et~al.(2014){Villanova}, {Geisler}, {Gratton}, and
  {Cassisi}}]{villanova14}
{Villanova} S, {Geisler} D, {Gratton} RG, {Cassisi} S (2014) {The Metallicity
  Spread and the Age-Metallicity Relation of {\ensuremath{\omega}} Centauri}.
  \apj 791(2):107, \doi{10.1088/0004-637X/791/2/107}, \eprint{1406.5069}

\bibitem[{{Voggel} et~al.(2016){Voggel}, {Hilker}, and {Richtler}}]{voggel16}
{Voggel} K, {Hilker} M, {Richtler} T (2016) {Globular cluster clustering and
  tidal features around ultra-compact dwarf galaxies in the halo of NGC 1399}.
  \aap 586:A102, \doi{10.1051/0004-6361/201527070}, \eprint{1510.05010}

\bibitem[{{Voggel} et~al.(2018){Voggel}, {Seth}, {Neumayer}, {Mieske},
  {Chilingarian}, {Ahn}, {Baumgardt}, {Hilker}, {Nguyen}, {Romanowsky},
  {Walsh}, {den Brok}, and {Strader}}]{voggel18}
{Voggel} KT, {Seth} AC, {Neumayer} N, {Mieske} S, {Chilingarian} I, {Ahn} C,
  {Baumgardt} H, {Hilker} M, {Nguyen} DD, {Romanowsky} AJ, {Walsh} JL, {den
  Brok} M, {Strader} J (2018) {Upper Limits on the Presence of Central Massive
  Black Holes in Two Ultra-compact Dwarf Galaxies in Centaurus A}. \apj
  858(1):20, \doi{10.3847/1538-4357/aabae5}, \eprint{1803.09750}

\bibitem[{{Voggel} et~al.(2019){Voggel}, {Seth}, {Baumgardt}, {Mieske},
  {Pfeffer}, and {Rasskazov}}]{voggel19}
{Voggel} KT, {Seth} AC, {Baumgardt} H, {Mieske} S, {Pfeffer} J, {Rasskazov} A
  (2019) {The Impact of Stripped Nuclei on the Supermassive Black Hole Number
  Density in the Local Universe}. \apj 871(2):159,
  \doi{10.3847/1538-4357/aaf735}

\bibitem[{{Volonteri}(2010)}]{volonteri10}
{Volonteri} M (2010) {Formation of supermassive black holes}. \aapr
  18(3):279--315, \doi{10.1007/s00159-010-0029-x}, \eprint{1003.4404}

\bibitem[{{Volonteri}(2012)}]{volonteri12}
{Volonteri} M (2012) {The Formation and Evolution of Massive Black Holes}.
  Science 337(6094):544, \doi{10.1126/science.1220843}, \eprint{1208.1106}

\bibitem[{{Walcher} et~al.(2005){Walcher}, {van der Marel}, {McLaughlin},
  {Rix}, {B{\"o}ker}, {H{\"a}ring}, {Ho}, {Sarzi}, and {Shields}}]{walcher05}
{Walcher} CJ, {van der Marel} RP, {McLaughlin} D, {Rix} HW, {B{\"o}ker} T,
  {H{\"a}ring} N, {Ho} LC, {Sarzi} M, {Shields} JC (2005) {Masses of Star
  Clusters in the Nuclei of Bulgeless Spiral Galaxies}. \apj 618:237--246,
  \doi{10.1086/425977}, \eprint{astro-ph/0409216}

\bibitem[{{Walcher} et~al.(2006){Walcher}, {B{\"o}ker}, {Charlot}, {Ho}, {Rix},
  {Rossa}, {Shields}, and {van der Marel}}]{walcher06}
{Walcher} CJ, {B{\"o}ker} T, {Charlot} S, {Ho} LC, {Rix} HW, {Rossa} J,
  {Shields} JC, {van der Marel} RP (2006) {Stellar Populations in the Nuclei of
  Late-Type Spiral Galaxies}. \apj 649:692--708, \doi{10.1086/505166},
  \eprint{astro-ph/0604138}

\bibitem[{{Wehner} and {Harris}(2006)}]{wehner06}
{Wehner} EH, {Harris} WE (2006) {From Supermassive Black Holes to Dwarf
  Elliptical Nuclei: A Mass Continuum}. \apjl 644(1):L17--L20,
  \doi{10.1086/505387}, \eprint{astro-ph/0603801}

\bibitem[{{Wevers} et~al.(2017){Wevers}, {van Velzen}, {Jonker}, {Stone},
  {Hung}, {Onori}, {Gezari}, and {Blagorodnova}}]{wevers17}
{Wevers} T, {van Velzen} S, {Jonker} PG, {Stone} NC, {Hung} T, {Onori} F,
  {Gezari} S, {Blagorodnova} N (2017) {Black hole masses of tidal disruption
  event host galaxies}. \mnras 471(2):1694--1708, \doi{10.1093/mnras/stx1703},
  \eprint{1706.08965}

\bibitem[{{Wevers} et~al.(2019){Wevers}, {Stone}, {van Velzen}, {Jonker},
  {Hung}, {Auchettl}, {Gezari}, {Onori}, {Mata S{\'a}nchez},
  {Kostrzewa-Rutkowska}, and {Casares}}]{wevers19}
{Wevers} T, {Stone} NC, {van Velzen} S, {Jonker} PG, {Hung} T, {Auchettl} K,
  {Gezari} S, {Onori} F, {Mata S{\'a}nchez} D, {Kostrzewa-Rutkowska} Z,
  {Casares} J (2019) {Black hole masses of tidal disruption event host galaxies
  II}. \mnras 487(3):4136--4152, \doi{10.1093/mnras/stz1602},
  \eprint{1902.04077}

\bibitem[{{Williams} et~al.(2017){Williams}, {Dolphin}, {Dalcanton}, {Weisz},
  {Bell}, {Lewis}, {Rosenfield}, {Choi}, {Skillman}, and
  {Monachesi}}]{williams17}
{Williams} BF, {Dolphin} AE, {Dalcanton} JJ, {Weisz} DR, {Bell} EF, {Lewis} AR,
  {Rosenfield} P, {Choi} Y, {Skillman} E, {Monachesi} A (2017) {PHAT. XIX. The
  Ancient Star Formation History of the M31 Disk}. \apj 846(2):145,
  \doi{10.3847/1538-4357/aa862a}

\bibitem[{{Wise} et~al.(2019){Wise}, {Regan}, {O'Shea}, {Norman}, {Downes}, and
  {Xu}}]{wise19}
{Wise} JH, {Regan} JA, {O'Shea} BW, {Norman} ML, {Downes} TP, {Xu} H (2019)
  {Formation of massive black holes in rapidly growing pre-galactic gas
  clouds}. \nat 566(7742):85--88, \doi{10.1038/s41586-019-0873-4},
  \eprint{1901.07563}

\bibitem[{{Woo} et~al.(2019){Woo}, {Cho}, {Gallo}, {Hodges-Kluck}, {Le},
  {Shin}, {Son}, and {Horst}}]{woo19}
{Woo} JH, {Cho} H, {Gallo} E, {Hodges-Kluck} E, {Le} HAN, {Shin} J, {Son} D,
  {Horst} JC (2019) {A 10,000-solar-mass black hole in the nucleus of a
  bulgeless dwarf galaxy}. Nature Astronomy 3:755--759,
  \doi{10.1038/s41550-019-0790-3}, \eprint{1905.00145}

\bibitem[{{Worthey}(1994)}]{worthey94}
{Worthey} G (1994) {Comprehensive Stellar Population Models and the
  Disentanglement of Age and Metallicity Effects}. \apjs 95:107,
  \doi{10.1086/192096}

\bibitem[{{Zhang} et~al.(2019){Zhang}, {Shao}, and {Zhu}}]{zhang19}
{Zhang} F, {Shao} L, {Zhu} W (2019) {Gravitational-wave Merging Events from the
  Dynamics of Stellar-mass Binary Black Holes around the Massive Black Hole in
  a Galactic Nucleus}. \apj 877(2):87, \doi{10.3847/1538-4357/ab1b28},
  \eprint{1903.02685}

\end{thebibliography}

\end{document}